\documentclass[twocolumn,times,tighten]{aastex631}
\usepackage{color}
\usepackage{amsmath}
\usepackage[caption=false]{subfig}

\accepted{April 16, 2023 by ApJ, in press}
\shorttitle{Cosmology from Clustering with the CAMELS-SAM suite}
\shortauthors{Perez, Genel, et al.}

\graphicspath{{./}{figures/}}

\begin{document}

\title{Constraining Cosmology with Machine Learning and Galaxy Clustering: the CAMELS-SAM Suite}

\correspondingauthor{Lucia A. Perez}
% \email{lucia.perez@asu.edu}
\email{lucia.perez.phd@gmail.com}

\author[0000-0002-8449-1956]{Lucia A. Perez}
\affiliation{Department of Astrophysical Sciences,
Princeton University,
4 Ivy Lane,
Princeton, NJ 08544, USA}
\affiliation{Center for Computational Astrophysics, Flatiron Institute, 162 5th Ave, New York, NY 10010, USA }
\affiliation{School of Earth and Space Exploration, Arizona State University, 781 Terrace Mall, Tempe, AZ 85287, USA}

\author[0000-0002-3185-1540]{Shy Genel}
\affiliation{Center for Computational Astrophysics, Flatiron Institute, 162 5th Ave, New York, NY 10010, USA }
\affiliation{Columbia Astrophysics Laboratory, Columbia University, 550 West 120th Street, New York, NY 10027, USA}

\author[0000-0002-4816-0455]{Francisco Villaescusa-Navarro}
\affiliation{Center for Computational Astrophysics, Flatiron Institute, 162 5th Ave, New York, NY 10010, USA }
\affiliation{Department of Astrophysical Sciences,
Princeton University,
4 Ivy Lane,
Princeton, NJ 08544, USA}

\author{Rachel S. Somerville}
\affiliation{Center for Computational Astrophysics, Flatiron Institute, 162 5th Ave, New York, NY 10010, USA }

\author[0000-0003-4295-3793]{Austen Gabrielpillai}
\affiliation{Institute for Astrophysics and Computational Sciences, Catholic University of America, USA}
\affiliation{Astrophysics Science Division, NASA GSFC, 8800 Greenbelt Rd, Greenbelt, MD 20771, USA }
\affiliation{Center for Research and Exploration in Space Science and Technology, NASA GSFC, 8800 Greenbelt Rd, Greenbelt, MD 20771, USA}

\author[0000-0001-5769-4945]{Daniel Angl\'es-Alc\'azar}
\affiliation{Department of Physics, University of Connecticut, 196 Auditorium Road, Storrs, CT 06269, USA}
\affiliation{Center for Computational Astrophysics, Flatiron Institute, 162 5th Ave, New York, NY 10010, USA }

\author[0000-0002-5854-8269]{Benjamin D. Wandelt}
\affiliation{Institut d’Astrophysique de Paris (IAP), UMR 7095, CNRS, Sorbonne Universit\'e, France}
\affiliation{Center for Computational Astrophysics, Flatiron Institute, 162 5th Ave, New York, NY 10010, USA }

\author[0000-0003-3466-035X]{L. Y. Aaron Yung}
\affiliation{Astrophysics Science Division, NASA GSFC, 8800 Greenbelt Rd, Greenbelt, MD 20771, USA }

\begin{abstract}

As the next generation of large galaxy surveys come online, it is becoming increasingly important to develop and understand the machine learning tools that analyze big astronomical data. Neural networks are powerful and capable of probing deep patterns in data, but must be trained carefully on large and representative data sets. We present a new `hump' of the Cosmology and Astrophysics with MachinE Learning Simulations (CAMELS) project: \textbf{CAMELS-SAM}, encompassing one thousand dark-matter only simulations of (100 $h^{-1}$ cMpc)$^3$ with different cosmological parameters ($\Omega_m$ and $\sigma_8$) and run through the Santa Cruz semi-analytic model for galaxy formation over a broad range of astrophysical parameters. As a proof-of-concept for the power of this vast suite of simulated galaxies in a large volume and broad parameter space, we probe the power of simple clustering summary statistics to marginalize over astrophysics and constrain cosmology using neural networks. We use the two-point correlation function, count-in-cells, and the Void Probability Function, and probe non-linear and linear scales across $0.68<$ R $<27\ h^{-1}$ cMpc. 
We find our neural networks can both marginalize over the uncertainties in astrophysics to constrain cosmology to 3-8\% error across various types of galaxy selections, while simultaneously learning about the SC-SAM astrophysical parameters.
This work encompasses vital first steps toward creating algorithms able to marginalize over the uncertainties in our galaxy formation models and measure the underlying cosmology of our universe.
CAMELS-SAM has been publicly released alongside the rest of CAMELS, and offers great potential to many applications of machine learning in astrophysics: \url{https://camels-sam.readthedocs.io}.

\end{abstract}


\section{Introduction} \label{sec:intro}

Since the earliest galaxy redshift surveys, it has been known that galaxies are not distributed randomly in space, but trace out vast structures, including walls, filaments, and voids. Dark matter (DM) makes up the majority of the mass content of the Universe, and is the dominant driver behind large-scale structure formation. The distribution of galaxies in space is heavily influenced by the clustering of dark matter halos, but also carries signatures of how galaxy properties map to the properties of these dark matter halos (\citealt{Peebles1980, WechslerTinker2018}). Galaxy clustering is a potential key probe of cosmology, yet accurately describing the baryonic physics that drives galaxy evolution, and determines this mapping between galaxy and DM halo properties, is a large ongoing area of research. Astrophysical processes such as cooling, chemical enrichment, star formation, stellar feedback, black hole growth and feedback, and galaxy mergers interact in highly non-linear ways. Active Galactic Nuclei (AGN) expel gas far beyond the center of the host galaxy, are a crucial element of the feedback cycle in galaxies, and may even affect the distribution of dark matter itself (e.g.\ \citealt{McKeeOstriker2007, Fabian2012, Kormendy2013, Netzger2015, Borrow2020}). Stellar feedback in the form of supernovae and radiation from massive stars drives galactic winds that are key to regulating star formation in galaxies (e.g\ \citealt{MadauDickinson2014, SomervilleRomeel2015, DanielAA2017}). 

Modern hydrodynamic simulations including these physical processes within the $\Lambda$CDM cosmological framework have been quite successful at reproducing many features of the large scale distribution of galaxies. For example, \citet{Springel2018} measured the matter power spectrum for the dark matter, gas, and stellar components in the IllustrisTNG simulations, explored the halo-galaxy connection, and found projected correlation functions that showed great consistency with diverse observations. More recently, the MillenniumTNG simulations expanded the model to enormous volumes and have refined our understanding of the galaxy-halo connection (\citealt{MillenniumTNG_HernandezAguayo, MillenniumTNG_Bose, MillenniumTNG_Contreras, MillenniumTNG_Hadzhiyska1halo, MillenniumTNG_Hadzhiyska2halo, MillenniumTNG_Barrera}).
However, although a broad narrative has developed for how these types of feedback tie into the formation and evolution of galaxies, there is still much uncertainty in our understanding of the details of how these physical processes operate to shape galaxy observables (e.g.\ \citealt{SteinhardtSpeagle2014, SomervilleRomeel2015, Naab2017, Forster2020}). One of the major open questions in astrophysics is how to disentangle the effects of cosmology and baryonic physics in order to realize the full potential of galaxies as probes of both cosmology and astrophysics.

The Cosmology and Astrophysics with MachinE Learning Simulations (\textbf{CAMELS}) project\footnote{\url{https://www.camel-simulations.org/}} \citep{CAMELSannouncement} posits: to constrain cosmology, we should leverage tools that can marginalize over uncertainties in baryonic physics, and thereby measure the underlying cosmology. Machine learning has great promise for this goal, as algorithms can learn relationships between features without requiring explicit functional forms or likelihoods. However, machine learning requires large data sets for accurate training and robust results. The CAMELS project created a large suite of simulations to specifically explore the potential of machine learning to constrain cosmology. The initial focus of CAMELS has been developing techniques that can constrain the density of matter in the universe, $\Omega_{\text{M}}$, and the amplitude of density fluctuations in the early universe, $\sigma_8$, under the broadly supported $\Lambda$CDM model of cosmology. The project created 4,000+ cosmological simulations of (25 $h^{-1}$ cMpc)$^3$ spanning thousands of cosmological models. Half the simulations are dark matter only, and the other half are run with the \textit{IllustrisTNG} (\citealt{Weinberger2017, Pillepich2018}) and \textit{SIMBA} \citep{Dave2019} hydrodynamic models of galaxy formation, creating 2,000+ simulations also spanning thousands of astrophysical models. All of CAMELS has been publicly released, as detailed in \citet{CAMELSpublic2022}\footnote{\url{https://camels.readthedocs.io/}}.

Various studies based on CAMELS have explored different methods of constraining cosmology or astrophysical models with diverse types of machine learning, computational tools, and astrophysical objects and phenomena. For example, \citet{Paco2021b_robust} obtain constraints on $\Omega_{\text{M}}$ with 3-4\% errors using neural networks trained on total matter density maps, providing robust predictions irrespective of galaxy formation physics implementation. Similarly, \citet{Nicola2022} used the electron density power spectrum $P_{ee}(k)$ to obtain cosmological constraints and to probe the strength of baryonic feedback through the mean baryon fraction ($\bar{f}_{\text{bar}}$). The $z=0\ P_{ee}(k)$ from the IllustrisTNG `hump' of CAMELS yields constraints with approximately $5-10\%$ error on $\Omega_{\text{M}}$ but no constraints on $\sigma_8$. However, their results are encouragingly robust across galaxy formation models: the same neural network is also able to predict $\Omega_{\text{M}}$ and $\bar{f}_{\text{bar}}$ when instead traind on simulations from the SIMBA hump. \citet{Nicola2022} also confirm that it is possible to obtain constraints on $\Omega_{\text{M}}$ with approximately $4\%$ errors using the CAMELS matter power spectrum $P_{mm}(k)$s, though no significant constraints on baryon fraction $\bar{f}_{\text{bar}}$ and $\sigma_8$ can be obtained with this method.

Though they represent a valuable resource for many science goals, the original CAMELS simulations, due to their small volumes and sensitivity to cosmic variance, are not well-suited for leveraging the most readily used summary statistics for measuring cosmology from observations: galaxy clustering. For example, \citet{Paco2021b_robust} find that the power spectra measured from CAMELS 2D maps of the matter density field provided constraints on $\Omega_{\text{M}}$ with 20\% error--a significant loss of information compared to using the full maps in their neural networks. The clear path to improve constraints from galaxy clustering statistics is to increase CAMELS' volume.  However, it is not currently computationally feasible to directly scale CAMELS and its full hydrodynamic simulations up to larger volumes. In this work, we present \textbf{CAMELS-SAM}, a third and larger `hump' of the CAMELS project to address the need for larger volumes of simulated galaxies, and use it to probe the power of clustering summary statistics towards CAMELS' goals.

Semi-analytic models are a well-established technique for simulating galaxy properties in a cosmological context, using simplified but physically motivated recipes. SAMs are very successful at reproducing a broad range of galaxy observables, and the predictions of SAMs for many global galaxy properties have been shown to be in good agreement with the predictions of hydrodynamic simulations over a broad range of cosmic time \citep{SomervilleRomeel2015}. However, SAMs are more computationally efficient than hydrodynamic simulations by many orders of magnitude. In this work, we make use of the well-established Santa Cruz SAM (SC-SAM; \citealt{SomervillePrimack1999,Somerville2008, Somerville2015, Somerville2021}). A detailed, halo by halo comparison of the predictions of the SC-SAM with the IllustrisTNG hydro simulations has recently been carried out by \citet{Gabrielpillai2022}. \citet{Hadzhiyska2021b} showed that the SC-SAM produces very similar predictions to IllustrisTNG for galaxy clustering, including two-point and higher order clustering statistics.

SAMs are set within the backbone of cosmological dark matter merger trees, which specify how halos grow over time via accretion and mergers. Like most SAMs, the SC-SAM implements treatments of cooling, partitioning of cold gas into a molecular, atomic, and ionized phase, star formation, stellar feedback, chemical enrichment, and black hole growth and feedback. Each of these processes contain free parameters that represent our incomplete understanding of the physical processes. Traditionally, these parameters are adjusted to reproduce a set of observational calibrations for nearby galaxies. In this work, we run the SC-SAMs within merger trees extracted from the suite of new DM-only simulations that we have created for the CAMELS-SAM hump. In addition, we re-run the SAMs for many different values of the parameters controlling stellar and AGN feedback, in a similar spirit to the CAMELS hydro humps. 

There has been much previous work describing galaxy clustering using simple mappings between galaxy and halo properties, often referred to as the Halo Occupation Distribution (HOD) framework
\citep{WechslerTinker2018}. HOD approaches often distinguish between the `one-halo' and `two-halo' terms, corresponding to small-scale clustering between galaxies in the same halo, and clustering between separate halos, respectively; these studies have yielded great understanding of how particular types of galaxies inhabit halos (e.g.\ \citealt{MillenniumTNG_Hadzhiyska1halo, MillenniumTNG_Hadzhiyska2halo}).
The HOD framework is central to the \textit{bias} formalism, which describes the relationship between galaxy clustering and dark matter halo clustering. Related methods to connect galaxies to halos include \textit{empirical models}, which map halo and galaxy properties using observational constraints into functional forms, often e.g.\ parametrizing galaxy star formation rates as a function of host halo masses, mass accretion rates, and redshifts \citep{Behroozi2020}. A notable subset of empirical models is the \textit{subhalo abundance matching} method (e.g.\ \citealt{Conroy2006, Guo2010, Contreras2021, Hearin2022}), which specifically maps the abundance of galaxies to the abundance of halos at a given redshift, and by definition accurately recreates the observed stellar mass function.
% The HOD and bias approach 
These approaches have the advantage of simplicity and great computational efficiency, and have been used in many studies that attempt to use galaxy clustering observations to constrain cosmological parameters such as $\Omega_{\text{M}}$ and $\sigma_8$. For example, \citet{MillenniumTNG_Contreras} use a SHAM atop the new MillenniumTNG simulations to build a very accurate emulator of galaxy clustering for a large range of scales, and \citet{Kwan2023} use an HOD atop the MiraTitan simulations for an accurate emulator of clustering that also uses the small scale galaxy-galaxy lensing signal.

`Classical' analyses with the HOD framework have, for example, leveraged theoretical descriptions of clustering statistics to find the best-fitting cosmology for simulations or observations \citep[e.g.][]{ReppSzapudi2020, Sugiyama2020}, or to create emulators or forward models for these statistics \citep[e.g.][]{Zhai2019-aemulus3, Wibking2020, Barriera2021, Kokron2021, Mead2021}. Several studies have also used the enormous Quijote simulation suite \citep{Quijote} to determine the cosmology dependence of various types of clustering using the Fisher matrix formalism \citep[e.g.][]{Uhlemann2020, Bayer2021, Hahn2021, Massara2021}.
However, most of these approaches focus on larger scales (R $\gtrsim 7\ h^{-1}$ cMpc or $k_{\max} < 1\ h$ cMpc $^{-1}$) due to the difficulty of capturing the complexity of clustering on strongly non-linear scales with these approaches\footnote{Some previous works have probed or tested clustering at mildly nonlinear scales up to $k_{\max} < 5\ h$ Mpc$^{-1}$ (R $> 1.26\ h^{-1}$ cMpc); e.g.\ \citealt{Arico2021}, and as part of verifying the UNIT simulations in \citealt{Chuang2019}.}. Smaller non-linear scales of clustering have been found to hold more cosmological information than larger scales (\citealt{Lange2022, MillenniumTNG_Contreras, Lange2023}), and are affected by the details of feedback and baryonic physics (seen acutely in the CAMELS matter power spectra in \citealt{Delgado2023}). Within the HOD framework and even for the thoroughly-studied power spectrum, the chosen prescription for the nonlinear regime strongly affect measured cosmological parameters \citep[e.g.\ 5$\sigma$ bias for a Euclid-like survey,][]{Safi2021}; this motivates approaches that inherently reproduce small-scale galaxy clustering while also incorporating baryonic effects. 

Another advantage of a SAM or hydro based approach is that it has the potential to provide direct insights into the astrophysical processes, which HOD/bias models bypass. Additionally, empirical and and SHAM models often have limited predictive or interpretive power, being tied to the observations they are tuned to match. It has also been shown that the most basic and widely used HOD-type models (which assume that the galaxy-halo mapping depends only on halo mass) do not accurately describe the clustering predictions of full hydrodynamic simulations \citep{Hadzhiyska2021a}. Galaxies in hydrodynamic simulations such as IllustrisTNG show a phenomenon called ``assembly bias'', which means that their clustering depends on halo properties in addition to mass. SC-SAM generated galaxies have been shown to closely reproduce the assembly bias signal seen in IllustrisTNG \citep{Hadzhiyska2021b}. Other works have also shown indications that HOD models may need more secondary parameters to accurately recreate the complexity of observed galaxy clustering (\citealt{Szewciw2021, Hahn2021}).

Machine learning enables the study of the galaxy-halo connection using nearly any type of galaxy property or feature, particularly those for which relationships are very hard to formulate or model (e.g.\ \citealt{deSanti2022a, ShaoM2022, Jo2022, Delgado2023, Rodrigues2023}). Machine learning also avoids some of the limitations of `classical' methods, and notably, has the ability to find constraints with fewer samples or over a larger parameter space than a Fisher formalism or a covariance matrix use (\citealt{Alsing2019, AlsingWandelt2019, deSanti2022b}), and for summary statistics for which theoretical descriptions and likelihoods do not exist \citep{Makinen2021}. Several works have leveraged machine learning to probe how galaxy clustering is influenced by cosmology and baryonic physics. 
\citet{Arico2021} created a multidimensional neural network emulator of the `baryonification' of the non-linear matter power spectrum atop the unique BACCO suite \citep{BACCO} that performs cosmological rescaling of N-body simulations. The emulator has been tuned to scales within $0.01 < k < 5\ h$ cMpc$^{-1}$ and $0 < z < 1.5$, and yields 1-2\% accuracy when tested against several dozens of hydrodynamic/N-body simulation pairs. Combined with the emulator of \citet{Contreras2020} for the dark matter power spectrum from BACCO, it is expected to give predictions for the non-linear matter power spectrum within 2-4\% accuracy. 
Additionally, \citet{Xu2021} used machine learning to predict the HOD, real-space 3D correlation function clustering, and assembly bias for the SAM of \citet{Guo2011}. \citet{Ntampaka2020} developed a hybrid deep machine learning-based technique for accurately measuring $\sigma_8$ and $\Omega_{\text{M}}$ (to within 3-4 \%) from mock galaxy redshift surveys built atop AbacusCosmos \citep{ABACUScode} with a HOD. \citet{Rodrigues2023} used neural networks to predict various galaxy properties (including clustering) in IllustrisTNG. Finally and notably, the SimBIG forward modeling framework \citep[][using the Quijote simulations and state-of-the-art supplemented HOD models]{Hahn2022a} finds very precise constraints for $\Omega_{\text{M}}$ and $\sigma_8$ using simulation-based inference with normalizing flows and the power spectrum ($k_{\rm max}=0.5 h$ Mpc$^{-1}$) for the BOSS CMASS sample.

The primarily deliverable of this work is the new CAMELS-SAM suite (publicly released in \citealt{CAMELSpublic2022}, found at: \url{https://camels-sam.readthedocs.io}). This work is also a proof-of-concept example of CAMELS-SAM's potential as a machine learning data set for cosmology and astrophysics. 
In this work, we probe how well galaxy clustering and neural networks can together: 1) marginalize over the uncertainties in astrophysics to constrain cosmology, and 2) also learn something about secondary effects of astrophysics on galaxy clustering. We also investigate how well neural networks constrain the astrophysics parameters that control stellar and AGN feedback with galaxy clustering.
Our work encompasses vital first steps toward being able to marginalize over the uncertainties in our galaxy formation models and measure the underlying cosmology of our universe. 
More specifically, the proof-of-concept work presented here with CAMELS-SAM makes several new contributions to the field, particularly in our use of neural networks for parameter inference: 

\begin{enumerate}
    \item We use a physics-based SAM to model galaxy formation, which innately provides predictions of galaxy clustering that agree well with hydrodynamic simulations and are more physically meaningful than HOD based approaches.
    \item With this SAM, we create an extensive suite of simulated galaxies in large dark matter-only simulations across a very wide range of both cosmological and astrophysical parameter space, specifically aimed at providing training data for machine learning. 
    \item Beyond two-point clustering, our neural networks use the information on non-linear and higher order clustering within counts-in-cells (CiC) and the less commonly used Void Probability Function (VPF) statistics. The VPF and CiC incorporate information on non-linear higher order clustering across many distance scales (e.g.\ \citealt{Croton2004}), but have been difficult to model in physically logical ways (e.g.\ \citealt{Yang2011, Hurtado-Gil2017}). Additionally, theoretical likelihoods do not yet exist for the VPF or have only recently been probed for CiC \citep{ReppSzapudi2020}, complicating `classical' approaches to cosmological constraints with them. 
    \item We probe smaller, more non-linear scales than most cosmological inference works have attempted ($k_{\max} <$ 8.5 $h$ cMpc$^{-1}$ for the two-point correlation function and $k_{\max} <$ 5.85 $h$ cMpc$^{-1}$ for the VPF and CiC), where baryonic effects are expected to be important to galaxy clustering and constraints on cosmology are stronger;
    \item Our neural networks are able to successfully marginalize over the large breadth of astrophysical models from the SAM, finding constraints on cosmology as low as 3-5\% with galaxy clustering.
    \item While constraining cosmology, our neural networks are \textit{also} able to learn the effects of individual astrophysical parameters in a physics-based model with galaxy clustering--a nontrivial result first seen here, made possible by the creation of CAMELS-SAM. In particular, we include astrophysical parameters in our parameter inference and leverage orthogonal information about cosmology they encode in galaxy clustering, rather than marginalizing over that information.
\end{enumerate}

The layout of this paper is as follows: in \textsection \ref{sec:CAMELS-SAM}, we describe the creation of the N-body simulations, the Santa Cruz SAM, and how we apply it to our simulations. In \textsection \ref{sec:Methods}, we explain how we measure galaxy clustering and our implementation of neural networks to infer the input cosmological and astrophysical parameters. In \textsection \ref{sec:CosmoConstraints}, we explore how well our neural networks constrain the cosmological parameters $\Omega_{\text{M}}$ and $\sigma_8$ across various experiments, such as: different galaxy selections, including or excluding a random down-sampling to fixed number density, and using a single redshift or combining several. In \textsection \ref{sec:SCSAM_constraints}, we explore how these experiments affect instead the SC-SAM feedback parameters. In \textsection \ref{sec:CompareStats}, we focus particularly on how each of the clustering statistics we measure perform independently. We discuss the CAMELS-SAM suite through a `meta'-lens in \textsection \ref{sec:Discussion}, comparing it to CAMELS and other simulation suites, and discussing the potential in our data release and our results. We conclude and summarize our results in \textsection \ref{sec:Conclusion}. Our Appendices hold additional explanatory figures for various parts of the project.

\section{The new CAMELS-SAM suite} \label{sec:CAMELS-SAM}

\begin{table*}[t]
    \begin{center}
    \caption{Description of all parameters varied in the CAMELS-SAM suite and their significance. The very broad ranges guarantee our neural networks will be unaffected by tight priors near the true expected values of these parameters \citep{Villaescusa-Navarro2020}. See \textsection \ref{subsec:SAMdetails} for more detailed descriptions of the parameters, and \citet{Somerville2008} and \citet{Somerville2015} for a full description of their role in the SC-SAM.}

	\begin{tabular}{p{0.7cm}p{12cm}p{1.25cm}p{0.5cm}p{1.5cm}} 
    Param. & Significance & IC Range & Fid. & Distribution \\
    \hline
		$\Omega_{\text{M}}$ & Fraction of universe's energy density in dark matter $\&$ baryons &  0.1, 0.5 & 0.3 & uniform \\
        $\sigma_8$ & Amplitude of the linear power spectrum on the scale of 8 $h^{-1}$ cMpc & 0.6, 1.0 & 0.8 & uniform \\
        A$_{\text{SN1}}$ & Multiplicative pre-factor to $\epsilon_{\text{SN}}=1.7$, amplitude of mass outflow rate due to SNe & 0.25, 4.0 & 1 & log uniform \\
        A$_{\text{SN2}}$ & Additive pre-factor to $\alpha_{rh}=3.0$, power law slope of mass outflow rate due to SNe & -2.0, +2.0 & 0 & uniform \\
        A$_{\text{AGN}}$ & Multiplicative pre-factor to $\kappa_{\mathrm{radio}}=0.002$, amplitude of mass ejection rate in AGN radio jets & 0.25, 4.0 & 1 & log uniform \\
    \hline \hline	
    % \\
    \label{table:ParamDescriptions}
	\end{tabular}
	\end{center}
\end{table*}

\subsection{Specifications for the Simulations}

The backbone of CAMELS-SAM consists of 1,005 N-body simulations of volume (100 $h^{-1}$ cMpc)$^3$ and N=640$^3$ particles, covering the broad cosmological space of $\Omega_{\text{M}}$=[0.1, 0.5] and $\sigma_8$=[0.6,1.0], and containing 100 snapshots between $20 \leq z \leq 0$. $\Omega_{\text{M}}$ and $\sigma_8$ are parameters central and strongly influential for large scale structure \citep{Dodelson2003}, and are among the least well-constrained cosmological parameters. For example, constraints on cosmology directly from galaxy clustering constrain the combined $S_8=\sigma_8 \sqrt{\Omega_{\text{M}}/0.3.} \approx 0.78 \pm 0.027$ \citep{Lange2023}, meaning constraints for $\Omega_{\text{M}}$ and $\sigma_8$ are muddled not just by uncertainties in the galaxy-halo connection, but also by degeneracies with each other.

We generated initial conditions with second order Lagrangian perturbation theory starting at $z=127$, and generated the linear power spectra with CAMB \citep{CAMB}. We specified a periodic box of volume (100 $h^{-1}$ cMpc)$^3$ and with N=$640^3$ dark matter particles. The N-body portion of CAMELS-SAM was run with a setup of \textsc{Arepo} similar to that which ran Illustris(TNG) (\citealt{AREPOog, Vogelsberger2013, Genel2014}). The resulting mass resolutions are between roughly $1-5\times 10^8\ h^{-1}$ M$_{\odot}$, and the gravitational softening length was fixed to 4 comoving kpc until $z=1$, after which it is fixed to a maximum of 2 physical kpc. 
Apart from varying $\Omega_{\text{M}}$ and $\sigma_8$ for our work, we assume a standard flat $\Lambda$CDM cosmology with $\Omega_{\rm b}=0.049$
\footnote{Some readers may wonder why $\Omega_{\rm b}$ was not varied in CAMELS or CAMELS-SAM, especially considering how sensitive galaxies are thought to be to the number of baryons in a halo. $\Omega_{\rm b}$ is more strongly constrained by studies of the Cosmic Microwave Background than $\Omega_{\text{M}}$ and $\sigma_8$ constraints. As derived in \citet{Planck2018cosmology}: $\Omega_{\rm b}h^2=0.02233 \pm 0.00015$ (0.67\%, dominated by $h$); $\Omega_{\text{M}}=0.3146 \pm 0.0074$ (2.4\%); $\sigma_8=0.8101 \pm 0.0061$ (0.75\%). However, the CAMELS team recently created another wing of $L=25\ h^{-1}$ cMpc hydrodynamic simulations that vary $\Omega_{\rm b}$ alongside $\Omega_{\text{M}}$ and $\sigma_8$, to fully account for its effects, using the Astrid model \citep{AstridCAMELS}. }
, $h=0.6711$, $n_s=0.9624$, $\sum m_\nu=0.0$ eV, and $w=-1$. We stored 100 snapshots between $27 \leq z \leq 0$ in the same spacing as IllustrisTNG\footnote{See the complete list within `Description of Simulations' in the documentation: \url{https://camels-sam.readthedocs.io/en/main/simulations.html}.}. 

These 100 snapshots were run through \textsc{Rockstar}\footnote{\citet{ROCKSTAR}: \url{https://bitbucket.org/gfcstanford/rockstar}} and \textsc{ConsistentTrees}\footnote{\citet{ConsistentTrees}: \url{https://bitbucket.org/pbehroozi/consistent-trees/src/ main/}} to identify dark matter halos and subhalos, and obtain merger trees.  Finally, each N-body simulation was run through multiple iterations of the SC-SAM, where we varied parameters for stellar and AGN feedback across a similarly broad hyperspace. Table \ref{table:ParamDescriptions} summarizes all the parameters we vary in CAMELS-SAM, and we discuss the physical meaning of these parameters in more detail in the next sub-section. The 1,000+ resulting SAM catalogs each contain hundreds of thousands to millions of halos and galaxies at each of the 100 redshift snapshots. We refer readers to \textsection 2.2-3.1 of \citet{Gabrielpillai2022} for a fuller explanation of how \textsc{Rockstar} and \textsc{ConsistentTrees} are used for the SC-SAM, as well as the full narrative of the SC-SAM astrophysics in the version used to create CAMELS-SAM.

Our implementation of the SC-SAM demands that `root' $z=0$ halos have at least 100 dark matter particles, and also prunes any parts of the tree that are less than 100 times the mass of the dark matter particles. This guarantees that all halos and the merger histories the SC-SAM uses are well resolved. Given our cosmological parameter range, these strong demands of the merger tree resolution translate to probing halos of at least M$_{\text{halo}} > 1-5 \times 10^{10}\ h^{-1}$ M$_{\odot}$. Additionally, the \textsc{ConsistentTrees} merger trees are post-processed to exclude subhalo trees, as the SC-SAM models the evolution of sub-halos internally. Central galaxies of M$_{\text{star}} > 10^{8}\ h^{-1}$ M$_{\odot}$ created by the SC-SAM have been shown to remain very similar throughout a large range of mass resolutions \citep{Gabrielpillai2022}. In practice, this means that key galaxy observables for SC-SAM galaxies remain converged beyond M$_{\text{star}} > 10^{8}\ h^{-1}$ M$_{\odot}$ and M$_{\text{halo}} > 10^{11}\ h^{-1}$ M$_{\odot}$, as we discuss in upcoming \textsection \ref{subsec:techdetails} and show in Appendix \ref{app:ExtraSAMverif}. Taken together, these specifications for the SC-SAM make us confident that the difference in each LH simulation's particle mass resolution does not artificially improve our cosmological constraints under our galaxy and halo selections.

Key CAMELS-SAM simulation data products, such as the halo catalogs, merger trees, and SC-SAM galaxy catalogs, have been publicly released alongside the CAMELS suites in \citet{CAMELSpublic2022}. We direct readers there for access details, and to the CAMELS-SAM documentation website for up-to-date information: \url{https://camels-sam.readthedocs.io}. 

\subsection{The Santa Cruz Semi-Analytic Model in Context}\label{subsec:SAMcontext}

Here we briefly  summarize the version of the Santa-Cruz semi-analytic model for galaxy formation used in this work. The core SC-SAM is similar to that used in \citet{Somerville2015}, and the small updates in the version used here are described in \citet{Gabrielpillai2022}. We also direct readers to the descriptions in \citet{Somerville2008} and \citet{Porter2014a}, as well as the recent applications to creating robust mock observations in \citet{Somerville2021} and \citet{Yung_i}. Our pipeline to go from \textsc{AREPO} N-body simulations to SC-SAM galaxy catalogs is nearly identical to that described in \citet{Gabrielpillai2022}, who probe how well the SC-SAM run upon the IllustrisTNG100-1 dark matter only simulation compares to the full hydrodynamic IllustrisTNG100-1 simulation.

Numerical simulations like IllustrisTNG and SIMBA explicitly solve equations of gravity, thermodynamics, (magneto-)hydrodynamics, etc.\ for discrete particles and fluid elements or cells representing dark matter, gas, stars, and black holes. However, they must adopt sub-grid recipes for small-scale processes that are not resolved nor fully understood in galaxy formation (e.g.\ stellar feedback). These simulations provide detailed information on the spatial distribution, kinematics, and composition of the baryonic component as galaxies evolve which allows for rich and diverse science. For example, the hydrodynamic core of CAMELS led to generated maps of neutral hydrogen in \citealt{Hassan2021} and detailed analyses of the circumgalactic medium in \citealt{Moser2022}. However, full hydrodynamic simulations are computationally expensive to run, often limited to either small volumes or low resolution. This makes it difficult to thoroughly explore the parameter space of the sub-grid recipes (though even CAMELS is helping to address this in \citealt{Jo2022}).

SAMs instead work from a dark matter halo merger tree that represents how halos grow and merge in the context of a chosen $\Lambda$CDM cosmology. Merger trees are either measured directly from N-body simulations, as we have done, or constructed from Extended Press-Schechter formalism (e.g.\ \citealt{SomervillePrimack1999,Yung_iv}). 
Within this framework, SAMs compute how mass and metals move between different reservoirs--intergalactic medium (IGM), circumgalactic medium (CGM), interstellar medium (ISM), stars, etc.--using a set of coupled ordinary differential equations. Nearly all SAMs assume that the rate at which gas is accreted into the CGM is proportional to the growth rate of the dark matter halo. Nearly all SAMs adopt a cooling model that determines how rapidly gas in the CGM cools and accretes into the ISM, often based on the one presented in \citet{WhiteFrenk1991}. The main differences between different SAMs lie in the specific equations and parameters that are adopted to describe star formation, stellar feedback, and black hole growth and AGN feedback. 
This approach offers flexibility; for example, the minimum resolution of a SC-SAM galaxy is not limited to the resolution of stellar mass particles. As noted previously, several of the key processes in SAMs contain adjustable parameters, similar to the parameters contained in the sub-grid recipes in hydrodynamic simulations. 

The SC-SAM is fairly typical in the general approach and functional forms used to describe these processes. For example, most SAMs specify the mass loading of stellar driven winds using a power law function similar the SC-SAM's, described in the next section. The implementation of star formation in the SC-SAM is somewhat different from that implemented in many other SAMs; the SC-SAM relies on partitioning gas into different phases (ionized, molecular, and atomic), and the adopted star formation relation is based on only the molecular gas density (rather than total gas density). However, \citet{Somerville2015} showed that this has a rather minor effect on most predictions of the SAM. The treatment of black hole growth and AGN feedback in the SC-SAM is also substantially different from the implementation in other SAMs, but this also turns out to have a rather minor effect on most galaxy properties. Models can also differ significantly in their treatment of satellite galaxies and orphans, but this should not strongly affect the relatively large scales considered in this work. There are many works comparing the predictions of different SAMs (e.g.\  \citealt{Lu2014,SomervilleRomeel2015, Knebe2018}). The overall takeaway from these studies is that, in spite of the many differences in implementation and choice of physical recipes, most SAMs produce similar predictions for key quantities such as the stellar mass function, suggesting a similar stellar to halo mass relationship \citep{SomervilleRomeel2015}. This implies that, to first order, these models will also make similar predictions for galaxy clustering. The predictions of SAMs for more ``second order" effects such as assembly bias has not yet been systematically studied. 

Finally, the fiducial SC-SAM (calibrated to $z=0$ observations) naturally predicts galaxy clustering consistent with both hydrodynamic simulations and galaxy observations across a wide span of redshifts. \citet{Yung_vi} robustly compared the SC-SAM predictions for future galaxy clustering observations in the process of creating forecasts for JWST. Their large lightcones\footnote{
Notable is the updated version of the L-GALAXIES SAM run atop the enormous MillenniumTNG lightcones in \citet{MillenniumTNG_Barrera}, who show the great accuracy of the produced two-point clustering of their galaxies in their improved SAM infrastructure.}, populated with SC-SAM galaxies and post-processing galaxy observables, produced projected two-point correlation functions in good agreement with those measured by PRIMUS and DEEP-2 in GOODS-N for $0.2<z<1.2$ galaxies, $1.25<z<4.5$ galaxies in CANDELS, and $3.5<z<7$ galaxies from \citet{Harikane2016}. \citet{Yung2023}, while making forecasts for the Roman Space Telescope, also compared their lightcones' SC-SAM clustering to those from lightcones made with the very different UniverseMachine \citep{Behroozi2020} and DREAM \citep{Drakos2022} models, and found great agreement across all models for the angular correlation function of rest-UV selected galaxies at $z>4$. Additionally, \citet{Hadzhiyska2021b} found SC-SAM galaxies generated atop the IllustrisTNG dark matter merger trees found two-point clustering statistics and galaxy assembly bias signatures very similar to first order to the IllustrisTNG complete hydrodynamic galaxies. Given these results, one can likely trust the SC-SAM clustering predictions for in both idealized simulation space (as our work is) and in a more realistic observational space. Finally, we remind readers that the SC-SAM was \textit{not} calibrated using galaxy clustering; these findings are predictions naturally resulting from the model.

\subsection{The Santa Cruz Semi-Analytic Model for CAMELS}\label{subsec:SAMdetails}

To mimic the stellar feedback parameter variation in CAMELS, we focus on the SC-SAM parameters $\epsilon_{\text{SN}}$ and $\alpha_{\rm rh}$, which control the mass outflow rate out of galaxies driven by supernovae and radiation from massive stars. Like other SAMs and numerical models, which expect winds to conserve energy and momentum in a galaxy, the SC-SAM assumes that the mass outflow rate due to stellar feedback scales with the depth of the galaxy's potential well. Specifically, the SC-SAM relates the mass outflow rate from stellar driven winds to the SFR of the galaxy and the circular velocity of the halo:

\begin{equation}
  \dot{m}_{\mathrm{out}}=\epsilon_{\mathrm{SN}} \Bigg( \frac{V_0}{V_c} \Bigg)^{\alpha_{\mathrm{rh}}} \dot{m}_*
\label{eq:SCSAM_mdotout}
\end{equation}

Here, $\dot{m}_*$ is the SFR; $V_0$ is a normalization constant set to 200 km s$^{-1}$; and $V_c$ is the maximum circular velocity of galaxy's disc (assumed to be the maximum rotational velocity of its host dark matter halo). The parameters $\epsilon_{\text{SN}}$ and $\alpha_{\rm rh}$ are adjustable. This is quite similar in spirit to the treatment of kinetic stellar driven winds in IllustrisTNG and SIMBA.

The SC-SAM, however, employs a somewhat different approach for implementing AGN feedback compared to the CAMELS TNG and SIMBA model variations. In the SAM, radiatively inefficient accretion onto black holes is assumed to cause heating of the hot halo gas via energetic radio jets. The rate of accretion onto the black hole from the hot halo is given by:

\begin{equation}
  \dot{m}_{\mathrm{radio}}=\kappa_{\mathrm{radio}} \Bigg[ \frac{kT}{\Lambda [T,Z_h]} \Bigg] \Bigg( \frac{M_{\mathrm{BH}}}{10^8 M_{\odot}} \Bigg)
\label{eq:SCSAM_mdotradio}
\end{equation}

Here, $kT$ is the temperature of gas within the Bondi accretion radius ($r_A \equiv 2GM_{\text{BH}}/c_s^2$); and $\Lambda[T,Z_h]$ is the temperature- and metallicity-dependent cooling function \citep{S-D1993}. This radio mode accretion heats the hot halo gas at a rate that is proportional to $\dot{m}_{\mathrm{radio}}$, and can partially or completely offset cooling and accretion into the ISM. Thus, we can control the strength of the feedback from the jet mode by varying the parameter $\kappa_{\mathrm{radio}}$. The AGN feedback mainly affects the most massive galaxies (e.g.\ Figure \ref{fig:SAMplots_AAGN} in Appendix \ref{app:ExtraSAMverif}).

In the SC-SAM, the default values for the above parameters are: $\epsilon_{\text{SN}}=1.7$, $\alpha_{rh}=3.0$, and $\kappa_{\mathrm{radio}}=0.002$. These values, and the other parameter values that go into the SAM, were selected by tuning the SAM ``by hand'' to reproduce a set of key observed relationships (\citealt{Somerville2008, Somerville2015, Somerville2021, Yung_i}), such as the stellar mass function, cold gas fraction, mass-metallicity relation for stars, and black hole mass vs. bulge mass relation (\citealt{Bernardi2013, Moustakas2013, Baldry2012, Rodriguez-Puebla2017, Catinella2018, Calette2018, Gallazzi2005, Kirby2011, McConnell2013, Kormendy2013}). We check that we reproduce these calibrations with our implementation of the SC-SAM within the CAMELS merger trees in Appendix \ref{app:ExtraSAMverif} Figure \ref{fig:SAM_verif_plotcCV0}, finding excellent agreement with \citet{Gabrielpillai2022}. The predictions of the SAM for these galaxy properties are quite insensitive to the resolution of the input dark matter merger trees (\citealt{Gabrielpillai2022} and Figure \ref{fig:SAM_verif_plotcCV0}). 

In this work, as in the CAMELS hydrodynamic humps, we vary pre-factors for each parameter over a fairly broad range. We assign the pre-factor of $\epsilon_{\text{SN}}$ as the \textit{multiplicative} A$_{\text{SN1}}$, and vary it between [$\frac{1}{4}$,4] with a default of unity. Similarly, the pre-factor for $\kappa_{\mathrm{radio}}$ is the multiplicative factor A$_{\text{AGN}}$, and varies between [$\frac{1}{4}$,4] with a default of unity. Both A$_{\text{SN1}}$ and A$_{\text{AGN}}$ are generated evenly in logarithmic space to compare the `order of magnitude' scale of effects\footnote{When we explore our neural networks' constraints for these parameters, A$_{\text{SN1}}$ and A$_{\text{AGN}}$ will therefore appear to bias toward smaller values on the plots' linear [1/4,4] $x$-axes.}. The parameter $\alpha_{\rm rh}$ appears in the exponent of a power law, and we therefore define an \textit{additive} factor A$_{\text{SN2}}$. We vary it between [-2,2] and generate it evenly in linear space with a default value of zero. 

Though this work selects three influential parameters of broad galaxy properties in the SC-SAM to mimic what CAMELS has done, there is ongoing work in both CAMELS and CAMELS-SAM to explore the full range of astrophysical parameters in their given galaxy models. The CAMELS team has begun creating an expanded TNG wing where all 28 parameters of the IllustrisTNG model are varied over a 1024 simulations using a Sobol sampling sequence \citep{Sobol1986}. The CAMELS-SAM team is beginning an exploration of all fifteen or so adjustable parameters in the SC-SAM atop our existing merger trees. These data sets will offer fascinating explorations of the full power of these models for galaxy formation, as well as push the boundaries of parameter inference from moderately sized data sets.

Table \ref{table:ParamDescriptions} gives a summary of our chosen CAMELS-SAM scaling parameters and their ranges. They appear within Equations \ref{eq:SCSAM_mdotout} and \ref{eq:SCSAM_mdotradio} in the following manner:

\begin{equation}
\begin{split}
  \dot{m}_{\mathrm{out}} = & \Big( \epsilon_{\mathrm{SN}}\times A_{\text{SN1}} \Big) \Bigg( \frac{V_0}{V_c} \Bigg)^{(\alpha_{\mathrm{rh}}+A_{\text{SN2}})} \dot{m}_* \\
  & \dot{m}_{\mathrm{radio}}= \Big( \kappa_{\mathrm{radio}}\times A_{\text{AGN}} \Big) \Bigg[ \frac{kT}{\Lambda [T,Z_h]} \Bigg]
\end{split}
\end{equation}

\subsection{LH, CV, and 1P Simulation Sets}

\begin{table*}[t]
	\begin{center}
    \caption{Description of the different simulation products across the CAMELS-SAM suite and their significance.}

	\begin{tabular}{p{0.8cm}p{9cm}p{1cm}p{5.5cm}} 
    Label & Parameters covered & Total & Products \\
    \hline
		LH & Latin hypercube across \{$\Omega_{\text{M}}, \sigma_8$, A$_{\text{SN1}}$, A$_{\text{SN2}}$, A$_{\text{AGN}}$\} = \{0.1:0.5, 0.6:1.0, 0.25:4, -2:2, 0.25:4\}, unique random seeds. &  1,000 & \textsc{ROCKSTAR} halo catalogs, \textsc{ConsistentTrees} merger trees, galaxy catalogs \\
        CV & Fiducial cosmology \{$\Omega_{\text{M}}, \sigma_8$, A$_{\text{SN1}}$, A$_{\text{SN2}}$, A$_{\text{AGN}}$\}=\{0.3, 0.8, 1, 0, 1\} and unique random seeds. & 5 & \textsc{ROCKSTAR} halo catalogs, \textsc{ConsistentTrees} merger trees, galaxy catalogs \\
        1P & For CV$\_$0 and CV$\_$1, the minimum and maximum values of \{A$_{\text{SN1}}$, A$_{\text{SN2}}$, A$_{\text{AGN}}$\} one at a time. All else held to fiducial values. & 12 & Additional galaxy catalogs \\
        Ex & `Ex'-treme cosmology and volume simulations of (205 $h^{-1}$ cMpc)$^3$ and (1240)$^3$ at $\Omega_M$=\{0.1, 0.5, 0.5, 0.1, 0.3\} and $\sigma_8$=\{0.6, 1.0, 0.6, 1.0, 0.8\} & 5 & Not currently shared; \textsc{ROCKSTAR} catalogs used in Figure \ref{fig:HMF_CVs}. \\
    \hline \hline	
    \label{table:SimSets}
	\end{tabular}
	\end{center}
\end{table*}

\begin{figure}
    \centering
    \includegraphics[width=0.48\textwidth]{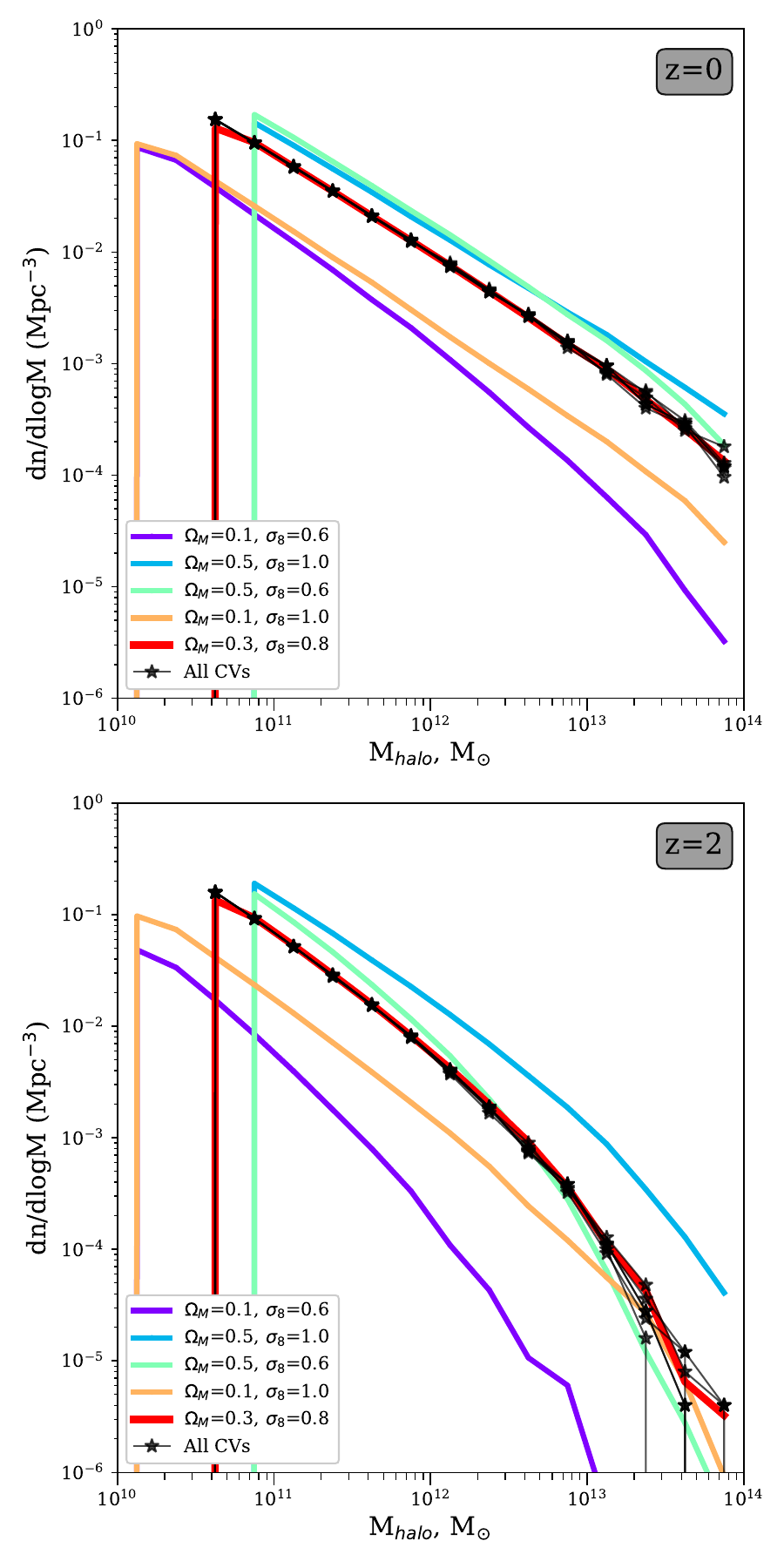}
    \caption{Halo mass functions (HMFs) at $z=0$ (top) and $z=2$ (bottom) for the CV simulations (black stars; fiducial cosmology and unique random seeds). HMFs of the (205 $h^{-1}$ cMpc)$^3$, N$=1280^3$ `extreme' cosmology volumes, where Ex\{0,1,2,3,4\} have $\Omega_{\text{M}}$=\{0.1, 0.5, 0.5, 0.1, 0.3\} and $\sigma_8$=\{0.6, 1.0, 0.6, 1.0, 0.8\}, are shown with \{purple, blue, cyan, orange, red\} lines. 
    % \textcolor{pink}{LAP note for Shy: corrected! CV and Ex4 z=0 HMF consistent with Murray et al 2013 Fig 1 comparisons of many fitting functions for the HMF.}
    }
    \label{fig:HMF_CVs}
\end{figure}

Table \ref{table:SimSets} describes all products within CAMELS-SAM and their significance and use. The core ``latin hypercube" (\textbf{LH}) set consists of 1,000 simulations, each with different values of $\Omega_{\text{M}}$, $\sigma_8$, A$_{\text{SN1}}$, A$_{\text{SN2}}$, and A$_{\text{AGN}}$. We first generated 1,000 N-body simulations with \textsc{Arepo} over a latin hypercube\footnote{
See \citet{Santner2003} and \citet{Fang2005} for review texts about this method, which originally was developed in ancient Rome to optimize agriculture, but allows for inference with sparse coverage of a high-dimensional parameter space. There has also been recent innovation on further reducing the number of instances needed for parameter space sampling in cosmological contexts \citep{Rogers2019}, or leveraging the latin hypercube for innovations in astrophysical computation \citep{Albers2019}.} of $\Omega_{\text{M}}$=[0.1,0.5] and $\sigma_8$=[0.6,1.0] (yielding individual dark matter particle masses of approximately $1.3-6\times10^8$ $h^{-1}$ M$_{\odot}$). The random phases of the initial conditions in the N-body simulations are allowed to vary. The parameters in the latin hypercube are randomly generated: linearly across $\Omega_{\text{M}}$=[0.1, 0.5], $\sigma_8$=[0.6, 1.0], and A$_{\text{SN2}}$=[-2, +2]; and logarithmically across both A$_{\text{SN1}}$ and A$_{\text{AGN}}$=[0.25, 4.0]. These SC-SAM pre-factor parameters are described in \textsection \ref{subsec:SAMdetails} and Table \ref{table:ParamDescriptions}.

In addition to the full CAMELS-SAM LH suite of 1,000 simulations, we also created a small set of 5 `cosmic variance' (\textbf{CV}) simulations, run with different random seeds for the initial conditions while using the same fiducial parameters of \{$\Omega_{\text{M}}, \sigma_8$, A$_{\text{SN1}}$, A$_{\text{SN2}}$, A$_{\text{AGN}}$\}=\{0.3, 0.8, 1, 0, 1\}. The CV set allows us to evaluate how cosmic variance in our (100 $h^{-1}$ Mpc)$^3$ volumes may affect clustering statistics.

Atop the first two CV N-body simulations, CV$\_$0 and CV$\_$1, we also created twelve min-max `one-parameter' (\textbf{1P}) simulations to serve a similar purpose as in the hydrodynamic CAMELS suites \citep{CAMELSpublic2022}. The 1P SC-SAM galaxy catalogs cover the minimum and maximum pre-factor parameter values, or \{A$_{\text{SN1}}$\}=\{0.25, 4.0\},  \{A$_{\text{AGN}}$\}=\{0.25, 4.0\}, and \{A$_{\text{SN2}}$\}=\{-2, 2\}. These simulations are useful to investigate each parameter's effect on our clustering statistics. For example, in Appendix \ref{app:ExtraSAMverif} (Figures \ref{fig:SAMplots_Asn1}, \ref{fig:SAMplots_Asn2}, and \ref{fig:SAMplots_AAGN}), we examine how the extreme ends of these parameters affect key galaxy summary statistics, including the stellar mass function and the ratio of cold gas fraction to stellar mass in galaxies.

We additionally created five `extreme' simulations with the same resolution but eight times larger volume. These simulations have box size length L=205 $h^{-1}$ cMpc and N=1280$^3$ particles. One was created with the fiducial cosmology of \{$\Omega_{\text{M}}$, $\sigma_8$\}=\{0.3, 0.8\}, and the other four were created at the corners of the full cosmological parameter space of $\Omega_{\text{M}}$=[0.1,0.5] and $\sigma_8$=[0.6,1.0]. 
We ran the SC-SAM on each of these ‘extreme’ simulations using the fiducial parameters that best recreate $z=0$ observations.  This allowed us to confirm that our selected galaxy clustering statistics show the expected influence of cosmological parameter variations.\footnote{At the time of writing, the `extreme'(-ly large) simulations are not released with the rest of CAMELS-SAM.}

\subsection{Verifying the Simulations} \label{subsec:techdetails}

As a confirmation of our N-body volumes and the \textsc{ROCKSTAR} products, we first examine the behavior of the halo mass functions (HMFs). Figure \ref{fig:HMF_CVs} shows $z=0$ and $z=2$ HMFs for the five CV simulations. We compare the CV set against the HMFs of the `extreme' simulations, 4 of which exist at the extreme corners of our cosmological parameter space and 1 of which is at the same fiducial cosmology as the CV set. These simulations show well converged halo statistics for our smaller fiducial volume and illustrate the broad range of conditions probed by CAMELS-SAM. We also note that the relevant HMFs of the IllustrisTNG300-1 and -2 simulations are very consistent with the `fiducial' cosmology Ex4 simulation and all the CV simulations (confirmed in Appendix \ref{app:ExtraSAMverif} Figure \ref{fig:SAM_verif_plotcCV0}). 

The SC-SAM has been tested using the largest and highest resolution IllustrisTNG simulations, with box side length of L=205 $h^{-1}$ cMpc and 2500$^3$ particles \citep{Gabrielpillai2022}. We confirmed the SC-SAM would still robustly match key observational statistics for high-mass halos at our lower resolution, similar to IllustrisTNG300-2 (of L$_{\text{box}}$=205 $h^{-1}$ Mpc and 1250$^3$ particles), with all parameters held to the IllustrisTNG cosmology and the best-fit SC-SAM parameters from \citet{Somerville2021}.

In Appendix \ref{app:ExtraSAMverif}, we confirm that our suite setup recreates key observed summary statistics under the fiducial model. We compare the result of our CV simulations to the SAM outputs of the two highest resolution IllustrisTNG300 volumes, and various near-universe observations for: the stellar mass function, the stellar mass-halo mass, the cold gas fraction vs. stellar mass of disk-dominated galaxies, stellar metallicity-stellar mass, and black hole mass-bulge mass relationships. The consistency with the larger IllustrisTNG300 SAM catalogs and the overall SC-SAM agreement with $z=0$ observations support the choice of volume and resolution for our CAMELS-SAM suite. In Appendix \ref{app:ExtraSAMverif}, we also probe these relationships for the `1P' galaxy catalogs, to further understand how each of the SC-SAM parameters we vary affects astrophysical summary statistics.

% --------------------------------------------------------------------------------

\section{Methodology} \label{sec:Methods}

In this section, we describe our methodology for constraining cosmology and astrophysics using clustering summary statistics and neural networks. In \textsection \ref{subsec:Clustering}, we describe the clustering statistics we test in this work, how they were measured, and how they are prepared for the neural network pipeline. In \textsection \ref{subsec:NNimplementation}, we describe our neural network architecture and process.

\subsection{Galaxy Clustering} \label{subsec:Clustering}

\begin{figure*}
    \centering
     \begin{subfloat}[VPF (left), CiC (middle), 2ptCF (right) clustering of $z=0$ galaxies with nonzero stellar mass, down-sampled to 0.005 ($h^{-1}$ cMpc)$^3$.]{
    \includegraphics[width=\textwidth]{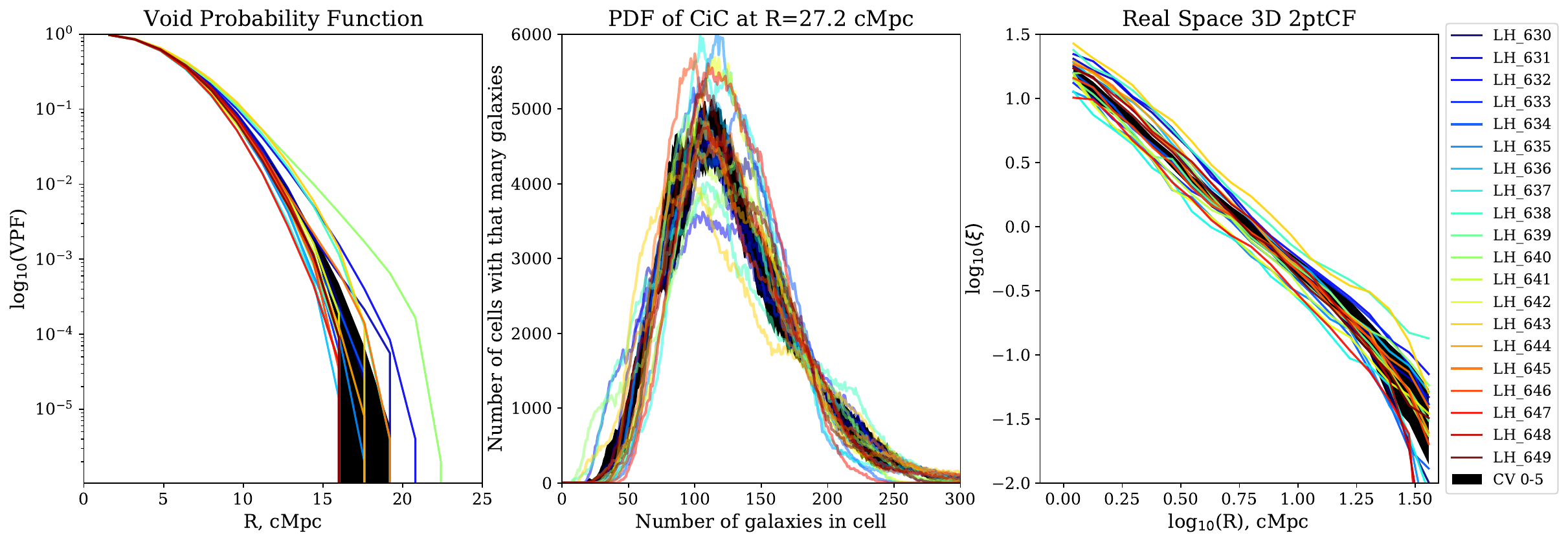}
    %  \caption{}
     \label{fig:clusteringex_wCV}
     }
     \end{subfloat}

     \begin{subfloat}[VPF (left), CiC (middle), 2ptCF (right) clustering of $z=0$ galaxies with SFR $>$ 1.25 M$_{\odot}$ yr$^{-1}$, down-sampled to 0.001 ($h^{-1}$ cMpc)$^3$.]{
    \includegraphics[width=\textwidth]{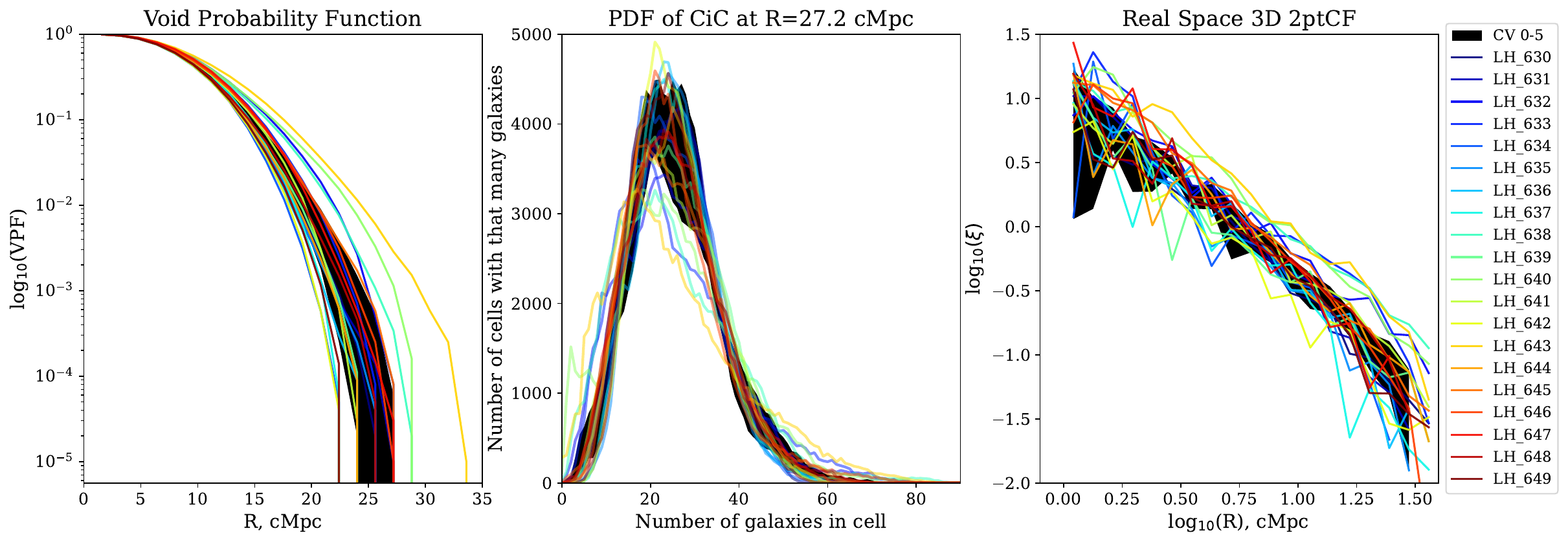}
    %  \caption{}
     \label{fig:clusteringex_nonzero}
     }
    \end{subfloat}
    \caption{Clustering statistics for CAMELS-SAM simulations 630 through 649 (each in a unique color). We show the Void Probability Function (VPF, left), count-in-cells at 27.2 cMpc (CiC, center), and real-space 3D two-point correlation function (2ptCF, right), for two different selections of $z=0$ galaxies (top: stellar mass, bottom: SFR). The shaded black area in the top figure indicates the clustering across 5 simulations using the `fiducial' cosmological and SAM parameters but generated with different random seeds, to show the effect of cosmic variance for these volumes.}
    \label{fig:example_Clustering}
\end{figure*}

\subsubsection{Introduction to Clustering Statistics} \label{subsubsec:BckgdClustering}

The spatial distribution of galaxies traces the structure of underlying dark matter, and carries signatures of both the cosmology as well as details of how galaxies interact with their environment and each other. There are many ways to measure the clustering of galaxies, each with unique strengths, uses, theoretical foundations, and connections to other physical concepts. In this proof of concept work, we use the Void Probability Function (VPF), count-in-cells (CiC), and (real space) two-point correlation function (2ptCF).

The widely used 2ptCF quantifies the probability of finding two galaxies within a certain distance from each other (compared to a random distribution; e.g.\ \citealt{Peebles1980, L-S1993}). CiC quantifies the number of galaxies within a randomly placed cell of a given size and theoretically includes all higher-order $n$-point clustering statistics \citep{Peebles1980}, but is the most computationally costly of these three statistics. The VPF is a less commonly used clustering statistic that simply asks: how likely is a sphere of a given size to contain zero objects for a given galaxy selection criterion? The VPF is simple and efficient to calculate, is tied to all higher order correlation functions as the 0$^{\text{th}}$ moment of count-in-cells, and encodes information from higher order clustering that is not captured in the 2ptCF (\citealt{Perez2021, Conroy2005, White1979})\footnote{
We emphasize here the VPF is separate from work with \textit{cosmic} voids, which are large underdense regions in the cosmic web that require large detailed sky surveys to map and catalog. The VPF is a simple statistical tool that counts empty circles/spheres, where cosmic voids can have complex shapes and contain rare and interesting galaxies \citep[e.g.][]{Habouzit2020}. Cosmic voids are being successfully used as alternate cosmological probes \citep{Pisani2015,Hamaus2016, GZhang2020}.} .

These clustering statistics are known to be powerful tools for constraining cosmology in observations and N-body simulations. CiC has been prominent and promising in several recent works constraining cosmology. \citet{Uhlemann2020} note the inclusion of CiC improved their constraints on $\Omega_{\text{M}}$ and $\sigma_8$ by factors of 5 and 2 respectively compared to the matter power spectrum alone, and also broke the degeneracy between massive neutrino mass and $\sigma_8$. \citet{Salvador2019} also probed how well CiC can constrain linear and higher-order galaxy bias, finding its constraints are consistent with measurements of the bias from galaxy–galaxy clustering, galaxy–galaxy lensing, cosmic microwave background lensing, and shear-clustering measurements. Excitingly, \citet{ReppSzapudi2020} developed a \textit{theoretical} description for the CiC distribution (enabling analyses requiring understanding of the variance and covariance), and found that CiC breaks the degeneracy between $\sigma_8$ and the bias parameter, and yields an 11\% error on their $\sigma_8$ measurement for the SDSS Main Galaxy Sample. \citet{Dantas2021} identified that the presence of baryons in IllustrisTNG affects the best-fitting theoretical model for CiC. And lastly, \citet{Wen2020} developed a new technique with the CiC PDF to probe different cosmological models with dark energy, focusing on dark matter halos in the DEUS simulations \citep{DEUS} at $0 < z < 4$ and scales $2-25\ h^{-1}$ cMpc.

Additionally, more uncommon types of the 2ptCF have been recently found to help constrain dark energy observables \citep{Zhai2019-cts, Zhai2019-aemulus3}. \citet{VanDaalen2016} also used the projected 2ptCF, creating a clustering estimator that improved constraints on astrophysics within SAMs (specifically, the Munich SAM of \citealt{Guo2013}). \citet{Wang2019} used the VPF and CiC in conjunction with the projected 2ptCF to help constrain galaxy assembly bias using halo occupation distributions, and \citet{McCullagh2017} and \citet{WalshTinker2019} used the VPF to refine generalized HOD fitting. Our work is among the first to use the VPF as the 2ptCF and CiC have been used, and to systematically compare the constraints that can be obtained from all three statistics. Part of the VPF's uncommon use can be attributed to how difficult it is to model for non-random galaxy distribution even for a narrow fiducial $\Lambda$CDM cosmology, often excluding it from `traditional' frameworks of cosmological parameter inference from galaxy clustering.

\subsubsection{Measuring Clustering in CAMELS-SAM}\label{subsubsec:MeasuringClustering}

We measure all clustering statistics between 1 and 40 cMpc (0.6711 to 26.84 $h^{-1}$ cMpc) using the \textsc{CORRFUNC} package \citep{CORRFUNC}. We constrain our analysis to distances larger than 1 cMpc due to known inaccuracies of the assignment of satellite positions in the SC-SAM relative to hydro simulations \citep{Hadzhiyska2021b}. This is because the SC-SAM does not use the subhalo positions within their host halo provided by the N-body simulation, but treats subhalo merging and tidal destruction using a semi-analytic recipe (see \textsection 2.2.1 in \citealt{Somerville2021}).
Moreover, it is expected that the presence of baryons will affect the orbits of satellites and the efficiency of their tidal heating and destruction, in a manner not currently modeled in the SC-SAM \citep{Jiang:2021}. Our analysis still probes smaller nonlinear scales than most similar studies, reaching the equivalent of $k_{\text{max}}<5.85 h$ cMpc$^{-1}$ (VPF/CiC) or $8.5\ h$ cMpc$^{-1}$ (2ptCF), and leverages additional information from non-linear terms of galaxy bias.

For the 2ptCF, we generate 20 distance scales evenly in logarithmic space between $10^0$ and $10^{1.6}$ cMpc, yielding 19 distance bins centered between 1.1 cMpc and 36.1 cMpc. With \textsc{CORRFUNC}, we measure the 2ptCF in three-dimensional real space using the \citet{L-S1993} estimator: $\xi(R)= (DD-2D\mathcal{R} +\mathcal{RR}) /\mathcal{RR}$, where $DD$ are the number of data-data pairs in the bin that encloses the distance scale $R$, $D\mathcal{R}$ data-random pairs, and $\mathcal{RR}$ random-random pairs. For the 2ptCF, we use 100 times as many random points as there are galaxies.  

For the VPF and CiC, we use \textsc{CORRFUNC} to perform the calculation for 25 distance scales with a maximum of 40 cMpc, yielding 25 linearly spaces radii between 1.6 and 40 cMpc. We measure the VPF by randomly dropping 100,000 or 500,000 spheres of each tested distance scale (for galaxy densities of 0.001 or 0.005 $h^{3}$ cMpc$^{-3}$, respectively), and counting those with no galaxies (see \citealt{Conroy2005} for a discussion on the effects of the number of dropped spheres). For CiC, we also drop 100,000 or 500,000 spheres randomly of each tested distance scale, and count how many have 0, 1, 2, ... $n=500$ galaxies inside them. Measuring to $n=500$ captures nearly the entire CiC distribution at these densities over almost all scales. The VPF is additionally calculated and verified\footnote{During this work, we identified a feature of how \textsc{CORRFUNC} generated the `random test spheres' for the VPF and CiC that led to inaccurate VPF measurements for small-to-medium sized galaxy samples. This has now been corrected fully, allowing the powerful \textsc{CORRFUNC} to be applied to even more samples than it was originally created for.} using the swift \textit{k}-nearest neighbor method of \citet{Banerjee2020}. Future expansions to this work would do well to include higher-order nearest neighbor statistics, shown to give additional cosmological constraints by \citet{Banerjee2020}.

Figure \ref{fig:example_Clustering} displays these clustering measurements for twenty of the LH galaxy catalogs, under two types of `selections' that vary the SC-SAM galaxy property, the threshold value, and number density down-sampling (described in the next section). For the twenty simulations shown (LH$\_630-649$), the span of covered parameters is: $\Omega_{\text{M}}=[0.131,0.479]$, $\sigma_{8}=[0.6342,0.9862]$, A$_{\text{SN1}}=[0.266, 2.45]$, A$_{\text{SN2}}=[-1.506,+1.998]$, A$_{\text{AGN}} = [0.26, 3.84]$, and therefore qualitatively representative of our parameter space. We also measured the clustering of the CV catalogs to confirm that the change in clustering caused by varying the parameters exceeded that caused by cosmic variance. 

\subsubsection{Context: Galaxy Selections within CAMELS-SAM}\label{subsubsec:Selections}

Before measuring any clustering statistic, we apply a selection across CAMELS-SAM to yield a coherent sample of galaxies across all cosmologies and feedback parameters. In this initial proof of concept study, galaxies are selected by: 

\begin{itemize}
    \item \textit{halo mass}, usually $\log_{10}(\text{M}_{\text{halo}}$/M$_{\odot}) >$ 10-12 %, for dark matter-only selections
    \item \textit{stellar mass}, usually $\log_{10}(\text{M}_{\text{stellar}}$/M$_{\odot}) >$ 9-10
    \item \textit{instantaneous star formation rate} (SFR), M$_{\odot}$ yr$^{-1}$
    \item or \textit{specific star formation rate} (sSFR = SFR / M$_{\text{stellar}}$), Gyr$^{-1}$.
\end{itemize}

\begin{table*}[t]
	\begin{center}
    \caption{Radii and bins of count-in-cells given to the neural network for our various tests. The spread of $n$ refers to which 'counts' we consider--e.g.\ cells with 0 to 50 galaxies only.  When doing a single redshift at a time, we broadened our choices to encompass the parts of the distribution with the most divergence across our simulations. To maximize the data we include for a single redshift, we skip every other $n$ value. Therefore, each simulation's single-redshift count-in-cells sampling yields 510 total data points for the neural network to work upon. For all redshifts combined, we sample the first 50 $n$ values at each distance scale, since the distance scales we selected encompass most of their variety in those regions. The all-redshifts CiC sampling yields 600 total data points.}

	\begin{tabular}{p{2cm}p{2cm}p{2cm}p{2cm}p{2cm}p{2cm}} 
	\hline \hline
	\multicolumn{2}{c}{All redshifts, $\mathcal{N}$=0.001 $h^3$ cMpc$^{-3}$} & \multicolumn{2}{c}{All redshifts, $\mathcal{N}$=0.005 $h^3$ cMpc$^{-3}$} & \multicolumn{2}{c}{Single redshift, any $\mathcal{N}$} \\
	\hline 
    Radius, cMpc & Spread of $n$ & Radius, cMpc & Spread of $n$ & Radius, cMpc & Spread of $n$ \\
    \hline
		11.2 & 0-50 & 16.0 & 0-50 & 16.0 & 0-50 \\
        16.0 & 0-50 & 22.4 & 0-50 & 25.6 & 0-170 \\
        20.8 & 0-50 & 28.8 & 0-50 & 30.4 & 50-250 \\
        -- & -- & -- & -- & 35.2 & 100-350 \\
        -- & -- & -- & -- & 40.0 & 250-500 \\
    \hline \hline	
    % \\
    \label{table:CiCRadii}
	\end{tabular}
	\end{center}
\end{table*}

These properties are predicted directly by the SC-SAM, and are relatively easy to compare directly to other simulations. Although they are not directly observable, there is extensive work in the literature to estimate stellar masses and SFR from observed samples using various methods (e.g.\ see reviews such as \citealt{BlantonMoustakas2009, MadauDickinson2014, SomervilleRomeel2015}). For example, the stellar mass of galaxies is often measured with spectral energy distribution fitting to broad or medium band photometry \citep{Walcher2011, Barro2013, Conroy2013, Duncan2014, Mobasher2015}, or spectra (e.g.\ \citealt{Brinchmann2004, Brammer2011}). Star formation rate can be measured with flux in several different emission lines or bands targeting different sources or tracers of star formation, though many are sensitive to dust or contamination, and all carry the uncertainties of assumed timescales (e.g.\ \citealt{Calzetti2013, Ellis2008}). Specific SFR gives deep insight into the process of galaxy evolution over cosmic history (e.g.\ Fig. 11 in \citealt{Speagle2014}), and is sometimes measured by proxy with emission lines or with careful analysis of the stellar mass function (e.g.\ \citealt{Davidzon2018}). Finally, the clear bimodal nature of galaxy colors (and the separation between star-forming vs. quiescent galaxies) is easily seen in the sSFR vs. stellar mass plane (e.g.\ \citealt{Muzzin2013}).

Ultimately, a natural end goal of this work would be to train a neural network capable of measuring the cosmology from observed galaxy clustering. This is an ambitious goal that requires much future work, such as identifying a galaxy sample that CAMELS-SAM is large enough to simulate, the generation of such realistic galaxies and/or their clustering over each CAMELS-SAM LH simulation, and a careful understanding of the observation's selection function and systematic errors. In \textsection \ref{subsec:FutureWorkIdeas}, we expand on the future work needed to apply neural networks like those we present in the next section. For this initial proof-of-concept work, however, we apply straightforward and idealized selections on fundamental galaxy properties.

Our simplistic selections on stellar mass, SFR, and sSFR are consistent with the results of past studies and upcoming surveys. For example, several studies have implemented or found similar halo mass limits as we use: \citet[log$_{10}$ M$_{\text{halo}} > 10.9-13\ h^{-1}$ M$_{\odot}$]{Springel2018}, \citet[log$_{10}$ M$_{\text{halo}} > 10.7\ h^{-1}$ M$_{\odot}$]{Xu2021}; \citet[log$_{10}$ M$_{\text{halo}} > 11.38\ h^{-1}$ M$_{\odot}$]{Wen2020}; \citet[log$_{10}$ M$_{\text{halo}} > 10.3\ h^{-1}$ M$_{\odot}$]{G-P2018}. Our probed SFR and stellar mass cuts are achievable for future surveys; for example, SFR $>0.1$ M$_{\odot}$ yr$^{-1}$ will likely be observable within $z<1$ for wide Roman and Euclid surveys, assuming surveys to (dust corrected) limiting magnitudes of 27 in Roman's WFI F062 filter at 62 $\mu$m and Euclid's VIS instrument (from Yung et al. (in prep.) Roman light cones created with the SC-SAM, and forecasts similar to those in \citealt{Yung_i, Yung_ii}). Our M$_{\text{star}}>10^9$ M$_{\odot}$ selection should similarly be widely detected with Roman within $z<1$, and to a similar extent with Euclid (though more easily at closer to $z\sim1$; stellar masses above $10^9$ M$_{\odot}$ will likely correspond to the brightest Euclid galaxies at $z<0.1$) Additionally, our higher stellar mass selections complement the limiting stellar masses derived for 14,000 deg$^2$ across various Legacy Surveys serving the DESI project: M$_{\text{star}}>10^{9.5}$ M$_{\odot}$ at $z \sim 0.1$, M$_{\text{star}}>10^{10.5-11}$ M$_{\odot}$ above $z>0.5$ \citep{Zou2019}. Finally, these selections are consistent with what the Vera Rubin Observatory will measure with LSST: M$_{\text{star}}>10^{9.5}$ M$_{\odot}$ at $z<0.5$, M$_{\text{star}}>10^{10}$ M$_{\odot}$ at $0.5<z<1.0$ \citep[Figure 6]{Riccio2021}.

When finalizing the selection criteria for our chosen galaxy or halo property, we confirm each criterion will be met by enough CAMELS-SAM simulations to obtain robust statistics across most of the 1,000 LH simulations. As we later discuss in \textsection \ref{subsubsec:NN_specifics}, we split up our 1,000 CAMELS-SAM LH simulations into roughly 70/15/15 percent training, validation, and testing sets, respectively. 
However, not all simulations yield a large enough sample to compute the clustering statistics with all selection criteria, and we choose to prioritize having at least 700 simulations in the training set, while still keeping moderately large validation and testing sets. 

For all halo or galaxy selections we present in this work, enough simulations meet the criteria to guarantee training sets of at least 700 samples, and validation and testing sets of at least 80-100 each. This range is still large enough to sample the parameter space well, while allowing the flexibility to use selection criteria that pick out rarer objects, potentially revealing unique relationships between the parameters and galaxy clustering. The simulations that do not meet the selection criteria tend to have little structure formation--very small $\Omega_{\text{M}}$ and $\sigma_8$--meaning they cluster in a specific corner of the parameter space. CAMELS and CAMELS-SAM intentionally cover such a large parameter space to avoid being affected by the distribution priors in the central, more realistic region of the parameter space. Therefore, missing a single corner will still yield robust constraints near the values of the cosmological parameters that are favored by observational constraints. 

The final aspect of our selection is whether or not we normalize to a specific number density of objects. Number density has a strong effect on the CiC and the VPF, and some stellar mass or SFR selections yield samples with very large numbers of objects, creating a computational bottleneck. We address this in this work by randomly ``down-sampling'' the objects that pass a given selection to either 0.001 $h^{3}$ cMpc$^{-3}$ or 0.005 $h^{3}$ cMpc$^{-3}$, corresponding to 1,000 or 5,000 galaxies in our (100 $h^{-1}$ cMpc)$^3$ volumes.
% \textcolo{green}{rachel: this is a bit misleading since these samples are not down-selected...so i'd instead relate the observational samples to stellar mass and SFR limits (and move to that section)} 
Additionally, these densities are large enough to mitigate Poisson noise while producing samples small enough to calculate the clustering statistics within a reasonable computational cost. In \textsection \ref{subsec:Downsampling}, we examine how our constraints on cosmology and SAM astrophysics change if we do not randomly down-sample to a fixed number density after applying our mass or SFR-based selections.

Finally, as a contextual example to compare against our selections, the \citet{Springel2018} clustering measurements of IllustrisTNG300 found thresholds for these selections at these densities: for a galaxy space density of 0.001 $h^3$ cMpc$^{-3}$, log$_{10}$M$_{\text{stellar}} > 10.18\ h^{-1}$ M$_{\odot}$, SFR $> 1.55$ M$_{\odot}$ yr$^{-1}$, sSFR $> 1.27\ h$ Gyr$^{-1}$. For a galaxy space density of 0.003 $h^3$ cMpc$^{-3}$, log$_{10}$M$_{\text{stellar}} > 10.49\ h^{-1}$ M$_{\odot}$, SFR $> 3.03$ M$_{\odot}$ yr$^{-1}$, sSFR $> 4.43\ h$ Gyr$^{-1}$. In observations, our chosen densities mimic the galaxy number densities many studies have measured or are expected to measure. Examples near $10^{-3-4}\ h^3$ cMpc$^{-3}$: SDSS red/blue galaxies with predicted M$_{\text{halo}} \sim 10^{12.7}\ h^{-1}$ M$_{\odot}$ \citep[Table 4]{Zehavi2005}; 
SDSS-IV BOSS for ELGs with measured M$_{\text{star}} \sim 10^{10.5}\ h^{-1}$ M$_{\odot}$ \citep[Figure 11]{Raichoor2017}, projected EUCLID galaxies within $z<1.5$ \citep[Table 3]{Amendola2018}; and a `dense' sample of red sequence galaxies in the Kilo Degree Survey with M$_{\text{stellar}} \sim 10^{10.8}\ h^{-1}$ M$_{\odot}$ \citep{Vakili2020}. However, we note these samples were not randomly down-sampled, and are often flux- or volume-limited.

In simulations and SAMs, the exact mapping of a given selection in stellar mass or SFR to the resulting galaxy number density is strongly model-dependent and can be at odds with other observational calibrations. This often leads to a different approach to fixing the number density of a galaxy sample in the literature. Instead of randomly down-sampling, studies will choose objects above a threshold in mass or SFR in order to obtain a desired number density, sometimes referred to as abundance matching. For example, \citet{Hadzhiyska2021a} (comparing IllustrisTNG and SC-SAM galaxies) select the most massive galaxies such that they reach their desired number density. This has a very different effect on clustering, as it is selecting a differently clustered/biased population of dark matter halos, while our random down-sampling selects halos with the same clustering properties, but just sparsely samples them to reduce computational load. 

\subsection{Neural Network Implementation} \label{subsec:NNimplementation}

\subsubsection{Preparing CAMELS-SAM clustering for neural networks}\label{subsubsec:NN_specifics}

We use a 70/15/15 percent split of the suite for training/validation/testing, meaning each network was trained on the first 700 of the CAMELS-SAM LH simulations, validated for performance on the next 150, and tested on the final 150. (See \textsection \ref{subsubsec:Selections} for what we do if not all 1,000 simulations are usable.) The `best' model is whichever has the lowest error value when applied to the validation set, though we often found that at least 5-10 models performed quite similarly, meaning these results are not tied to a specific architecture. In the figures that follow, we show the performance of these best models on the test set of simulations.

In this fiducial case of \textit{`all' clustering}, the data given to the neural networks consists of: for $z=\{0, 0.1, 0.5, 1.0\}$, the 2ptCF and the VPF between roughly 1 and 40 cMpc, and the CiC probability distribution for several distance scales (between 11-21 cMpc for our lower density samples, and between 16-30 cMpc for the larger density samples; see Table \ref{table:CiCRadii}). The clustering measurements at all four redshifts are strung into a 1D array\footnote{Though explained later in the relevant sections, when we are \textit{not} looking at `all' clustering statistics at all four redshifts, the neural network is instead given in a 1D array: all statistics at a single redshift (\textsection \ref{subsec:RedshiftChoices}), or an individual clustering statistic at the four redshifts (\textsection \ref{sec:CompareStats}).}; all statistics are measured at exactly the same distance scales across all simulations that pass the selection, so the radius of each measurement is irrelevant for the neural network. Additionally, we randomly re-sample to our selected number density for each redshift\footnote{There is no way to fully know how a neural network learns what it does. However, our choice to re-sample at each redshift, and the additional noise introduced, may help prevent the networks from focusing upon the growth factor \citep{Hamilton2001} if it e.g.\ decides to try taking ratios between clustering statistics. In future work, shuffling the order of different redshifts' clustering when training may help further counteract this.}. See \textsection \ref{subsec:RedshiftChoices} for an exploration of how our results change if we use only one redshift at a time, and \textsection \ref{sec:CompareStats} for the performance when the clustering statistics are used separately.

Neural networks perform best when trained on normalized data, where the mean is about zero and the standard deviation is one. We normalize the clustering 1D array so: for each value (corresponding to e.g.\ the 2ptCF at $R\approx$ 10 cMpc), we take the base-ten logarithm across all 1000 values, subtract the mean from each, and then divide each by the standard deviation. For galaxy selections that yield values of zero for a given simulation (e.g.\ a particularly unclustered simulation that finds no voids at large scales), the value is set to $10^{-12}$ before normalizing. For galaxy selections where all 1000 simulations yielded zero (rare, and often at the largest distance scales for extreme selections), we set the first simulation to a value of $2\times10^{-12}$ to guarantee the normalization does not fail, and continue. The cosmological and astrophysical parameters are normalized in almost the same way---by their mean and standard deviation in linear space---and are returned by the neural network as 1D output arrays.

\subsubsection{Loss Functions}\label{subsubsec:NNlosses}

In this work, we use the CAMELS-SAM simulations to do parameter inference with galaxy clustering statistics: how do the input parameter(s) relate to the statistics, and how do we measure the input parameter(s) given only the statistics? This \textit{marginal posterior} we seek to learn is $p(\vec{\theta}|\vec{C})$, which relates the parameter space $\vec{\theta}$ to the 1D array of clustering measurements $\vec{C}$. In our work, we often have $\vec{\theta}$\ =\ \{$\Omega_{\text{M}}$, $\sigma_8$, $A_{\text{SN1}}$, $A_{\text{SN2}}$, $A_{\text{AGN}}$\} for all 5 input parameters of each simulation. 

The marginal posterior over a single parameter $\theta_i$ (out of e.g.\ all 5 we constrain) can be defined as:
\begin{equation}
    p(\theta_i|\vec{C}) =  \int_{\theta} p(\theta_1, \theta_2, ...\theta_n|\vec{C})  d\theta_1...d\theta_{i-1}d\theta_{i+1}...d\theta_n
\end{equation}

The marginal posterior describes the probability that a simulation (and its array of clustering measurements) were created with a particular combination of parameters $\vec{\theta}$. Here with CAMELS-SAM and in many applications of CAMELS, we give the network completely flat priors, or no prior knowledge of the underlying distribution of parameters (e.g.\ the measurement of cosmology from \citealt{Planck2016}). The estimated mean of the marginal posterior for a given parameter $\theta_i$ is: 

\begin{equation}
    \mu_i(\vec{C}) = \int_{\theta_i} p(\theta_i|\vec{C}) \theta_i d\theta_i
\end{equation}

The estimated standard deviation of the marginal posterior for a given parameter $\theta_i$ is: 

\begin{equation}
    \sigma_i(\vec{C}) = \int_{\theta_i} p(\theta_i|\vec{C}) (\theta_i - \mu_i)^2 d\theta_i
\end{equation}

The goal of a neural network is to learn the posterior accurately enough that the mean $\mu_i$ and standard deviation $\sigma_i$ it predicts are consistent with estimated posterior values given the input parameters. Our neural network here assumes a single-peak posterior with one mean and standard deviation. The actual marginal posterior may not have these properties; for example, there may be a degenerate combination of our parameters that yields very similar clustering measurements. In these cases, we can expect a large standard deviation measurement that will attempt to cover the multiple peaks in the posterior.

For a neural network to learn a posterior, it requires a loss function to measure its performance (i.e.\ calculate the gradients it uses to update the weights between neurons in order to eventually converge on values closest to the true ones). We perform both parameter regression with a standard \textit{mean-squared error} (MSE) validation criterion, and \textit{likelihood-free inference} (LFI) with the method from \citet{MomentNetworks}, updated for CAMELS in \citet{CAMELS_CMD_announcement} and featured in \citet{Jo2022}. Our parameter regression is a fast and straightforward way to measure the mean of the posterior and therefore roughly approximate the network's accuracy, while the latter trains for longer to also measure the posterior's standard deviation. Finally, our loss functions are assessed over a given batch of input data, with a batch size of N$_{\text{batch}}=64$ as default, meaning that sixty-four random simulations in the training set are trained at a time per node worker.

The MSE loss function\footnote{A default option in \textit{Pytorch} \citep{Pytorch}} we use simply measures the mean squared error between each element of the neural network's predictions and its true value, and is built to approximate the mean of the marginal posterior (even if non-Gaussian). For a given batch size, the MSE loss on the model's predictions is:

\begin{equation}
    \mathcal{L}_{\text{MSE}} = \sum_{i=1}^{5} \Bigg( \frac{1}{\text{N}_{\text{batch}}} \sum_{j\in {\text{N}_{\text{batch}}}} (\theta_{i,j} - \mu_{i,j})^2 \Bigg) ,
    \label{eq:MSEloss}
\end{equation}

where $\mu_{i,j}$ is the network's prediction for the mean of parameter $i$'s posterior for simulation $j$; and $\theta_{i,j}$ is the true input value of parameter $i$ for simulation $j$. We note that this loss sums over all the cosmological and astrophysical parameters being constrained, meaning it attempts to measure the marginal posterior means for each parameter at once.

With the LFI loss function, our goal is to instead get the neural network output to converge on both the mean and standard deviation of the marginal posterior. With slight modifications to the loss function presented in \citet{MomentNetworks}, one can define the loss function so that the neural network outputs converge over both $\mu_i$ and $\sigma_i$ (see \citealt{CAMELS_CMD_announcement} \textsection 3.1.2):

\begin{equation}
\begin{split}
  \mathcal{L}_{\text{LFI}} = & \sum_{i=1}^{5} \log \Bigg( \sum_{j\in {\text{N}_{\text{batch}}}} (\theta_{i,j} - \mu_{i,j})^2 \Bigg) \\
  & + \sum_{i=1}^{5} \log \Bigg( \sum_{j\in {\text{N}_{\text{batch}}}} \Big( (\theta_{i,j} - \mu_{i,j})^2 -\sigma_{i,j}^2 \Big)^2 \Bigg)
\end{split}
\end{equation}

As described in \citet{CAMELS_CMD_announcement}, the effect of this loss function is that the neural network ignores noisy parameters until it has learned the well-determined parameters first. This is important in circumstances where a feature is very sensitive to specific parameters and not at all for others (e.g.\ the CAMELS dark matter density field, which very mildly detects TNG/SIMBA astrophysical parameter variations).
This LFI loss function removes the dependence on the overall scale in the scatter of a parameter with the included logarithm, therefore inverse-variance weighting the combination of gradients from the different terms compared to the MSE loss in Eq. \ref{eq:MSEloss}. The clustering of galaxies is dominated by that of dark matter halos, meaning the influence of the cosmological parameters will likely be much stronger than any of the SC-SAM astrophysical parameters. Therefore, we prioritize the use of the LFI loss method throughout this work. 

We use the MSE loss only on the single-redshift exploration in \textsection \ref{subsec:RedshiftChoices} due to its faster training time, and LFI everywhere else for more detailed constraints. We generally find the MSE and LFI results are broadly consistent, though the LFI are often slightly more precise, likely due to the introduction of the logarithm discussed above.

We do, however, leverage the \textit{root mean square error} (rMSE) to gauge the accuracy of our neural network predictions. In Tables \ref{table:oneatatime}-\ref{table:ClusteringStats_CosmoConstraints} reporting our results, we list two types of errors, `rMSE' and $\bar{\sigma}$. $\bar{\sigma}_i$ is the mean across the test set's $\sigma_{i}$ values after going through the best-performing neural network under the LFI loss, and reflects the actual 1$\sigma$ error on the constraint. `rMSE' refers to the rMSE calculated on the mean $\mu_{i}$ values from the relevant loss function, and roughly measures the accuracy of the predictions. Following the common definitions, the rMSE and $\bar{\sigma}_i$ for the predicted parameter $\theta_i$ across the test set of length N$_{\text{test}}$ are:

\begin{equation}
   \text{rMSE}_i= \sqrt{ \frac{1}{\text{N}_{\text{test}}} \sum_{j\in {\text{N}_{\text{test}}}} (\theta_{i,j} - \mu_{i,j})^2 }
\end{equation}

\begin{equation}
    \bar{\sigma}_{i} = \frac{1}{\text{N}_{\text{test}}} \sum_{j\in {\text{N}_{\text{test}}}} \sigma_{i,j} 
\end{equation}

\subsubsection{Tests for accuracy}\label{subsubsec:Chi2}

We generally find that the rMSE calculated on the LFI-loss predicted means is often very close to the LFI-loss predicted error, though that often rMSE $>\bar{\sigma}$. To confirm the general behavior of our LFI-loss networks is accurate and is not under-predicting the error, we implement a few simple quantitative tests. 

First, we include a calculation of each individual test simulation's `$Z$-value' to exclude the rare outlier(s), according to the behavior of the entire test set's LFI predicted means. We begin by calculating the absolute value difference between actual and predicted parameter $i$ values across the whole test set of length $t$:

\begin{equation}
    \vec{\phi}_i=\{\phi_{i,1}, \phi_{i,2}, ..., \phi_{i,t}\}, \quad \phi_{i,j} = | \theta_{i,j} - \mu_{i,j} |
    \label{eq:phi_def_acc}
\end{equation}

Then for each simulation $j$ in the test set, the `$Z$ value' for a parameter $i$ is: 

\begin{equation}
    Z_{i,j}= \frac{ \Big( \phi_{i,j} - \textrm{mean}(\vec{\phi}) \Big)} {(\text{std}(\vec{\phi}))}
    \label{eq:Zvalue}
\end{equation}

We remove simulations in the test set with $Z > 6$ before calculating the rMSE on the LFI predicted means that are listed in future tables; visually, these are the extreme outliers clear to the eye whose predictions are often unrealistic or beyond the bounds of the parameters (e.g.\ see A$_{\text{AGN}}$ for Figure \ref{fig:z0p5_mstar}). Though very rare to find an outlier of $Z>6$, none of the selections we have implemented had more than a single outlier in the test set, and we have found it is nearly always the same LH simulation. We remind readers that the best-performing neural network minimizes the loss across all simulations in the given set; therefore, the network as a whole still performs well across the parameter space even with the outlier. This simple `$Z$-value' test helps generate rMSE assessments for the networks that are reflective of their visual performance in our Figures.

Next, we examine whether our neural networks are under- or over-estimating the $\bar{\sigma}$ errors, as we find the rMSE is often slightly larger. As did \citet{Jeffrey2022} for their moment networks, we define for a given parameter $\theta_i$ and simulation $j$ $\chi_j (\theta_i)$ as:

\begin{equation}
    \chi_j (\theta_i) = \frac{\theta_{i,j\ \text{true}}\ - \mu_{i,j}}{\sigma_{i,j}}
    \label{eq:chi_def}
\end{equation}

The distribution of $\chi(\theta_i)$ can help qualify the accuracy of the neural network. Parameters that find accurate constraints--or, that follow the 1:1 slope on our figures, regardless of how closely--have $\chi(\theta_i)$ roughly consistent to a Gaussian centered at zero with a variance of 1. For parameters that cannot be constrained by the network, e.g.\ the SC-SAM parameters in halo-mass selections, the distributions of $\chi(\theta_i)$ are either flat distributions within $-2<\chi(\theta_i)<2$, or whose peak is $\chi(\theta_i)\approx \pm 1.5$. 

In Appendix \ref{app:chi_LFI}'s Figure \ref{fig:example_chiHistograms}, we plot the distributions of $\chi(\theta_i)$ for the two types of mass-selected clustering in Figures \ref{fig:DMonlyBest} and \ref{fig:SAMmstarBest} across all 5 parameters. We find that our good constraints, even if their $\bar{\sigma}$ trends slightly smaller than rMSE, have $\chi(\theta_i)$ consistent with neural networks that are accurate and not over- or under- predicted. In Tables where relevant, we list both the rMSE and $\bar{\sigma}$ calculated for various parameters; and we show only the rMSE errors in summary Figures \ref{fig:Summary_Allclust_Constraints} - \ref{fig:Summ_Clustering} for clarity.

\subsubsection{Architecture}\label{subsubsec:NNarch}

We used the \textsc{Optuna} package \citep{optuna_2019} to quickly train and validate 1D:1D fully-connected neural networks, all while identifying the best-performing hyperparameters (e.g.\ number of hidden layers, neurons per layer, learning rate, etc.). When using \textsc{Optuna} to create neural networks, we limit it to: 
\begin{itemize}
    \item take in a 1D array of normalized clustering values for a given galaxy selection, and predict the 5 parameters of {$\Omega_{\text{M}}$, $\sigma_8$, A$_{\text{SN1}}$, A$_{\text{SN2}}$, A$_{\text{AGN}}$} (or a single one, in the specific experiment of \textsection \ref{subsec:SCSAMfocus})
    \item have no more than 5 layers total, each with no more than 1,000 neurons 
    \item assess 250-1,000 trials (i.e.\ sample neural networks within the \textit{hyper}parameter space), each with 500 training epochs per trial
    % \item use single GPU node, often with 20GB of reserved memory
    \item use with Leaky ReLu activation in each layer
    \item and use the Adam optimizer with $\beta$ parameters equal to {0.5, 0.999}.
\end{itemize}

We often find 3-4 layers of 500-700 neurons do very well, testing more than 250 trials was often unnecessary, and most trials converge within 300 epochs and after approximately 2-3 GPU days. The MSE loss training and hyper-parameter selection often took less than 24 GPU hours, while the LFI loss averaged 3-3.5 GPU days per trained network.

% --------------------------------------------------------------------------------

\section{Constraining Cosmology with Clustering using CAMELS-SAM} \label{sec:CosmoConstraints}

\begin{figure*}
	\begin{center}
\includegraphics[width=\textwidth]{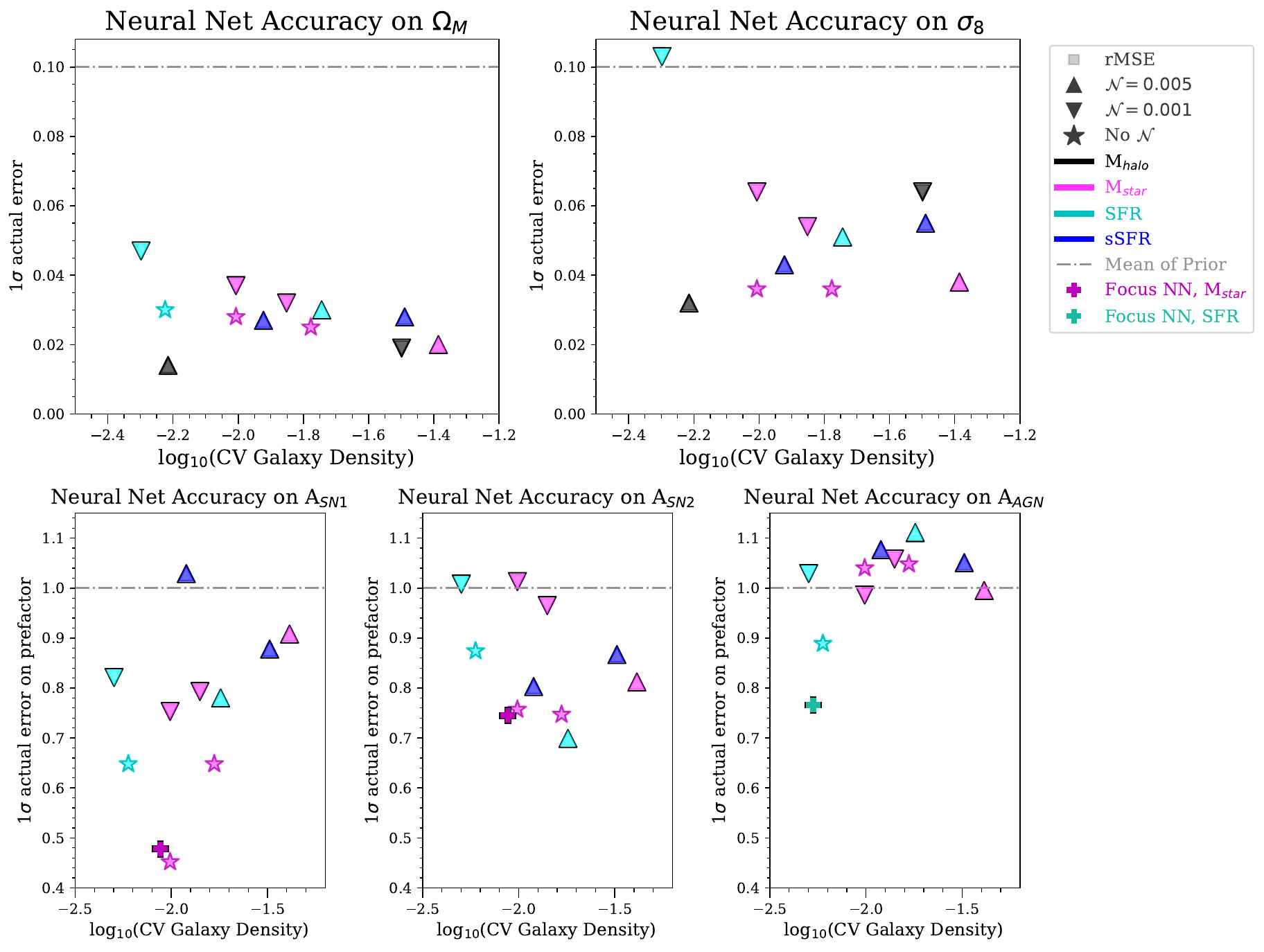}
	\caption{A summary of the $1\sigma$ rMSE errors we find on our five parameters when using `all' clustering (the 2ptCF, VPF, and CiC at $z=\{0.0, 0.1, 0.5, 1.0\}$) across our various galaxy selections. The color of the points indicates the property we selected by: halo mass (black), stellar mass (magenta), SFR (cyan), sSFR (blue). The shape of the marker indicates what density we down-sampled to after the selection: 0.001 $h^3$ cMpc$^{-3}$ (downward pointing triangle), 0.005 $h^3$ cMpc$^{-3}$ (upward pointing triangle), or no density down-sampling (star). The plus sign markers in deeper colors indicate the rMSE constraints found by having a neural network focus only on the single parameter (see Table \ref{table:oneatatime}); these symbols have been slightly shifted left for clarity. The grey dash-dotted lines indicate the mean of the parameter's prior space; errors around this value indicate poor to no constraints found by the neural networks. The density plotted on the $x$-axis is the $z=0$ value averaged across the 5 CV simulations after applying each selection, but before randomly down-sampling. See Tables \ref{table:GlxSelections_Constraints}, \ref{table:GlxSelections_noDS}, and \ref{table:oneatatime} for complete details.}
	\label{fig:Summary_Allclust_Constraints}
	\end{center}
\end{figure*}

In this section, we explore the constraints on the cosmological parameters $\Omega_{\text{M}}$ and $\sigma_8$ that we obtain using clustering statistics from the CAMELS-SAM simulation suite. In \textsection \ref{subsec:CosmoConstraints_byselection}, we explore how the constraints respond to various galaxy selections, especially between halo and stellar mass. In \textsection \ref{subsec:Downsampling}, we test how our cosmological constraints respond to including density down-sampling or not. Finally, in \textsection \ref{subsec:RedshiftChoices}, we examine the constraints at each of the individual redshifts we probe between $0<z<1$. For ease of reading, the bulk of the quantitative details, figures, and tabulated results can be found in Appendix \ref{app:CosmoConstraints}.

Throughout the rest of this work, our Figures will often contain the results for constraints on the SC-SAM astrophysical parameters; those will be reported and discussed independently in the following \textsection \ref{sec:SCSAM_constraints}. For the best comparison, we compare only the results of `all' clustering, which includes the VPF, 2ptCF, and CiC at $z={0, 0.1, 0.5, 1.0}$. The 2ptCF and VPF are measured between roughly $1<R<40$ cMpc, the CiC at select distance scales within that range (see Table \ref{table:CiCRadii}). See \textsection \ref{subsec:RedshiftChoices} for redshift-by-redshift comparisons, and \textsection \ref{sec:CompareStats} for statistic-by-statistic comparisons. When we report errors as percentages, we calculate them against the fiducial values (i.e.\ $\Omega_{\text{M}}=0.3$, $\sigma_8=0.8$).

\subsection{Cosmological Constraints: Comparing Galaxy Selections} \label{subsec:CosmoConstraints_byselection}

The SC-SAM properties we select upon are halo mass, stellar mass, % (given in units of M$_{\odot}$)
star formation rate,  % (given in units of M$_{\odot}$ yr$^{-1}$)
and specific star formation rate. % (derived from the former, with values yielding units of Gyr$^{-1}$)
In this section, we specifically focus on cosmological constraints. In \textsection \ref{sec:SCSAM_constraints}, we shift to discussing the constraints on the SC-SAM astrophysical parameters. 
Throughout both sections, we will briefly summarize the most relevant results of our experiments and primarily focus on discussing their meaning and significance. The detailed quantitative results are tabulated in Appendices \ref{app:CosmoConstraints}-\ref{app:Zs_Clust_Focus}.

Figure \ref{fig:Summary_Allclust_Constraints} summarizes (and Table \ref{table:GlxSelections_Constraints} in Appendix \ref{app:CosmoConstraints} details) the various selections applied (the SC-SAM galaxy property, the cut-off value applied, and the down-sampling density), and results for all 5 parameters (the rMSE error across the entire test set, and the mean $1\bar{\sigma}$ error estimation from the LFI loss method across the test set). Good fits are much smaller than the mean of the distribution (dash-dotted lines) and show a 1:1 relationship between the predicted and actual parameters values. Parameters with no constraints beyond the prior will appear centered around the mean of the prior distribution. Finally, we note that Figure \ref{fig:Summary_Allclust_Constraints} includes constraints that will be discussed in \textsection \ref{subsec:Downsampling}, where we do not randomly down-sample to a specific number density before measuring the galaxies' clustering.

\subsubsection{Dark Matter Halo vs. Stellar Mass Selections} \label{subsubsec:MassConstraints}

\begin{figure}
    \centering
     \includegraphics[width=0.49\textwidth]{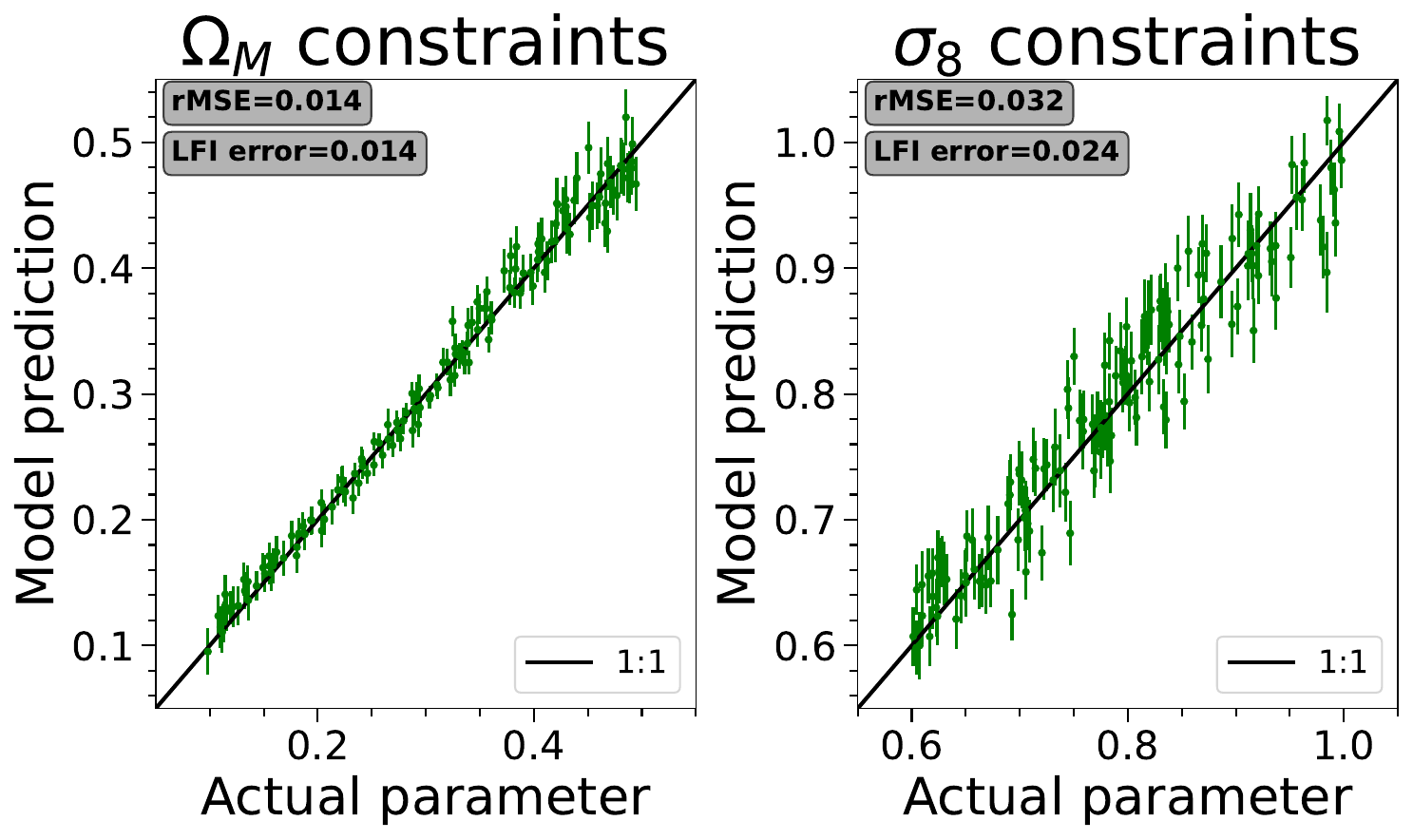}
     \caption{Our best cosmological constraints with CAMELS-SAM, using `all' N-body only \textit{\textbf{halo clustering}} and the LFI loss function. Here we selected halos with mass greater than $2\times 10^{11}$ M$_{\odot}$, down-sampled to a density of 0.005 $h^{3}$ cMpc$^{-3}$ at $z=\{0.0, 0.1, 0.5, 1.0\}$.}
    \label{fig:DMonlyBest}
\end{figure}

Figure \ref{fig:DMonlyBest} presents the best constraints our neural networks find on $\Omega_{\text{M}}$ and $\sigma_8$, using a moderate halo mass selection and when randomly down-sampling to a relatively high density. The best-performing neural network produces $\Omega_{\text{M}}$ predictions accurate to 4.7\% about the fiducial value $\Omega_{\text{M}}=0.3$, under both the rMSE and LFI loss regimes. For $\sigma_8$, this same neural networks predicts $\sigma_8$ to 3-4\% accuracy (LFI, rMSE respectively). These are the `best' possible constraints we refer to throughout the rest of this work.

% In Figure \ref{fig:DMonlyBest}, we show the best constraints we find on $\Omega_{\text{M}}$ and $\sigma_8$, from `all' clustering of halos with halo mass greater than $2\times10^{11}$ M$_{\odot}$ (or log$_{10}$ M$_{\text{halo}} > 11.3$ M$_{\odot}$), randomly down-sampled to a density of 0.005 $h^{3}$ cMpc$^{-3}$. 
% The best-performing neural network produces $\Omega_{\text{M}}$ predictions accurate to rMSE = 0.014, or approximately 5\% about the fiducial value $\Omega_{\text{M}}=0.3$. Our LFI method also measures an average 1$\sigma$ standard deviation error of $\bar{\sigma}=0.014$ on $\Omega_{\text{M}}$. For $\sigma_8$, the rMSE error of the mean values across the test set is 0.032 (4\% for $\sigma_8=0.8$), while $\bar{\sigma}$ is 0.024 (3\%). --> into appendix!
% We also probe a much higher halo mass selection of $1.2\times 10^{12}$ M$_{\odot}$ with a lower density down-sampling of 0.001 $h^{3}$ cMpc$^{-3}$, and find comparable though slightly worse constraints (likely due to Poisson noise introduced from the lower density). We take the tightest constraints from our dark matter selections as the `best' throughout this work.

\begin{figure}
    \centering
        \includegraphics[width=0.49\textwidth]{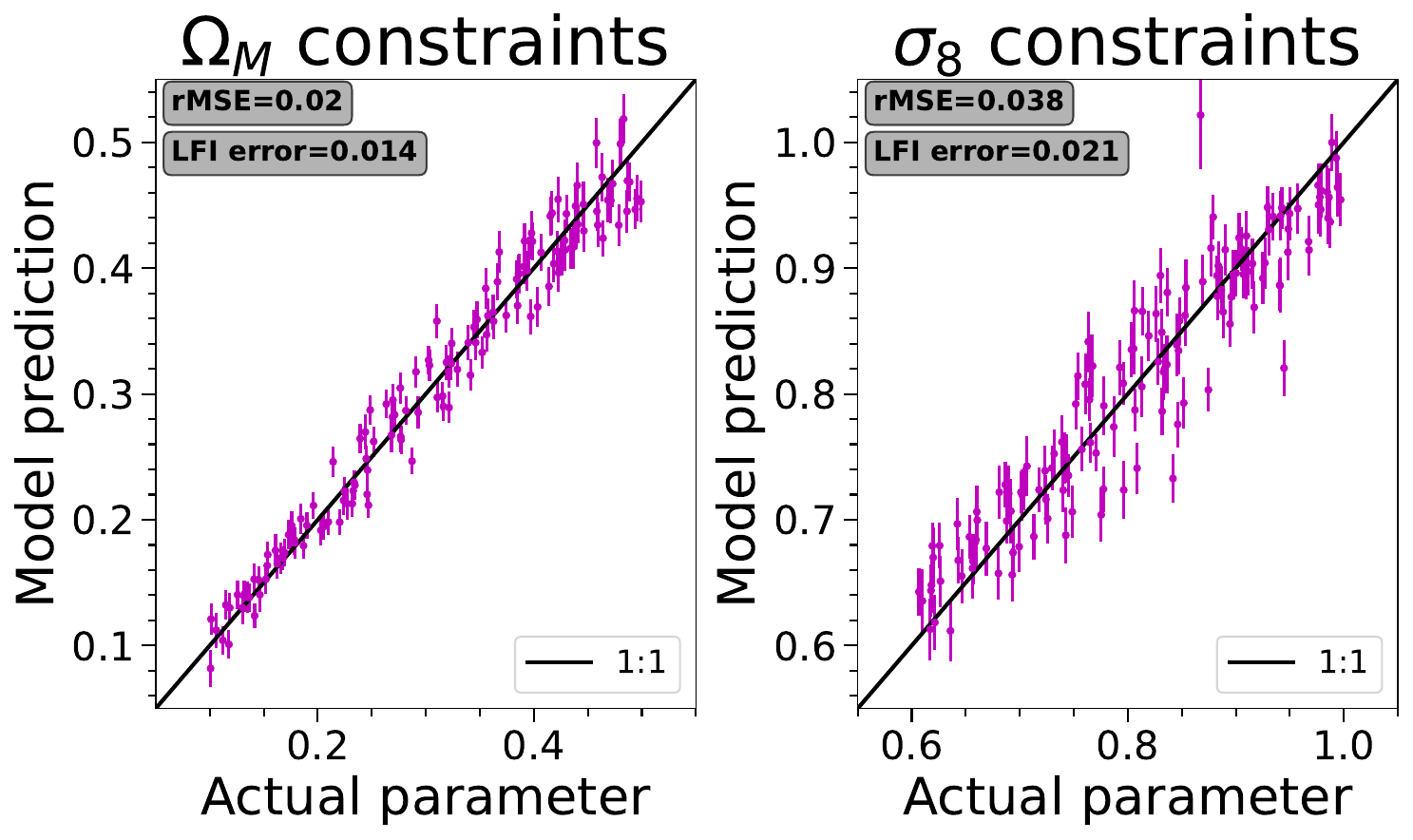}
         \caption{Our best cosmological constraints using a \textit{\textbf{stellar mass selected sample}} and the LFI loss function. Here, we selected SAM galaxies with stellar mass greater than $10^{9}$ M$_{\odot}$, down-sampled to a density of 0.005 $h^{3}$ cMpc$^{-3}$ at $z=\{0.0, 0.1, 0.5, 1.0\}$. }
    \label{fig:SAMmstarBest}
\end{figure}

Of the SC-SAM galaxy property selections we test, stellar mass can be expected to give the tightest cosmological constraints. The stellar and halo masses of galaxies are the most correlated of the selections we probe, meaning the tight constraints from halo mass clustering will likely also be seen somewhat in stellar mass. As with halo mass, a low-threshold mass selection and higher down-sampling density yields the tightest constraints on cosmology, and nearly exactly what the halo mass `best-case' selection found: errors of 4.7-6.7\% for $\Omega_{\text{M}}$ and 3-5\% for $\sigma_8$ (LFI, rMSE respectively). Figure \ref{fig:SAMmstarBest} presents these best constraints our neural networks find on $\Omega_{\text{M}}$ and $\sigma_8$, using stellar mass selections.

% In Figure \ref{fig:SAMmstarBest}, we show the constraints on $\Omega_{\text{M}}$ and $\sigma_8$ from `all' clustering at $z=\{0.0, 0.1, 0.5, 1.0\}$ of galaxies with stellar mass greater than $1\times10^9$ M$_{\odot}$, randomly down-sampled to 0.005 $h^{3}$ cMpc$^{-3}$. Other stellar mass selections we also tested include stellar mass greater than $1\times10^{10}$ M$_{\odot}$ to a density of 0.001 $h^{3}$ cMpc$^{-3}$; and stellar mass greater than $2\times10^{10}$ M$_{\odot}$ to a density of 0.001 $h^{3}$ cMpc$^{-3}$. We found the lower-threshold and higher down-sampling density selection yielded the tightest constraints on cosmology.

% The best-performing neural network, using the clustering of galaxies with stellar mass greater than $1\times10^9$ M$_{\odot}$, produces $\Omega_{\text{M}}$ predictions accurate to rMSE = 0.02, or approximately 7\% about the fiducial value $\Omega_{\text{M}}=0.3$. The LFI loss measures an average 1$\sigma$ standard deviation error of $\bar{\sigma}=0.014$ on $\Omega_{\text{M}}$, the same value as it found with dark-matter only clustering. For $\sigma_8$, the rMSE error of the mean values across the test set is 0.038 (approximately 5\% for $\sigma_8=0.8$), while $\bar{\sigma}=0.021$ (3\%). 

\subsubsection{SFR and sSFR Selections}
\label{subsubsec:ConstrCosmo_s_SFR}

Next, we explore the constraints obtained when galaxies are selected via their star formation rate (SFR) and specific star formation rate (sSFR, or SFR divided by stellar mass). In this Section, we focus on cosmological constraints when using these selections; see \textsection \ref{subsec:SCSAM_selections} for the constraints on the SC-SAM parameters controlling baryonic processes of stellar and AGN feedback. 

Much like with stellar mass, our experiments here test both the selection threshold, and the number density to which we randomly down-sample. Here,
% with `all' clustering statistics at $z=\{0.0, 0.1, 0.5, 1.0\}$ and random down-sampling to a fixed number density, 
we probe SC-SAM galaxies with SFR$> 0.2$ or 1.25 M$_{\odot}$ yr$^{-1}$ at a high and low density, respectively.
% $> 0.2$ M$_{\odot}$ yr$^{-1}$ randomly sampled to $\mathcal{N}=0.005\ h^{3}$ cMpc$^{-3}$, and also SFR $> 1.25$ M$_{\odot}$ yr$^{-1}$ randomly sampled to $\mathcal{N}=0.001\ h^{3}$ cMpc$^{-3}$. 
For sSFR, we probe SC-SAM galaxies with sSFR $> 0.1$ or 0.2 Gyr$^{-1}$ with a high density down-sampling.
% Gyr$^{-1}$ and sSFR $> 0.2$ Gyr$^{-1}$, both randomly sampled to $\mathcal{N}=0.005\ h^{3}$ cMpc$^{-3}$. 
For SFR and sSFR, errors on the cosmological parameters span 9-15\% for $\Omega_{\text{M}}$, and 5-8\% for $\sigma_{8}$ (though we note the high threshold, low density SFR selection finds essentially no constraints). The tightest constraints on cosmology consistently come from stellar mass.

\subsubsection{Discussion: Selections for Cosmological Constraints}
\label{subsubsec:ConstrCosmo_Discussion}

The good accuracy and precision found on $\Omega_{\text{M}}$ (5-7\%) and $\sigma_8$ (3-5\%) with SC-SAM stellar mass-selected clustering are quite encouraging. First, we reach accuracy comparable to the dark matter only clustering (5\% on $\Omega_{\text{M}}$, 3-4\% on $\sigma_8$), even with the introduction of an astrophysical model. Secondly, it is heartening to see that though this neural network was trained on simulations with a broad range of feedback from the SAM baryonic prescriptions, and was asked to constrain all 5 parameters at once, it is still able to obtain robust cosmological constraints and focus on their strong influence on clustering. 

Of the basic galaxy properties we select by, stellar mass tends to give slightly better constraints across the board. This is expected, as dark matter halo mass clustering would be expected to give the best constraints, and stellar mass tracks halo mass the most closely of the selections we probe. SFR provides good constraints on $\Omega_{\text{M}}$, but interestingly much worse constraints on $\sigma_8$, even obtaining \textit{no} constraint at the strongest SFR selection at the lower density. sSFR shows good constraints on both cosmological parameters, likely benefiting from the simple combination of stellar mass and star formation. 

We also find that more extreme selections on these parameters have mixed effects, occasionally occluded by the necessity of a lower down-sampling density and the introduction of more Poisson noise. For example, doubling the stellar mass threshold 
at the lower down-sampling density
% $\mathcal{N}=0.001\ h^{3}$ cMpc$^{-3}$ from $1\times10^{10}$ M$_{\odot}$ to $2\times10^{10}$ M$_{\odot}$ 
has inconclusive effects on $\Omega_{\text{M}}$ (increasing rMSE but decreasing $\bar{\sigma}$), and worsens $\sigma_8$ by approximately 0.1\%. However, doubling the sSFR threshold 
% at $\mathcal{N}=0.005\ h^{3}$ cMpc$^{-3}$ from 0.1 Gyr$^{-1}$ to 0.2 Gyr$^{-1}$ 
improves cosmological constraints by 0.1-1 percentage points ($\Omega_{\text{M}}$, $\sigma_8$ respectively). 

Finally, there is a possible trend of obtaining better constraints for simulations with parameter values that yield higher densities across the suite, notably $\Omega_{\text{M}}$ and A$_{\text{SN2}}$. Why would $\Omega_{\text{M}}$ be better constrained with the clustering of galaxies under a selection that yields higher densities, especially for stellar mass? A lower stellar mass threshold combined with our density down-sampling means that the objects whose clustering we measure will tend to be lower mass, and therefore may better probe the general range of structure formation within the simulations. Additionally, poorer constraints at higher stellar mass may come from a combination of more sensitivity to Poisson noise (at the lower density), as well as a possible degeneracy with $\sigma_8$ that may mask the effect of $\Omega_{\text{M}}$.

\subsubsection{Information in Galaxy vs. Halo Clustering}
\label{subsubsec:MstarintoMhalo_Discussion}

\begin{figure*}
	\begin{center}
\includegraphics[width=.9\textwidth]{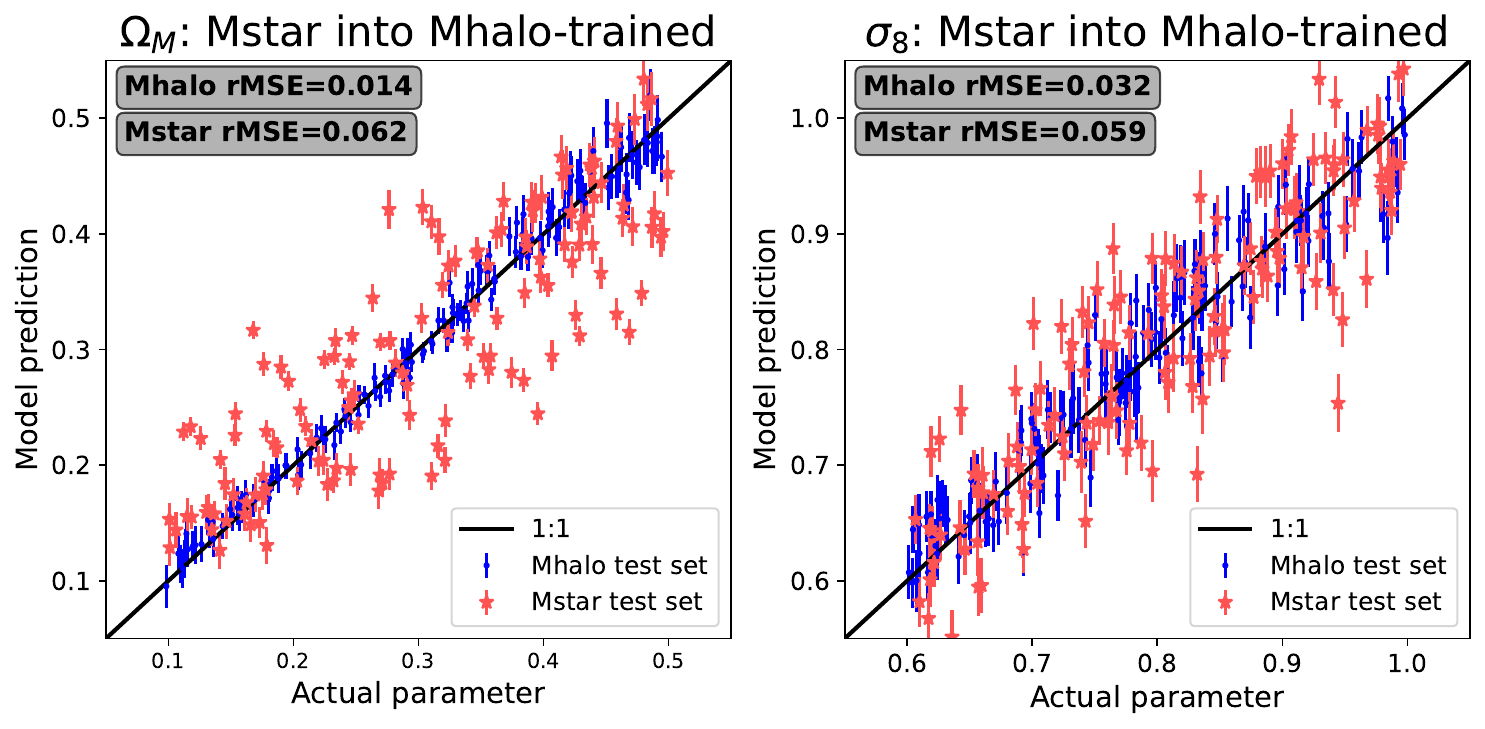}
	\caption{Comparing how well a neural network trained on the clustering of a dark matter halo mass selected sample can predict cosmological parameters for a stellar mass-selected sample.
	This network was trained with `all' clustering at $z=\{0.0, 0.1, 0.5, 1.0\}$ for dark matter halos with \textit{dark matter halo mass} greater than $2\times10^{11}$ M$_{\odot}$, down-sampled to a density of 0.005 $h^{3}$ cMpc$^{-3}$.
	Blue circles show the results of giving the network the test set for the same type of clustering with which it was trained and validated.
	Red stars show the results from instead evaluating the best-performing model on the test set of simulations using the clustering of \textit{SAM galaxies of stellar mass} greater than $1\times10^9$ M$_{\odot}$, down-sampled to the same density of 0.005 $h^{3}$ cMpc$^{-3}$.}
	\label{fig:trainDMgiveMstar}
	\end{center}
\end{figure*}

Some readers may wonder: exactly how much information about cosmology is contained within the clustering of SAM galaxies? We can probe the relationship between halo and galaxy clustering in our sample with a simple experiment: if we \textit{trained} a neural network to measure cosmology with only the clustering of dark matter halos, how well can it predict cosmology when \textit{tested} using galaxy clustering instead?

Figure \ref{fig:trainDMgiveMstar} shows the results of this experiment. We first train a neural network with `all' clustering for SAM galaxies with host \textit{dark matter halo mass} greater than $2\times10^{11}$ M$_{\odot}$, down-sampled to a density of 0.005 $h^{3}$ cMpc$^{-3}$. As expected, the network finds tight constraints if it is tested with the same type of dark matter only clustering. Notably, the LFI loss gives very precise and accurate predictions, 
% very small 1$\sigma$ error measurements that are 
consistent with the predictions on the means, for dark matter only clustering: 4.7\% for $\Omega_{\text{M}}$ and 4\% for $\sigma_{8}$.

We then have the best-performing model try to measure the cosmology when given only the clustering of SAM galaxies of \textit{stellar mass} greater than $1\times10^9$ M$_{\odot}$, down-sampled to the same density of 0.005 $h^{3}$ cMpc$^{-3}$. We note that this stellar mass selection was chosen because it performed the best of those we tried, and not through any connection between the halo mass above and the stellar mass of galaxies within them (making this a more challenging problem for the neural network). We carry out exactly the same calculations, only changing the objects in the test set. The constraints are less precise, but still fairly accurate: $\Omega_{\text{M}}$ rMSE = 0.062 (21\%) and $\sigma_8$ rMSE = 0.059 (7.4\%). The LFI errors remain very small, consistent with what the network achieved on its training set. 

This exercise of training on halo mass clustering and then testing on stellar mass clustering is helpful in two ways. First, it humbly reminds us that the precision of machine learning results is only as good as the training samples. The LFI error predictions underestimate the true error on the stellar mass clustering constraints, as the network simply assumes the data it is given works the same way that the dark matter only clustering did. Second, this exercise emphasizes the strength of using clustering: having networks trained only on dark matter halos (which we theoretically understand well and can simulate cheaply) \textit{can} yield accurate (if imprecise) constraints when given galaxy clustering. The neural network is retaining some information, even when being tested on the `wrong' thing, precisely because galaxies generally follow the clustering of dark matter.

However, this is not to understate the significance of how much our cosmological constraints improve when training directly with SAM galaxy clustering (as in Figure \ref{fig:SAMmstarBest}). The clustering of dark matter halos dominates the signal of galaxy clustering, but it does not completely describe it. Without the additional information of the SC-SAM galaxy formation model, and without allowing the neural network to marginalize over the effects and uncertainties they introduce, there would be no improving the dark-matter-only constraints and no path forward in this field. Galaxies and halos have complicated relationships that are not well understood, yet even with the many variations of the SC-SAM models that we include and the broad parameter space that our suite covers, there is still more information to be learned than exists in a simplistic galaxy-halo model. Information about cosmology is lost when the clustering of galaxies is assumed to be related to dark matter in an overly simplistic way, and much of it may be gained back when including a robust galaxy formation model.

This exercise therefore reinforces the importance of one of CAMELS' central tenets as a project: to teach neural networks to \textit{marginalize} over the uncertain galaxy formation physics. Neural networks are able to learn the complicated ways that galaxy physics affect clustering statistics, and yield constraints within 3-10\% on cosmology from galaxy clustering. Finally, this exercise also strongly motivates implementing additional methods of linking galaxy properties to dark matter (e.g.\ various HODs, more SAMs, sub-halo abundance matching), to more fully explore the marginalizing power of the neural networks.

\subsection{Cosmological Constraints: Effect of Number Density Down-Sampling}\label{subsec:Downsampling}

Next, we examine how the choice to down-sample to a given density affects our cosmological constraints. CiC and VPF are sensitive to number density and encode its effect very strongly, so we chose to correct for its influence to pull out the effect our parameters have on the large-scale structure more clearly. Additionally, randomly down-sampling after a selection means the same underlying clustering is maintained, and reduces the computational load of clustering measurements for our 1000+ simulations. However, how do constraints change if we allow number density to vary, and therefore allow the neural networks to use that information for predictions?

To test constraints without our density down-sampling, we applied high threshold cuts for select SC-SAM properties and measured the clustering on all resulting galaxies. To reduce the computational load while still maintaining robust number statistics for the sparsest volumes, we chose high value cuts for our simulations that yielded at least several hundred to a few thousand galaxies in more than 95\% of the simulations. These selections yield galaxy densities mostly between $10^{-4}-10^{-1}\ h^{3}$ cMpc$^{-3}$, which corresponds to as much as fifty to one hundred thousand galaxies in a volume across the entire breadth of our parameter space. For more grounded context, the CV simulations with the fiducial parameters yield ten to twenty thousand galaxies under these selections. These selections therefore likely have minimal Poisson error in the bulk of their clustering measurements. 

The cosmological constraints found for these three selections are detailed in Appendix \ref{app:CosmoConstraints} Table \ref{table:GlxSelections_noDS} and are included in the summary Figure \ref{fig:Summary_Allclust_Constraints} for direct comparison with earlier selections. We discuss the effect of not down-sampling on the SC-SAM astrophysical parameters in \textsection \ref{subsec:SCSAM_densityredshift}. Appendix \ref{app:CosmoConstraints} Figure \ref{fig:noDS_Sampled} highlights some of the best-performing neural networks' results.

Without the normalization to a fixed number density across all simulations and models, SFR- and stellar mass- selected clustering yields slightly improved cosmological constraints. For the stellar mass selections, $\Omega_{\text{M}}$ constraints stay the same or improve slightly, while $\sigma_8$ constraints notably improve by a few percent (dropping from approximately 6.3\% to 4-5.5\%). For SFR selection, $\Omega_{\text{M}}$ constraints are about as good as when down-sampling is applied (under the LFI loss). However, removing the down-sampling greatly improves the $\sigma_8$ constraints from the strong SFR selection, finding 8.1\% errors. This is 
% The clustering of all galaxies with SFR$> 1$ M$_{\odot}$ yr$^{-1}$ yields $\bar{\sigma}=0.065$ (8.1\% error), 
much improved over the essentially unconstrained $\bar{\sigma}=0.103$ when taking a similarly strong SFR cut and then randomly down-sampling. Though we cannot confirm just how much of the improvement comes from reduced Poisson noise, our other results suggest that some of this improvement might be attributed to the additional information contained in the predicted number density values.

The results of this experiment, at least with regard to cosmology, can be interpreted in a mixed way: constraints are often improved by including number density, especially for $\sigma_8$ when SFR selected samples are used, though not a significant amount of information about $\Omega_{\text{M}}$ is lost when randomly down-sampling to a fixed number density. The computational effort required to measure so many galaxies' clustering is not negligible, meaning that perhaps for an initial assessment of cosmology, randomly down-sampling would be a practical choice. We revisit this assessment when analyzing the constraints on the SC-SAM astrophysical parameters in \ref{subsec:SCSAM_densityredshift}.

\subsection{Cosmological Constraints: Effect of Redshift Choice} \label{subsec:RedshiftChoices}

Next, we briefly examine the effect of combining multiple redshifts when training our neural networks\footnote{We note that e.g.\ \citet{Nicola2022} and \citet{Uhlemann2020} have also included a few redshifts at a time in their analyses.}. With real galaxy observations, it would be difficult to make identical galaxy selections at multiple redshifts, especially those as observationally different as $z=0$ vs. $z=1$. Therefore, we create single-redshift samples at $z=\{0.0, 0.1, 0.5, 1.0\}$ to examine the constraints at single redshifts for a few types of selections.

At each redshift, we randomly down-sample to our selected number density, meaning each redshift measures the clustering of a different randomly selected sample of objects. We also include more radii in the CiC distribution, as using one redshift at a time allows for more spatial scales to be included (see Table \ref{table:CiCRadii} for full details; in essence, we sample more distance scales for more $n$ count-in-cells measurements). We use the same VPF and 2ptCF as in previous sections.

\begin{figure*}[t]
	\begin{center}
\includegraphics[width=\textwidth]{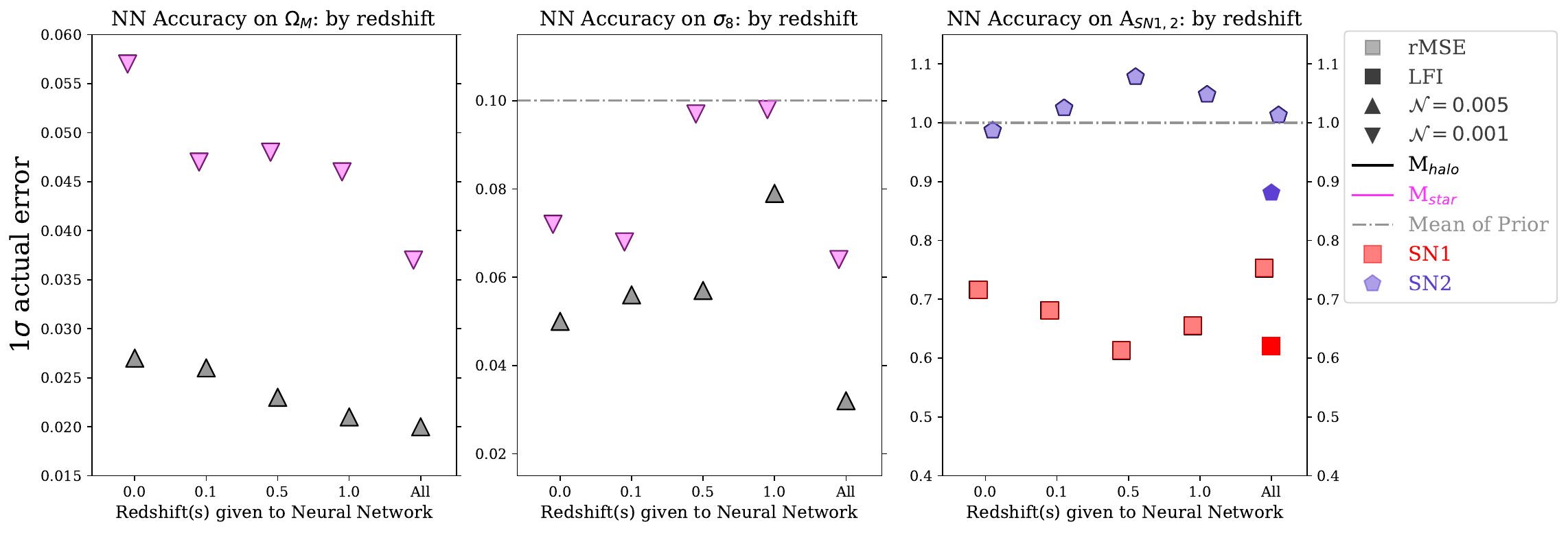}
	\caption{Similar to Figure \ref{fig:Summary_Allclust_Constraints}, but instead summarizing the constraints of \textsection \ref{subsec:RedshiftChoices}, Table \ref{table:SingleRedshift_rMSE}, and Appendix \ref{app:Zs_Clust_Focus}, illustrating the effects of using clustering constraints at a single redshift vs. four redshifts combined (All). }
	\label{fig:Summ_Redshift}
	\end{center}
\end{figure*}

Summary Figure \ref{fig:Summ_Redshift} presents the calculated rMSE errors across the test set for all selections probed; the detailed results are tabulated in Appendix \ref{app:Zs_Clust_Focus} Table \ref{table:SingleRedshift_rMSE}.
We discuss the SC-SAM parameter constraints when data from a single redshift at a time is used in \textsection \ref{subsubsec:AstroConstr_Redshift}.  
Additionally, Appendix \ref{app:Zs_Clust_Focus} Figures \ref{fig:DMonlyBestOm_redshift} and \ref{fig:DMonlyBests8_redshift} focus on results from experiments using halo mass and stellar mass selections at a high and low density, respectively.

As expected, combining the information from multiple redshifts improves the constraints. Networks trained on clustering from a single redshift find slightly worse constraints, with fractional errors increasing by at least 5\% on $\Omega_{\text{M}}$ and 1-5\% on $\sigma_8$. We note that one or two extreme outlier test simulations are more common in the neural networks trained on single redshifts (see the visible red star outliers in several of the SC-SAM parameters in Figures \ref{fig:z0p1_mstar} and \ref{fig:z0p5_mstar}). Constraints on  $\Omega_{\text{M}}$ and $\sigma_8$ worsen by at least a few percent within the two types of mass selection we probe for individual redshifts. The individual redshifts have comparable rMSE errors across all parameters. There may be a degradation in $\Omega_{\text{M}}$ constraints with decreasing redshift, but the pattern is not conclusive.

We remind readers that, for the single redshift experiment, the VPF and 2ptCF were kept identical, but the CiC distributions were expanded to use the longer arrays the neural network can easily handle (\textsection \ref{subsubsec:NN_specifics} and Table \ref{table:CiCRadii}). Interestingly, we note here that including clustering at more redshifts improves the constraints much more than including more of the CiC information at a single redshift. As Table \ref{table:SingleRedshift_rMSE} and summary Figure \ref{fig:Summ_Redshift} show, the constraints from `all' clustering at four redshifts always improve for $\sigma_8$ and nearly always improve for $\Omega_{\text{M}}$, often dropping several percentage points. However, this may be due to the neural network independently learning the growth factor (a factor that determines the growth of density perturbations as a function of cosmology, especially $\Omega_{\text{M}}$). We do not find strong evidence for a particular redshift outperforming another. 

Though we have not run experiments for each iteration and combination of clustering selections, our initial results lend credence to prioritizing getting samples of the same type of galaxies at different redshifts rather than measuring more detailed clustering statistics for a single sample, or measuring clustering across a broader range of scales. For example, a galaxy sample for which this is likely feasible in the near future is H$\alpha$ and [OIII] emission line galaxies confirmed with photometric redshifts, either from large scale structure surveys like DES \citep{DES}, or narrowband surveys like HiZELS \citep{Khostovan2018} and LAGER \citep{Khostovan2020}. 

Finally, we note that each CAMELS-SAM simulation has halo and galaxy information at 100 snapshots between $20 < z < 0$, and that we have only probed $z < 1$ in this initial work. We also re-emphasize the caveat that we did not use exactly the same CiC for the single redshift training as we did with all four redshifts (see Table \ref{table:CiCRadii}): we include more distributions at a slightly larger range, in an attempt to leverage as much information as we could give the simple 1D:1D neural network at a single redshift. 
% \textcolor{cyan}{However, the strong correlation between distance bins for CiC, combined with the continued good performance of our NNs, might indicate that fewer well-spaced distance scales for the CiC yields similar amounts of information as using most of the distribution.}

%------------------------------------------------------------------------------------

\section{Constraining SC-SAM parameters for stellar and AGN feedback}\label{sec:SCSAM_constraints}

Next, we explore in detail the constraints our neural networks obtain on the SC-SAM parameters controlling baryonic processes related to stellar and AGN feedback. All selections we tested are included in Figures \ref{fig:Summary_Allclust_Constraints}, and \ref{fig:Summ_Redshift}, and are detailed in Appendix \ref{app:CosmoConstraints} and \ref{app:Zs_Clust_Focus} Tables \ref{table:GlxSelections_Constraints}, \ref{table:GlxSelections_noDS}, and \ref{table:SingleRedshift_rMSE}.

\subsection{Astrophysical Constraints: Comparing Galaxy Selections}\label{subsec:SCSAM_selections}

First, we compare how the selection on galaxy property affects how well our neural networks are able to constrain astrophysical parameters from galaxy clustering.
See upcoming \textsection \ref{subsec:SCSAM_densityredshift} for a discussion about how constraints on the SC-SAM parameters change with no down-selection and at individual redshifts, and \textsection \ref{subsec:CompStats} for how individual clustering statistics perform for these parameters.

As introduced in \textsection \ref{subsec:CosmoConstraints_byselection}, we select SC-SAM galaxies based on stellar mass, star formation rate, and specific star formation rate. Here, we examine the constraints that our experiments find on the SC-SAM astrophysical parameters A$_{\text{SN1}}$, A$_{\text{SN2}}$, and A$_{\text{AGN}}$. We remind readers that all neural networks have been asked to constrain all five parameters at the same time (with the exception of the `focused' experiment in the upcoming \textsection \ref{subsec:SCSAMfocus}), and that our LFI loss function (described in \textsection \ref{subsubsec:NNlosses}) has been used specifically for its strength in pulling out the influence of weaker parameters. The results for constraints on the SC-SAM astrophysical parameters are summarized in Figure \ref{fig:Summary_Allclust_Constraints}; detailed in Appendix \ref{app:CosmoConstraints} Table \ref{table:GlxSelections_Constraints}; and select neural network constraint examples can be seen in Appendix \ref{app:CosmoConstraints} Figure \ref{fig:GlxsSelections_all5}. We note that poor fits on the SC-SAM parameters correspond to rMSE and $\bar{\sigma}$ errors near 1.

Of the core selections with `all' clustering with density down-sampling, we find all selections find good constraints in different circumstances. Stellar mass and sSFR do well at constraining the parameter A$_{\text{SN2}}$ ($0.5<\bar{\sigma}<0.8$), and find moderate ($0.6<\bar{\sigma}<0.8$, stellar mass) or very poor ($\bar{\sigma}=1$, sSFR) constraints on A$_{\text{SN1}}$. Both are outperformed by SFR selections, especially at high density down-samplings. All selections in this category find poor to no constraints on A$_{\text{AGN}}$. As we will explore later, removing the density down-sampling vastly improves constraints on the SC-SAM parameters.

The high-density SFR selection does well at constraining both SC-SAM A$_{\text{SN}}$ parameters--not surprisingly, since those parameters control the normalization and slope of the mass outflow rate driven by stellar feedback, regulating star formation in galaxies. Specific SFR, as the simple combination of stellar mass and star formation, unfortunately does not show ``the best of both worlds" and does not improve on the results from stellar mass or SFR selections separately (and, as explored in \textsection \ref{subsubsec:ConstrCosmo_s_SFR}, does not stand out for cosmological constraints). 

The poor to non-existent constraints on A$_{\text{AGN}}$ may perhaps be due to the fact that it only affects the properties of galaxies that are much more massive than our selection limit (see Figure \ref{fig:SAMplots_AAGN} in Appendix \ref{app:ExtraSAMverif}). In \textsection \ref{subsec:SCSAMfocus}, we attempt to improve these astrophysical constraints by having the neural networks learn one parameter at a time.

Finally, we note some trends on the constraints obtained with different galaxy selections. As seen in Figure \ref{fig:Summary_Allclust_Constraints}, the parameter A$_{\text{SN2}}$ (like $\Omega_{\text{M}}$) is better constrained with lower stellar mass and lower SFR selections (higher number density), and the dependence on mass or SFR is fairly strong. A$_{\text{SN1}}$ appears to show the inverse behavior, with better constraints from higher mass selections that yield fewer galaxies (but less of a clear trend with SFR selection). Marginally better constraints on A$_{\text{AGN}}$ may also be obtained with higher mass (lower density) selections, likely because this parameter has the greatest effect on the highest mass, and therefore most star-forming, galaxies (e.g.\ Appendix \ref{app:ExtraSAMverif} Figure \ref{fig:SAMplots_AAGN}).

\subsection{Astrophysical Constraints: Effect of Number Density and Redshift} \label{subsec:SCSAM_densityredshift}

Next, we report the effects of using clustering at one redshift, and not down-sampling to a single number density, on the constraints on the astrophysical parameters A$_{\text{SN1}}$, A$_{\text{SN2}}$, and A$_{\text{AGN}}$.

\subsubsection{One Redshift vs. Multiple} \label{subsubsec:AstroConstr_Redshift}

Our work has found there is no strong trend on the quality of the constraint on the astrophysics parameters with redshift (Appendix \ref{app:Zs_Clust_Focus} Table \ref{table:SingleRedshift_rMSE} and Figure \ref{fig:Mstar_redshift}), much like what was found with the cosmological parameters in \textsection \ref{subsec:RedshiftChoices} . A$_{\text{SN2}}$ constraints are non-informative (i.e. they are close to the mean of the prior) at the individual redshifts, and only somewhat constrained with all four redshifts combined. The constraints on A$_{\text{AGN}}$ remain very poor at individual redshifts. A$_{\text{SN1}}$ is still decently constrained across redshifts. There is an interesting phenomenon of `all' four redshifts combined finding slightly worse rMSE constraints on A$_{\text{SN1}}$ than the individual redshifts, though we note the LFI loss prediction finds $\bar{\sigma}=0.621$, slightly better than three of the four constraints. For the SC-SAM supernova parameters, there therefore may be no loss of constraining power when focusing on a single redshift. %Further analysis with other selections can confirm this is not due to changes in Poisson noise due to the lower number density down-sampling. 

\subsubsection{Effect of Density Down-Sampling} \label{subsubsec:AstroConstr_DensityResults}

Throughout our process of narrowing down what galaxy selections we would undertake, we found that varying the SC-SAM parameters for stellar feedback created large variations in the number of galaxies that pass a given stellar mass or star formation rate cut. This is not surprising, especially with e.g.\ the strong influence the A$_{\text{SN}}$ parameters have on the stellar mass functions, as seen in Appendix \ref{app:ExtraSAMverif}. However, how much do they affect the clustering of galaxies? 

We revisit summary Figure \ref{fig:Summary_Allclust_Constraints} and Appendix \ref{app:CosmoConstraints} Figure \ref{fig:noDS_Sampled} and Table \ref{table:GlxSelections_noDS} to assess the constraints found on the SC-SAM parameters A$_{\text{SN1}}$, A$_{\text{SN2}}$, and A$_{\text{AGN}}$. Without down-selecting to fixed number density, we find significantly improved constraints on the SC-SAM parameters for both SFR and stellar mass selected samples: often decreasing $\bar{\sigma}$ by a tenth or more. Both types of selections get imprecise but not inaccurate constraints on A$_{\text{SN2}}$. Stellar mass selected samples produce better constraints on A$_{\text{SN1}}$, while stellar mass and SFR selected samples show the first signs of finding any constraint on the elusive A$_{\text{AGN}}$ parameter. 

Appendix \ref{app:ExtraSAMverif} uses the 1P catalogs to explore how the SC-SAM parameters affect key galaxy relationships at either very low or high values. We find that the SN parameters show strong influence on relationships involving stellar mass, especially the stellar mass function (SMF) and stellar mass-halo mass function. The AGN parameter has the strongest influence at high halo masses. Therefore, it is not surprising that including the additional information on galaxy number density would improve the SC-SAM parameter constraints. 

The VPF and CiC are quite sensitive to number density (the VPF especially, essentially constraining it alone). Including the number density as additional information for a neural network to leverage gives it a lot of knowledge about the SMF, and therefore allows it to more easily learn the effect of the SC-SAM parameters. This is perhaps why A$_{\text{AGN}}$ is finally being constrained at all: the VPF and CiC sense whatever small influence it may have on galaxies' stellar mass and sSFR, and the neural network has more information to learn its relationships. 

\subsection{Astrophysical Constraints: Focused Neural Networks}\label{subsec:SCSAMfocus}

We have explored how various galaxy selections and choices around clustering statistics affect constraints on cosmological parameters and SC-SAM astrophysics parameters when all are constrained at once. However, how much improvement can we find on our constraints for the SC-SAM parameters around stellar and AGN feedback by focusing on them one at a time? 

These experiments function just like those already described, but the neural networks are asked to predict only one parameter at a time given some array of galaxy clustering. We first confirm that this method works by isolating the cosmological parameters; our constraints do not improve when a neural network focuses on $\Omega_{\text{M}}$ or $\sigma_8$ alone given dark-matter only or stellar mass-selected galaxy clustering. This likely indicates that the full 5-parameter predictions are accurately pulling out their influence. See Appendix \ref{app:Zs_Clust_Focus} Table \ref{table:oneatatime} and Figure \ref{fig:SCSAMparamsalone} for detailed comparisons.

Next, from our various experiments, we determine which galaxy and clustering selections may give the most useful constraints on the SC-SAM parameters. For the A$_{\text{SN}}$ parameters, we select a strong stellar mass selection with no density down-sampling. For A$_{\text{AGN}}$, we select a strong star formation rate with no down-sampling. For both, we use `all' clustering and all four redshifts.

Our best constraints for the SC-SAM parameters when training networks to focus on them alone are detailed in Figure \ref{fig:SCSAMparamsalone} and Appendix \ref{app:Zs_Clust_Focus} Table \ref{table:oneatatime}. They are also included in summary Figure \ref{fig:Summary_Allclust_Constraints} for easier comparison against the neural networks that constrained all parameters at once (slightly darker plus signs). The rMSE constraints A$_{\text{SN1}}$ and A$_{\text{SN2}}$ are only slightly improved by individualized training. Similarly, the rMSE constraints on A$_{\text{AGN}}$ are slightly improved with the focused approach, which we partially attribute to this parameter being the most subtle of the SC-SAM parameters with weak influence (see Appendix \ref{app:ExtraSAMverif}'s Figure \ref{fig:SAMplots_AAGN}). The LFI-predicted errors $\bar{\sigma}$ when constraining all five parameters at once give very similar or slightly better constraints than the focused parameter rMSEs.

Though the focused neural networks did not yield the desired outcome---stronger constraints on the SC-SAM astrophysical parameters---this exercise communicates the strength of our neural network implementation. The `Focused NN' results we summarize in Figure \ref{fig:Summary_Allclust_Constraints} are all either nearly identical to or slightly worse than the $\bar{\sigma}$ that our LFI loss predicts when predicting all five parameters at the same time. In theory, focusing the entire breadth of a neural network to predict only one parameter will lead to the tightest constraints the architecture is able to find. This result therefore indicates that the LFI loss is behaving exactly as advertised---it can, indeed, learn noisy and less sensitive parameters even in the presence of strongly influential parameters. 

As we describe in \textsection \ref{subsubsec:NNlosses}, the LFI loss function removes from the scatter of a parameter the dependence of the overall space, which in practice in CAMELS and CAMELS-SAM means it is able to learn parameters with relatively subtle effects, such as A$_{\text{AGN}}$, while also learning $\Omega_{\text{M}}$ and $\sigma_8$. Finally, we also ran focused neural networks for the cosmological parameters, and found their constraints showed very little difference from those obtained when constraining all five parameters, additionally confirming the LFI loss method has optimized the constraints on the strongest parameters.

% --------------------------------------------------------------------------------

\section{Comparing Constraints from Different Statistics: 2ptCF vs. VPF vs. CiC} \label{sec:CompareStats}

Throughout this work, we have obtained constraints using the combined results of the two-point correlation function (2ptCF), count-in-cells (CiC), and the Void Probability Function (VPF). This leads to natural questions: how much is each statistic contributing? Is one better than the others for specific constraints? What statistics are worth investing computational time into?

In this section, we compare the constraining power of each of the clustering statistics that we have used. We train neural networks keeping all choices but the clustering statistics the same. We test several galaxy selections, making sure to keep the density, the radii tested, and redshifts selected ($z={0.0, 0.1, 0.5, 1.0}$) the same for each clustering statistic. We compare each clustering statistic's constraints against the combination of `all' clustering to determine which statistic may be dominating a given constraint for a given selection.

Summary Figure \ref{fig:Summ_Clustering} shows the 1$\sigma$ error predictions for $\Omega_{\text{M}}$ and $\sigma_8$ (top half), and A$_{\text{SN1}}$ and A$_{\text{SN2}}$ (bottom half). Appendix \ref{app:Zs_Clust_Focus} Tables \ref{table:ClusteringStats_CosmoConstraints} and \ref{table:ClusteringStats_AsnConstraints} detail all test results. 

\subsection{Clustering Statistic Experiment Setup} \label{subsec:CompStats_Experiment}

We test the constraints from individual clustering statistics across several halo and galaxy selections: two thresholds of halo mass, three thresholds of stellar mass, and one of SFR, all randomly down-sampled to either $\mathcal{N}$=0.001 $h^3$ cMpc$^{-3}$ or 0.005 $h^3$ cMpc$^{-3}$. The exact details of how we measure and prepare the 2ptCF, CiC, and VPF for our neural networks are explained in \textsection \ref{subsubsec:MeasuringClustering}-\ref{subsubsec:NN_specifics}. In this experiment, when giving a neural network the independent clustering statistics, we pass exactly the same values that go into `all' clustering, allowing for fair comparison.

% As representative examples, we plot the constraints from different clustering statistics under two mass selections:  the clustering of dark matter halos with halo mass greater than $2\times 10^{11}$ M$_{\odot}$, randomly sampled to a density of 0.005 $h^{3}$ cMpc$^{-3}$; and the clustering of SAM galaxies with stellar mass greater than $1\times 10^{9}$ M$_{\odot}$, randomly sampled to a density of 0.005 $h^{3}$ cMpc$^{-3}$. Figures \ref{fig:DMonlyBestOm_ClusteringStat}, \ref{fig:DMonlyBests8_ClusteringStat}, and \ref{fig:Mstar_ClusteringCompare} showing these constraints are shown in Appendix \ref{app:Zs_Clust_Focus}, as the exact constraints closely resemble Figures already shown. 

\subsection{Results: Constraints by Statistic} \label{subsec:CompStats}

\begin{figure*}
	\begin{center}
\includegraphics[width=\textwidth]{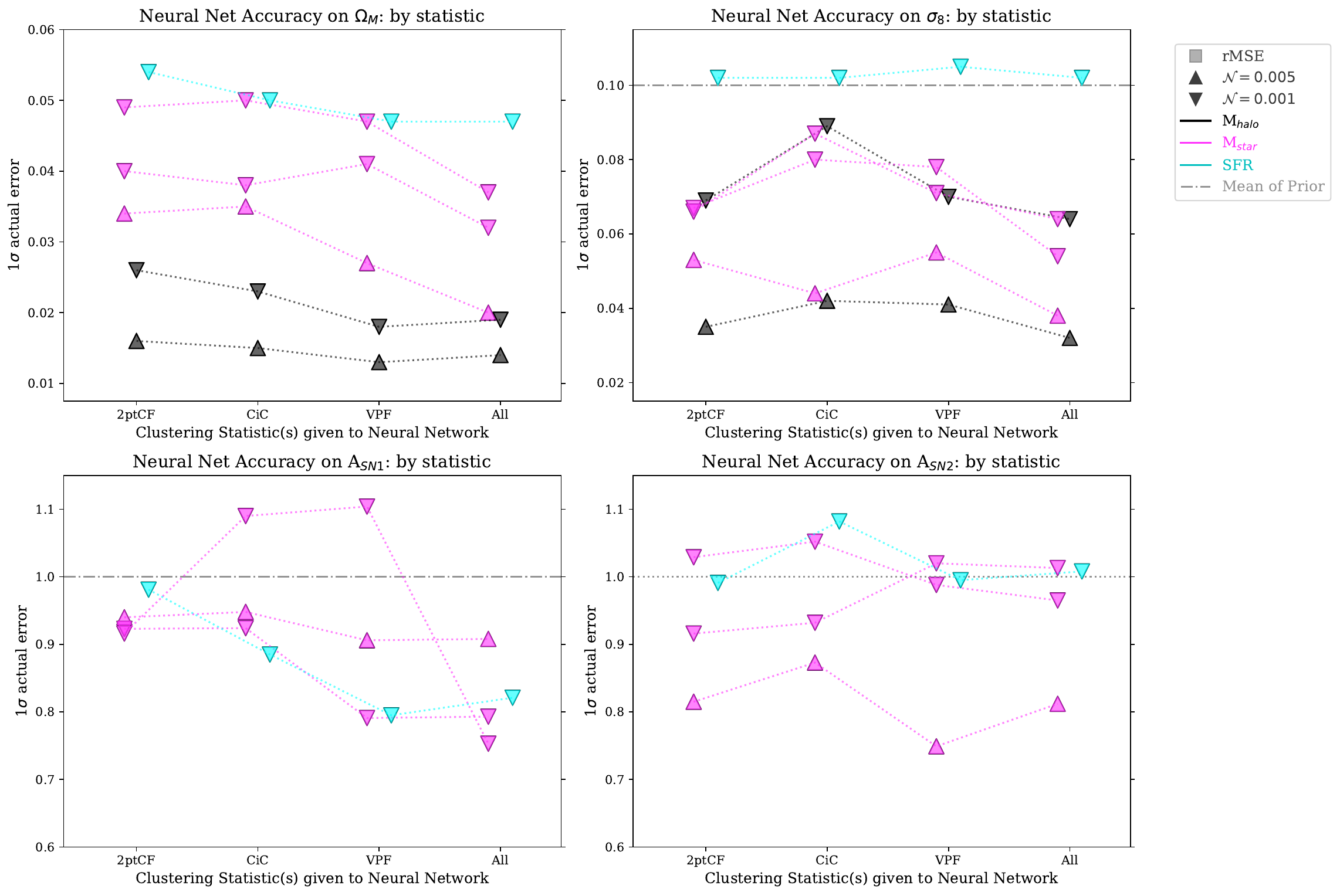}
	\caption{Similar to Figures \ref{fig:Summary_Allclust_Constraints} and \ref{fig:Summ_Redshift}, instead showing the constraints obtained when different clustering statistics are used separately  (see \textsection \ref{sec:CompareStats}, Tables \ref{table:ClusteringStats_CosmoConstraints} and \ref{table:ClusteringStats_AsnConstraints}, and Appendix \ref{app:Zs_Clust_Focus}). Dotted lines connect the same galaxy selection scenarios for ease of comparison. The dash-dotted lines indicate the mean of the prior, where constraints are not informative.}
	\label{fig:Summ_Clustering}
	\end{center}
\end{figure*}

We summarize how the 2ptCF, CiC, and VPF constrain $\Omega_{\text{M}}$ and $\sigma_8$ in the upper half of summary Figure \ref{fig:Summ_Clustering}; detailed results can be found in Appendix \ref{app:Zs_Clust_Focus}  Table \ref{table:ClusteringStats_CosmoConstraints}. The lower half of  Figure \ref{fig:Summ_Clustering} (and Appendix \ref{app:Zs_Clust_Focus}  Table \ref{table:ClusteringStats_AsnConstraints}) focuses instead on the SC-SAM parameters controlling stellar feedback in Equation \ref{eq:SCSAM_mdotout}. 
We note that the AGN feedback parameter A$_{\text{AGN}}$ from Equation \ref{eq:SCSAM_mdotradio} is generally difficult to constrain, so that each clustering statistic alone (or even all combined) cannot constrain it beyond the mean of the prior under the selections we probe.

Across the many selections we attempt, patterns emerge in how cosmological constraints are affected by the choice of clustering statistic. First, as expected, the combination of multiple clustering statistics nearly always improves the constraints found, especially for the galaxy selections. The clustering statistics often find similar constraints on $\sigma_8$. The 2ptCF tends to do best at $\sigma_8$ for lower density down-sampling selections, which is not unexpected: the 2ptCF, via its connection to the power spectrum, is a close measurement of the density fluctuations of the universe. Interestingly, the VPF often yields better constraints on $\Omega_{\text{M}}$ than the 2ptCF and CiC; our thoughts on why are discussed in \ref{subsec:DiscussCompClustStat}.

Unlike the cosmological parameters, there is notable improvement in constraining A$_{\text{SN1}}$ and A$_{\text{SN2}}$ when all clustering statistics are combined. There is some evidence across the galaxy selections that the VPF may drive the bulk of the constraints on A$_{\text{SN1}}$ and A$_{\text{SN2}}$. CiC and 2ptCF mostly perform comparably to each other, and the VPF tends to also perform similarly for A$_{\text{SN2}}$ across the bulk of our selections.

Examination of the neural network results in Appendix \ref{app:Zs_Clust_Focus}  Figure \ref{fig:Mstar_ClusteringCompare} helps give important context to some of the constraints we find. That specific neural network predicted low-value A$_{\text{SN1}}$ slightly more accurately than the high A$_{\text{SN1}}$ values, while maintaining a still generally flat dependence across the whole parameter space; this leads to deceptively lower errors given the performance. For a visual example of this phenomenon, compare this parameter's constraints in Figure \ref{fig:VPF_mstargt1N5k} against the others. Table \ref{table:ClusteringStats_AsnConstraints} indicates that the VPF yields the best constraints on the A$_{\text{SN1}}$ and A$_{\text{SN2}}$ feedback parameters (and therefore likely dominates the `all' clustering neural network results), but this is not an evident pattern in the Figures themselves and could be a statistical anomaly.

\subsection{Discussion: Comparing Clustering Statistics} \label{subsec:DiscussCompClustStat}

The clustering statistics we probe often find similar constraints on $\sigma_8$, with some evidence that the 2ptCF tends to do best at $\sigma_8$ for lower density down-sampling selections. As the Fourier transform of the power spectrum, which describes the amplitude of density fluctuations across distance scales, the 2ptCF is indeed expected to constrain $\sigma_8$ well, even despite higher Poisson noise at the lower density. We also find that the VPF often results in better constraints on $\Omega_{\text{M}}$ than the 2ptCF and CiC. The VPF also appears to drive the bulk of the weak constraints on the SAM SN parameters. Why might this be?

First, let us consider our approach to CiC. CiC in its entirety contains all information from all orders of correlations \citep{Uhlemann2020}, and might therefore be expected to give the most robust constraints (e.g.\ \citealt{Samushia2021}). However, due to the limited amount of data the neural network can take in and still promptly converge on a solution, we cannot use the full range of our CiC measurements. The input 1D array of clustering should have fewer than 1000 values, so that our fully-connected neural networks do not spiral into an unwieldy size with each new layer of neurons. For example, a fully-connected network of 4 layers with 1,000 neurons with an input of 10,000 elements and output of 5 elements would have:  10,000$\times$1,000$^4\times$5 internal relationships to consider when optimizing the network. This can be feasible with powerful GPUs and patience, but the amount of time it takes to train grows very quickly. In our work, even sampling to cells with fewer than 50 galaxies ($n=50$ in the terminology of Table \ref{table:CiCRadii}) for 10 distance scales between 1 and 40 cMpc yields 500 data points for a single redshift.

The default of our current setup uses four redshifts between $0<z<1$, and in this scenario we instead sample CiC at three distance scales where $n=0-50$, taking care to choose distance scales that show the differences between the simulations, as well as aiming to sample evenly across approximately $5-40$ cMpc. Additionally, taking distance scales that are far apart helps reduce the correlation between measurements (e.g.\ see \citealt{Gangolli2021} for an examination of correlations within VPF measurements, and \citealt{Uhlemann2020} for CiC). Table \ref{table:CiCRadii} fully describes how we sampled a small part of the full CiC distribution due to computational limitations, both for our default four-redshift networks, as well as when focusing on one redshift at a time in \textsection \ref{subsec:RedshiftChoices} and \ref{subsec:SCSAM_densityredshift}. Though we have sampled a small part of the full CiC distribution to reduce computational load, we still find the CiC inputs offer competitive constraints and sometimes outperform the 2ptCF.

We note here the relevant study of \citet{Wang2019} that reached  similar conclusions within a different framework. Using the `decorated' HOD \citep{Hearin2016} that includes galaxy assembly bias, they used a complete Fisher matrix analysis to probe how robustly various clustering statistics constrain assembly bias. They compared the projected 2ptCF (which projects the 3D 2ptCF into the dimensions of galaxy observations in RA, Dec, and degrees); the VPF; the galaxy-galaxy lensing signal; and variations on CiC such as count-in-cylinders and -annuli and probability distributions of them. Their work strongly motivates including CiC statistics alongside the popular projected 2ptCF and lensing signal for efficient constraints on galaxy assembly bias. Specifically, the VPF was good at constraining the number of central galaxies (not surprisingly, since it is a binary statistic that finds only empty test spheres, and will be less sensitive to satellites), while the varied CiC refined the number of centrals and satellites well.

Our handling of CiC may give an explanation for why the VPF performs well. The VPF is the 0$^{\text{th}}$ moment of counts-in-cells, the $n=0$ measurement. Because it yields a single value at each distance scale, we are able to include many more distance scales when preparing our data for a neural network. Therefore, the VPF serves as a sampling of the CiC and all the higher-order moments at many distance scales \citep{White1979}. Future work may therefore benefit from including the VPF and similar statistics (e.g.\ the $k$-NN statistics proposed by \citealt{Banerjee2020}) alongside the popular 2ptCF. 

%----------------------------------------------------------------------

\section{CAMELS-SAM Discussion and Implications}\label{sec:Discussion}
 
This work is a proof of concept, both for the power of the CAMELS-SAM simulation suite and
the use of SAM-generated galaxy catalogs, galaxy clustering, and neural networks to constrain cosmology and galaxy formation. In this section, we discuss implications of our analyses.

\subsection{Comparison with original hydrodynamic CAMELS}

Some readers may wonder if the work carried out in this work can be done to any extent with the original CAMELS suites. 
Important to the conception of CAMELS-SAM was the difficulty of applying galaxy selections across the entire CAMELS hydrodynamic suite. For example, if one tries to select \textit{any} objects with non-zero stellar mass, the SIMBA `hump' will often easily find several thousand galaxies in the (25 $h^{-1}$ cMpc)$^3$ volumes, while 10 or more percent of the IllustrisTNG `hump' would struggle to get more than a hundred galaxies\footnote{As explored in \citet{Perez2021}, when, where, and how the VPF clustering should be measured depends on galaxy density and total covered volume. This logic can be extended to CiC comfortably, and confirms much of the common-sense logic of the 2ptCF in the literature.}. 
We found there is \textit{no} basic galaxy property selection that yields enough objects for acceptable Poisson noise and a large enough training sets across both the SIMBA and TNG humps, even if not down-sampling to a single density\footnote{
Worrisome, since the accuracy of `deep learning' algorithms like our neural networks roughly scales with the size of the training set (e.g.\ \citealt{Hestness2017}).}. 
We are therefore unable to robustly explore how well galaxy clustering marginalizes over multiple hydrodynamic astrophysics models to measure cosmology.

Among the most stringent CAMELS selection we are able to make across both hydrodynamic `humps' is a \textit{halo mass} cut of M$_{\text{halo}} > 2\times10^{10} h^{-1}$ M$_{\odot}$. We make this cut and then randomly down-sample to $\mathcal{N}=0.064 h^3$ cMpc$^{-3}$ (1000 galaxies in each volume). We apply this to both the IllustrisTNG-DM and SIMBA-DM humps, and update our clustering to account for the smaller volume (and therefore allowed distance scales)\footnote{The VPF radii are 10 between with R$=0.8-8\ h^{-1}$ cMpc; the 2ptCF radii are 19 bins whose edges are evenly log-spaced between 0.68 and 9.3 $h^{-1}$ cMpc; and we include CiC to $n=50$ at R $=\{3.2, 5.6, 7.2\}$ cMpc. We combine the two `humps' for training under the MSE loss function criterion, yielding a total combined suite of approximately 1400/300/300 training/validation/testing simulations.}. Using the clustering of these CAMELS-DM halos, we find rMSE constraints of 0.031 for $\Omega_{\text{M}}$ and 0.081 for $\sigma_8$ (10\% atop the fiducial $\Omega_{\text{M}}=0.3$ and $\sigma_8=0.8$).

Cosmic variance in CAMELS---or the variance due to creating cosmological volumes of just (25 $h^{-1}$ cMpc)$^3$ for each combination of parameters---results in noisy clustering statistics, and makes the neural network predictions less accurate. Our choice to create much larger volumes at lower mass resolution improved the predictive power of our neural networks, both in terms of decreasing the effect of cosmic variance and expanding the galaxy selection and scale of clustering measurements we are able to carry out. This exercise further motivates the creation of larger CAMELS hydrodynamic wings, and the development of other techniques such as ``next generation" SAMs, which are specifically designed to emulate the results of specific hydrodynamic simulations. 

\subsection{The influence of astrophysics on cosmological parameter inference}

Our work with the clustering in CAMELS-SAM reaches two key conclusions: first, that we are able to successfully marginalize over astrophysics to constrain cosmology to precision nearly as good as with dark matter only clustering; and second, that we are able to learn something about the astrophysical parameters at the same time. Some readers may wonder just how much influence the presence of astrophysical variations in CAMELS-SAM affects our constraints.

First, we can explore how much information is lost to astrophysics in our CAMELS-SAM clustering analysis. We compare the results presented earlier in this work--where neural networks work with the clustering of galaxy catalogs generated using parameters across a 5D latin hypercube--with the results for a neural network trained on the clustering of galaxy catalogs whose generating parameters cover only a 2D cosmological space. Appendix \ref{app:infolostAstro} details this experiment.

e find that the cosmological constraints stay the same or improve when trained on the clustering of galaxy catalogs run with only the fiducial SC-SAM prescriptions. The rMSE errors slightly worsen for both $\Omega_{\text{M}}$ and $\sigma_{8}$, but the LFI loss errors remain the name for $\Omega_{\text{M}}$ and slightly improve for $\sigma_{8}$. We conservatively believe that some of the improvement of $\sigma_{8}$ constraints are partially due to the SC-SAM reinforcing or encoding information from the merger trees. However, this experiment shows that including the astrophysical parameters in the inference does \textit{not} lead to lost information on cosmology from galaxy clustering.

In a secondary experiment, we test the constraints when instead using the clustering of all galaxies above a high stellar mass threshold (and therefore introducing variations in number density for the neural networks to use). This scenario finds remarkable improvements in both cosmological parameters, though especially in $\Omega_{\text{M}}$ (not surprisingly, as increasing the total mass density will increase the total number of halos formed). We attribute this to the neural networks not having to work around degenerate effects on the number density that the SC-SAM parameters cause. However, the two experiments taken together emphasize the importance of including variations in astrophysics in cosmological parameter inference: the effects of astrophysics are many and not well-understood, and not marginalizing over them might lead to spuriously optimistic constraints.

One might also wonder: how robust is the cosmological inference to the number of SC-SAM parameters varied? This work only begins to probe this question, comparing constraints with 0 vs. 3 SC-SAM parameters varied. As shown in earlier in this section and in Appendix \ref{app:infolostAstro}, cosmological constraints remain quite robust under 3 SC-SAM parameters, with the LFI loss maintaining the precision and accuracy found within our dark matter only clustering tests. We believe this is due to the properties of the LFI loss function our networks use (see \textsection \ref{subsubsec:NNlosses}), and that we have a large enough latin hypercube so that the networks can learn the behavior of all five CAMELS-SAM parameters. Projects in progress within CAMELS-SAM and CAMELS will test this completely with the SC-SAM and the IllustrisTNG models, by varying all available parameters in an expanded latin hypercube or Sobol sequence \citep{Sobol1986} respectively.

\subsection{CAMELS-SAM among other simulations} \label{subsec:OtherSims}

Next, we discuss CAMELS-SAM in context with other large simulation suites, and some future work and possible experiments that CAMELS-SAM enables.

The N-body portion of CAMELS-SAM straddles a unique point between the limitations of computing power, data storage capacity, and useful scientific application, especially for machine learning. This part of CAMELS-SAM comprises over 1000 unique simulations of a moderately large volume, with moderate mass resolution (sufficient to robustly resolve galaxy properties relevant to upcoming observational samples), with 100 snapshots saved throughout $20<z<0$, and sampling of very broad cosmological parameter space in $\Omega_{\text{M}}$ and $\sigma_8$. Individually, other simulation suites may be comparable to or superior to each of these aspects, but they are combined in such a way to fill a unique role.

There exist N-body simulation suites that are much larger and/or higher resolution---e.g.\ BACCO \citep{BACCO}, Aemulus \citep{AemulusI}, ABACUS Cosmo/Summit (\citealt{ABACUScode, ABACUSSUMMIT}), Uchuu \citep{Ishiyama2021}, Dark Quest \citep{Nishimichi2019}--but that may not be as well suited for training neural networks or for running semi-analytic models. For example, comparable simulation suites often contain significantly lower numbers of realizations (e.g.~between several dozen to several hundred), which risks providing a small training set\footnote{It is worth noting, however, that these suites were \textit{not} created for machine learning training, and they excel at creating e.g.\ accurate and robust emulators of various phenomena. We acknowledge our somewhat unfair comparison.}. Next, many of these suites cover a narrower cosmological parameter space, which can risk neural network results too tightly focused around the priors (\citealt{Villaescusa-Navarro2020}; e.g.\ Figure 6  of \citealt{Ntampaka2020} with AbacusCosmo, who had to restrict their cosmological space in $\sigma_8$ due to biasing at the edges). Additionally, many of these suites solved the volume-resolution-data storage balance by saving a small number of snapshots (often 10-50, with as many as 65 or as few as 5). Though ideal for their specific science goals, this limits the possibility of running SAMs to generate galaxies, as they require densely sampled merger tree histories. Finally, we also point out N-body simulation suites that do not vary cosmological parameters, but which are well-suited to study other key features of large-scale structure and cosmology: Indra from \citet{Indra}, several hundred large volumes and many snapshots for excellent statistics; and UNIT from \citet{Chuang2019}, several hundred large volumes at excellent resolution for non-linear statistics. 

Most comparable to CAMELS-SAM (beyond the hydrodynamic CAMELS `humps') is the Quijote project \citep{Quijote}. The Quijote suite is unique in that it covers an even broader cosmological parameter space than all of CAMELS: 7,000 unique models over 6 cosmological parameters, including massive neutrinos. Quijote also has much larger volumes of 1 ($h^{-1}$ Gpc)$^3$ (though at lower resolution), and an astounding 44,000 simulations. However, its 5 stored snapshots and lower mass resolution make it unsuitable for applications using merger trees, and therefore from being the backbone for CAMELS-SAM. Though dark matter only, the vast Quijote simulation suite may be able to answer questions this work has not or cannot; for example, probing larger scale clustering, using more clustering statistics, implementing more sophisticated statistical tools like the Fisher matrix, expanding the cosmological models probed, etc. 

\subsection{CAMELS-SAM Data Release \& Possibilities} \label{subsec:FutureWorkIdeas}

Like the previous CAMELS `humps', CAMELS-SAM was created to serve as a data set to train machine learning tools to measure and analyze cosmology and uncertain aspects of astrophysics models. With the addition of the Santa Cruz SAM-generated galaxies atop large N-body only volumes, CAMELS-SAM offers a completely unique data set for machine learning. As part of the CAMELS Public Data Release in \citet{CAMELSpublic2022}\footnote{\url{https://camels.readthedocs.io/}, camels.simulations@gmail.com}, we release:

\begin{itemize}
\item The halo catalogues from \textsc{Rockstar}.
\item The merger trees generated from \textsc{Consistent trees}.
\item The galaxy catalogues from the Santa Cruz SAM.
\item Documentation at \url{https://camels-sam.readthedocs.io/}
\end{itemize}

Those wishing to directly analyze the raw simulation snapshots of CAMELS-SAM should reach out to the CAMELS team. The raw data (full N-body snapshots across redshifts) has been stored on tape and may be retrieved upon request.

An intuitive next step with this work is to leverage other types of neural networks, and use all available information rather than just summary statistics. \citet{Ntampaka2020} incorporated \textit{convolutional} neural networks (cNNs; \citealt{cnn1982, LeCun1999}) for cosmological inference on maps of HOD-simulated galaxies, finding few-percentage constraints on $\Omega_{\text{M}}$ and $\sigma_8$. cNNs can extract image features, like edges, shapes, and textures form 2D or 3D images, and have been shown to remarkably improve cosmological constraints compared to standard statistical approaches in the context of cosmology. For example,
% \footnote{Other examples in cosmology: \citet{Fluri2018} show their cNN of noisy convergence maps (from weak lensing) does much better than standard power spectrum analysis using a covariance matrix and likelihood. Similarly, \citet{Ribli2019} show their cNN finds much better and unbiased estimates of $\Omega_{\text{M}}$ and $\sigma_8$ from weak lensing maps.} 
\citet{Ravanbakhsh2017} fed their cNN the full dark matter distribution and compared to the classic method of maximum likelihood fitting, finding significantly better constraints with the cNN even with much smaller volumes. Later, \citet{Pan2020} further tested cNN constraints and their robustness and biases when using 3D dark matter distributions to measure $\Omega_{\text{M}}$ and $\sigma_8$. With its relatively large volumes and SAM-generated galaxies, CAMELS-SAM offers a great opportunity to train cNNs to constrain cosmology from galaxy maps. 

The CAMELS collaboration has recently found great success in cosmological parameter inference with \textit{graph} neural networks (gNNs). gNNs operate on data structured into nodes (the objects themselves) and their edges (how they interact), and are able to leverage global and local relationships in the data (\citealt{Xu2018, Battaglia2018, Zhou2018gnn, hamilton2020graph, Naidoo2020}). 
% Benefits of gNNs over cNNs are the ability to handle irregular data of any density, without requiring grid-like structure and the consideration of pixels or voxels. 
First in CAMELS to use gNNs for cosmological inference was \citet{PabloPaco2022}, who trained gNNs to perform likelihood-free inference of $\Omega_{\text{M}}$ at the galaxy-field level. With only $z=0$ real-space galaxy positions, their gNNs were able to accurately predict the power spectrum of CAMELS galaxies and moderately constrain $\Omega_{\text{M}}$; by including galaxy properties, the gNNs find significantly tighter cosmological constraints (though unfortunately only when training and testing on the same hydrodynamic suite, leaving improved robustness to astrophysics as a goal for the team). Next, \citet{Shao2022a} were able to robustly constrain $\Omega_{\text{M}}$ and $\sigma_8$ across several types of N-body codes and the CAMELS TNG and SIMBA hydrodynamic suites using information about dark matter halos (indicating perhaps some fundamental relation, as they probe and identify in \citealt{Shao2022b}). de Santi et al. (in prep) are training gNNs to predict $\Omega_{\text{M}}$ given galaxy properties, finding great robustness to the various CAMELS astrophysical models.
With its larger volume and separate model for galaxy physics, CAMELS-SAM promises to yield great results using the full halo or galaxy distributions with gNNs. Beyond CAMELS but relevant to CAMELS-SAM, gNNs have also been used to accurately map the complete hyperplane and dispersion of galaxy properties onto merger trees, offering a path to more precisely and efficiently emulate the output of SAMs \citep{Jespersen2022}. Additionally, \citet{Makinen2022} utilized information maximizing neural networks \citep{Charnock2018} atop a gNN architecture of DM halo catalogs to constrain cosmology far beyond what two-point galaxy clustering could, even in a noisy survey.

Finally, what of the possibilities of doing the work of this project with observed galaxy clustering? If we hope to use neural networks to constrain for our universe the value of $\Omega_{\text{M}}$ and $\sigma_8$, and astrophysical feedback in the context of the SC-SAM, how must this analysis progress? A neural network's predictions are only as good as the data on which it is trained, meaning an analysis like we have presented will only succeed on observations if the training data accurately reflects them. Therefore, expanding this project to observed galaxy clustering requires first robustly generating realistic galaxies from the properties the SC-SAM produces. This would entail selecting a galaxy sample that CAMELS-SAM is large and resolved enough to simulate (e.g.\ perhaps the 2D clustering of emission line galaxies in a small but well-studied field like COSMOS; \citealt{Khostovan2018}), and a detailed understanding of the selection function and observational systematics for the sample (as, e.g.\ \citealt{Hahn2022a, Hahn2022b} include in their forward model for BOSS CMASS galaxies). Additionally, expanding to other SAMs (and their unique parametrization of galaxy physics) could improve the neural networks' ability to marginalize over many forms of astrophysical prescriptions, and better constrain $\Omega_{\text{M}}$ and $\sigma_8$ for similar galaxy selections. A large variety of future work is still required within CAMELS-SAM, as well as for the cosmological inference field as a whole, before directly working with observed galaxies.

% \textcolor{red}{mention LtU? synthetic obs?}
% Within the realm of our work with CAMELS-SAM, future progress includes: expanding the clustering statistics and methods applied, varying more parameters in the SC-SAM, exploring more methods for tightening our cosmological constraints and marginalization over astrophysics, and implementing or creating more realistic galaxy selections to yield realistic mock galaxy catalogs. \textcolor{red}{Expand on what it would take to apply to observations here; and make this less stagnated}

\section{Conclusion}\label{sec:Conclusion}

In this work, we present and show an initial proof-of-concept project of CAMELS-SAM, a new and bigger `hump' of the CAMELS project. CAMELS-SAM is composed of more than 1000 unique N-body only simulations of volume ($100\ h^{-1}$ Mpc)$^{3}$ and N=$640^3$ particles each. The N-body simulations were generated across a broad cosmological parameter space of $\Omega_{\text{M}}$=[0.1,0.5] and $\sigma_8$=[0.6,1.0], with 100 stored snapshots between $20\leq z \leq 0$. Each of these N-body simulations has associated \textsc{Rockstar} halo catalogs and \textsc{ConsistentTrees} merger trees. Finally, each N-body simulation was run through a unique iteration of the Santa Cruz SAM for galaxy formation, covering a broad range of the parameters controlling feedback from massive stars, supernovae, and AGN radio jets. 

These halo catalogs, merger trees, and galaxy catalogs have been publicly released for the community to use for a variety of science applications:  \url{https://camels-sam.readthedocs.io}  \citep{CAMELSpublic2022}. As a proof-of-concept for the capabilities of this simulation suite, we used galaxy clustering statistics to constrain cosmology, marginalize over astrophysics, and probe astrophysical feedback with simple neural networks. 

We give a brief summary of our proof-of-concept work with CAMELS-SAM:

\begin{itemize}
    \item We measure and analyze the Void Probability Function, counts-in-cells, and real-space 3D two-point correlation function of halos and SC-SAM galaxies. We compare clustering across selections by halo mass, stellar mass, star formation rate, and specific star formation rate.
    \item We leverage simple 1D:1D neural networks and a likelihood-free inference method to measure the mean and standard deviation of each parameter's marginal posterior. We leverage CAMELS-SAM's large suite size for a split of 700/150/150 training/validation/testing simulation sets.
    \item We explore how the accuracy and precision of our parameter inference varies with different halo/galaxy selections and density down-samplings, and compare the choice of combining vs. keeping separate the redshift and clustering statistics used. \textsection \ref{sec:CosmoConstraints} focuses on the cosmological parameters \{$\Omega_{\text{M}}$, $\sigma_8$\}, while \textsection \ref{sec:SCSAM_constraints} focuses on the SC-SAM feedback parameters.
    
    \item Our neural networks are indeed able to marginalize over astrophysics to accurately constrain cosmology. The tightest constraints on cosmology we find are with the clustering by \textit{dark matter halo mass}. Our cosmological constraints for \{$\Omega_{\text{M}}$, $\sigma_8$\} find fractional errors of \{4.7\%, 3\%\} about their fiducial values of \{0.3, 0.8\}, respectively.
    With the clustering of \textit{SC-SAM galaxies by stellar mass}, we predict \{$\Omega_{\text{M}}$, $\sigma_8$\} with errors of \{4.7-6.7\%, 3-5\%\}.
    \item We find that our other selections based on \textit{stellar mass} yield predictions for \{$\Omega_{\text{M}}$, $\sigma_8$\} with errors of \{8.6-12.3\%, 5.4-8\%\}. Selecting on \textit{instantaneous star formation rate} often yields predictions for \{$\Omega_{\text{M}}$, $\sigma_8$\} with errors of \{10-15.6\%, 6.4-12.8\%\}. Selecting by \textit{specific star formation rate} often yields predictions for \{$\Omega_{\text{M}}$, $\sigma_8$\} with errors of \{8.3-9.3\%, 4.6-6.9\%\}. 
    
    %rssf this contradicts figure 3 -- revising
    %   \item We find tighter constraints on both cosmological parameters and astrophysics parameters when we do \emph{not} randomly down-sample to a fixed number density. 
    %lucia's update in response:
    \item Our neural networks are also able to constrain the astrophysical parameters alongside the cosmological parameters. We are able to learn some information about each of the SN and AGN feedback parameters of the SC-SAM, reaching constraints as good as $30\%$. We tend to find tighter constraints on both cosmological parameters and astrophysics parameters when we do \emph{not} randomly down-sample to a fixed number density. However, some galaxy property selections perform comparably well after down-sampling.
      
    \item We find better constraints when we combine clustering statistics from several redshifts. When using one redshift at a time, we do not find evidence for strong redshift dependence on the quality of the constraints.
    
    \item In \textsection \ref{sec:CompareStats}, we compare the constraints that each clustering statistics finds independently.  
    When comparing each of the clustering statistics--VPF, CiC, 2ptCF--for various mass-threshold clustering samples, we find that all the statistics find similar constraints for $\sigma_8$. 
    \item We find that CiC and VPF often slightly out-perform the 2ptCF alone for $\Omega_{\text{M}}$. We also find the VPF often drives the bulk of the constraints found on the SC-SAM parameters. 
      
    % \item We find stellar mass-selected galaxy clustering with no density down-sampling best constrains A$_{\text{SN1}}$ and A$_{\text{SN2}}$ ($\pm$0.5/0.7 or 13.3\%/17.5\% about fiducial values of 1/0, respectively).
    % \item The subtle A$_{\text{AGN}}$ parameter finds moderate constraints ($\pm$0.8 or 21.3\% about a fiducial value of 1) with star formation rate-selected galaxy clustering with no density down-sampling.
\end{itemize}

Finally, these are key implications of our work and CAMELS-SAM:
\begin{itemize}
    \item This is the first work to leverage machine learning and adjacent tools to not only constrain parameters of a SAM (e.g.\ \citealt{VanDaalen2016}), but to simultaneously constrain cosmology and improve the information a neural network is able to learn from galaxy clustering. 
    \item Our work with CAMELS-SAM includes smaller non-linear scales than most have probed for cosmology (reaching R$>$1.1-1.6 cMpc, or $k_{\text{max}} < 5.85-8.15\ h$ Mpc$^{-1}$). 
    \item Our work contributes to the growing literature that advocates the inclusion of count-in-cells, the VPF, and related statistics alongside the 2ptCF for constraining cosmology (e.g.\ \citealt{Wang2019, Uhlemann2020, Samushia2021}).
    \item By implementing a robust model for galaxies and their complex astrophysics, our neural networks are able to learn the unique ways that the SC-SAM affects galaxy bias, and significantly improve constraints on cosmology than if we were to simply assume galaxies follow the clustering of dark matter halos.
    \item The halo products from CAMELS-SAM inhabit a unique position in simulation volume and resolution, excellent snapshot and redshift coverage, number of simulations, and vast cosmological parameter space, ideal for several applications across machine learning, galaxy modeling, and dark matter analysis.

\end{itemize}

% \begin{acknowledgments}
\textit{This work was supported by the Flatiron Institute's Center for Computational Astrophysics Pre-Doctoral Program for the fall of 2020. CAMELS-SAM was run, stored, and remains accessible through the Flatiron Institute Scientific Computing Hub. LAP is supported by the Future Faculty in the Physical Sciences Fellowship at Princeton University. DAA was supported in part by NSF grants AST-2009687 and AST-2108944.} 

\textit{We gratefully thank M. Hirschmann for valuable conversations and explorations about simulating galaxies for clustering early in the project; R. Angulo for helpful discussion about the \textsc{BACCO} suite and its cosmological rescaling methods; A. Banerjee for fascinating discussion about their k-NN method; and L. Garrison for support and helpful context around \textsc{CORRFUNC} and the Abacus code. We thank P. Behroozi for creating and sharing \textsc{ROCKSTAR} and \textsc{ConsistentTrees}.
We also thank the Flatiron Institute's Scientific Computing team for helping make this project feasible and enduring, and especially to J. Tischio for support with the enormous data space. }

% \end{acknowledgments}

\bibliography{Paper}{}

\begin{thebibliography}{}
\expandafter\ifx\csname natexlab\endcsname\relax\def\natexlab#1{#1}\fi
\providecommand{\url}[1]{\href{#1}{#1}}
\providecommand{\dodoi}[1]{doi:~\href{http://doi.org/#1}{\nolinkurl{#1}}}
\providecommand{\doeprint}[1]{\href{http://ascl.net/#1}{\nolinkurl{http://ascl.net/#1}}}
\providecommand{\doarXiv}[1]{\href{https://arxiv.org/abs/#1}{\nolinkurl{https://arxiv.org/abs/#1}}}

\bibitem[{{Abbott} {et~al.}(2018){Abbott}, {Abdalla}, {Alarcon}, {Aleksi{\'c}},
  {Allam}, {Allen}, {Amara}, {Annis}, {Asorey}, {Avila}, {Bacon}, {Balbinot},
  {Banerji}, {Banik}, {Barkhouse}, {Baumer}, {Baxter}, {Bechtol}, {Becker},
  {Benoit-L{\'e}vy}, {Benson}, {Bernstein}, {Bertin}, {Blazek}, {Bridle},
  {Brooks}, {Brout}, {Buckley-Geer}, {Burke}, {Busha}, {Campos}, {Capozzi},
  {Carnero Rosell}, {Carrasco Kind}, {Carretero}, {Castander}, {Cawthon},
  {Chang}, {Chen}, {Childress}, {Choi}, {Conselice}, {Crittenden}, {Crocce},
  {Cunha}, {D'Andrea}, {da Costa}, {Das}, {Davis}, {Davis}, {De Vicente},
  {DePoy}, {DeRose}, {Desai}, {Diehl}, {Dietrich}, {Dodelson}, {Doel},
  {Drlica-Wagner}, {Eifler}, {Elliott}, {Elsner}, {Elvin-Poole}, {Estrada},
  {Evrard}, {Fang}, {Fernandez}, {Fert{\'e}}, {Finley}, {Flaugher}, {Fosalba},
  {Friedrich}, {Frieman}, {Garc{\'\i}a-Bellido}, {Garcia-Fernandez}, {Gatti},
  {Gaztanaga}, {Gerdes}, {Giannantonio}, {Gill}, {Glazebrook}, {Goldstein},
  {Gruen}, {Gruendl}, {Gschwend}, {Gutierrez}, {Hamilton}, {Hartley}, {Hinton},
  {Honscheid}, {Hoyle}, {Huterer}, {Jain}, {James}, {Jarvis}, {Jeltema},
  {Johnson}, {Johnson}, {Kacprzak}, {Kent}, {Kim}, {King}, {Kirk}, {Kokron},
  {Kovacs}, {Krause}, {Krawiec}, {Kremin}, {Kuehn}, {Kuhlmann}, {Kuropatkin},
  {Lacasa}, {Lahav}, {Li}, {Liddle}, {Lidman}, {Lima}, {Lin}, {MacCrann},
  {Maia}, {Makler}, {Manera}, {March}, {Marshall}, {Martini}, {McMahon},
  {Melchior}, {Menanteau}, {Miquel}, {Miranda}, {Mudd}, {Muir}, {M{\"o}ller},
  {Neilsen}, {Nichol}, {Nord}, {Nugent}, {Ogando}, {Palmese}, {Peacock},
  {Peiris}, {Peoples}, {Percival}, {Petravick}, {Plazas}, {Porredon}, {Prat},
  {Pujol}, {Rau}, {Refregier}, {Ricker}, {Roe}, {Rollins}, {Romer}, {Roodman},
  {Rosenfeld}, {Ross}, {Rozo}, {Rykoff}, {Sako}, {Salvador}, {Samuroff},
  {S{\'a}nchez}, {Sanchez}, {Santiago}, {Scarpine}, {Schindler}, {Scolnic},
  {Secco}, {Serrano}, {Sevilla-Noarbe}, {Sheldon}, {Smith}, {Smith}, {Smith},
  {Soares-Santos}, {Sobreira}, {Suchyta}, {Tarle}, {Thomas}, {Troxel},
  {Tucker}, {Tucker}, {Uddin}, {Varga}, {Vielzeuf}, {Vikram}, {Vivas},
  {Walker}, {Wang}, {Wechsler}, {Weller}, {Wester}, {Wolf}, {Yanny}, {Yuan},
  {Zenteno}, {Zhang}, {Zhang}, {Zuntz}, \& {Dark Energy Survey
  Collaboration}}]{DES}
{Abbott}, T.~M.~C., {Abdalla}, F.~B., {Alarcon}, A., {et~al.} 2018, \prd, 98,
  043526, \dodoi{10.1103/PhysRevD.98.043526}

\bibitem[{Akiba {et~al.}(2019)Akiba, Sano, Yanase, Ohta, \&
  Koyama}]{optuna_2019}
Akiba, T., Sano, S., Yanase, T., Ohta, T., \& Koyama, M. 2019, in Proceedings
  of the 25rd {ACM} {SIGKDD} International Conference on Knowledge Discovery
  and Data Mining

\bibitem[{{Albers} {et~al.}(2019){Albers}, {Fidler}, {Lesgourgues},
  {Sch{\"o}neberg}, \& {Torrado}}]{Albers2019}
{Albers}, J., {Fidler}, C., {Lesgourgues}, J., {Sch{\"o}neberg}, N., \&
  {Torrado}, J. 2019, \jcap, 2019, 028, \dodoi{10.1088/1475-7516/2019/09/028}

\bibitem[{{Alimi} {et~al.}(2012){Alimi}, {Bouillot}, {Rasera}, {Reverdy},
  {Corasaniti}, {Balmes}, {Requena}, {Delaruelle}, \& {Richet}}]{DEUS}
{Alimi}, J.-M., {Bouillot}, V., {Rasera}, Y., {et~al.} 2012, arXiv e-prints,
  arXiv:1206.2838.
\newblock \doarXiv{1206.2838}

\bibitem[{{Alsing} {et~al.}(2019){Alsing}, {Charnock}, {Feeney}, \&
  {Wandelt}}]{Alsing2019}
{Alsing}, J., {Charnock}, T., {Feeney}, S., \& {Wandelt}, B. 2019, \mnras, 488,
  4440, \dodoi{10.1093/mnras/stz1960}

\bibitem[{{Alsing} \& {Wandelt}(2019)}]{AlsingWandelt2019}
{Alsing}, J., \& {Wandelt}, B. 2019, \mnras, 488, 5093,
  \dodoi{10.1093/mnras/stz1900}

\bibitem[{{Amendola} {et~al.}(2018){Amendola}, {Appleby}, {Avgoustidis},
  {Bacon}, {Baker}, {Baldi}, {Bartolo}, {Blanchard}, {Bonvin}, {Borgani},
  {Branchini}, {Burrage}, {Camera}, {Carbone}, {Casarini}, {Cropper}, {de
  Rham}, {Dietrich}, {Di Porto}, {Durrer}, {Ealet}, {Ferreira}, {Finelli},
  {Garc{\'\i}a-Bellido}, {Giannantonio}, {Guzzo}, {Heavens}, {Heisenberg},
  {Heymans}, {Hoekstra}, {Hollenstein}, {Holmes}, {Hwang}, {Jahnke},
  {Kitching}, {Koivisto}, {Kunz}, {La Vacca}, {Linder}, {March}, {Marra},
  {Martins}, {Majerotto}, {Markovic}, {Marsh}, {Marulli}, {Massey}, {Mellier},
  {Montanari}, {Mota}, {Nunes}, {Percival}, {Pettorino}, {Porciani},
  {Quercellini}, {Read}, {Rinaldi}, {Sapone}, {Sawicki}, {Scaramella},
  {Skordis}, {Simpson}, {Taylor}, {Thomas}, {Trotta}, {Verde}, {Vernizzi},
  {Vollmer}, {Wang}, {Weller}, \& {Zlosnik}}]{Amendola2018}
{Amendola}, L., {Appleby}, S., {Avgoustidis}, A., {et~al.} 2018, Living Reviews
  in Relativity, 21, 2, \dodoi{10.1007/s41114-017-0010-3}

\bibitem[{{Angl{\'e}s-Alc{\'a}zar} {et~al.}(2017){Angl{\'e}s-Alc{\'a}zar},
  {Faucher-Gigu{\`e}re}, {Kere{\v{s}}}, {Hopkins}, {Quataert}, \&
  {Murray}}]{DanielAA2017}
{Angl{\'e}s-Alc{\'a}zar}, D., {Faucher-Gigu{\`e}re}, C.-A., {Kere{\v{s}}}, D.,
  {et~al.} 2017, \mnras, 470, 4698, \dodoi{10.1093/mnras/stx1517}

\bibitem[{{Angulo} {et~al.}(2021){Angulo}, {Zennaro}, {Contreras}, {Aric{\`o}},
  {Pellejero-Iba{\~n}ez}, \& {St{\"u}cker}}]{BACCO}
{Angulo}, R.~E., {Zennaro}, M., {Contreras}, S., {et~al.} 2021, \mnras, 507,
  5869, \dodoi{10.1093/mnras/stab2018}

\bibitem[{{Aric{\`o}} {et~al.}(2021){Aric{\`o}}, {Angulo}, {Contreras},
  {Ondaro-Mallea}, {Pellejero-Iba{\~n}ez}, \& {Zennaro}}]{Arico2021}
{Aric{\`o}}, G., {Angulo}, R.~E., {Contreras}, S., {et~al.} 2021, \mnras, 506,
  4070, \dodoi{10.1093/mnras/stab1911}

\bibitem[{{Baldry} {et~al.}(2012){Baldry}, {Driver}, {Loveday}, {Taylor},
  {Kelvin}, {Liske}, {Norberg}, {Robotham}, {Brough}, {Hopkins}, {Bamford},
  {Peacock}, {Bland-Hawthorn}, {Conselice}, {Croom}, {Jones}, {Parkinson},
  {Popescu}, {Prescott}, {Sharp}, \& {Tuffs}}]{Baldry2012}
{Baldry}, I.~K., {Driver}, S.~P., {Loveday}, J., {et~al.} 2012, \mnras, 421,
  621, \dodoi{10.1111/j.1365-2966.2012.20340.x}

\bibitem[{{Banerjee} \& {Abel}(2020)}]{Banerjee2020}
{Banerjee}, A., \& {Abel}, T. 2020, arXiv e-prints, arXiv:2007.13342.
\newblock \doarXiv{2007.13342}

\bibitem[{{Barreira} {et~al.}(2021){Barreira}, {Lazeyras}, \&
  {Schmidt}}]{Barriera2021}
{Barreira}, A., {Lazeyras}, T., \& {Schmidt}, F. 2021, \jcap, 2021, 029,
  \dodoi{10.1088/1475-7516/2021/08/029}

\bibitem[{{Barrera} {et~al.}(2022){Barrera}, {Springel}, {White},
  {Hern{\'a}ndez-Aguayo}, {Hernquist}, {Frenk}, {Pakmor}, {Ferlito},
  {Hadzhiyska}, {Delgado}, {Kannan}, \& {Bose}}]{MillenniumTNG_Barrera}
{Barrera}, M., {Springel}, V., {White}, S., {et~al.} 2022, arXiv e-prints,
  arXiv:2210.10419, \dodoi{10.48550/arXiv.2210.10419}

\bibitem[{{Barro} {et~al.}(2013){Barro}, {Faber}, {P{\'e}rez-Gonz{\'a}lez},
  {Koo}, {Williams}, {Kocevski}, {Trump}, {Mozena}, {McGrath}, {van der Wel},
  {Wuyts}, {Bell}, {Croton}, {Ceverino}, {Dekel}, {Ashby}, {Cheung},
  {Ferguson}, {Fontana}, {Fang}, {Giavalisco}, {Grogin}, {Guo}, {Hathi},
  {Hopkins}, {Huang}, {Koekemoer}, {Kartaltepe}, {Lee}, {Newman}, {Porter},
  {Primack}, {Ryan}, {Rosario}, {Somerville}, {Salvato}, \& {Hsu}}]{Barro2013}
{Barro}, G., {Faber}, S.~M., {P{\'e}rez-Gonz{\'a}lez}, P.~G., {et~al.} 2013,
  \apj, 765, 104, \dodoi{10.1088/0004-637X/765/2/104}

\bibitem[{{Battaglia} {et~al.}(2018){Battaglia}, {Hamrick}, {Bapst},
  {Sanchez-Gonzalez}, {Zambaldi}, {Malinowski}, {Tacchetti}, {Raposo},
  {Santoro}, {Faulkner}, {Gulcehre}, {Song}, {Ballard}, {Gilmer}, {Dahl},
  {Vaswani}, {Allen}, {Nash}, {Langston}, {Dyer}, {Heess}, {Wierstra}, {Kohli},
  {Botvinick}, {Vinyals}, {Li}, \& {Pascanu}}]{Battaglia2018}
{Battaglia}, P.~W., {Hamrick}, J.~B., {Bapst}, V., {et~al.} 2018, arXiv
  e-prints, arXiv:1806.01261.
\newblock \doarXiv{1806.01261}

\bibitem[{{Bayer} {et~al.}(2021){Bayer}, {Villaescusa-Navarro}, {Massara},
  {Liu}, {Spergel}, {Verde}, {Wandelt}, {Viel}, \& {Ho}}]{Bayer2021}
{Bayer}, A.~E., {Villaescusa-Navarro}, F., {Massara}, E., {et~al.} 2021, \apj,
  919, 24, \dodoi{10.3847/1538-4357/ac0e91}

\bibitem[{{Behroozi} {et~al.}(2020){Behroozi}, {Conroy}, {Wechsler}, {Hearin},
  {Williams}, {Moster}, {Yung}, {Somerville}, {Gottl{\"o}ber}, {Yepes}, \&
  {Endsley}}]{Behroozi2020}
{Behroozi}, P., {Conroy}, C., {Wechsler}, R.~H., {et~al.} 2020, \mnras, 499,
  5702, \dodoi{10.1093/mnras/staa3164}

\bibitem[{{Behroozi} {et~al.}(2013{\natexlab{a}}){Behroozi}, {Wechsler}, \&
  {Wu}}]{ROCKSTAR}
{Behroozi}, P.~S., {Wechsler}, R.~H., \& {Wu}, H.-Y. 2013{\natexlab{a}}, \apj,
  762, 109, \dodoi{10.1088/0004-637X/762/2/109}

\bibitem[{{Behroozi} {et~al.}(2013{\natexlab{b}}){Behroozi}, {Wechsler}, {Wu},
  {Busha}, {Klypin}, \& {Primack}}]{ConsistentTrees}
{Behroozi}, P.~S., {Wechsler}, R.~H., {Wu}, H.-Y., {et~al.} 2013{\natexlab{b}},
  \apj, 763, 18, \dodoi{10.1088/0004-637X/763/1/18}

\bibitem[{{Bernardi} {et~al.}(2013){Bernardi}, {Meert}, {Sheth}, {Vikram},
  {Huertas-Company}, {Mei}, \& {Shankar}}]{Bernardi2013}
{Bernardi}, M., {Meert}, A., {Sheth}, R.~K., {et~al.} 2013, \mnras, 436, 697,
  \dodoi{10.1093/mnras/stt1607}

\bibitem[{{Blanton} \& {Moustakas}(2009)}]{BlantonMoustakas2009}
{Blanton}, M.~R., \& {Moustakas}, J. 2009, \araa, 47, 159,
  \dodoi{10.1146/annurev-astro-082708-101734}

\bibitem[{{Borrow} {et~al.}(2020){Borrow}, {Angl{\'e}s-Alc{\'a}zar}, \&
  {Dav{\'e}}}]{Borrow2020}
{Borrow}, J., {Angl{\'e}s-Alc{\'a}zar}, D., \& {Dav{\'e}}, R. 2020, \mnras,
  491, 6102, \dodoi{10.1093/mnras/stz3428}

\bibitem[{{Bose} {et~al.}(2022){Bose}, {Hadzhiyska}, {Barrera}, {Delgado},
  {Ferlito}, {Frenk}, {Hern{\'a}ndez-Aguayo}, {Hernquist}, {Kannan}, {Pakmor},
  {Springel}, \& {White}}]{MillenniumTNG_Bose}
{Bose}, S., {Hadzhiyska}, B., {Barrera}, M., {et~al.} 2022, arXiv e-prints,
  arXiv:2210.10065, \dodoi{10.48550/arXiv.2210.10065}

\bibitem[{{Brammer} {et~al.}(2011){Brammer}, {Whitaker}, {van Dokkum},
  {Marchesini}, {Franx}, {Kriek}, {Labb{\'e}}, {Lee}, {Muzzin}, {Quadri},
  {Rudnick}, \& {Williams}}]{Brammer2011}
{Brammer}, G.~B., {Whitaker}, K.~E., {van Dokkum}, P.~G., {et~al.} 2011, \apj,
  739, 24, \dodoi{10.1088/0004-637X/739/1/24}

\bibitem[{{Brinchmann} {et~al.}(2004){Brinchmann}, {Charlot}, {White},
  {Tremonti}, {Kauffmann}, {Heckman}, \& {Brinkmann}}]{Brinchmann2004}
{Brinchmann}, J., {Charlot}, S., {White}, S.~D.~M., {et~al.} 2004, \mnras, 351,
  1151, \dodoi{10.1111/j.1365-2966.2004.07881.x}

\bibitem[{{Calette} {et~al.}(2018){Calette}, {Avila-Reese},
  {Rodr{\'\i}guez-Puebla}, {Hern{\'a}ndez-Toledo}, \&
  {Papastergis}}]{Calette2018}
{Calette}, A.~R., {Avila-Reese}, V., {Rodr{\'\i}guez-Puebla}, A.,
  {Hern{\'a}ndez-Toledo}, H., \& {Papastergis}, E. 2018, \rmxaa, 54, 443.
\newblock \doarXiv{1803.07692}

\bibitem[{{Calzetti}(2013)}]{Calzetti2013}
{Calzetti}, D. 2013, {Star Formation Rate Indicators}, ed.
  J.~{Falc{\'o}n-Barroso} \& J.~H. {Knapen}, 419

\bibitem[{{Catinella} {et~al.}(2018){Catinella}, {Saintonge}, {Janowiecki},
  {Cortese}, {Dav{\'e}}, {Lemonias}, {Cooper}, {Schiminovich}, {Hummels},
  {Fabello}, {Ger{\'e}b}, {Kilborn}, \& {Wang}}]{Catinella2018}
{Catinella}, B., {Saintonge}, A., {Janowiecki}, S., {et~al.} 2018, \mnras, 476,
  875, \dodoi{10.1093/mnras/sty089}

\bibitem[{{Charnock} {et~al.}(2018){Charnock}, {Lavaux}, \&
  {Wandelt}}]{Charnock2018}
{Charnock}, T., {Lavaux}, G., \& {Wandelt}, B.~D. 2018, \prd, 97, 083004,
  \dodoi{10.1103/PhysRevD.97.083004}

\bibitem[{{Chuang} {et~al.}(2019){Chuang}, {Yepes}, {Kitaura},
  {Pellejero-Ibanez}, {Rodr{\'\i}guez-Torres}, {Feng}, {Metcalf}, {Wechsler},
  {Zhao}, {To}, {Alam}, {Banerjee}, {DeRose}, {Giocoli}, {Knebe}, \&
  {Reyes}}]{Chuang2019}
{Chuang}, C.-H., {Yepes}, G., {Kitaura}, F.-S., {et~al.} 2019, \mnras, 487, 48,
  \dodoi{10.1093/mnras/stz1233}

\bibitem[{{Conroy}(2013)}]{Conroy2013}
{Conroy}, C. 2013, \araa, 51, 393, \dodoi{10.1146/annurev-astro-082812-141017}

\bibitem[{{Conroy} {et~al.}(2006){Conroy}, {Wechsler}, \&
  {Kravtsov}}]{Conroy2006}
{Conroy}, C., {Wechsler}, R.~H., \& {Kravtsov}, A.~V. 2006, \apj, 647, 201,
  \dodoi{10.1086/503602}

\bibitem[{{Conroy} {et~al.}(2005){Conroy}, {Coil}, {White}, {Newman}, {Yan},
  {Cooper}, {Gerke}, {Davis}, \& {Koo}}]{Conroy2005}
{Conroy}, C., {Coil}, A.~L., {White}, M., {et~al.} 2005, \apj, 635, 990,
  \dodoi{10.1086/497682}

\bibitem[{{Contreras} {et~al.}(2021){Contreras}, {Angulo}, \&
  {Zennaro}}]{Contreras2021}
{Contreras}, S., {Angulo}, R.~E., \& {Zennaro}, M. 2021, \mnras, 508, 175,
  \dodoi{10.1093/mnras/stab2560}

\bibitem[{{Contreras} {et~al.}(2020){Contreras}, {Angulo}, {Zennaro},
  {Aric{\`o}}, \& {Pellejero-Iba{\~n}ez}}]{Contreras2020}
{Contreras}, S., {Angulo}, R.~E., {Zennaro}, M., {Aric{\`o}}, G., \&
  {Pellejero-Iba{\~n}ez}, M. 2020, \mnras, 499, 4905,
  \dodoi{10.1093/mnras/staa3117}

\bibitem[{{Contreras} {et~al.}(2022){Contreras}, {Angulo}, {Springel}, {White},
  {Hadzhiyska}, {Hernquist}, {Pakmor}, {Kannan}, {Hern{\'a}ndez-Aguayo},
  {Barrera}, {Ferlito}, {Delgado}, {Bose}, \&
  {Frenk}}]{MillenniumTNG_Contreras}
{Contreras}, S., {Angulo}, R.~E., {Springel}, V., {et~al.} 2022, arXiv
  e-prints, arXiv:2210.10075, \dodoi{10.48550/arXiv.2210.10075}

\bibitem[{{Croton} {et~al.}(2004){Croton}, {Gazta{\~n}aga}, {Baugh}, {Norberg},
  {Colless}, {Baldry}, {Bland-Hawthorn}, {Bridges}, {Cannon}, {Cole},
  {Collins}, {Couch}, {Dalton}, {De Propris}, {Driver}, {Efstathiou}, {Ellis},
  {Frenk}, {Glazebrook}, {Jackson}, {Lahav}, {Lewis}, {Lumsden}, {Maddox},
  {Madgwick}, {Peacock}, {Peterson}, {Sutherland}, \& {Taylor}}]{Croton2004}
{Croton}, D.~J., {Gazta{\~n}aga}, E., {Baugh}, C.~M., {et~al.} 2004, \mnras,
  352, 1232, \dodoi{10.1111/j.1365-2966.2004.08017.x}

\bibitem[{{Dantas}(2021)}]{Dantas2021}
{Dantas}, C.~C. 2021, \mnras, 502, 5495, \dodoi{10.1093/mnras/stab445}

\bibitem[{{Dav{\'e}} {et~al.}(2019){Dav{\'e}}, {Angl{\'e}s-Alc{\'a}zar},
  {Narayanan}, {Li}, {Rafieferantsoa}, \& {Appleby}}]{Dave2019}
{Dav{\'e}}, R., {Angl{\'e}s-Alc{\'a}zar}, D., {Narayanan}, D., {et~al.} 2019,
  \mnras, 486, 2827, \dodoi{10.1093/mnras/stz937}

\bibitem[{{Davidzon} {et~al.}(2018){Davidzon}, {Ilbert}, {Faisst}, {Sparre}, \&
  {Capak}}]{Davidzon2018}
{Davidzon}, I., {Ilbert}, O., {Faisst}, A.~L., {Sparre}, M., \& {Capak}, P.~L.
  2018, \apj, 852, 107, \dodoi{10.3847/1538-4357/aaa19e}

\bibitem[{{de Santi} \& {Abramo}(2022)}]{deSanti2022b}
{de Santi}, N. S.~M., \& {Abramo}, L.~R. 2022, \jcap, 2022, 013,
  \dodoi{10.1088/1475-7516/2022/09/013}

\bibitem[{{de Santi} {et~al.}(2022){de Santi}, {Rodrigues}, {Montero-Dorta},
  {Abramo}, {Tucci}, \& {Artale}}]{deSanti2022a}
{de Santi}, N. S.~M., {Rodrigues}, N. V.~N., {Montero-Dorta}, A.~D., {et~al.}
  2022, \mnras, 514, 2463, \dodoi{10.1093/mnras/stac1469}

\bibitem[{{Delgado} {et~al.}(2023){Delgado}, {Angles-Alcazar}, {Thiele},
  {Ntampaka}, {Pandey}, {Lehman}, {Somerville}, {Genel}, \&
  {Villaescusa-Navarro}}]{Delgado2023}
{Delgado}, A.~M., {Angles-Alcazar}, D., {Thiele}, L., {et~al.} 2023, arXiv
  e-prints, arXiv:2301.02231.
\newblock \doarXiv{2301.02231}

\bibitem[{{DeRose} {et~al.}(2019){DeRose}, {Wechsler}, {Tinker}, {Becker},
  {Mao}, {McClintock}, {McLaughlin}, {Rozo}, \& {Zhai}}]{AemulusI}
{DeRose}, J., {Wechsler}, R.~H., {Tinker}, J.~L., {et~al.} 2019, \apj, 875, 69,
  \dodoi{10.3847/1538-4357/ab1085}

\bibitem[{{Dodelson}(2003)}]{Dodelson2003}
{Dodelson}, S. 2003, {Modern Cosmology}

\bibitem[{{Drakos} {et~al.}(2022){Drakos}, {Villasenor}, {Robertson}, {Hausen},
  {Dickinson}, {Ferguson}, {Furlanetto}, {Greene}, {Madau}, {Shapley}, {Stark},
  \& {Wechsler}}]{Drakos2022}
{Drakos}, N.~E., {Villasenor}, B., {Robertson}, B.~E., {et~al.} 2022, \apj,
  926, 194, \dodoi{10.3847/1538-4357/ac46fb}

\bibitem[{{Duncan} {et~al.}(2014){Duncan}, {Conselice}, {Mortlock}, {Hartley},
  {Guo}, {Ferguson}, {Dav{\'e}}, {Lu}, {Ownsworth}, {Ashby}, {Dekel},
  {Dickinson}, {Faber}, {Giavalisco}, {Grogin}, {Kocevski}, {Koekemoer},
  {Somerville}, \& {White}}]{Duncan2014}
{Duncan}, K., {Conselice}, C.~J., {Mortlock}, A., {et~al.} 2014, \mnras, 444,
  2960, \dodoi{10.1093/mnras/stu1622}

\bibitem[{{Ellis}(2008)}]{Ellis2008}
{Ellis}, R.~S. 2008, {Observations of the High Redshift Universe}, ed.
  A.~{Loeb}, A.~{Ferrara}, \& R.~S. {Ellis}, 259--364,
  \dodoi{10.1007/978-3-540-74163-3\_3}

\bibitem[{{Fabian}(2012)}]{Fabian2012}
{Fabian}, A.~C. 2012, \araa, 50, 455,
  \dodoi{10.1146/annurev-astro-081811-125521}

\bibitem[{{Falck} {et~al.}(2021){Falck}, {Wang}, {Jenkins}, {Lemson},
  {Medvedev}, {Neyrinck}, \& {Szalay}}]{Indra}
{Falck}, B., {Wang}, J., {Jenkins}, A., {et~al.} 2021, \mnras, 506, 2659,
  \dodoi{10.1093/mnras/stab1823}

\bibitem[{{Fang} {et~al.}(2005){Fang}, {Li}, \& {Sudjianto}}]{Fang2005}
{Fang}, K.-T., {Li}, R., \& {Sudjianto}, A. 2005, {Design and Modeling for
  Computer Experiments}, \dodoi{10.1201/9781420034899}

\bibitem[{{F{\"o}rster Schreiber} \& {Wuyts}(2020)}]{Forster2020}
{F{\"o}rster Schreiber}, N.~M., \& {Wuyts}, S. 2020, \araa, 58, 661,
  \dodoi{10.1146/annurev-astro-032620-021910}

\bibitem[{{Fukushima} \& {Miyake}(1982)}]{cnn1982}
{Fukushima}, K., \& {Miyake}, S. 1982, Pattern Recognition, 15, 455,
  \dodoi{10.1016/0031-3203(82)90024-3}

\bibitem[{{Gabrielpillai} {et~al.}(2022){Gabrielpillai}, {Somerville}, {Genel},
  {Rodriguez-Gomez}, {Pandya}, {Yung}, \& {Hernquist}}]{Gabrielpillai2022}
{Gabrielpillai}, A., {Somerville}, R.~S., {Genel}, S., {et~al.} 2022, \mnras,
  517, 6091, \dodoi{10.1093/mnras/stac2297}

\bibitem[{{Gallazzi} {et~al.}(2005){Gallazzi}, {Charlot}, {Brinchmann},
  {White}, \& {Tremonti}}]{Gallazzi2005}
{Gallazzi}, A., {Charlot}, S., {Brinchmann}, J., {White}, S. D.~M., \&
  {Tremonti}, C.~A. 2005, \mnras, 362, 41,
  \dodoi{10.1111/j.1365-2966.2005.09321.x}

\bibitem[{{Gangolli} {et~al.}(2021){Gangolli}, {D'Aloisio}, {Nasir}, \&
  {Zheng}}]{Gangolli2021}
{Gangolli}, N., {D'Aloisio}, A., {Nasir}, F., \& {Zheng}, Z. 2021, \mnras, 501,
  5294, \dodoi{10.1093/mnras/staa3843}

\bibitem[{{Garrison} {et~al.}(2021){Garrison}, {Eisenstein}, {Ferrer},
  {Maksimova}, \& {Pinto}}]{ABACUScode}
{Garrison}, L.~H., {Eisenstein}, D.~J., {Ferrer}, D., {Maksimova}, N.~A., \&
  {Pinto}, P.~A. 2021, \mnras, 508, 575, \dodoi{10.1093/mnras/stab2482}

\bibitem[{{Genel} {et~al.}(2014){Genel}, {Vogelsberger}, {Springel}, {Sijacki},
  {Nelson}, {Snyder}, {Rodriguez-Gomez}, {Torrey}, \& {Hernquist}}]{Genel2014}
{Genel}, S., {Vogelsberger}, M., {Springel}, V., {et~al.} 2014, \mnras, 445,
  175, \dodoi{10.1093/mnras/stu1654}

\bibitem[{{Gonzalez-Perez} {et~al.}(2018){Gonzalez-Perez}, {Comparat},
  {Norberg}, {Baugh}, {Contreras}, {Lacey}, {McCullagh}, {Orsi}, {Helly}, \&
  {Humphries}}]{G-P2018}
{Gonzalez-Perez}, V., {Comparat}, J., {Norberg}, P., {et~al.} 2018, \mnras,
  474, 4024, \dodoi{10.1093/mnras/stx2807}

\bibitem[{{Guo} {et~al.}(2013){Guo}, {White}, {Angulo}, {Henriques}, {Lemson},
  {Boylan-Kolchin}, {Thomas}, \& {Short}}]{Guo2013}
{Guo}, Q., {White}, S., {Angulo}, R.~E., {et~al.} 2013, \mnras, 428, 1351,
  \dodoi{10.1093/mnras/sts115}

\bibitem[{{Guo} {et~al.}(2010){Guo}, {White}, {Li}, \&
  {Boylan-Kolchin}}]{Guo2010}
{Guo}, Q., {White}, S., {Li}, C., \& {Boylan-Kolchin}, M. 2010, \mnras, 404,
  1111, \dodoi{10.1111/j.1365-2966.2010.16341.x}

\bibitem[{{Guo} {et~al.}(2011){Guo}, {White}, {Boylan-Kolchin}, {De Lucia},
  {Kauffmann}, {Lemson}, {Li}, {Springel}, \& {Weinmann}}]{Guo2011}
{Guo}, Q., {White}, S., {Boylan-Kolchin}, M., {et~al.} 2011, \mnras, 413, 101,
  \dodoi{10.1111/j.1365-2966.2010.18114.x}

\bibitem[{{Habouzit} {et~al.}(2020){Habouzit}, {Pisani}, {Goulding}, {Dubois},
  {Somerville}, \& {Greene}}]{Habouzit2020}
{Habouzit}, M., {Pisani}, A., {Goulding}, A., {et~al.} 2020, \mnras, 493, 899,
  \dodoi{10.1093/mnras/staa219}

\bibitem[{{Hadzhiyska} {et~al.}(2021{\natexlab{a}}){Hadzhiyska}, {Liu},
  {Somerville}, {Gabrielpillai}, {Bose}, {Eisenstein}, \&
  {Hernquist}}]{Hadzhiyska2021b}
{Hadzhiyska}, B., {Liu}, S., {Somerville}, R.~S., {et~al.} 2021{\natexlab{a}},
  \mnras, 508, 698, \dodoi{10.1093/mnras/stab2564}

\bibitem[{{Hadzhiyska} {et~al.}(2021{\natexlab{b}}){Hadzhiyska}, {Tacchella},
  {Bose}, \& {Eisenstein}}]{Hadzhiyska2021a}
{Hadzhiyska}, B., {Tacchella}, S., {Bose}, S., \& {Eisenstein}, D.~J.
  2021{\natexlab{b}}, \mnras, 502, 3599, \dodoi{10.1093/mnras/stab243}

\bibitem[{{Hadzhiyska} {et~al.}(2022{\natexlab{a}}){Hadzhiyska}, {Hernquist},
  {Eisenstein}, {Delgado}, {Bose}, {Kannan}, {Pakmor}, {Springel}, {Contreras},
  {Barrera}, {Ferlito}, {Hern{\'a}ndez-Aguayo}, {White}, \&
  {Frenk}}]{MillenniumTNG_Hadzhiyska1halo}
{Hadzhiyska}, B., {Hernquist}, L., {Eisenstein}, D., {et~al.}
  2022{\natexlab{a}}, arXiv e-prints, arXiv:2210.10068,
  \dodoi{10.48550/arXiv.2210.10068}

\bibitem[{{Hadzhiyska} {et~al.}(2022{\natexlab{b}}){Hadzhiyska}, {Eisenstein},
  {Hernquist}, {Pakmor}, {Bose}, {Delgado}, {Contreras}, {Kannan}, {White},
  {Springel}, {Frenk}, {Hern{\'a}ndez-Aguayo}, {Ferlito}, \&
  {Barrera}}]{MillenniumTNG_Hadzhiyska2halo}
{Hadzhiyska}, B., {Eisenstein}, D., {Hernquist}, L., {et~al.}
  2022{\natexlab{b}}, arXiv e-prints, arXiv:2210.10072,
  \dodoi{10.48550/arXiv.2210.10072}

\bibitem[{{Hahn} \& {Villaescusa-Navarro}(2021)}]{Hahn2021}
{Hahn}, C., \& {Villaescusa-Navarro}, F. 2021, \jcap, 2021, 029,
  \dodoi{10.1088/1475-7516/2021/04/029}

\bibitem[{{Hahn} {et~al.}(2022{\natexlab{a}}){Hahn}, {Eickenberg}, {Ho}, {Hou},
  {Lemos}, {Massara}, {Modi}, {Moradinezhad Dizgah}, {R{\'e}galdo-Saint
  Blancard}, \& {Abidi}}]{Hahn2022a}
{Hahn}, C., {Eickenberg}, M., {Ho}, S., {et~al.} 2022{\natexlab{a}}, arXiv
  e-prints, arXiv:2211.00723, \dodoi{10.48550/arXiv.2211.00723}

\bibitem[{{Hahn} {et~al.}(2022{\natexlab{b}}){Hahn}, {Eickenberg}, {Ho}, {Hou},
  {Lemos}, {Massara}, {Modi}, {Moradinezhad Dizgah}, {R{\'e}galdo-Saint
  Blancard}, \& {Abidi}}]{Hahn2022b}
---. 2022{\natexlab{b}}, arXiv e-prints, arXiv:2211.00660,
  \dodoi{10.48550/arXiv.2211.00660}

\bibitem[{{Hamaus} {et~al.}(2016){Hamaus}, {Pisani}, {Sutter}, {Lavaux},
  {Escoffier}, {Wandelt}, \& {Weller}}]{Hamaus2016}
{Hamaus}, N., {Pisani}, A., {Sutter}, P.~M., {et~al.} 2016, \prl, 117, 091302,
  \dodoi{10.1103/PhysRevLett.117.091302}

\bibitem[{{Hamilton}(2001)}]{Hamilton2001}
{Hamilton}, A.~J.~S. 2001, \mnras, 322, 419,
  \dodoi{10.1046/j.1365-8711.2001.04137.x}

\bibitem[{Hamilton(2020)}]{hamilton2020graph}
Hamilton, W. 2020, Graph Representation Learning, Synthesis lectures on
  artificial intelligence and machine learning (Morgan \& Claypool Publishers).
\newblock \url{https://books.google.com/books?id=V6HnzQEACAAJ}

\bibitem[{{Harikane} {et~al.}(2016){Harikane}, {Ouchi}, {Ono}, {More}, {Saito},
  {Lin}, {Coupon}, {Shimasaku}, {Shibuya}, {Price}, {Lin}, {Hsieh}, {Ishigaki},
  {Komiyama}, {Silverman}, {Takata}, {Tamazawa}, \& {Toshikawa}}]{Harikane2016}
{Harikane}, Y., {Ouchi}, M., {Ono}, Y., {et~al.} 2016, \apj, 821, 123,
  \dodoi{10.3847/0004-637X/821/2/123}

\bibitem[{{Hassan} {et~al.}(2021){Hassan}, {Villaescusa-Navarro}, {Wandelt},
  {Spergel}, {Angl{\'e}s-Alc{\'a}zar}, {Genel}, {Cranmer}, {Bryan}, {Dav{\'e}},
  {Somerville}, {Eickenberg}, {Narayanan}, {Ho}, \&
  {Andrianomena}}]{Hassan2021}
{Hassan}, S., {Villaescusa-Navarro}, F., {Wandelt}, B., {et~al.} 2021, arXiv
  e-prints, arXiv:2110.02983.
\newblock \doarXiv{2110.02983}

\bibitem[{{Hearin} {et~al.}(2022){Hearin}, {Ramachandra}, {Becker}, \&
  {DeRose}}]{Hearin2022}
{Hearin}, A.~P., {Ramachandra}, N., {Becker}, M.~R., \& {DeRose}, J. 2022, The
  Open Journal of Astrophysics, 5, 3, \dodoi{10.21105/astro.2112.08423}

\bibitem[{{Hearin} {et~al.}(2016){Hearin}, {Zentner}, {van den Bosch},
  {Campbell}, \& {Tollerud}}]{Hearin2016}
{Hearin}, A.~P., {Zentner}, A.~R., {van den Bosch}, F.~C., {Campbell}, D., \&
  {Tollerud}, E. 2016, \mnras, 460, 2552, \dodoi{10.1093/mnras/stw840}

\bibitem[{{Hern{\'a}ndez-Aguayo} {et~al.}(2022){Hern{\'a}ndez-Aguayo},
  {Springel}, {Pakmor}, {Barrera}, {Ferlito}, {White}, {Hernquist},
  {Hadzhiyska}, {Delgado}, {Kannan}, {Bose}, \&
  {Frenk}}]{MillenniumTNG_HernandezAguayo}
{Hern{\'a}ndez-Aguayo}, C., {Springel}, V., {Pakmor}, R., {et~al.} 2022, arXiv
  e-prints, arXiv:2210.10059, \dodoi{10.48550/arXiv.2210.10059}

\bibitem[{{Hestness} {et~al.}(2017){Hestness}, {Narang}, {Ardalani}, {Diamos},
  {Jun}, {Kianinejad}, {Patwary}, {Yang}, \& {Zhou}}]{Hestness2017}
{Hestness}, J., {Narang}, S., {Ardalani}, N., {et~al.} 2017, arXiv e-prints,
  arXiv:1712.00409.
\newblock \doarXiv{1712.00409}

\bibitem[{{Hurtado-Gil} {et~al.}(2017){Hurtado-Gil}, {Mart{\'\i}nez},
  {Arnalte-Mur}, {Pons-Border{\'\i}a}, {Pareja-Flores}, \&
  {Paredes}}]{Hurtado-Gil2017}
{Hurtado-Gil}, L., {Mart{\'\i}nez}, V.~J., {Arnalte-Mur}, P., {et~al.} 2017,
  \aap, 601, A40, \dodoi{10.1051/0004-6361/201629097}

\bibitem[{{Ishiyama} {et~al.}(2021){Ishiyama}, {Prada}, {Klypin}, {Sinha},
  {Metcalf}, {Jullo}, {Altieri}, {Cora}, {Croton}, {de la Torre},
  {Mill{\'a}n-Calero}, {Oogi}, {Ruedas}, \&
  {Vega-Mart{\'\i}nez}}]{Ishiyama2021}
{Ishiyama}, T., {Prada}, F., {Klypin}, A.~A., {et~al.} 2021, \mnras, 506, 4210,
  \dodoi{10.1093/mnras/stab1755}

\bibitem[{{Jeffrey} {et~al.}(2022){Jeffrey}, {Boulanger}, {Wandelt},
  {Regaldo-Saint Blancard}, {Allys}, \& {Levrier}}]{Jeffrey2022}
{Jeffrey}, N., {Boulanger}, F., {Wandelt}, B.~D., {et~al.} 2022, \mnras, 510,
  L1, \dodoi{10.1093/mnrasl/slab120}

\bibitem[{{Jeffrey} \& {Wandelt}(2020)}]{MomentNetworks}
{Jeffrey}, N., \& {Wandelt}, B.~D. 2020, arXiv e-prints, arXiv:2011.05991.
\newblock \doarXiv{2011.05991}

\bibitem[{{Jespersen} {et~al.}(2022){Jespersen}, {Cranmer}, {Melchior}, {Ho},
  {Somerville}, \& {Gabrielpillai}}]{Jespersen2022}
{Jespersen}, C.~K., {Cranmer}, M., {Melchior}, P., {et~al.} 2022, \apj, 941, 7,
  \dodoi{10.3847/1538-4357/ac9b18}

\bibitem[{{Jiang} {et~al.}(2021){Jiang}, {Dekel}, {Freundlich}, {van den
  Bosch}, {Green}, {Hopkins}, {Benson}, \& {Du}}]{Jiang:2021}
{Jiang}, F., {Dekel}, A., {Freundlich}, J., {et~al.} 2021, \mnras, 502, 621,
  \dodoi{10.1093/mnras/staa4034}

\bibitem[{{Jo} {et~al.}(2022){Jo}, {Genel}, {Wandelt}, {Somerville},
  {Villaescusa-Navarro}, {Bryan}, {Angles-Alcazar}, {Foreman-Mackey}, {Nelson},
  \& {Kim}}]{Jo2022}
{Jo}, Y., {Genel}, S., {Wandelt}, B., {et~al.} 2022, arXiv e-prints,
  arXiv:2211.16461, \dodoi{10.48550/arXiv.2211.16461}

\bibitem[{{Khostovan} {et~al.}(2018){Khostovan}, {Sobral}, {Mobasher}, {Best},
  {Smail}, {Matthee}, {Darvish}, {Nayyeri}, {Hemmati}, \&
  {Stott}}]{Khostovan2018}
{Khostovan}, A.~A., {Sobral}, D., {Mobasher}, B., {et~al.} 2018, \mnras, 478,
  2999, \dodoi{10.1093/mnras/sty925}

\bibitem[{{Khostovan} {et~al.}(2020){Khostovan}, {Malhotra}, {Rhoads}, {Jiang},
  {Wang}, {Wold}, {Zheng}, {Barrientos}, {Coughlin}, {Harish}, {Hu}, {Infante},
  {Perez}, {Pharo}, {Valdes}, {Walker}, \& {Yang}}]{Khostovan2020}
{Khostovan}, A.~A., {Malhotra}, S., {Rhoads}, J.~E., {et~al.} 2020, \mnras,
  493, 3966, \dodoi{10.1093/mnras/staa175}

\bibitem[{{Kirby} {et~al.}(2011){Kirby}, {Lanfranchi}, {Simon}, {Cohen}, \&
  {Guhathakurta}}]{Kirby2011}
{Kirby}, E.~N., {Lanfranchi}, G.~A., {Simon}, J.~D., {Cohen}, J.~G., \&
  {Guhathakurta}, P. 2011, \apj, 727, 78, \dodoi{10.1088/0004-637X/727/2/78}

\bibitem[{{Knebe} {et~al.}(2018){Knebe}, {Pearce}, {Gonzalez-Perez}, {Thomas},
  {Benson}, {Asquith}, {Blaizot}, {Bower}, {Carretero}, {Castander},
  {Cattaneo}, {Cora}, {Croton}, {Cui}, {Cunnama}, {Devriendt}, {Elahi}, {Font},
  {Fontanot}, {Gargiulo}, {Helly}, {Henriques}, {Lee}, {Mamon}, {Onions},
  {Padilla}, {Power}, {Pujol}, {Ruiz}, {Srisawat}, {Stevens}, {Tollet},
  {Vega-Mart{\'\i}nez}, \& {Yi}}]{Knebe2018}
{Knebe}, A., {Pearce}, F.~R., {Gonzalez-Perez}, V., {et~al.} 2018, \mnras, 475,
  2936, \dodoi{10.1093/mnras/stx3274}

\bibitem[{{Kokron} {et~al.}(2021){Kokron}, {DeRose}, {Chen}, {White}, \&
  {Wechsler}}]{Kokron2021}
{Kokron}, N., {DeRose}, J., {Chen}, S.-F., {White}, M., \& {Wechsler}, R.~H.
  2021, \mnras, 505, 1422, \dodoi{10.1093/mnras/stab1358}

\bibitem[{{Kormendy} \& {Ho}(2013)}]{Kormendy2013}
{Kormendy}, J., \& {Ho}, L.~C. 2013, \araa, 51, 511,
  \dodoi{10.1146/annurev-astro-082708-101811}

\bibitem[{{Kwan} {et~al.}(2023){Kwan}, {Saito}, {Leauthaud}, {Heitmann},
  {Habib}, {Frontiere}, {Guo}, {Huang}, {Pope}, \&
  {Rodr{\'\i}guez-Torres}}]{Kwan2023}
{Kwan}, J., {Saito}, S., {Leauthaud}, A., {et~al.} 2023, arXiv e-prints,
  arXiv:2302.12379.
\newblock \doarXiv{2302.12379}

\bibitem[{{Landy} \& {Szalay}(1993)}]{L-S1993}
{Landy}, S.~D., \& {Szalay}, A.~S. 1993, \apj, 412, 64, \dodoi{10.1086/172900}

\bibitem[{{Lange} {et~al.}(2022){Lange}, {Hearin}, {Leauthaud}, {van den
  Bosch}, {Guo}, \& {DeRose}}]{Lange2022}
{Lange}, J.~U., {Hearin}, A.~P., {Leauthaud}, A., {et~al.} 2022, \mnras, 509,
  1779, \dodoi{10.1093/mnras/stab311110.48550/arXiv.2101.12261}

\bibitem[{{Lange} {et~al.}(2023){Lange}, {Hearin}, {Leauthaud}, {van den
  Bosch}, {Xhakaj}, {Guo}, {Wechsler}, \& {DeRose}}]{Lange2023}
---. 2023, arXiv e-prints, arXiv:2301.08692.
\newblock \doarXiv{2301.08692}

\bibitem[{LeCun {et~al.}(1999)LeCun, Haffner, Bottou, \& Bengio}]{LeCun1999}
LeCun, Y., Haffner, P., Bottou, L., \& Bengio, Y. 1999, in Shape, Contour and
  Grouping in Computer Vision (Berlin, Heidelberg: Springer-Verlag), 319

\bibitem[{{Lewis} {et~al.}(2000){Lewis}, {Challinor}, \& {Lasenby}}]{CAMB}
{Lewis}, A., {Challinor}, A., \& {Lasenby}, A. 2000, \apj, 538, 473,
  \dodoi{10.1086/309179}

\bibitem[{{Lu} {et~al.}(2014){Lu}, {Wechsler}, {Somerville}, {Croton},
  {Porter}, {Primack}, {Behroozi}, {Ferguson}, {Koo}, {Guo}, {Safarzadeh},
  {Finlator}, {Castellano}, {White}, {Sommariva}, \& {Moody}}]{Lu2014}
{Lu}, Y., {Wechsler}, R.~H., {Somerville}, R.~S., {et~al.} 2014, \apj, 795,
  123, \dodoi{10.1088/0004-637X/795/2/123}

\bibitem[{{Madau} \& {Dickinson}(2014)}]{MadauDickinson2014}
{Madau}, P., \& {Dickinson}, M. 2014, \araa, 52, 415,
  \dodoi{10.1146/annurev-astro-081811-125615}

\bibitem[{{Makinen} {et~al.}(2021){Makinen}, {Charnock}, {Alsing}, \&
  {Wandelt}}]{Makinen2021}
{Makinen}, T.~L., {Charnock}, T., {Alsing}, J., \& {Wandelt}, B.~D. 2021,
  \jcap, 2021, 049, \dodoi{10.1088/1475-7516/2021/11/049}

\bibitem[{{Makinen} {et~al.}(2022){Makinen}, {Charnock}, {Lemos}, {Porqueres},
  {Heavens}, \& {Wandelt}}]{Makinen2022}
{Makinen}, T.~L., {Charnock}, T., {Lemos}, P., {et~al.} 2022, The Open Journal
  of Astrophysics, 5, 18,
  \dodoi{10.21105/astro.2207.0520210.48550/arXiv.2207.05202}

\bibitem[{{Maksimova} {et~al.}(2021){Maksimova}, {Garrison}, {Eisenstein},
  {Hadzhiyska}, {Bose}, \& {Satterthwaite}}]{ABACUSSUMMIT}
{Maksimova}, N.~A., {Garrison}, L.~H., {Eisenstein}, D.~J., {et~al.} 2021,
  \mnras, 508, 4017, \dodoi{10.1093/mnras/stab2484}

\bibitem[{{Massara} {et~al.}(2021){Massara}, {Villaescusa-Navarro}, {Ho},
  {Dalal}, \& {Spergel}}]{Massara2021}
{Massara}, E., {Villaescusa-Navarro}, F., {Ho}, S., {Dalal}, N., \& {Spergel},
  D.~N. 2021, \prl, 126, 011301, \dodoi{10.1103/PhysRevLett.126.011301}

\bibitem[{{McConnell} \& {Ma}(2013)}]{McConnell2013}
{McConnell}, N.~J., \& {Ma}, C.-P. 2013, \apj, 764, 184,
  \dodoi{10.1088/0004-637X/764/2/184}

\bibitem[{{McCullagh} {et~al.}(2017){McCullagh}, {Norberg}, {Cole},
  {Gonzalez-Perez}, {Baugh}, \& {Helly}}]{McCullagh2017}
{McCullagh}, N., {Norberg}, P., {Cole}, S., {et~al.} 2017, arXiv e-prints,
  arXiv:1705.01988.
\newblock \doarXiv{1705.01988}

\bibitem[{{McKee} \& {Ostriker}(2007)}]{McKeeOstriker2007}
{McKee}, C.~F., \& {Ostriker}, E.~C. 2007, \araa, 45, 565,
  \dodoi{10.1146/annurev.astro.45.051806.110602}

\bibitem[{{Mead} {et~al.}(2021){Mead}, {Brieden}, {Tr{\"o}ster}, \&
  {Heymans}}]{Mead2021}
{Mead}, A.~J., {Brieden}, S., {Tr{\"o}ster}, T., \& {Heymans}, C. 2021, \mnras,
  502, 1401, \dodoi{10.1093/mnras/stab082}

\bibitem[{{Mobasher} {et~al.}(2015){Mobasher}, {Dahlen}, {Ferguson},
  {Acquaviva}, {Barro}, {Finkelstein}, {Fontana}, {Gruetzbauch}, {Johnson},
  {Lu}, {Papovich}, {Pforr}, {Salvato}, {Somerville}, {Wiklind}, {Wuyts},
  {Ashby}, {Bell}, {Conselice}, {Dickinson}, {Faber}, {Fazio}, {Finlator},
  {Galametz}, {Gawiser}, {Giavalisco}, {Grazian}, {Grogin}, {Guo}, {Hathi},
  {Kocevski}, {Koekemoer}, {Koo}, {Newman}, {Reddy}, {Santini}, \&
  {Wechsler}}]{Mobasher2015}
{Mobasher}, B., {Dahlen}, T., {Ferguson}, H.~C., {et~al.} 2015, \apj, 808, 101,
  \dodoi{10.1088/0004-637X/808/1/101}

\bibitem[{{Moser} {et~al.}(2022){Moser}, {Battaglia}, {Nagai}, {Lau}, {Machado
  Poletti Valle}, {Villaescusa-Navarro}, {Amodeo}, {Angles-Alcazar}, {Bryan},
  {Dave}, {Hernquist}, \& {Vogelsberger}}]{Moser2022}
{Moser}, E., {Battaglia}, N., {Nagai}, D., {et~al.} 2022, arXiv e-prints,
  arXiv:2201.02708.
\newblock \doarXiv{2201.02708}

\bibitem[{{Moustakas} {et~al.}(2013){Moustakas}, {Coil}, {Aird}, {Blanton},
  {Cool}, {Eisenstein}, {Mendez}, {Wong}, {Zhu}, \& {Arnouts}}]{Moustakas2013}
{Moustakas}, J., {Coil}, A.~L., {Aird}, J., {et~al.} 2013, \apj, 767, 50,
  \dodoi{10.1088/0004-637X/767/1/50}

\bibitem[{{Muzzin} {et~al.}(2013){Muzzin}, {Marchesini}, {Stefanon}, {Franx},
  {McCracken}, {Milvang-Jensen}, {Dunlop}, {Fynbo}, {Brammer}, {Labb{\'e}}, \&
  {van Dokkum}}]{Muzzin2013}
{Muzzin}, A., {Marchesini}, D., {Stefanon}, M., {et~al.} 2013, \apj, 777, 18,
  \dodoi{10.1088/0004-637X/777/1/18}

\bibitem[{{Naab} \& {Ostriker}(2017)}]{Naab2017}
{Naab}, T., \& {Ostriker}, J.~P. 2017, \araa, 55, 59,
  \dodoi{10.1146/annurev-astro-081913-040019}

\bibitem[{{Naidoo} {et~al.}(2020){Naidoo}, {Whiteway}, {Massara}, {Gualdi},
  {Lahav}, {Viel}, {Gil-Mar{\'\i}n}, \& {Font-Ribera}}]{Naidoo2020}
{Naidoo}, K., {Whiteway}, L., {Massara}, E., {et~al.} 2020, \mnras, 491, 1709,
  \dodoi{10.1093/mnras/stz3075}

\bibitem[{{Netzer}(2015)}]{Netzger2015}
{Netzer}, H. 2015, \araa, 53, 365, \dodoi{10.1146/annurev-astro-082214-122302}

\bibitem[{{Ni} {et~al.}(2023){Ni}, {Genel}, {Angl{\'e}s-Alc{\'a}zar},
  {Villaescusa-Navarro}, {Jo}, {Bird}, {Di Matteo}, {Croft}, {Chen}, {de
  Santi}, {Gebhardt}, {Shao}, {Pandey}, {Hernquist}, \& {Dave}}]{AstridCAMELS}
{Ni}, Y., {Genel}, S., {Angl{\'e}s-Alc{\'a}zar}, D., {et~al.} 2023, arXiv
  e-prints, arXiv:2304.02096, \dodoi{10.48550/arXiv.2304.02096}

\bibitem[{{Nicola} {et~al.}(2022){Nicola}, {Villaescusa-Navarro}, {Spergel},
  {Dunkley}, {Angl{\'e}s-Alc{\'a}zar}, {Dav{\'e}}, {Genel}, {Hernquist},
  {Nagai}, {Somerville}, \& {Wandelt}}]{Nicola2022}
{Nicola}, A., {Villaescusa-Navarro}, F., {Spergel}, D.~N., {et~al.} 2022, arXiv
  e-prints, arXiv:2201.04142.
\newblock \doarXiv{2201.04142}

\bibitem[{{Nishimichi} {et~al.}(2019){Nishimichi}, {Takada}, {Takahashi},
  {Osato}, {Shirasaki}, {Oogi}, {Miyatake}, {Oguri}, {Murata}, {Kobayashi}, \&
  {Yoshida}}]{Nishimichi2019}
{Nishimichi}, T., {Takada}, M., {Takahashi}, R., {et~al.} 2019, \apj, 884, 29,
  \dodoi{10.3847/1538-4357/ab3719}

\bibitem[{{Ntampaka} {et~al.}(2020){Ntampaka}, {Eisenstein}, {Yuan}, \&
  {Garrison}}]{Ntampaka2020}
{Ntampaka}, M., {Eisenstein}, D.~J., {Yuan}, S., \& {Garrison}, L.~H. 2020,
  \apj, 889, 151, \dodoi{10.3847/1538-4357/ab5f5e}

\bibitem[{{Pan} {et~al.}(2020){Pan}, {Liu}, {Forero-Romero}, {Sabiu}, {Li},
  {Miao}, \& {Li}}]{Pan2020}
{Pan}, S., {Liu}, M., {Forero-Romero}, J., {et~al.} 2020, Science China
  Physics, Mechanics, and Astronomy, 63, 110412,
  \dodoi{10.1007/s11433-020-1586-3}

\bibitem[{Paszke {et~al.}(2019)Paszke, Gross, Massa, Lerer, Bradbury, Chanan,
  Killeen, Lin, Gimelshein, Antiga, Desmaison, Kopf, Yang, DeVito, Raison,
  Tejani, Chilamkurthy, Steiner, Fang, Bai, \& Chintala}]{Pytorch}
Paszke, A., Gross, S., Massa, F., {et~al.} 2019, in Advances in Neural
  Information Processing Systems 32, ed. H.~Wallach, H.~Larochelle,
  A.~Beygelzimer, F.~d\textquotesingle Alch\'{e}-Buc, E.~Fox, \& R.~Garnett
  (Curran Associates, Inc.), 8024--8035

\bibitem[{{Peebles}(1980)}]{Peebles1980}
{Peebles}, P.~J.~E. 1980, {The large-scale structure of the universe}

\bibitem[{{Perez} {et~al.}(2021){Perez}, {Malhotra}, {Rhoads}, \&
  {Tilvi}}]{Perez2021}
{Perez}, L.~A., {Malhotra}, S., {Rhoads}, J.~E., \& {Tilvi}, V. 2021, \apj,
  906, 58, \dodoi{10.3847/1538-4357/abc88b}

\bibitem[{{Pillepich} {et~al.}(2018){Pillepich}, {Springel}, {Nelson}, {Genel},
  {Naiman}, {Pakmor}, {Hernquist}, {Torrey}, {Vogelsberger}, {Weinberger}, \&
  {Marinacci}}]{Pillepich2018}
{Pillepich}, A., {Springel}, V., {Nelson}, D., {et~al.} 2018, \mnras, 473,
  4077, \dodoi{10.1093/mnras/stx2656}

\bibitem[{{Pisani} {et~al.}(2015){Pisani}, {Sutter}, {Hamaus}, {Alizadeh},
  {Biswas}, {Wandelt}, \& {Hirata}}]{Pisani2015}
{Pisani}, A., {Sutter}, P.~M., {Hamaus}, N., {et~al.} 2015, \prd, 92, 083531,
  \dodoi{10.1103/PhysRevD.92.083531}

\bibitem[{{Planck Collaboration} {et~al.}(2016){Planck Collaboration}, {Ade},
  {Aghanim}, {Arnaud}, {Ashdown}, {Aumont}, {Baccigalupi}, {Banday},
  {Barreiro}, {Bartlett}, {Bartolo}, {Battaner}, {Battye}, {Benabed},
  {Beno{\^\i}t}, {Benoit-L{\'e}vy}, {Bernard}, {Bersanelli}, {Bielewicz},
  {Bock}, {Bonaldi}, {Bonavera}, {Bond}, {Borrill}, {Bouchet}, {Boulanger},
  {Bucher}, {Burigana}, {Butler}, {Calabrese}, {Cardoso}, {Catalano},
  {Challinor}, {Chamballu}, {Chary}, {Chiang}, {Chluba}, {Christensen},
  {Church}, {Clements}, {Colombi}, {Colombo}, {Combet}, {Coulais}, {Crill},
  {Curto}, {Cuttaia}, {Danese}, {Davies}, {Davis}, {de Bernardis}, {de Rosa},
  {de Zotti}, {Delabrouille}, {D{\'e}sert}, {Di Valentino}, {Dickinson},
  {Diego}, {Dolag}, {Dole}, {Donzelli}, {Dor{\'e}}, {Douspis}, {Ducout},
  {Dunkley}, {Dupac}, {Efstathiou}, {Elsner}, {En{\ss}lin}, {Eriksen},
  {Farhang}, {Fergusson}, {Finelli}, {Forni}, {Frailis}, {Fraisse},
  {Franceschi}, {Frejsel}, {Galeotta}, {Galli}, {Ganga}, {Gauthier}, {Gerbino},
  {Ghosh}, {Giard}, {Giraud-H{\'e}raud}, {Giusarma}, {Gjerl{\o}w},
  {Gonz{\'a}lez-Nuevo}, {G{\'o}rski}, {Gratton}, {Gregorio}, {Gruppuso},
  {Gudmundsson}, {Hamann}, {Hansen}, {Hanson}, {Harrison}, {Helou},
  {Henrot-Versill{\'e}}, {Hern{\'a}ndez-Monteagudo}, {Herranz}, {Hildebrandt},
  {Hivon}, {Hobson}, {Holmes}, {Hornstrup}, {Hovest}, {Huang}, {Huffenberger},
  {Hurier}, {Jaffe}, {Jaffe}, {Jones}, {Juvela}, {Keih{\"a}nen}, {Keskitalo},
  {Kisner}, {Kneissl}, {Knoche}, {Knox}, {Kunz}, {Kurki-Suonio}, {Lagache},
  {L{\"a}hteenm{\"a}ki}, {Lamarre}, {Lasenby}, {Lattanzi}, {Lawrence}, {Leahy},
  {Leonardi}, {Lesgourgues}, {Levrier}, {Lewis}, {Liguori}, {Lilje},
  {Linden-V{\o}rnle}, {L{\'o}pez-Caniego}, {Lubin}, {Mac{\'\i}as-P{\'e}rez},
  {Maggio}, {Maino}, {Mandolesi}, {Mangilli}, {Marchini}, {Maris}, {Martin},
  {Martinelli}, {Mart{\'\i}nez-Gonz{\'a}lez}, {Masi}, {Matarrese}, {McGehee},
  {Meinhold}, {Melchiorri}, {Melin}, {Mendes}, {Mennella}, {Migliaccio},
  {Millea}, {Mitra}, {Miville-Desch{\^e}nes}, {Moneti}, {Montier}, {Morgante},
  {Mortlock}, {Moss}, {Munshi}, {Murphy}, {Naselsky}, {Nati}, {Natoli},
  {Netterfield}, {N{\o}rgaard-Nielsen}, {Noviello}, {Novikov}, {Novikov},
  {Oxborrow}, {Paci}, {Pagano}, {Pajot}, {Paladini}, {Paoletti}, {Partridge},
  {Pasian}, {Patanchon}, {Pearson}, {Perdereau}, {Perotto}, {Perrotta},
  {Pettorino}, {Piacentini}, {Piat}, {Pierpaoli}, {Pietrobon}, {Plaszczynski},
  {Pointecouteau}, {Polenta}, {Popa}, {Pratt}, {Pr{\'e}zeau}, {Prunet},
  {Puget}, {Rachen}, {Reach}, {Rebolo}, {Reinecke}, {Remazeilles}, {Renault},
  {Renzi}, {Ristorcelli}, {Rocha}, {Rosset}, {Rossetti}, {Roudier},
  {Rouill{\'e} d'Orfeuil}, {Rowan-Robinson}, {Rubi{\~n}o-Mart{\'\i}n},
  {Rusholme}, {Said}, {Salvatelli}, {Salvati}, {Sandri}, {Santos},
  {Savelainen}, {Savini}, {Scott}, {Seiffert}, {Serra}, {Shellard}, {Spencer},
  {Spinelli}, {Stolyarov}, {Stompor}, {Sudiwala}, {Sunyaev}, {Sutton},
  {Suur-Uski}, {Sygnet}, {Tauber}, {Terenzi}, {Toffolatti}, {Tomasi},
  {Tristram}, {Trombetti}, {Tucci}, {Tuovinen}, {T{\"u}rler}, {Umana},
  {Valenziano}, {Valiviita}, {Van Tent}, {Vielva}, {Villa}, {Wade}, {Wandelt},
  {Wehus}, {White}, {White}, {Wilkinson}, {Yvon}, {Zacchei}, \&
  {Zonca}}]{Planck2016}
{Planck Collaboration}, {Ade}, P.~A.~R., {Aghanim}, N., {et~al.} 2016, \aap,
  594, A13, \dodoi{10.1051/0004-6361/201525830}

\bibitem[{{Planck Collaboration} {et~al.}(2020){Planck Collaboration},
  {Aghanim}, {Akrami}, {Ashdown}, {Aumont}, {Baccigalupi}, {Ballardini},
  {Banday}, {Barreiro}, {Bartolo}, {Basak}, {Battye}, {Benabed}, {Bernard},
  {Bersanelli}, {Bielewicz}, {Bock}, {Bond}, {Borrill}, {Bouchet}, {Boulanger},
  {Bucher}, {Burigana}, {Butler}, {Calabrese}, {Cardoso}, {Carron},
  {Challinor}, {Chiang}, {Chluba}, {Colombo}, {Combet}, {Contreras}, {Crill},
  {Cuttaia}, {de Bernardis}, {de Zotti}, {Delabrouille}, {Delouis}, {Di
  Valentino}, {Diego}, {Dor{\'e}}, {Douspis}, {Ducout}, {Dupac}, {Dusini},
  {Efstathiou}, {Elsner}, {En{\ss}lin}, {Eriksen}, {Fantaye}, {Farhang},
  {Fergusson}, {Fernandez-Cobos}, {Finelli}, {Forastieri}, {Frailis},
  {Fraisse}, {Franceschi}, {Frolov}, {Galeotta}, {Galli}, {Ganga},
  {G{\'e}nova-Santos}, {Gerbino}, {Ghosh}, {Gonz{\'a}lez-Nuevo}, {G{\'o}rski},
  {Gratton}, {Gruppuso}, {Gudmundsson}, {Hamann}, {Handley}, {Hansen},
  {Herranz}, {Hildebrandt}, {Hivon}, {Huang}, {Jaffe}, {Jones}, {Karakci},
  {Keih{\"a}nen}, {Keskitalo}, {Kiiveri}, {Kim}, {Kisner}, {Knox},
  {Krachmalnicoff}, {Kunz}, {Kurki-Suonio}, {Lagache}, {Lamarre}, {Lasenby},
  {Lattanzi}, {Lawrence}, {Le Jeune}, {Lemos}, {Lesgourgues}, {Levrier},
  {Lewis}, {Liguori}, {Lilje}, {Lilley}, {Lindholm}, {L{\'o}pez-Caniego},
  {Lubin}, {Ma}, {Mac{\'\i}as-P{\'e}rez}, {Maggio}, {Maino}, {Mandolesi},
  {Mangilli}, {Marcos-Caballero}, {Maris}, {Martin}, {Martinelli},
  {Mart{\'\i}nez-Gonz{\'a}lez}, {Matarrese}, {Mauri}, {McEwen}, {Meinhold},
  {Melchiorri}, {Mennella}, {Migliaccio}, {Millea}, {Mitra},
  {Miville-Desch{\^e}nes}, {Molinari}, {Montier}, {Morgante}, {Moss}, {Natoli},
  {N{\o}rgaard-Nielsen}, {Pagano}, {Paoletti}, {Partridge}, {Patanchon},
  {Peiris}, {Perrotta}, {Pettorino}, {Piacentini}, {Polastri}, {Polenta},
  {Puget}, {Rachen}, {Reinecke}, {Remazeilles}, {Renzi}, {Rocha}, {Rosset},
  {Roudier}, {Rubi{\~n}o-Mart{\'\i}n}, {Ruiz-Granados}, {Salvati}, {Sandri},
  {Savelainen}, {Scott}, {Shellard}, {Sirignano}, {Sirri}, {Spencer},
  {Sunyaev}, {Suur-Uski}, {Tauber}, {Tavagnacco}, {Tenti}, {Toffolatti},
  {Tomasi}, {Trombetti}, {Valenziano}, {Valiviita}, {Van Tent}, {Vibert},
  {Vielva}, {Villa}, {Vittorio}, {Wandelt}, {Wehus}, {White}, {White},
  {Zacchei}, \& {Zonca}}]{Planck2018cosmology}
{Planck Collaboration}, {Aghanim}, N., {Akrami}, Y., {et~al.} 2020, \aap, 641,
  A6, \dodoi{10.1051/0004-6361/201833910}

\bibitem[{{Porter} {et~al.}(2014){Porter}, {Somerville}, {Primack}, \&
  {Johansson}}]{Porter2014a}
{Porter}, L.~A., {Somerville}, R.~S., {Primack}, J.~R., \& {Johansson}, P.~H.
  2014, \mnras, 444, 942, \dodoi{10.1093/mnras/stu1434}

\bibitem[{{Raichoor} {et~al.}(2017){Raichoor}, {Comparat}, {Delubac}, {Kneib},
  {Y{\`e}che}, {Dawson}, {Percival}, {Dey}, {Lang}, {Schlegel}, {Gorgoni},
  {Bautista}, {Brownstein}, {Mariappan}, {Seo}, {Tinker}, {Ross}, {Wang},
  {Zhao}, {Moustakas}, {Palanque-Delabrouille}, {Jullo}, {Newmann}, {Prada}, \&
  {Zhu}}]{Raichoor2017}
{Raichoor}, A., {Comparat}, J., {Delubac}, T., {et~al.} 2017, \mnras, 471,
  3955, \dodoi{10.1093/mnras/stx1790}

\bibitem[{{Ravanbakhsh} {et~al.}(2017){Ravanbakhsh}, {Oliva}, {Fromenteau},
  {Price}, {Ho}, {Schneider}, \& {Poczos}}]{Ravanbakhsh2017}
{Ravanbakhsh}, S., {Oliva}, J., {Fromenteau}, S., {et~al.} 2017, arXiv
  e-prints, arXiv:1711.02033.
\newblock \doarXiv{1711.02033}

\bibitem[{{Repp} \& {Szapudi}(2020)}]{ReppSzapudi2020}
{Repp}, A., \& {Szapudi}, I. 2020, \mnras, 498, L125,
  \dodoi{10.1093/mnrasl/slaa139}

\bibitem[{{Riccio} {et~al.}(2021){Riccio}, {Ma{\l}ek}, {Nanni}, {Boquien},
  {Buat}, {Burgarella}, {Donevski}, {Hamed}, {Hurley}, {Shirley}, \&
  {Pollo}}]{Riccio2021}
{Riccio}, G., {Ma{\l}ek}, K., {Nanni}, A., {et~al.} 2021, \aap, 653, A107,
  \dodoi{10.1051/0004-6361/202140854}

\bibitem[{{Rodrigues} {et~al.}(2023){Rodrigues}, {de Santi}, {Montero-Dorta},
  \& {Abramo}}]{Rodrigues2023}
{Rodrigues}, N. V.~N., {de Santi}, N. S.~M., {Montero-Dorta}, A.~D., \&
  {Abramo}, L.~R. 2023, arXiv e-prints, arXiv:2301.06398,
  \dodoi{10.48550/arXiv.2301.06398}

\bibitem[{{Rodr{\'\i}guez-Puebla} {et~al.}(2017){Rodr{\'\i}guez-Puebla},
  {Primack}, {Avila-Reese}, \& {Faber}}]{Rodriguez-Puebla2017}
{Rodr{\'\i}guez-Puebla}, A., {Primack}, J.~R., {Avila-Reese}, V., \& {Faber},
  S.~M. 2017, \mnras, 470, 651, \dodoi{10.1093/mnras/stx1172}

\bibitem[{{Rogers} {et~al.}(2019){Rogers}, {Peiris}, {Pontzen}, {Bird},
  {Verde}, \& {Font-Ribera}}]{Rogers2019}
{Rogers}, K.~K., {Peiris}, H.~V., {Pontzen}, A., {et~al.} 2019, \jcap, 2019,
  031, \dodoi{10.1088/1475-7516/2019/02/031}

\bibitem[{{Safi} \& {Farhang}(2021)}]{Safi2021}
{Safi}, S., \& {Farhang}, M. 2021, \apj, 914, 65,
  \dodoi{10.3847/1538-4357/abfa18}

\bibitem[{{Salvador} {et~al.}(2019){Salvador}, {S{\'a}nchez}, {Pagul},
  {Garc{\'\i}a-Bellido}, {Sanchez}, {Pujol}, {Frieman}, {Gaztanaga}, {Ross},
  {Sevilla-Noarbe}, {Abbott}, {Allam}, {Annis}, {Avila}, {Bertin}, {Brooks},
  {Burke}, {Carnero Rosell}, {Carrasco Kind}, {Carretero}, {Castander},
  {Cunha}, {De Vicente}, {Diehl}, {Doel}, {Evrard}, {Fosalba}, {Gruen},
  {Gruendl}, {Gschwend}, {Gutierrez}, {Hartley}, {Hollowood}, {James}, {Kuehn},
  {Kuropatkin}, {Lahav}, {Lima}, {March}, {Marshall}, {Menanteau}, {Miquel},
  {Romer}, {Roodman}, {Scarpine}, {Schindler}, {Smith}, {Soares-Santos},
  {Sobreira}, {Suchyta}, {Swanson}, {Tarle}, {Thomas}, {Vikram}, {Walker}, \&
  {DES Collaboration}}]{Salvador2019}
{Salvador}, A.~I., {S{\'a}nchez}, F.~J., {Pagul}, A., {et~al.} 2019, \mnras,
  482, 1435, \dodoi{10.1093/mnras/sty2802}

\bibitem[{{Samushia} {et~al.}(2021){Samushia}, {Slepian}, \&
  {Villaescusa-Navarro}}]{Samushia2021}
{Samushia}, L., {Slepian}, Z., \& {Villaescusa-Navarro}, F. 2021, \mnras, 505,
  628, \dodoi{10.1093/mnras/stab1199}

\bibitem[{{Santner} {et~al.}(2003){Santner}, {Williams}, \&
  {Notz}}]{Santner2003}
{Santner}, T.~J., {Williams}, B.~J., \& {Notz}, W.~I. 2003, {The Design and
  Analysis Computer Experiments}, \dodoi{10.1007/9781475737998}

\bibitem[{{Shao} {et~al.}(2022{\natexlab{a}}){Shao}, {Villaescusa-Navarro},
  {Villanueva-Domingo}, {Teyssier}, {Garrison}, {Gatti}, {Inman}, {Ni},
  {Steinwandel}, {Kulkarni}, {Visbal}, {Bryan}, {Angles-Alcazar}, {Castro},
  {Hernandez-Martinez}, \& {Dolag}}]{Shao2022a}
{Shao}, H., {Villaescusa-Navarro}, F., {Villanueva-Domingo}, P., {et~al.}
  2022{\natexlab{a}}, arXiv e-prints, arXiv:2209.06843.
\newblock \doarXiv{2209.06843}

\bibitem[{{Shao} {et~al.}(2022{\natexlab{b}}){Shao}, {Villaescusa-Navarro},
  {Genel}, {Spergel}, {Angl{\'e}s-Alc{\'a}zar}, {Hernquist}, {Dav{\'e}},
  {Narayanan}, {Contardo}, \& {Vogelsberger}}]{Shao2022b}
{Shao}, H., {Villaescusa-Navarro}, F., {Genel}, S., {et~al.}
  2022{\natexlab{b}}, \apj, 927, 85, \dodoi{10.3847/1538-4357/ac4d30}

\bibitem[{{Shao} {et~al.}(2022{\natexlab{c}}){Shao}, {Anbajagane}, \&
  {Chang}}]{ShaoM2022}
{Shao}, M., {Anbajagane}, D., \& {Chang}, C. 2022{\natexlab{c}}, arXiv
  e-prints, arXiv:2212.05964, \dodoi{10.48550/arXiv.2212.05964}

\bibitem[{{Sinha} \& {Garrison}(2020)}]{CORRFUNC}
{Sinha}, M., \& {Garrison}, L.~H. 2020, \mnras, 491, 3022,
  \dodoi{10.1093/mnras/stz3157}

\bibitem[{Sobol(1967)}]{Sobol1986}
Sobol, I. 1967, USSR Computational Mathematics and Mathematical Physics, 7, 86,
  \dodoi{https://doi.org/10.1016/0041-5553(67)90144-9}

\bibitem[{{Somerville} \& {Dav{\'e}}(2015)}]{SomervilleRomeel2015}
{Somerville}, R.~S., \& {Dav{\'e}}, R. 2015, \araa, 53, 51,
  \dodoi{10.1146/annurev-astro-082812-140951}

\bibitem[{{Somerville} {et~al.}(2008){Somerville}, {Hopkins}, {Cox},
  {Robertson}, \& {Hernquist}}]{Somerville2008}
{Somerville}, R.~S., {Hopkins}, P.~F., {Cox}, T.~J., {Robertson}, B.~E., \&
  {Hernquist}, L. 2008, \mnras, 391, 481,
  \dodoi{10.1111/j.1365-2966.2008.13805.x}

\bibitem[{{Somerville} {et~al.}(2015){Somerville}, {Popping}, \&
  {Trager}}]{Somerville2015}
{Somerville}, R.~S., {Popping}, G., \& {Trager}, S.~C. 2015, \mnras, 453, 4337,
  \dodoi{10.1093/mnras/stv1877}

\bibitem[{{Somerville} \& {Primack}(1999)}]{SomervillePrimack1999}
{Somerville}, R.~S., \& {Primack}, J.~R. 1999, \mnras, 310, 1087,
  \dodoi{10.1046/j.1365-8711.1999.03032.x}

\bibitem[{{Somerville} {et~al.}(2021){Somerville}, {Olsen}, {Yung}, {Pacifici},
  {Ferguson}, {Behroozi}, {Osborne}, {Wechsler}, {Pandya}, {Faber}, {Primack},
  \& {Dekel}}]{Somerville2021}
{Somerville}, R.~S., {Olsen}, C., {Yung}, L.~Y.~A., {et~al.} 2021, \mnras, 502,
  4858, \dodoi{10.1093/mnras/stab231}

\bibitem[{{Speagle} {et~al.}(2014){Speagle}, {Steinhardt}, {Capak}, \&
  {Silverman}}]{Speagle2014}
{Speagle}, J.~S., {Steinhardt}, C.~L., {Capak}, P.~L., \& {Silverman}, J.~D.
  2014, \apjs, 214, 15, \dodoi{10.1088/0067-0049/214/2/15}

\bibitem[{{Springel}(2010)}]{AREPOog}
{Springel}, V. 2010, \mnras, 401, 791, \dodoi{10.1111/j.1365-2966.2009.15715.x}

\bibitem[{{Springel} {et~al.}(2018){Springel}, {Pakmor}, {Pillepich},
  {Weinberger}, {Nelson}, {Hernquist}, {Vogelsberger}, {Genel}, {Torrey},
  {Marinacci}, \& {Naiman}}]{Springel2018}
{Springel}, V., {Pakmor}, R., {Pillepich}, A., {et~al.} 2018, \mnras, 475, 676,
  \dodoi{10.1093/mnras/stx3304}

\bibitem[{{Steinhardt} \& {Speagle}(2014)}]{SteinhardtSpeagle2014}
{Steinhardt}, C.~L., \& {Speagle}, J.~S. 2014, \apj, 796, 25,
  \dodoi{10.1088/0004-637X/796/1/25}

\bibitem[{{Sugiyama} {et~al.}(2020){Sugiyama}, {Takada}, {Kobayashi},
  {Miyatake}, {Shirasaki}, {Nishimichi}, \& {Park}}]{Sugiyama2020}
{Sugiyama}, S., {Takada}, M., {Kobayashi}, Y., {et~al.} 2020, \prd, 102,
  083520, \dodoi{10.1103/PhysRevD.102.083520}

\bibitem[{{Sutherland} \& {Dopita}(1993)}]{S-D1993}
{Sutherland}, R.~S., \& {Dopita}, M.~A. 1993, \apjs, 88, 253,
  \dodoi{10.1086/191823}

\bibitem[{{Szewciw} {et~al.}(2021){Szewciw}, {Beltz-Mohrmann}, {Berlind}, \&
  {Sinha}}]{Szewciw2021}
{Szewciw}, A.~O., {Beltz-Mohrmann}, G.~D., {Berlind}, A.~A., \& {Sinha}, M.
  2021, arXiv e-prints, arXiv:2110.03701.
\newblock \doarXiv{2110.03701}

\bibitem[{{Uhlemann} {et~al.}(2020){Uhlemann}, {Friedrich},
  {Villaescusa-Navarro}, {Banerjee}, \& {Codis}}]{Uhlemann2020}
{Uhlemann}, C., {Friedrich}, O., {Villaescusa-Navarro}, F., {Banerjee}, A., \&
  {Codis}, S. 2020, \mnras, 495, 4006, \dodoi{10.1093/mnras/staa1155}

\bibitem[{{Vakili} {et~al.}(2020){Vakili}, {Hoekstra}, {Bilicki}, {Fortuna},
  {Kuijken}, {Wright}, {Asgari}, {Brown}, {Dombrovskij}, {Erben}, {Giblin},
  {Heymans}, {Hildebrandt}, {Johnston}, {Joudaki}, \& {Kannawadi}}]{Vakili2020}
{Vakili}, M., {Hoekstra}, H., {Bilicki}, M., {et~al.} 2020, arXiv e-prints,
  arXiv:2008.13154.
\newblock \doarXiv{2008.13154}

\bibitem[{{van Daalen} {et~al.}(2016){van Daalen}, {Henriques}, {Angulo}, \&
  {White}}]{VanDaalen2016}
{van Daalen}, M.~P., {Henriques}, B. M.~B., {Angulo}, R.~E., \& {White}, S.
  D.~M. 2016, \mnras, 458, 934, \dodoi{10.1093/mnras/stw405}

\bibitem[{{Villaescusa-Navarro}
  {et~al.}(2020{\natexlab{a}}){Villaescusa-Navarro}, {Wandelt},
  {Angl{\'e}s-Alc{\'a}zar}, {Genel}, {Zorrilla Mantilla}, {Ho}, \&
  {Spergel}}]{Villaescusa-Navarro2020}
{Villaescusa-Navarro}, F., {Wandelt}, B.~D., {Angl{\'e}s-Alc{\'a}zar}, D.,
  {et~al.} 2020{\natexlab{a}}, arXiv e-prints, arXiv:2011.05992.
\newblock \doarXiv{2011.05992}

\bibitem[{{Villaescusa-Navarro}
  {et~al.}(2020{\natexlab{b}}){Villaescusa-Navarro}, {Hahn}, {Massara},
  {Banerjee}, {Delgado}, {Ramanah}, {Charnock}, {Giusarma}, {Li}, {Allys},
  {Brochard}, {Uhlemann}, {Chiang}, {He}, {Pisani}, {Obuljen}, {Feng},
  {Castorina}, {Contardo}, {Kreisch}, {Nicola}, {Alsing}, {Scoccimarro},
  {Verde}, {Viel}, {Ho}, {Mallat}, {Wandelt}, \& {Spergel}}]{Quijote}
{Villaescusa-Navarro}, F., {Hahn}, C., {Massara}, E., {et~al.}
  2020{\natexlab{b}}, \apjs, 250, 2, \dodoi{10.3847/1538-4365/ab9d82}

\bibitem[{{Villaescusa-Navarro}
  {et~al.}(2021{\natexlab{a}}){Villaescusa-Navarro}, {Angl{\'e}s-Alc{\'a}zar},
  {Genel}, {Spergel}, {Somerville}, {Dave}, {Pillepich}, {Hernquist}, {Nelson},
  {Torrey}, {Narayanan}, {Li}, {Philcox}, {La Torre}, {Maria Delgado}, {Ho},
  {Hassan}, {Burkhart}, {Wadekar}, {Battaglia}, {Contardo}, \&
  {Bryan}}]{CAMELSannouncement}
{Villaescusa-Navarro}, F., {Angl{\'e}s-Alc{\'a}zar}, D., {Genel}, S., {et~al.}
  2021{\natexlab{a}}, \apj, 915, 71, \dodoi{10.3847/1538-4357/abf7ba}

\bibitem[{{Villaescusa-Navarro}
  {et~al.}(2021{\natexlab{b}}){Villaescusa-Navarro}, {Genel}, {Angles-Alcazar},
  {Spergel}, {Li}, {Wandelt}, {Thiele}, {Nicola}, {Zorrilla Matilla}, {Shao},
  {Hassan}, {Narayanan}, {Dave}, \& {Vogelsberger}}]{Paco2021b_robust}
{Villaescusa-Navarro}, F., {Genel}, S., {Angles-Alcazar}, D., {et~al.}
  2021{\natexlab{b}}, arXiv e-prints, arXiv:2109.10360.
\newblock \doarXiv{2109.10360}

\bibitem[{{Villaescusa-Navarro}
  {et~al.}(2021{\natexlab{c}}){Villaescusa-Navarro}, {Genel}, {Angles-Alcazar},
  {Thiele}, {Dave}, {Narayanan}, {Nicola}, {Li}, {Villanueva-Domingo},
  {Wandelt}, {Spergel}, {Somerville}, {Zorrilla Matilla}, {Mohammad}, {Hassan},
  {Shao}, {Wadekar}, {Eickenberg}, {Wong}, {Contardo}, {Jo}, {Moser}, {Lau},
  {Machado Poletti Valle}, {Perez}, {Nagai}, {Battaglia}, \&
  {Vogelsberger}}]{CAMELS_CMD_announcement}
---. 2021{\natexlab{c}}, arXiv e-prints, arXiv:2109.10915.
\newblock \doarXiv{2109.10915}

\bibitem[{{Villaescusa-Navarro} {et~al.}(2022){Villaescusa-Navarro}, {Genel},
  {Angl{\'e}s-Alc{\'a}zar}, {Perez}, {Villanueva-Domingo}, {Wadekar}, {Shao},
  {Mohammad}, {Hassan}, {Moser}, {Lau}, {Machado Poletti Valle}, {Nicola},
  {Thiele}, {Jo}, {Philcox}, {Oppenheimer}, {Tillman}, {Hahn}, {Kaushal},
  {Pisani}, {Gebhardt}, {Delgado}, {Caliendo}, {Kreisch}, {Wong}, {Coulton},
  {Eickenberg}, {Parimbelli}, {Ni}, {Steinwandel}, {La Torre}, {Dave},
  {Battaglia}, {Nagai}, {Spergel}, {Hernquist}, {Burkhart}, {Narayanan},
  {Wandelt}, {Somerville}, {Bryan}, {Viel}, {Li}, {Irsic}, {Kraljic}, \&
  {Vogelsberger}}]{CAMELSpublic2022}
{Villaescusa-Navarro}, F., {Genel}, S., {Angl{\'e}s-Alc{\'a}zar}, D., {et~al.}
  2022, arXiv e-prints, arXiv:2201.01300.
\newblock \doarXiv{2201.01300}

\bibitem[{{Villanueva-Domingo} \& {Villaescusa-Navarro}(2022)}]{PabloPaco2022}
{Villanueva-Domingo}, P., \& {Villaescusa-Navarro}, F. 2022, \apj, 937, 115,
  \dodoi{10.3847/1538-4357/ac8930}

\bibitem[{{Vogelsberger} {et~al.}(2013){Vogelsberger}, {Genel}, {Sijacki},
  {Torrey}, {Springel}, \& {Hernquist}}]{Vogelsberger2013}
{Vogelsberger}, M., {Genel}, S., {Sijacki}, D., {et~al.} 2013, \mnras, 436,
  3031, \dodoi{10.1093/mnras/stt1789}

\bibitem[{{Walcher} {et~al.}(2011){Walcher}, {Groves}, {Budav{\'a}ri}, \&
  {Dale}}]{Walcher2011}
{Walcher}, J., {Groves}, B., {Budav{\'a}ri}, T., \& {Dale}, D. 2011, \apss,
  331, 1, \dodoi{10.1007/s10509-010-0458-z}

\bibitem[{{Walsh} \& {Tinker}(2019)}]{WalshTinker2019}
{Walsh}, K., \& {Tinker}, J. 2019, \mnras, 488, 470,
  \dodoi{10.1093/mnras/stz1351}

\bibitem[{{Wang} {et~al.}(2019){Wang}, {Mao}, {Zentner}, {van den Bosch},
  {Lange}, {Schafer}, {Villarreal}, {Hearin}, \& {Campbell}}]{Wang2019}
{Wang}, K., {Mao}, Y.-Y., {Zentner}, A.~R., {et~al.} 2019, \mnras, 488, 3541,
  \dodoi{10.1093/mnras/stz1733}

\bibitem[{{Wechsler} \& {Tinker}(2018)}]{WechslerTinker2018}
{Wechsler}, R.~H., \& {Tinker}, J.~L. 2018, \araa, 56, 435,
  \dodoi{10.1146/annurev-astro-081817-051756}

\bibitem[{{Weinberger} {et~al.}(2017){Weinberger}, {Springel}, {Hernquist},
  {Pillepich}, {Marinacci}, {Pakmor}, {Nelson}, {Genel}, {Vogelsberger},
  {Naiman}, \& {Torrey}}]{Weinberger2017}
{Weinberger}, R., {Springel}, V., {Hernquist}, L., {et~al.} 2017, \mnras, 465,
  3291, \dodoi{10.1093/mnras/stw2944}

\bibitem[{{Wen} {et~al.}(2020){Wen}, {Kemball}, \& {Saslaw}}]{Wen2020}
{Wen}, D., {Kemball}, A.~J., \& {Saslaw}, W.~C. 2020, \apj, 890, 160,
  \dodoi{10.3847/1538-4357/ab6d6f}

\bibitem[{{White}(1979)}]{White1979}
{White}, S.~D.~M. 1979, \mnras, 186, 145, \dodoi{10.1093/mnras/186.2.145}

\bibitem[{{White} \& {Frenk}(1991)}]{WhiteFrenk1991}
{White}, S. D.~M., \& {Frenk}, C.~S. 1991, \apj, 379, 52,
  \dodoi{10.1086/170483}

\bibitem[{{Wibking} {et~al.}(2020){Wibking}, {Weinberg}, {Salcedo}, {Wu},
  {Singh}, {Rodr{\'\i}guez-Torres}, {Garrison}, \& {Eisenstein}}]{Wibking2020}
{Wibking}, B.~D., {Weinberg}, D.~H., {Salcedo}, A.~N., {et~al.} 2020, \mnras,
  492, 2872, \dodoi{10.1093/mnras/stz3423}

\bibitem[{{Xu} {et~al.}(2018){Xu}, {Hu}, {Leskovec}, \& {Jegelka}}]{Xu2018}
{Xu}, K., {Hu}, W., {Leskovec}, J., \& {Jegelka}, S. 2018, arXiv e-prints,
  arXiv:1810.00826.
\newblock \doarXiv{1810.00826}

\bibitem[{{Xu} {et~al.}(2021){Xu}, {Kumar}, {Zehavi}, \& {Contreras}}]{Xu2021}
{Xu}, X., {Kumar}, S., {Zehavi}, I., \& {Contreras}, S. 2021, \mnras, 507,
  4879, \dodoi{10.1093/mnras/stab2464}

\bibitem[{{Yang} \& {Saslaw}(2011)}]{Yang2011}
{Yang}, A., \& {Saslaw}, W.~C. 2011, \apj, 729, 123,
  \dodoi{10.1088/0004-637X/729/2/123}

\bibitem[{{Yung} {et~al.}(2019{\natexlab{a}}){Yung}, {Somerville},
  {Finkelstein}, {Popping}, \& {Dav{\'e}}}]{Yung_i}
{Yung}, L.~Y.~A., {Somerville}, R.~S., {Finkelstein}, S.~L., {Popping}, G., \&
  {Dav{\'e}}, R. 2019{\natexlab{a}}, \mnras, 483, 2983,
  \dodoi{10.1093/mnras/sty3241}

\bibitem[{{Yung} {et~al.}(2020){Yung}, {Somerville}, {Finkelstein}, {Popping},
  {Dav{\'e}}, {Venkatesan}, {Behroozi}, \& {Ferguson}}]{Yung_iv}
{Yung}, L.~Y.~A., {Somerville}, R.~S., {Finkelstein}, S.~L., {et~al.} 2020,
  \mnras, 496, 4574, \dodoi{10.1093/mnras/staa1800}

\bibitem[{{Yung} {et~al.}(2019{\natexlab{b}}){Yung}, {Somerville}, {Popping},
  {Finkelstein}, {Ferguson}, \& {Dav{\'e}}}]{Yung_ii}
{Yung}, L.~Y.~A., {Somerville}, R.~S., {Popping}, G., {et~al.}
  2019{\natexlab{b}}, \mnras, 490, 2855, \dodoi{10.1093/mnras/stz2755}

\bibitem[{{Yung} {et~al.}(2022){Yung}, {Somerville}, {Ferguson}, {Finkelstein},
  {Gardner}, {Dav{\'e}}, {Bagley}, {Popping}, \& {Behroozi}}]{Yung_vi}
{Yung}, L.~Y.~A., {Somerville}, R.~S., {Ferguson}, H.~C., {et~al.} 2022,
  \mnras, 515, 5416, \dodoi{10.1093/mnras/stac2139}

\bibitem[{{Yung} {et~al.}(2023){Yung}, {Somerville}, {Finkelstein}, {Behroozi},
  {Dav{\'e}}, {Ferguson}, {Gardner}, {Popping}, {Malhotra}, {Papovich},
  {Rhoads}, {Bagley}, {Hirschmann}, \& {Koekemoer}}]{Yung2023}
{Yung}, L.~Y.~A., {Somerville}, R.~S., {Finkelstein}, S.~L., {et~al.} 2023,
  \mnras, 519, 1578, \dodoi{10.1093/mnras/stac3595}

\bibitem[{{Zehavi} {et~al.}(2005){Zehavi}, {Zheng}, {Weinberg}, {Frieman},
  {Berlind}, {Blanton}, {Scoccimarro}, {Sheth}, {Strauss}, {Kayo}, {Suto},
  {Fukugita}, {Nakamura}, {Bahcall}, {Brinkmann}, {Gunn}, {Hennessy},
  {Ivezi{\'c}}, {Knapp}, {Loveday}, {Meiksin}, {Schlegel}, {Schneider},
  {Szapudi}, {Tegmark}, {Vogeley}, {York}, \& {SDSS
  Collaboration}}]{Zehavi2005}
{Zehavi}, I., {Zheng}, Z., {Weinberg}, D.~H., {et~al.} 2005, \apj, 630, 1,
  \dodoi{10.1086/431891}

\bibitem[{{Zhai} {et~al.}(2019{\natexlab{a}}){Zhai}, {Benson}, {Wang}, {Yepes},
  \& {Chuang}}]{Zhai2019-cts}
{Zhai}, Z., {Benson}, A., {Wang}, Y., {Yepes}, G., \& {Chuang}, C.-H.
  2019{\natexlab{a}}, \mnras, 490, 3667, \dodoi{10.1093/mnras/stz2844}

\bibitem[{{Zhai} {et~al.}(2019{\natexlab{b}}){Zhai}, {Tinker}, {Becker},
  {DeRose}, {Mao}, {McClintock}, {McLaughlin}, {Rozo}, \&
  {Wechsler}}]{Zhai2019-aemulus3}
{Zhai}, Z., {Tinker}, J.~L., {Becker}, M.~R., {et~al.} 2019{\natexlab{b}},
  \apj, 874, 95, \dodoi{10.3847/1538-4357/ab0d7b}

\bibitem[{{Zhang} {et~al.}(2020){Zhang}, {Li}, {Liu}, {Spergel}, {Kreisch},
  {Pisani}, \& {Wandelt}}]{GZhang2020}
{Zhang}, G., {Li}, Z., {Liu}, J., {et~al.} 2020, \prd, 102, 083537,
  \dodoi{10.1103/PhysRevD.102.083537}

\bibitem[{{Zhou} {et~al.}(2018){Zhou}, {Cui}, {Hu}, {Zhang}, {Yang}, {Liu},
  {Wang}, {Li}, \& {Sun}}]{Zhou2018gnn}
{Zhou}, J., {Cui}, G., {Hu}, S., {et~al.} 2018, arXiv e-prints,
  arXiv:1812.08434.
\newblock \doarXiv{1812.08434}

\bibitem[{{Zou} {et~al.}(2019){Zou}, {Gao}, {Zhou}, \& {Kong}}]{Zou2019}
{Zou}, H., {Gao}, J., {Zhou}, X., \& {Kong}, X. 2019, \apjs, 242, 8,
  \dodoi{10.3847/1538-4365/ab1847}

\end{thebibliography}
\bibliographystyle{aasjournal}

\appendix

\section{Verifying our SC-SAM runs against observations and IllustrisTNG300} \label{app:ExtraSAMverif}

\begin{figure*}[b]
    \centering
    \includegraphics[width=\textwidth]{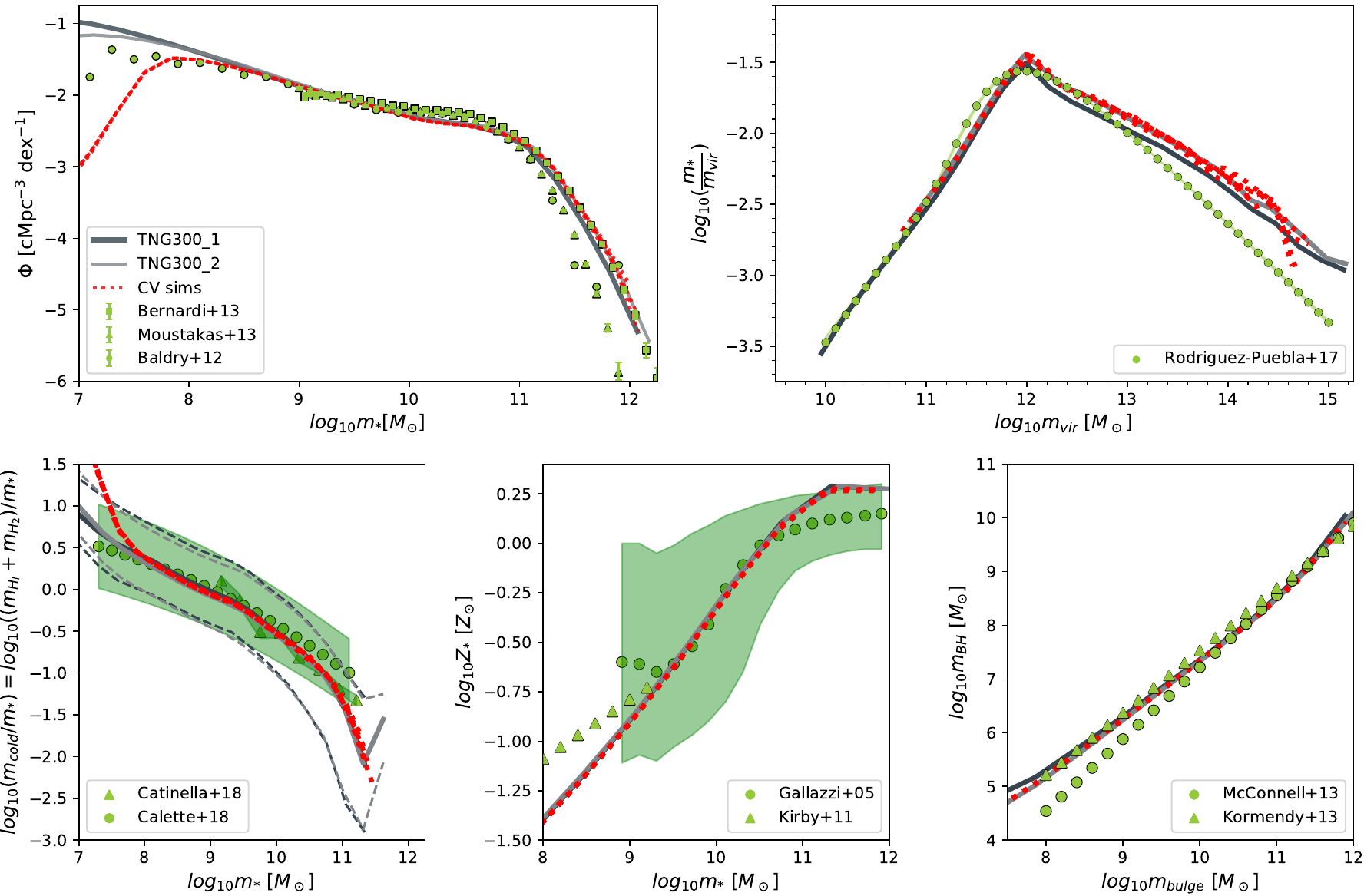}
    \caption{Verification of our fiducial model and set-up of the SC-SAM at $z=0$. The 5 CAMELS-SAM `cosmic variance' (CV) galaxy catalogs (red dotted lines) are compared against relevant $z\sim0$ observations (light green shapes) and the default SC-SAM run through IllustrisTNG300-1 (dark gray solid line; N=2500$^3$) and IllustrisTNG300-2 (light gray solid line; N=1250$^3$) dark matter-only volumes.
    Top row: stellar mass function (left), stellar mass-halo mass relationship (right). Bottom row: cold gas fraction vs. stellar mass for disk-dominated galaxies (left), stellar metallicity vs. stellar mass (middle), black hole mass vs. bulge mass (right). The observational data (green) used for the calibration of the SC-SAM are detailed in \citet{Somerville2015} and \citet{Yung_i}. The CV simulations were each generated with a different random initial seed, and all other parameters set to the fiducial value of \{$\Omega_{\text{M}}$, $\sigma_8$, A$_{\text{SN1}}$, A$_{\text{SN2}}$, A$_{\text{AGN}}$\}=\{0.3, 0.8, 1, 0, 1,\}.}
    \label{fig:SAM_verif_plotcCV0}
\end{figure*}

To confirm that our implementation of the SC-SAM is running correctly, we computed various useful summary statistics for the CV simulations and compared them to larger-volume SC-SAM runs and observations. In Figure \ref{fig:SAM_verif_plotcCV0}, we compare the SAM output of the CV simulations, the SAM output from the two highest resolution IllustrisTNG300 volumes, and various local universe observations for: the stellar mass function, the stellar mass-halo mass relation, the cold gas fraction vs. stellar mass,  stellar metallicity-stellar mass, and black hole mass-bulge mass relationships. The local universe observations were compiled in \citet{Yung_i} and include: \{\citealt{Bernardi2013, Moustakas2013, Baldry2012}\} for the stellar mass function; \{\citealt{Rodriguez-Puebla2017}\} for stellar mass-halo mass; \{\citealt{Catinella2018, Calette2018}\} for cold gas-stellar mass for disk-dominated galaxies (bulge-to-stellar mass ratio is less than 40\%); {\citealt{Gallazzi2005, Kirby2011}} for stellar metallicity vs. stellar mass; and {\citealt{McConnell2013, Kormendy2013}} for black hole mass vs. bulge mass relationships.

The great consistency with the larger IllustrisTNG300 SAM catalogs, and all SAM's agreement with local observations, support our use of the SC-SAM in this volume and resolution. We direct readers to the Figure 3 in \citet{Gabrielpillai2022} for more details for how the SC-SAM was run over IllustrisTNG300-1 (the highest resolution 300 volume), additional comparisons to even higher resolution volumes, and error bounds for most of the relations (removed here to minimize visual clutter). We note that IllustrisTNG used the cosmology of \citet[Table 4]{Planck2016}, with $\Omega_M=0.3089$, $\sigma_8=0.8159$, and $h=0.6774$. Though slightly different from the `fiducial' cosmology we used for the CV simulations, the SC-SAM still finds excellent agreement for these galaxy observables without retuning.

\begin{figure*}
    \centering
    \includegraphics[width=\textwidth]{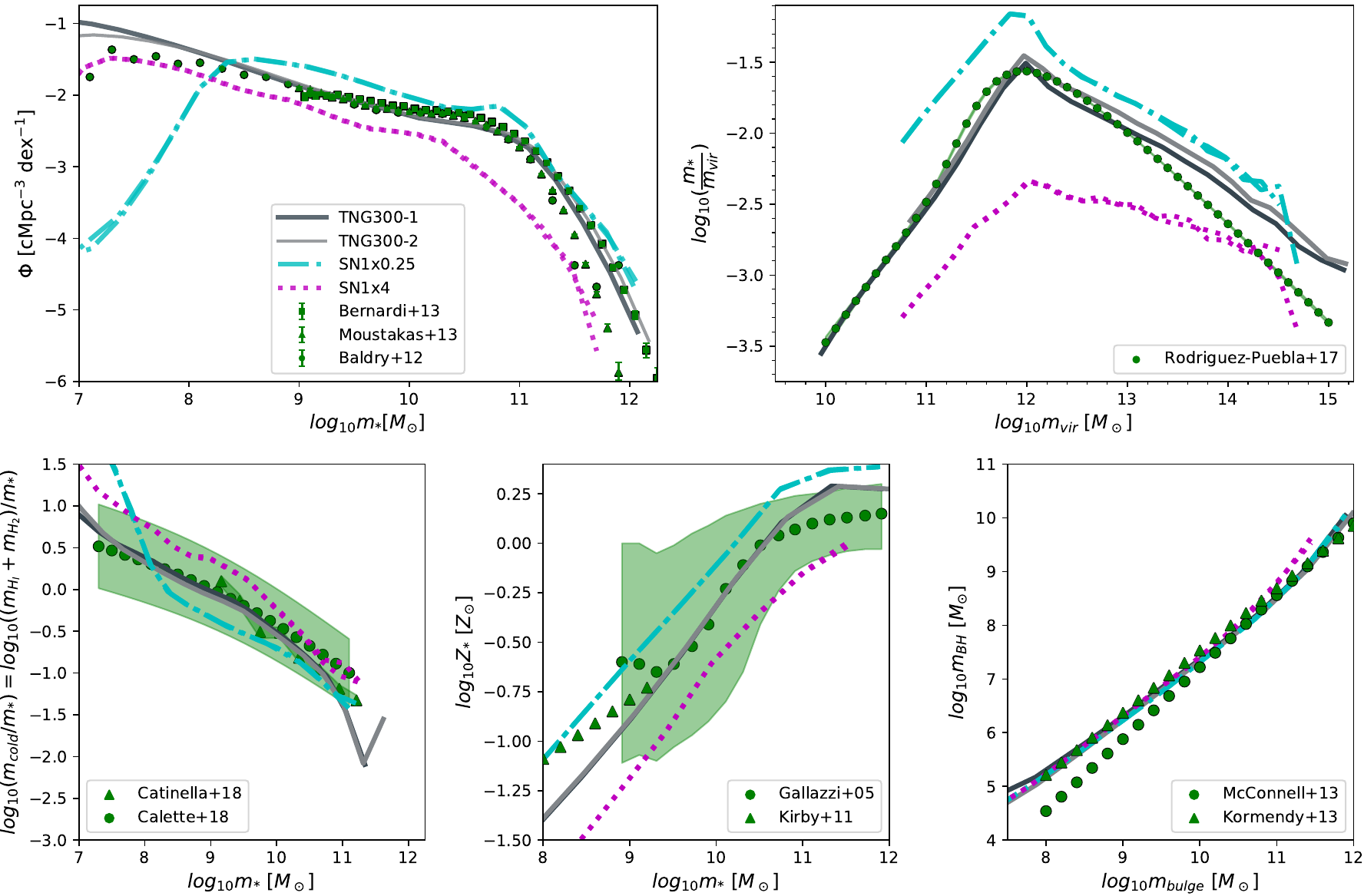}
     \caption{Key galaxy relationships for the 1P pair of  A$_{\text{SN1}}=0.25\times \epsilon_{\text{SN0}}$ (cyan) and A$_{\text{SN1}}=4.0\times \epsilon_{\text{SN0}}$ (magenta). See Fig.~\protect\ref{fig:SAM_verif_plotcCV0} for explanation of plotted quantities. }
     \label{fig:SAMplots_Asn1}
\end{figure*}

\begin{figure*}
    \centering
    \includegraphics[width=0.85\textwidth]{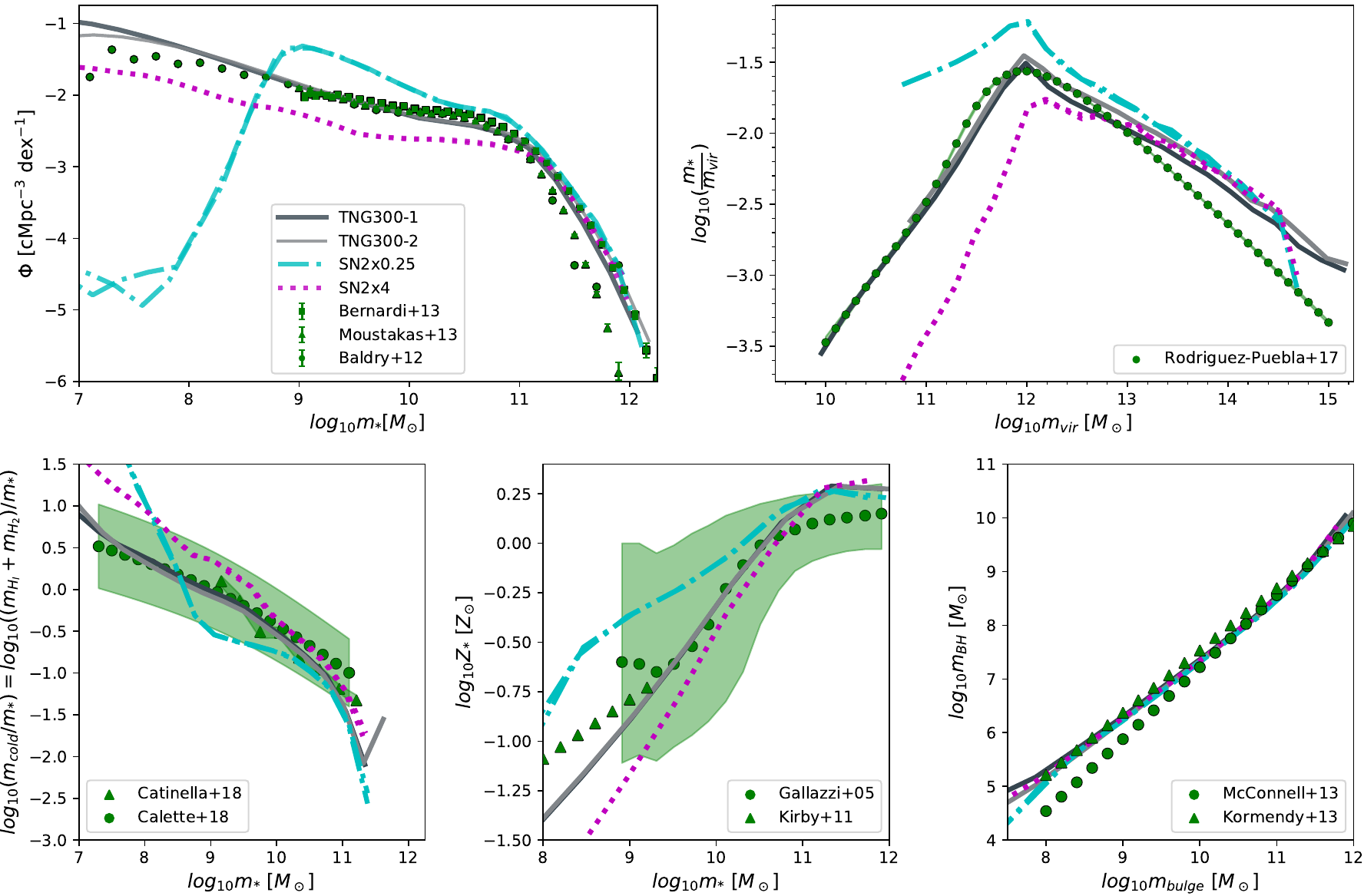}
     \caption{Key galaxy relationships for the 1P pair of A$_{\text{SN2}}=-2 + \alpha_{\text{rh}}$ (cyan) and A$_{\text{SN2}}=+2 + \alpha_{\text{rh}}$ (magenta). See Fig.~\protect\ref{fig:SAM_verif_plotcCV0} for explanation of plotted quantities.}
     \label{fig:SAMplots_Asn2}
\end{figure*}

\begin{figure*}
    \centering
    \includegraphics[width=0.85\textwidth]{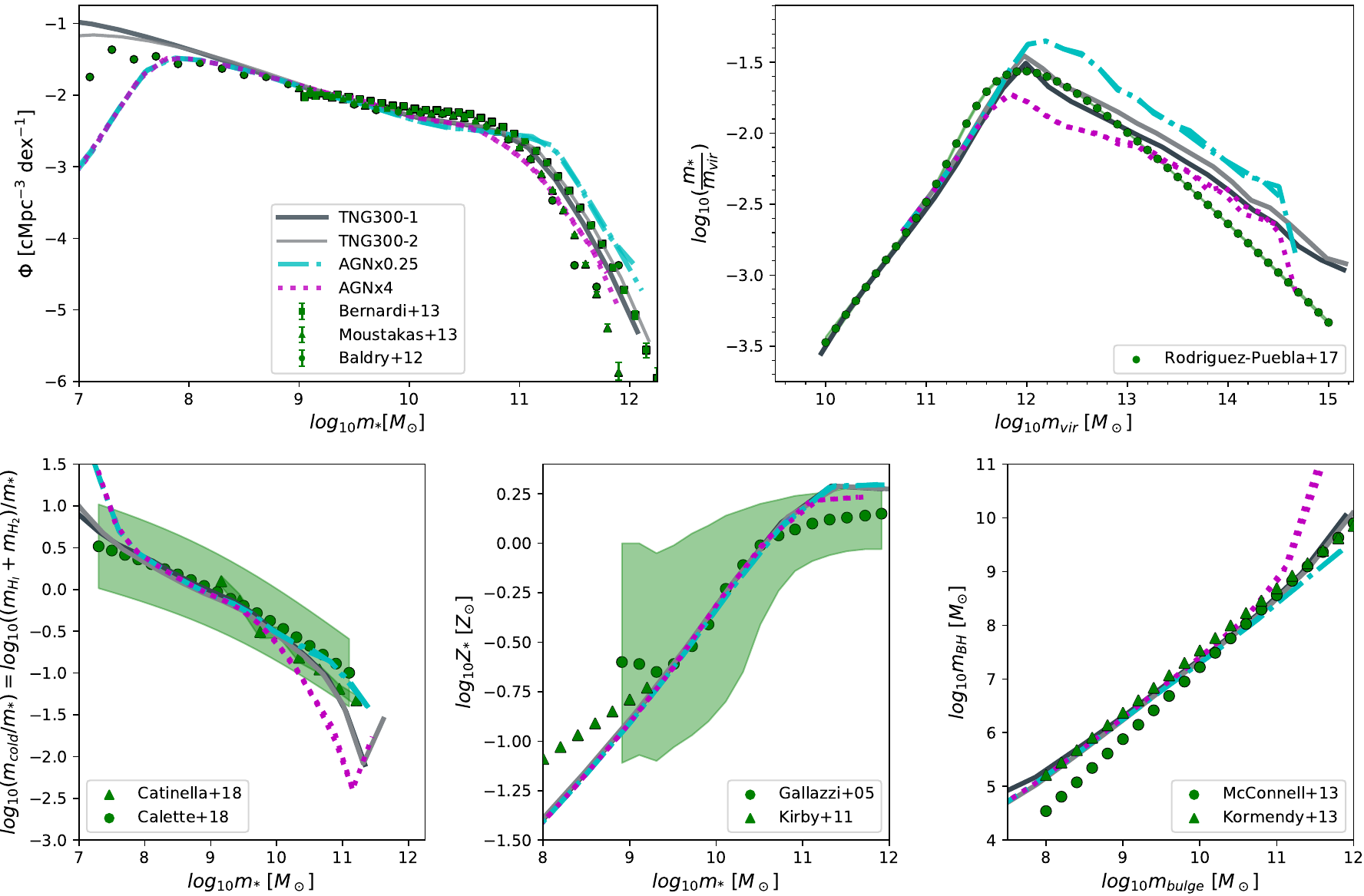}
     \caption{Key galaxy relationships for the 1P pair of A$_{\text{AGN}}=0.25\times \kappa_{\text{radio}}$ (cyan) and A$_{\text{AGN}}=4.0\times \kappa_{\text{radio}}$ (magenta). See Fig.~\protect\ref{fig:SAM_verif_plotcCV0} for explanation of plotted quantities.}
     \label{fig:SAMplots_AAGN}
\end{figure*}

Next, we examine the behavior of the ``1P" suite for the SC-SAM parameters across these galaxy relationships. In CAMELS-SAM, the ``1P" set is 12 galaxy catalogs that probe the minimum and maximum values of the SC-SAM parameters in our suite. We ran the SC-SAM with one of the SAM pre-factors at a time set to the minimum or maximum value atop the `fiducial' cosmology simulations CV$\_$0 and CV$\_$1. This creates three pairs of SC-SAM galaxy catalogs for each of the 2 N-body simulations. Figures \ref{fig:SAMplots_Asn1}, \ref{fig:SAMplots_Asn2}, and \ref{fig:SAMplots_AAGN} show key SC-SAM verification relationships for the 1P galaxy catalogs for A$_{\text{SN1}}$, A$_{\text{SN2}}$, and A$_{\text{AGN}}$ respectively. 

The stellar mass function and stellar mass-halo mass relation shows great sensitivity to A$_{\text{SN1}}$ and A$_{\text{SN2}}$, each with unique effects. Lower values of A$_{\text{SN1}}$ (weaker feedback) lift and increase the low-mass cutoff of the SMF slightly, while lower A$_{\text{SN2}}$ much more strongly sharpens the shape and increases the mass of most galaxies. High values of A$_{\text{SN1}}$ very strongly lowers the normalization of the stellar mass-halo mass relationship everywhere, whereas high A$_{\text{SN2}}$ makes the low-mass end of the stellar mass-halo mass relationship steeper. Finally, A$_{\text{SN1}}$ mostly shifts the stellar metallicity-stellar mass relationship up (weak feedback) and down (strong feedback), while A$_{\text{SN2}}$ affects the slope at all but the highest masses. Both A$_{\text{SN}}$ parameters affect the low-mass half of the cold gas fraction vs. stellar mass relationship in similar ways: lower values create a sharper drop and lower valleys.

A$_{\text{AGN}}$, on the other hand, has much more subtle effects. It has nearly no effect on stellar metallicity vs. stellar mass relationship, and shows only very mild effects on the SMF and black hole vs. bulge mass relationship at the highest stellar masses. Unlike the A$_{\text{SN}}$ parameters, its effect on the relationship of cold gas fraction vs. stellar mass is on the higher mass end. It does, however, show strong effects on the stellar mass-halo mass relationship on the right/higher-mass half of the `mountain', suppressing the stellar mass values as its effect is strengthened. 

\clearpage

\section{Tests for neural network accuracy, continued} \label{app:chi_LFI}

As discussed in \textsection \ref{subsubsec:Chi2}, part of our verification of our neural network's accuracy under the likelihood-free inference involves examining the distribution of $\chi_j(\theta_i)$ for the test set. This confirmation of the LFI is nearly identical to that in Figure 3 of \citet{Jeffrey2022}. The authors developed the moment networks framework at the heart of our LFI loss in \citet{MomentNetworks}, and apply it in the context of detecting the primordial B-mode of the Cosmic Microwave Background. Even with a model with more than $10^5$ parameters, the moment networks gave good results with accurately estimated errors. We apply the same test to probe how underestimated or inaccurate our errors may be.

\begin{figure*}[b]
     \centering
     \begin{subfloat}[`All' clustering of dark matter halos with mass greater than $2\times10^{11}$ M$_{\odot}$ and randomly down-sampled to 0.005 $h^{3}$ cMpc$^{-3}$.]{
    \includegraphics[width=\textwidth]{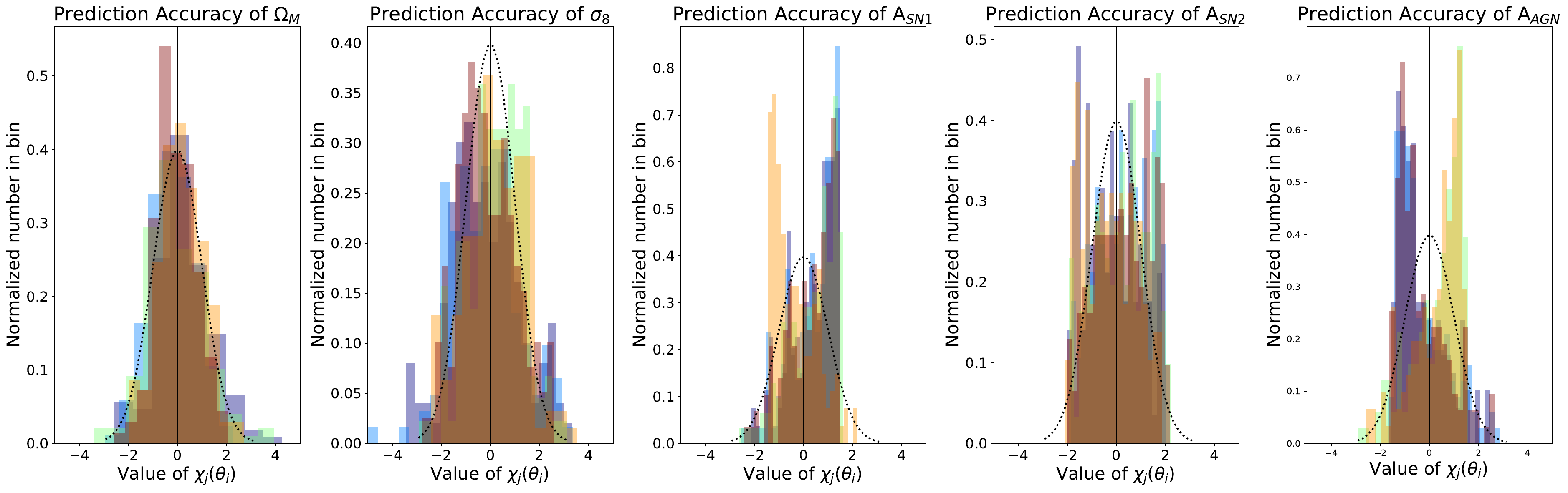}
     \label{fig:chi_hist_DMO}
     }
     \end{subfloat}
     \begin{subfloat}[`All' clustering of SC-SAM galaxies with stellar mass greater than $1\times10^{9}$ M$_{\odot}$ and randomly down-sampled to 0.005 $h^{3}$ cMpc$^{-3}$.]{
    \includegraphics[width=\textwidth]{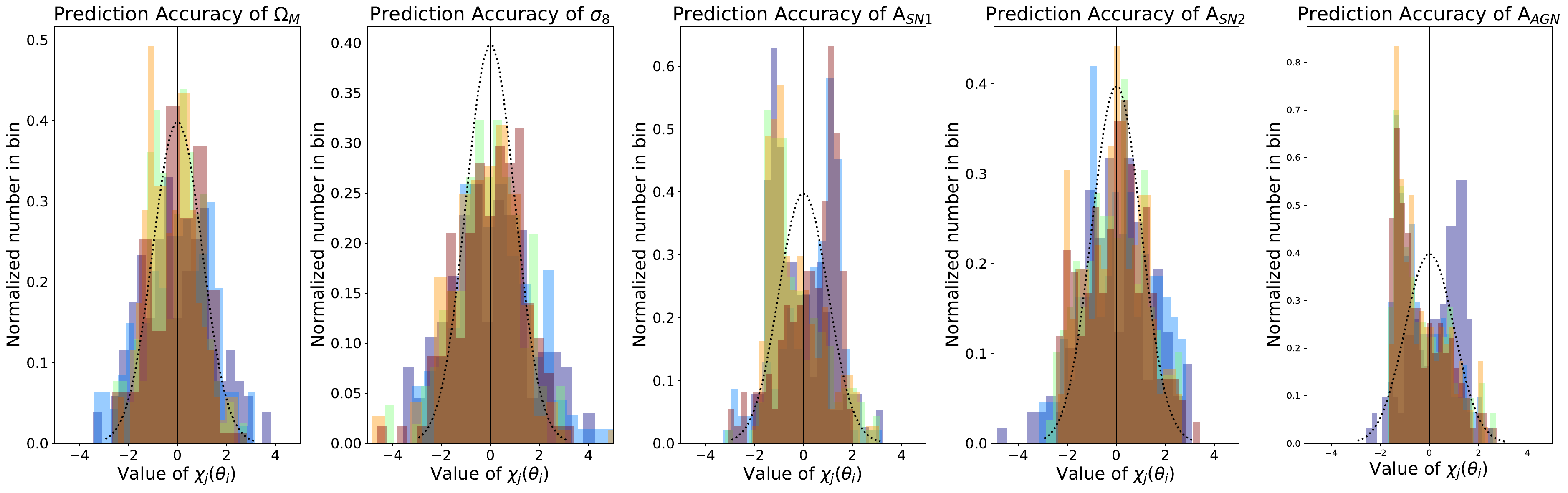}
     \label{fig:chi_hist_mstar}
     }
    \end{subfloat}
    \caption{The $\chi_i$ distribution across the test set for the top 5 best performing networks (each in a unique color). The dotted grey curve is a Gaussian whose center is at zero (black vertical line) and has a variance of one. }
    \label{fig:example_chiHistograms}
\end{figure*}

Figure \ref{fig:example_chiHistograms} shows the distribution of $\chi_j$ (for each $j$ simulation in the test set) for two of our clustering selections' neural network predictions each parameter $\theta_i$. We show the top 5 best-performing neural networks as a unique color. As discussed in \textsection \ref{subsec:NNimplementation}, the top few best-performing neural networks are all quite similar in overall behavior, so this highlights the general behavior of the LFI predictions for the given selections. If the LFI loss and moment network are working as they should to compute the mean and variances of the posteriors, the $\chi(\theta_i)$ distributions will have a mean of 0 and variance of 1. A pattern emerges among the $\chi(\theta_i)$ distributions we create: parameters that get decent constraints (i.e.\ about the 1:1 relationship and whose error bars look visually consistent) appear as rough Gaussian distributions with a mean of 0 and variance of 1. Unconstrained parameters appear flat within a similar range or peak at one of the extreme ends, though visually still appear to have a mean of 0 and variance of 1, indicating the moment network still performed as prescribed.
% When the LFI loss leads to accurate predictions, the distribution of $\chi(\theta_i)$ will approximate a Gaussian. Unconstrained parameters will appear flat within a similar range or peak at one of the extreme ends.

Figure \ref{fig:chi_hist_DMO} shows the $\chi_j$ distribution for the marginal posteriors given `all' clustering of dark matter halos with mass greater than $2\times10^{11}$ M$_{\odot}$ and randomly down-sampled to 0.005 $h^{3}$ cMpc$^{-3}$. The cosmological constraints tend to be very good, so their $\chi_j$ distributions are consistent with a Gaussian distribution centered at zero and with variance of one. The constraints on the SC-SAM astrophysical parameters are nonexistent (just around the mean of the prior and with large errors), which show up as $\chi_j$ distributions that are flat along the range or that peak at the Gaussian tails. Figure \ref{fig:chi_hist_mstar} shows the $\chi_j$ distribution for the marginal posteriors given `all' clustering of SC-SAM galaxies with stellar mass greater than $1\times10^{9}$ M$_{\odot}$ and randomly down-sampled to 0.005 $h^{3}$ cMpc$^{-3}$. This selection gives good constraints on the cosmological parameters and A$_{\text{SN2}}$. Even though the constraints are not as good on A$_{\text{SN2}}$ (i.e. spread around the 1:1 relationship with large error bars), the errors are not underestimated.

\clearpage

\section{Detailed Results: Constraining Cosmology \& Astrophysics with Clustering} \label{app:CosmoConstraints}

Here we share the the detailed experiments and results from \textsection \ref{sec:CosmoConstraints}.

\begin{table*}[t]
	\begin{center}
    \caption{Constraints for cosmological and astrophysical parameters across different SC-SAM galaxy selections with `all' galaxy clustering statistics at $z=\{0.0, 0.1, 0.5, 1.0\}$, when down sampling to a single number density (in $h^3$ cMpc $^{-3}$). }
	\begin{tabular}{p{2.4cm}p{1.cm}p{1.35cm}p{0.75cm}p{0.825cm}p{0.75cm}p{1cm}p{0.75cm}p{0.825cm}p{0.75cm}p{0.825cm}p{0.75cm}p{0.75cm}} 
	\hline \hline
	\multicolumn{3}{c}{Galaxy Selections} & \multicolumn{2}{c}{$\Omega_{\text{M}}$} & \multicolumn{2}{c}{$\sigma_8$} & \multicolumn{2}{c}{A$_{\text{SN1}}$} & \multicolumn{2}{c}{A$_{\text{SN2}}$} & \multicolumn{2}{c}{A$_{\text{AGN}}$} \\
    \hline
    Property & Value & Density & rMSE & $\bar{\sigma}$ & rMSE & $\bar{\sigma}$ & rMSE & $\bar{\sigma}$ & rMSE & $\bar{\sigma}$ & rMSE & $\bar{\sigma}$ \\
    \hline
		M$_{\text{halo}}$, log$_{10}$M$_{\odot}$& 11.3 & 0.005 & 0.014 & 0.014 & 0.032 & 0.024 & N/A & N/A & N/A & N/A & N/A & N/A \\
        ... & 12.08 & 0.001 & 0.019 & 0.016 & 0.064 & 0.040 & N/A & N/A & N/A & N/A & N/A & N/A \\
        M$_{\text{stellar}}$, log$_{10}$M$_{\odot}$ & 9.0 & 0.005 & 0.02 & 0.014 & 0.038 & 0.021 & 0.908 & 0.777 & 0.812 & 0.522 & 0.995 & 0.962 \\
        ... & 10.0 & 0.001 & 0.032 & 0.03 & 0.054 & 0.049 & 0.793 & 0.717 & 0.965 & 0.969 & 1.058 & 1.003 \\
        ... & 10.3 & 0.001 & 0.037 & 0.026 & 0.064 & 0.053 & 0.753 & 0.621 & 1.013 & 0.881 & 0.986 & 0.923 \\
		SFR, M$_{\odot}$ yr$^{-1}$ & 0.2 & 0.005 &  0.03 & 0.033 & 0.057 & 0.051 & 0.78 & 0.562 & 0.699 & 0.621 & 1.111 & 1.112 \\
        ... & 1.25 & 0.001 & 0.047 & 0.03 & 0.102 & 0.103 & 0.821 & 0.721 & 1.008 & 0.932 & 1.029 & 0.963 \\
        sSFR, log$_{10}$yr & -10.0 & 0.005 & 0.028 & 0.027 & 0.055 & 0.043 & 0.878 & 0.956 & 0.867 & 0.826 & 1.051 & 1.073 \\
        ... & -9.7 & 0.005 & 0.027 & 0.025 & 0.043 & 0.037 & 1.029 & 0.936 & 0.803 & 0.570 & 1.077 & 1.024 \\ %$10^{-9}$ yr
    \hline \hline	
    \label{table:GlxSelections_Constraints}
    \end{tabular}
	\end{center}
\end{table*}

\textbf{Halo Mass Clustering:} In \textsection \ref{subsubsec:MassConstraints} and Figure \ref{fig:DMonlyBest}, we show the best constraints we find on $\Omega_{\text{M}}$ and $\sigma_8$, from `all' clustering of halos with halo mass greater than $2\times10^{11}$ M$_{\odot}$ (or log$_{10}$ M$_{\text{halo}} > 11.3$ M$_{\odot}$), randomly down-sampled to a density of 0.005 $h^{3}$ cMpc$^{-3}$. The best-performing neural network produces $\Omega_{\text{M}}$ predictions accurate to rMSE = 0.014, or approximately 5\% about the fiducial value $\Omega_{\text{M}}=0.3$. Our LFI method also measures an average 1$\sigma$ standard deviation error of $\bar{\sigma}=0.014$ on $\Omega_{\text{M}}$. For $\sigma_8$, the rMSE error of the mean values across the test set is 0.032 (4\% for $\sigma_8=0.8$), while $\bar{\sigma}$ is 0.024 (3\%). We also probe a much higher halo mass selection of $1.2\times 10^{12}$ M$_{\odot}$ with a lower density down-sampling of 0.001 $h^{3}$ cMpc$^{-3}$, and find comparable though slightly worse constraints (likely due to Poisson noise introduced from the lower density). We take the tightest constraints from our dark matter selections as the `best' throughout this work.

\textbf{Stellar Mass Clustering:}
In Figure \ref{fig:SAMmstarBest}, we showed the constraints on $\Omega_{\text{M}}$ and $\sigma_8$ from `all' clustering at $z=\{0.0, 0.1, 0.5, 1.0\}$ of galaxies with stellar mass greater than $1\times10^9$ M$_{\odot}$, randomly down-sampled to 0.005 $h^{3}$ cMpc$^{-3}$. Other stellar mass selections we also tested include stellar mass greater than $1\times10^{10}$ M$_{\odot}$ to a density of 0.001 $h^{3}$ cMpc$^{-3}$; and stellar mass greater than $2\times10^{10}$ M$_{\odot}$ to a density of 0.001 $h^{3}$ cMpc$^{-3}$. We found the lower-threshold and higher down-sampling density selection yielded the tightest constraints on cosmology. The best-performing neural network, using the clustering of galaxies with stellar mass greater than $1\times10^9$ M$_{\odot}$, produces $\Omega_{\text{M}}$ predictions accurate to rMSE = 0.02, or approximately 7\% about the fiducial value $\Omega_{\text{M}}=0.3$. The LFI loss measures an average 1$\sigma$ standard deviation error of $\bar{\sigma}=0.014$ on $\Omega_{\text{M}}$, the same value as it found with dark-matter only clustering. For $\sigma_8$, the rMSE error of the mean values across the test set is 0.038 (approximately 5\% for $\sigma_8=0.8$), while $\bar{\sigma}=0.021$ (3\%). 

\begin{figure*}
    \centering
    \begin{subfloat}[Constraints from `all' clustering at $z=\{0.0, 0.1, 0.5, 1.0\}$ for galaxies with \textbf{\textit{stellar mass}} greater than $ 2\times10^{10}$ M$_{\odot}$ and down-sampled to a density of 0.001 $h^{3}$ cMpc$^{-3}$.]{
        \includegraphics[width=\textwidth]{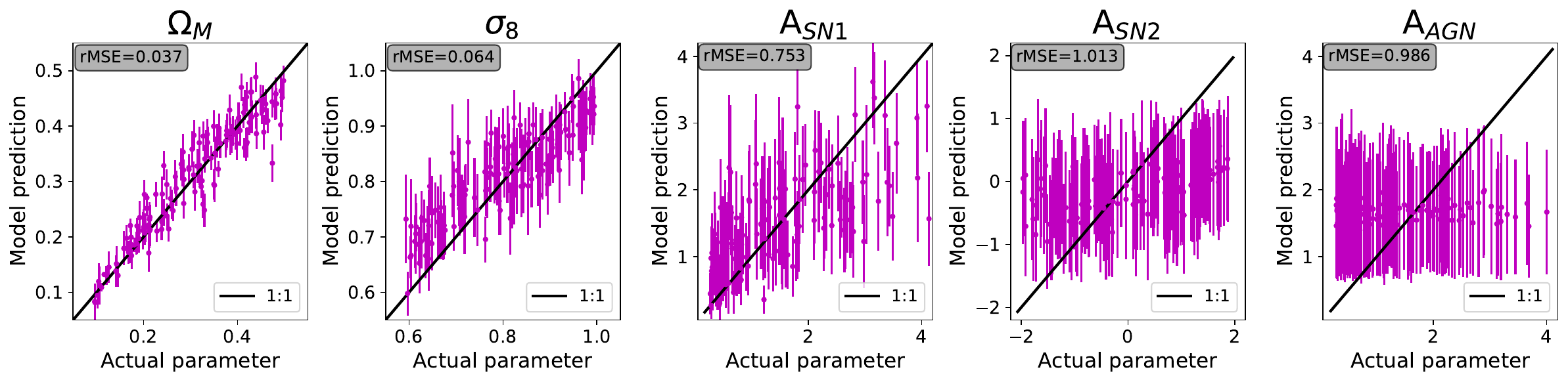}
    	\label{fig:mstargt2N1k_Sampled_LFI}
    	}
    \end{subfloat}
    \begin{subfloat}[Constraints from `all' clustering at $z=\{0.0, 0.1, 0.5, 1.0\}$ for galaxies with \textbf{\textit{star formation rate}} greater than 1.25 M$_{\odot}$ yr$^{-1}$ and down-sampled to a density of 0.001 $h^{3}$ cMpc$^{-3}$.]{
        \includegraphics[width=\textwidth]{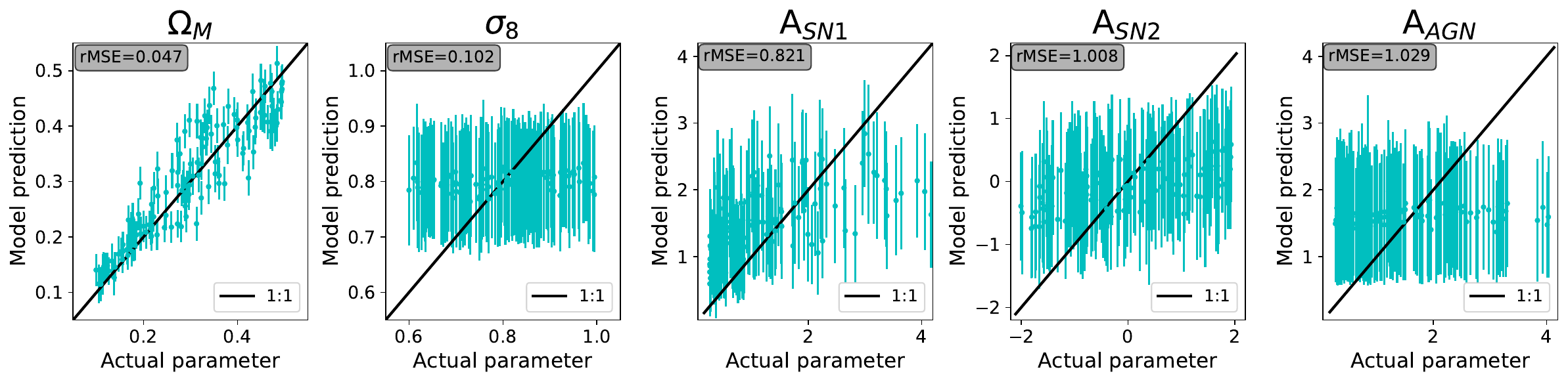}
    	\label{fig:sfrgt1p25N1k_Sampled_LFI}
    	}
    \end{subfloat}
    % \hfill
    \begin{subfloat}[Constraints from `all' clustering at $z=\{0.0, 0.1, 0.5, 1.0\}$ for galaxies with \textbf{\textit{specific star formation rate}} greater than 0.2 Gyr$^{-1}$ and down-sampled to a density of 0.005 $h^{3}$ cMpc$^{-3}$.]{
        \includegraphics[width=\textwidth]{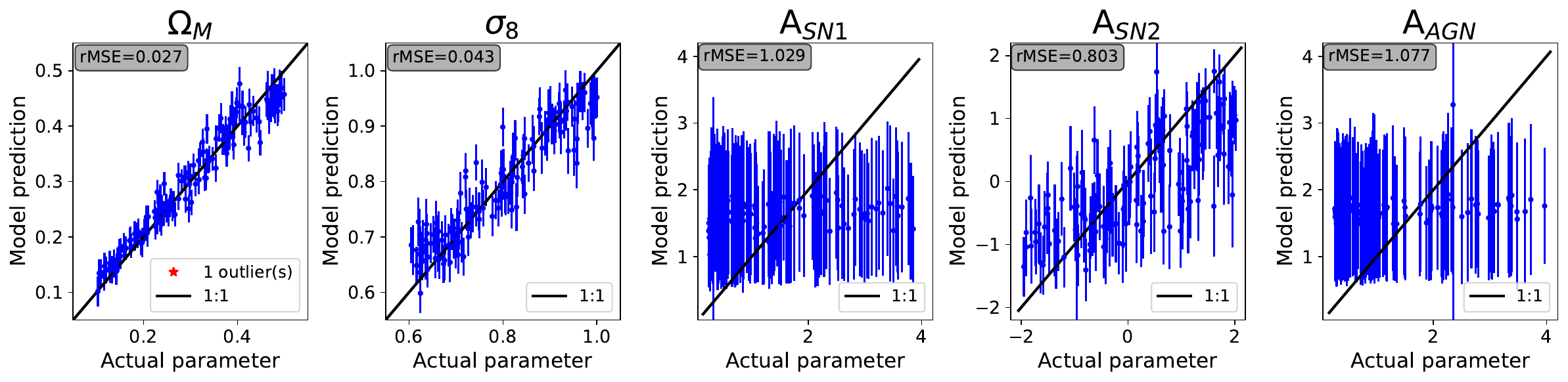}
        \label{fig:ssfrgt0p2N5k_Sampled_LFI}
        }
    \end{subfloat}
    \caption{Comparing how selections on basic galaxy properties generated by the SC-SAM affect the predictive power of a neural network trained on galaxy clustering: (a) stellar mass, (b) star formation rate, and (c) specific star formation rate. Detailed quantitative comparisons can be found in Table \ref{table:GlxSelections_Constraints}. Outliers (red stars, often out of range) are simulations in the test set whose `$Z$-value' (Eq. \ref{eq:Zvalue}) are greater than 6; see \textsection \ref{subsubsec:Chi2} for more details.}
	\label{fig:GlxsSelections_all5}
\end{figure*}

\textbf{SFR and sSFR:}  Under the set of choices we have adopted so far, with `all' clustering statistics at $z=\{0.0, 0.1, 0.5, 1.0\}$ and random down-sampling to a fixed number density, we probe SC-SAM galaxies with SFR$> 0.2$ M$_{\odot}$ yr$^{-1}$ randomly sampled to $\mathcal{N}=0.005\ h^{3}$ cMpc$^{-3}$, and also SFR $> 1.25$ M$_{\odot}$ yr$^{-1}$ randomly sampled to $\mathcal{N}=0.001\ h^{3}$ cMpc$^{-3}$. 
For sSFR, we probe SC-SAM galaxies with sSFR $> 0.1$ and Gyr$^{-1}$ and sSFR $> 0.2$ Gyr$^{-1}$, both randomly sampled to $\mathcal{N}=0.005\ h^{3}$ cMpc$^{-3}$. Figures \ref{fig:sfrgt1p25N1k_Sampled_LFI} and \ref{fig:ssfrgt0p2N5k_Sampled_LFI} show two of the neural networks trained on SFR- and sSFR- selected galaxy samples.

\textbf{Density Down-sampling:} The only repeated selection from \textsection \ref{subsec:CosmoConstraints_byselection} is M$_{\text{stellar}}>2 \times 10^{10}$ M$_{\odot}$. We also include M$_{\text{stellar}}>7 \times 10^{9}$ M$_{\odot}$ as a small test of threshold sensitivity, as well as SFR$
>1$ M$_{\odot}$ yr$^{-1}$. 

\begin{figure*}
        \centering
    \begin{subfloat}[Constraints from `all' clustering at $z=\{0.0, 0.1, 0.5, 1.0\}$ of SC-SAM galaxies with \textit{\textbf{stellar mass}} $> 2\times10^{10} h^{-1}$ M$_{\odot}$, with \textbf{\textit{no density down-sampling}}.]{
        \includegraphics[width=\textwidth]{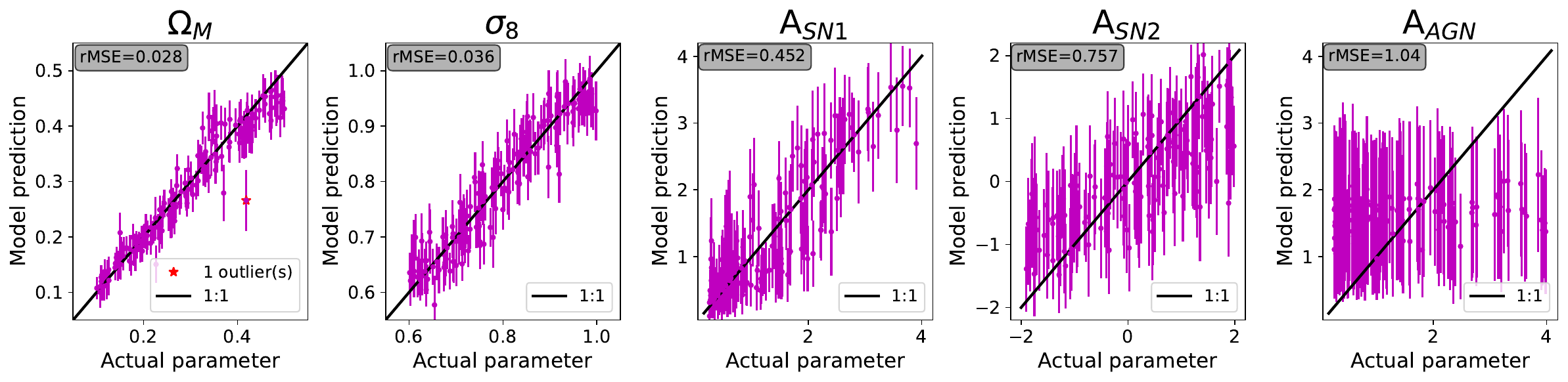}
    	\label{fig:mstargt20noDS_Sampled}
    	}
    \end{subfloat}
    \hfill
    \begin{subfloat}[Constraints from `all' clustering at $z=\{0.0, 0.1, 0.5, 1.0\}$ of SC-SAM galaxies with \textit{\textbf{star formation rate}} $> 1$ M$_{\odot}$ yr$^{-1}$, with \textbf{\textit{no density down-sampling}}.]{
        \includegraphics[width=\textwidth]{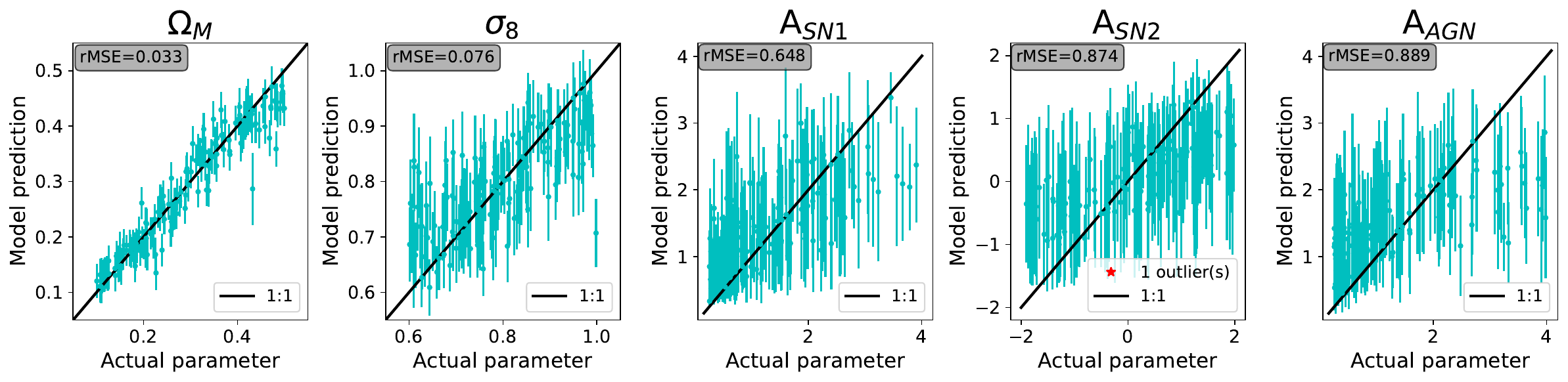}
        \label{fig:sfrgt1noDS_Sampled}
        }
    \end{subfloat}
    \caption{Exploring how \textit{not }down-sampling to a fixed density affects constraints on cosmology and the SC-SAM feedback parameters, for `all' clustering of  galaxies selected by (a) stellar mass (magenta) or (b) star formation rate (cyan). Detailed quantitative comparisons can be found in Table \ref{table:GlxSelections_noDS}. Outliers (red stars, often out of range) are simulations in the test set whose `$Z$-value' are greater than 6 (\textsection \ref{subsubsec:Chi2}).}
	\label{fig:noDS_Sampled}
\end{figure*}

\begin{table*}
	\begin{center}
    \caption{Constraints for cosmological and astrophysical parameters from different SAM galaxy selections with `all' galaxy clustering statistics at $z=\{0.0, 0.1, 0.5, 1.0\}$, when \textbf{\textit{not} correcting to a fixed number density}. The density ranges give a rough idea of the number of galaxies passing each selection across the LH suite, and are in units of $h^{3}$ cMpc$^{-3}$. }
	\begin{tabular}{p{2.4cm}p{1.1cm}p{2cm}p{0.75cm}p{0.75cm}p{0.75cm}p{0.75cm}p{0.75cm}p{0.75cm}p{0.75cm}p{0.75cm}p{0.75cm}p{0.75cm}} 
	\hline \hline
	\multicolumn{3}{c}{Galaxy Selections} & \multicolumn{2}{c}{$\Omega_{\text{M}}$} & \multicolumn{2}{c}{$\sigma_8$} & \multicolumn{2}{c}{A$_{\text{SN1}}$} & \multicolumn{2}{c}{A$_{\text{SN2}}$} & \multicolumn{2}{c}{A$_{\text{AGN}}$} \\
    \hline
    Property & Value & Density range & rMSE & $\bar{\sigma}$ & rMSE & $\bar{\sigma}$ & rMSE & $\bar{\sigma}$ & rMSE & $\bar{\sigma}$ & rMSE & $\bar{\sigma}$ \\
    \hline
		SFR, M$_{\odot}$ yr$^{-1}$ & 1.0 & 1.9e-4 to 1.9e-2 &  0.033 & 0.03 & 0.076 & 0.065 & 0.648 & 0.676 & 0.874 & 0.886 & 0.889 & 0.758 \\
        M$_{\text{stellar}}$, log$_{10}$M$_{\odot}$ & 9.845 & 2.4e-4 to 2.2e-2 & 0.025 & 0.021 & 0.036 & 0.030 & 0.648 & 0.549 & 0.747 & 0.612 & 1.048 & 0.857 \\
        ... & 10.3 & 1e-4 to 1e-2 & 0.028 & 0.029 & 0.036 & 0.043 & 0.452 & 0.478 & 0.757 & 0.717 & 1.04 & 0.978 \\
    \hline \hline	
    % \\
    \label{table:GlxSelections_noDS}
	\end{tabular}
	\end{center}
\end{table*}

\clearpage

% \section{Detailed Results: Testing Effect of Redshift on Constraints} \label{app:Zs_Clust_Focus}

\section{Detailed Results: Effect of Redshifts, Clustering Statistics, and Focused NNs} 
 \label{app:Zs_Clust_Focus}

\begin{table}
	\begin{center}
    \caption{Constraints from clustering statistics at a single redshift; using the same VPF and 2ptCF as with all four redshifts, but more of the CiC distribution (see \textsection \ref{subsubsec:NN_specifics}, \textsection \ref{subsec:RedshiftChoices}, and Table \ref{table:CiCRadii}). 
    %We update the clustering to use more of CiC. 
    Masses are log$_{10}$M$_{\odot}$, and densities for each selection are indicated with superscript symbols: $^*$ for 0.001, $^{\dag}$ for 0.005 $h^3$ cMpc $^{-3}$. `All' here indicates the four combined redshifts. }
	\begin{tabular}{p{1.0cm}p{1.1cm}p{0.5cm}p{0.7cm}p{0.7cm}p{0.7cm}p{0.7cm}}
	\hline \hline
	\multicolumn{3}{c}{Galaxy Selections} & \multicolumn{1}{c}{$\Omega_{\text{M}}$} & \multicolumn{1}{c}{$\sigma_8$} & \multicolumn{1}{c}{A$_{\text{SN1}}$} & \multicolumn{1}{c}{A$_{\text{SN2}}$}  \\
    \hline
    Property & Value & $z$ & rMSE & rMSE & rMSE & rMSE \\
    \hline
        M$_{\text{stellar}}$ & 10.3$^*$ & 0.0 & 0.057 & 0.072 & 0.716 & 0.987 \\
        ...                  & 10.3$^*$ & 0.1 & 0.047 & 0.075 & 0.681 & 1.025 \\
        ...                  & 10.3$^*$ & 0.5 & 0.048 & 0.097 & 0.613 & 1.069 \\
        ...                  & 10.3$^*$ & 1.0 & 0.046 & 0.098 & 0.655 & 1.078 \\
        ...                  & 10.3$^*$ & All & 0.037 & 0.064 & 0.753 & 1.013 \\
        M$_{\text{halo}}$ & 11.3$^{\dag}$ & 0.0 & 0.027 & 0.05 & N/A & N/A \\
        ...               & 11.3$^{\dag}$ & 0.1 & 0.026 & 0.056 & N/A & N/A \\
        ...               & 11.3$^{\dag}$ & 0.5 & 0.023 & 0.057 & N/A & N/A \\
        ...               & 11.3$^{\dag}$ & 1.0 & 0.021 & 0.079 & N/A & N/A \\
        ...               & 11.3$^{\dag}$ & All & 0.020 & 0.032 & N/A & N/A \\
    \hline \hline	
    \label{table:SingleRedshift_rMSE}
	\end{tabular}
	\end{center}
\end{table}

\textbf{Comparing Redshifts:} Throughout this work, our default input to our neural networks has been clustering statistics from $z=\{0.0, 0.1, 0.5, 1.0\}$ combined together. This, for example, mimics possible future experiments leveraging similarly selected galaxy populations at different redshifts. However, in \textsection \ref{subsec:RedshiftChoices} and \ref{subsec:SCSAM_densityredshift}, we considered  how might 

We describe our clustering methodology in \textsection \ref{subsec:Clustering}, and specifically note Table \ref{table:CiCRadii} for the slight adjustments we made to what clustering we give to the single redshift neural network. We note that a more fair comparison would have followed the example of the clustering statistic tests in \textsection \ref{sec:CompareStats}, where the only adjustment we made was splitting the data up by statistic. However, we expand the CiC distributions given to the neural network at each redshift to answer: \textit{how good might our constraints be if we focus in on a single redshift, and give a neural network as much data as it can easily handle?}

\begin{figure*}[b]
    \centering
     \begin{subfloat}[$z=1.0$.]{
        \includegraphics[width=0.23\textwidth]{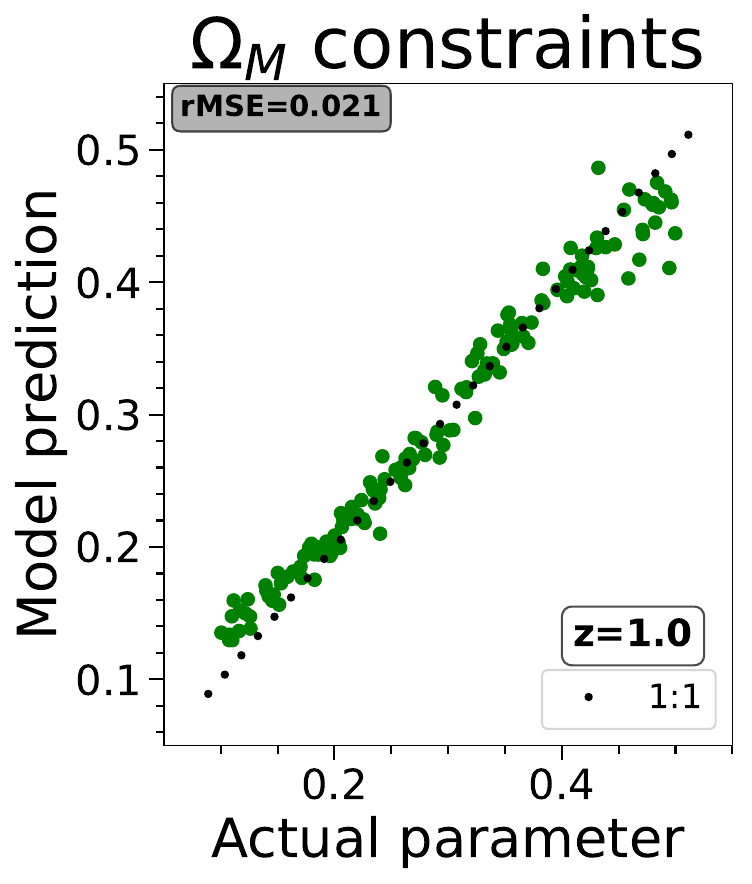}
         \label{fig:mhal_z1p0_Om}
         }
     \end{subfloat}
     \begin{subfloat}[$z=0.5$.]{
        \includegraphics[width=0.23\textwidth]{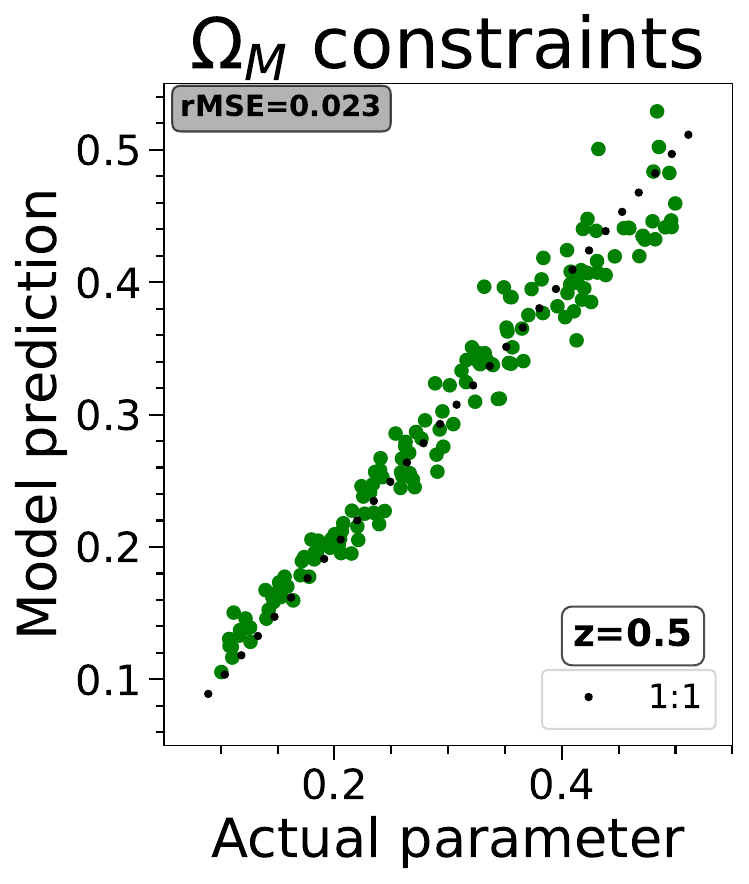}
         \label{fig:mhalo_z0p5_Om}
         }
     \end{subfloat}
    \begin{subfloat}[$z=0.1$.]{
        \includegraphics[width=0.23\textwidth]{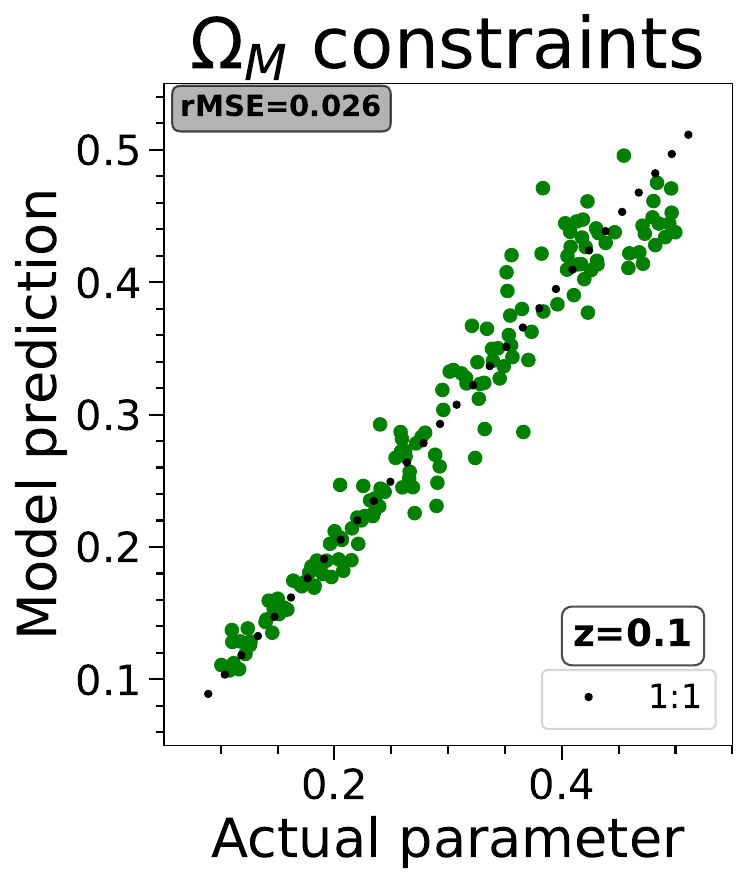}
         \label{fig:mhalo_z0p1_Om}
         }
     \end{subfloat}
    \begin{subfloat}[$z=0.0$.]{
        \includegraphics[width=0.23\textwidth]{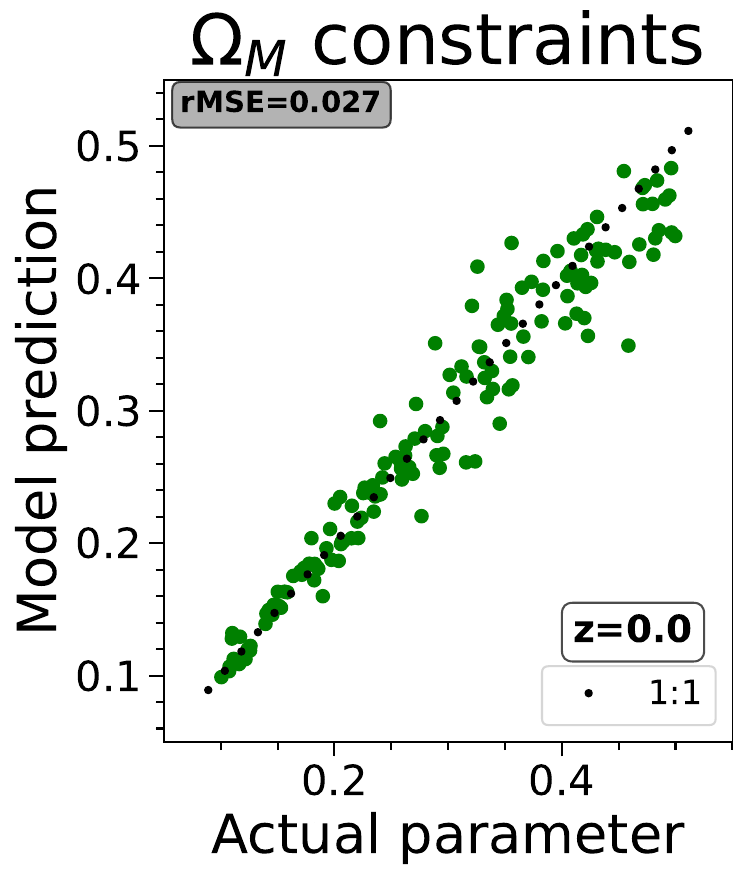}
         \label{fig:mhalo_z0p0_Om}
         }
     \end{subfloat}
     \caption{Examining redshift dependence on constraints for the the cosmological parameter $\Omega_{\text{M}}$, using the clustering of halos with mass greater than $2\times 10^{11}$ M$_{\odot}$ down-sampled to a density 0.005 $h^{3}$ cMpc$^{-3}$. See Table \ref{table:CiCRadii} for details for what exact distance scales and values were used for CiC for `all' clustering measured for this training.}
    \label{fig:DMonlyBestOm_redshift}
\end{figure*}

\begin{figure*}[b]
    \centering
     \begin{subfloat}[$z=1.0$]{
        \includegraphics[width=0.23\textwidth]{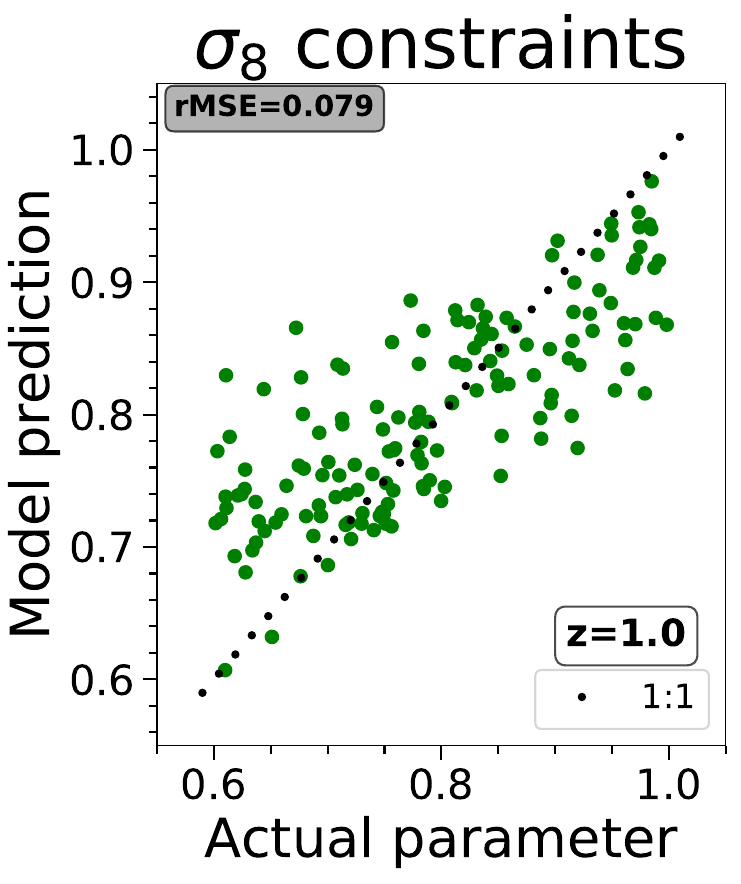}
         \label{fig:mhal_z1p0_s8}
         }
     \end{subfloat}
     \begin{subfloat}[$z=0.5$]{
        \includegraphics[width=0.23\textwidth]{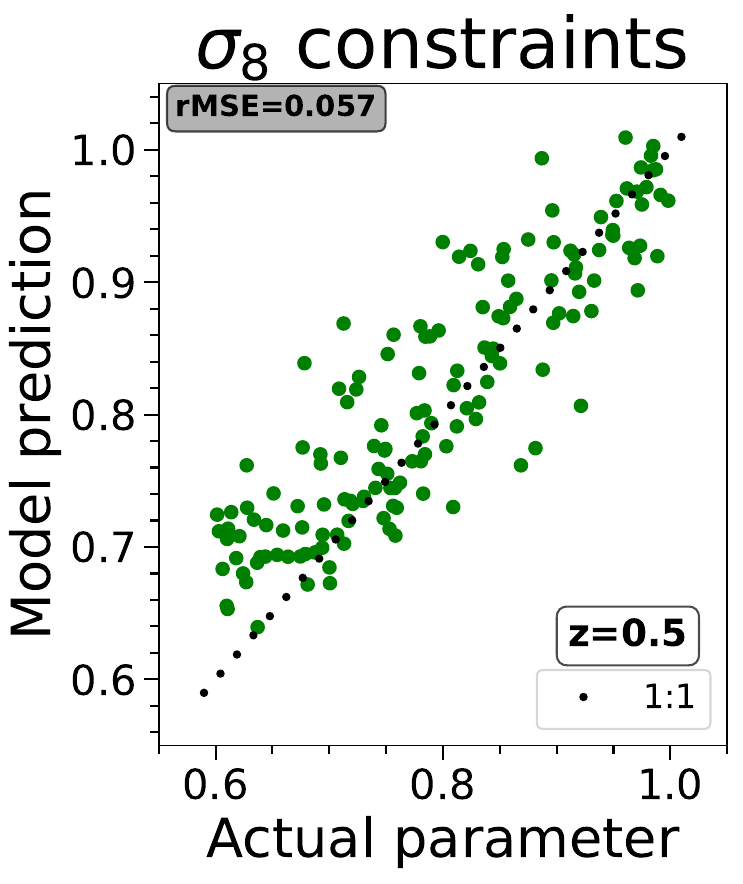}
         \label{fig:mhalo_z0p5_s8}
         }
     \end{subfloat}
    \begin{subfloat}[$z=0.1$]{
        \includegraphics[width=0.23\textwidth]{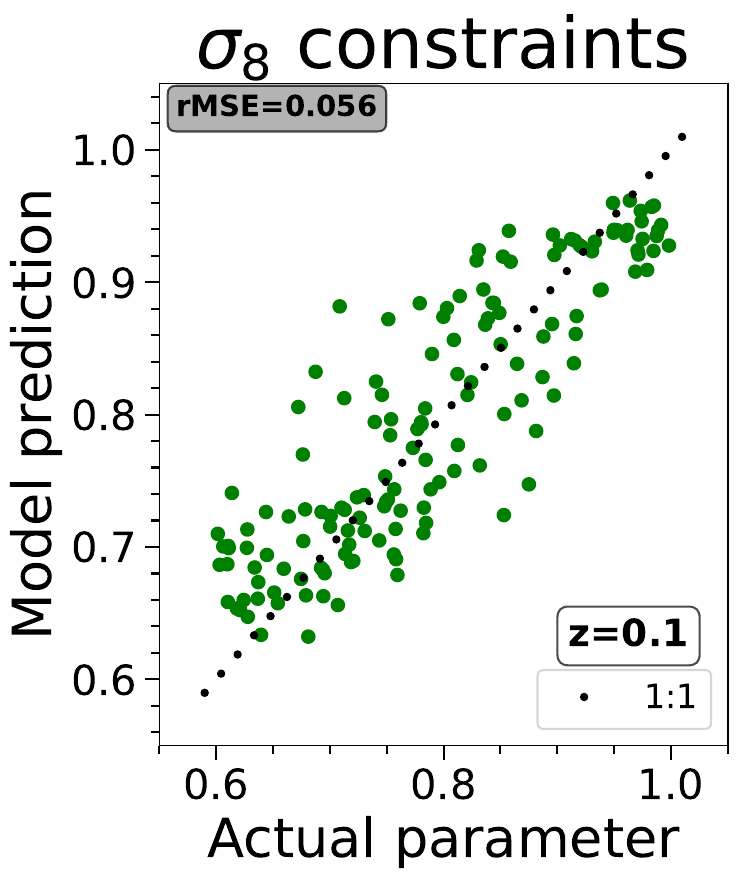}
         \label{fig:mhalo_z0p1_s8}
         }
     \end{subfloat}
    \begin{subfloat}[$z=0.0$]{
        \includegraphics[width=0.23\textwidth]{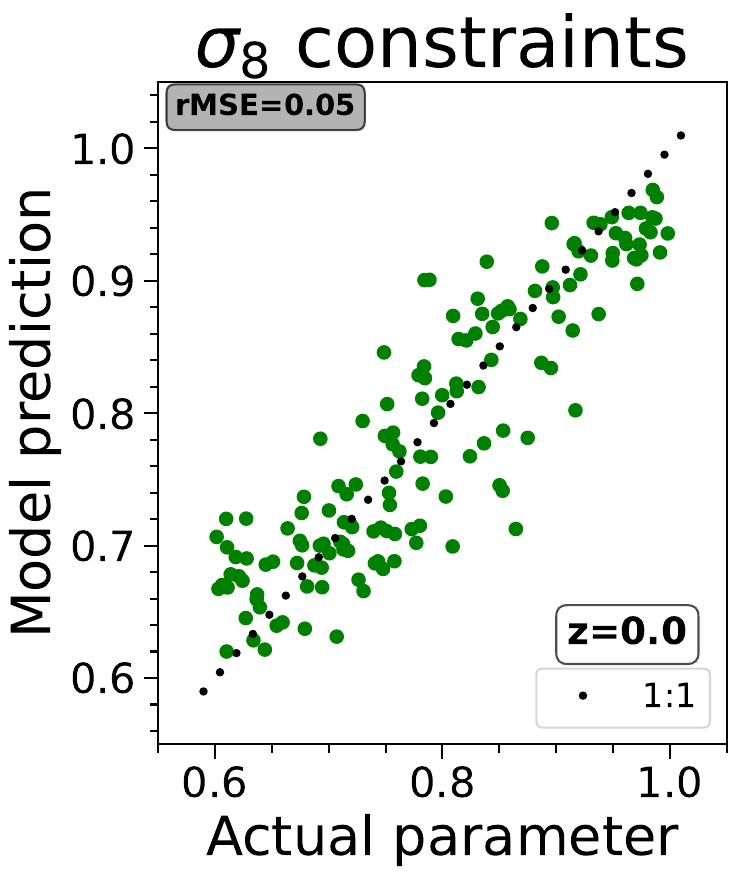}
         \label{fig:mhalo_z0p0_s8}
         }
     \end{subfloat}
     \caption{Similar to Figure \ref{fig:DMonlyBestOm_redshift}, but for      $\sigma_{8}$.}
    \label{fig:DMonlyBests8_redshift}
\end{figure*}

\begin{figure*}
    \centering
     \begin{subfloat}[Constraints with $z=1.0$ clustering.]{
        \includegraphics[width=0.95\textwidth]{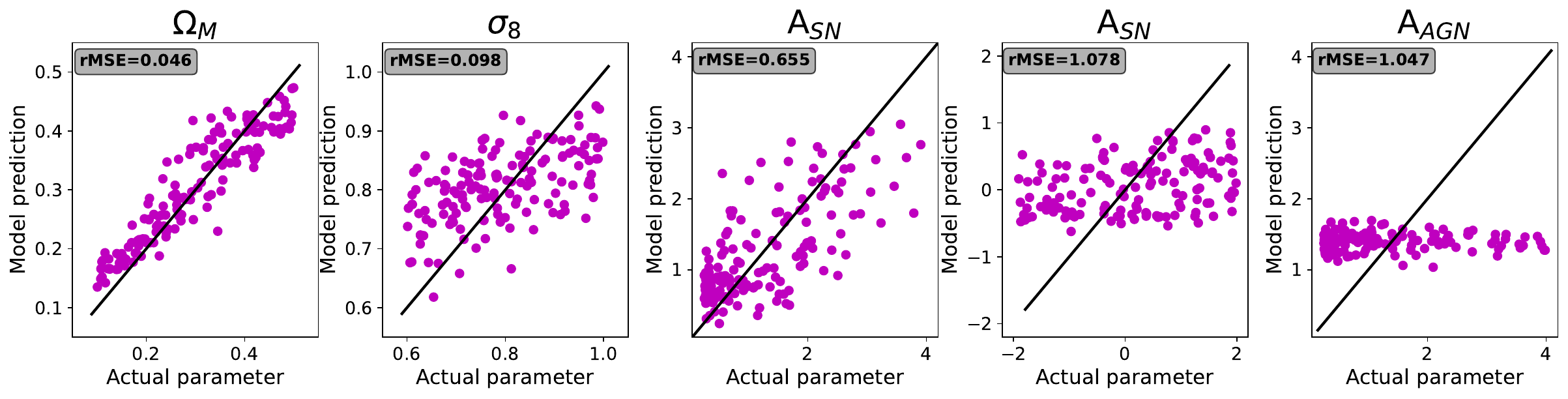}
         \label{fig:z1p0_mstar}
         }
     \end{subfloat}
     \begin{subfloat}[Constraints with $z=0.5$ clustering.]{
        \includegraphics[width=0.95\textwidth]{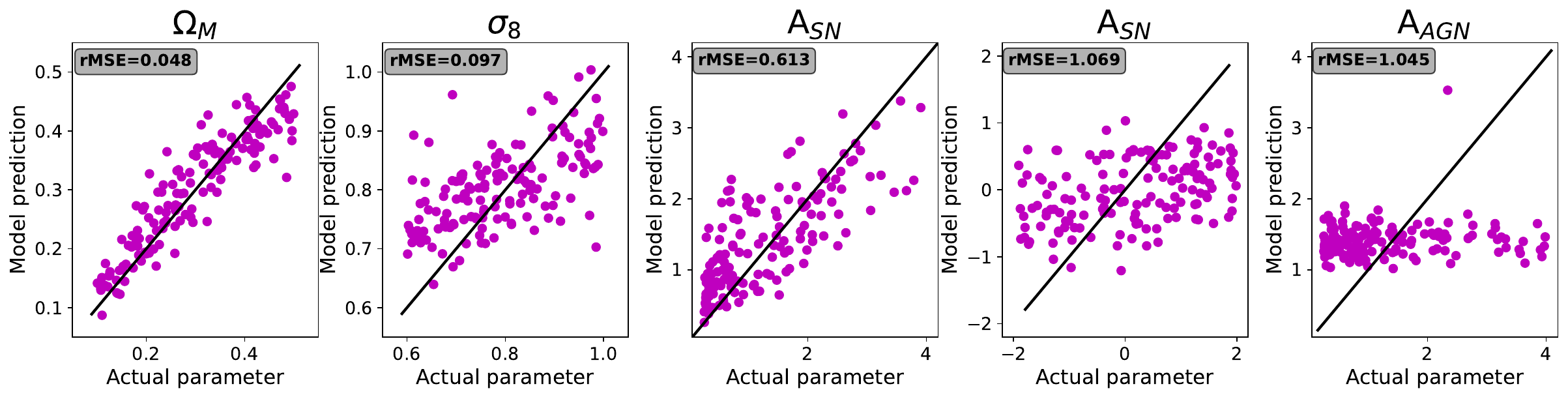}
         \label{fig:z0p5_mstar}
         }
     \end{subfloat}

    \begin{subfloat}[Constraints with $z=0.1$ clustering.]{
        \includegraphics[width=0.95\textwidth]{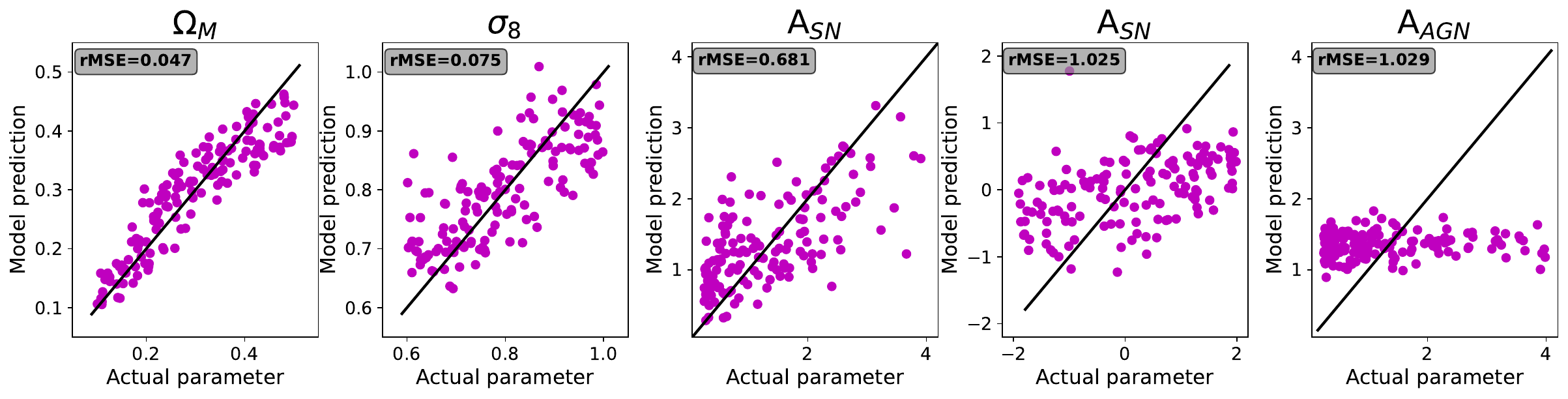}
         \label{fig:z0p1_mstar}
         }
     \end{subfloat}
    \begin{subfloat}[Constraints with $z=0$ clustering.]{
        \includegraphics[width=0.95\textwidth]{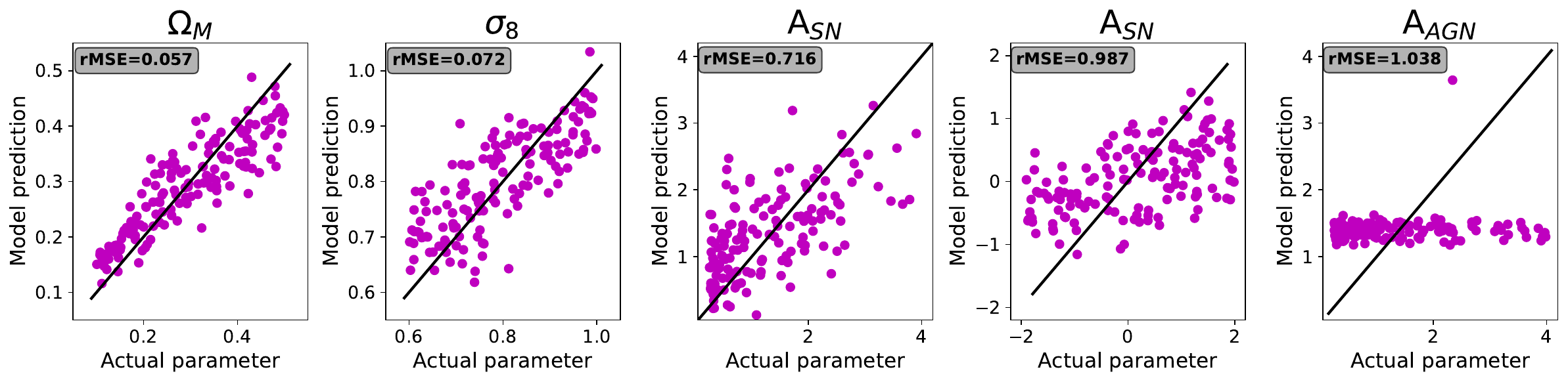}
         \label{fig:z0p0_mstar}
         }
     \end{subfloat}
     \caption{Examining redshift dependence on cosmological constraints for the clustering statistics of SAM galaxies with stellar mass greater than $2\times 10^{10}$ M$_{\odot}$, down-sampled to a density of 0.001 $h^{3}$ cMpc$^{-3}$. See Table \ref{table:CiCRadii} for details for what exact distance scales and values were used for CiC for `all' clustering measured for this training.}
    \label{fig:Mstar_redshift}
\end{figure*}

We share two representative examples of these neural networks in this appendix, both down-sampled to a density of 0.005 $h^{3}$ cMpc$^{-3}$: cosmology-only constraints for a high-density halo mass selection of $2\times 10^{11}$ M$_{\odot}$ (Figures \ref{fig:DMonlyBestOm_redshift} and \ref{fig:DMonlyBests8_redshift}), and constraints on all 5 parameters at once for a low-density stellar mass selection of $1\times 10^{9}$ M$_{\odot}$ (Figure \ref{fig:Mstar_redshift}).
% Figures \ref{fig:DMonlyBestOm_redshift} and \ref{fig:DMonlyBests8_redshift} focus on each redshift's (expanded) `all' clustering for dark matter halos with halo mass greater than $2\times 10^{11}$ M$_{\odot}$ and down-sampled to 0.005 $h^{3}$ cMpc$^{-3}$; and Appendix \ref{app:Zs_Clust_Focus} Figure \ref{fig:Mstar_redshift} focuses on each redshift's (expanded) `all' clustering for galaxies with stellar mass greater than $2\times 10^{10}$ M$_{\odot}$ and down-sampled to 0.001 $h^{3}$ cMpc$^{-3}$.

% \clearpage

\begin{table*}
    \begin{center}
    \caption{We compare the \textbf{cosmology} ($\Omega_{\text{M}}$ and $\sigma_8$) constraints from the best-performing neural networks \textbf{across clustering statistics} in Figures \ref{fig:DMonlyBestOm_ClusteringStat}, \ref{fig:DMonlyBests8_ClusteringStat}, and \ref{fig:Mstar_ClusteringCompare}. For $z=\{0.0, 0.1, 0.5, 1.0\}$, we use either the two-point correlation function, count-in-cells, the Void Probability Function, or `all' combined. The densities for each selection are indicated with superscript symbols: $^*$ means a density of 0.001 $h^3$ cMpc $^{-3}$, $^{\dag}$ means 0.005 $h^3$ cMpc $^{-3}$. We note that for these parameters, a rMSE on the LFI-loss means or the mean standard deviation $\bar{\sigma}$ around 0.1 indicate imprecise and inaccurate constraints, with error bars that span half the parameter space and predictions that are flat and around the mean of the prior. 
    }
    \footnotesize
	\begin{tabular}{p{2.05cm}p{0.7cm}p{0.6cm}p{0.55cm}p{0.6cm}p{0.55cm}p{0.6cm}p{0.55cm}p{0.6cm}p{0.55cm}p{0.6cm}p{0.55cm}p{0.6cm}p{0.55cm}p{0.6cm}p{0.55cm}p{0.6cm}p{0.55cm}} 
	\hline \hline
	\multicolumn{2}{c}{Clustering Statistic:} & \multicolumn{4}{c}{2pt Correlation Function} & \multicolumn{4}{c}{Count-in-Cells} &  \multicolumn{4}{c}{Void Probability Function} & \multicolumn{4}{c}{All Statistics}\\
    \hline
	\multicolumn{2}{c}{Galaxy Selections} & \multicolumn{2}{c}{$\Omega_{\text{M}}$} & \multicolumn{2}{c}{$\sigma_8$} & \multicolumn{2}{c}{$\Omega_{\text{M}}$} & \multicolumn{2}{c}{$\sigma_8$} & \multicolumn{2}{c}{$\Omega_{\text{M}}$} & \multicolumn{2}{c}{$\sigma_8$} & \multicolumn{2}{c}{$\Omega_{\text{M}}$} & \multicolumn{2}{c}{$\sigma_8$} \\
    \hline
    Property & Value & rMSE & $\bar{\sigma}$ & rMSE & $\bar{\sigma}$ & rMSE & $\bar{\sigma}$ & rMSE & $\bar{\sigma}$ & rMSE & $\bar{\sigma}$ & rMSE & $\bar{\sigma}$ & rMSE & $\bar{\sigma}$ & rMSE & $\bar{\sigma}$ \\
    \hline
        M$_{\text{halo}}$, $\log_{10}$M$_{\odot}$ 
        % $\log_{10}(\text{M}_{\text{halo}}/\text{M}_{\odot})$ 
        & 11.3$^{\dag}$ & 0.016 & 0.016 & 0.035 & 0.035 & 0.015 & 0.016 & 0.042 & 0.031 & 0.013 & 0.019 & 0.041 & 0.039 & 0.014 & 0.014 & 0.032 & 0.024 \\
        ... & 12.08$^*$ & 0.026 & 0.022 & 0.069 & 0.071 & 0.023 & 0.024 & 0.089 & 0.08 & 0.018 & 0.016 & 0.07 & 0.062 & 0.019 & 0.016 & 0.064 & 0.04 \\
        M$_{\text{stellar}}$, $\log_{10}$M$_{\odot}$ & 9.0$^{\dag}$ & 0.034 & 0.03 & 0.053 & 0.045 & 0.035 & 0.032 & 0.044 & 0.041 & 0.027 & 0.023 & 0.055 & 0.033 & 0.02 & 0.014 & 0.038 & 0.021 \\
        ... & 10.0$^*$ & 0.04 & 0.04 & 0.067 & 0.06 & 0.038 & 0.042 & 0.08 & 0.068 & 0.041 & 0.038 & 0.078 & 0.066  & 0.032 & 0.03 & 0.054 & 0.049 \\
        ... & 10.3$^*$ & 0.049 & 0.047 & 0.066 & 0.061 & 0.05 & 0.042 & 0.087 & 0.075 & 0.047 & 0.042 & 0.071 & 0.07 & 0.037 & 0.026 & 0.064 & 0.053 \\
        SFR, M$_{\odot}$ yr$^{-1}$ & 1.25$^*$ & 0.054 & 0.044 & 0.102 & 0.109 & 0.05 & 0.036 & 0.102 & 0.104 & 0.047 & 0.041 & 0.105 & 0.113 & 0.047 & 0.03 & 0.102 & 0.103 \\
    \hline \hline	
    \label{table:ClusteringStats_CosmoConstraints}
	\end{tabular}
	\end{center}
\end{table*}

\begin{table*}
	\begin{center}
    \caption{We compare the \textbf{SC-SAM supernova parameter} constraints from the best-performing neural networks \textbf{across clustering statistics} in Figure \ref{fig:Mstar_ClusteringCompare}. For $z=\{0.0, 0.1, 0.5, 1.0\}$, we use either the two-point correlation function, count-in-cells, the Void Probability Function, or `all' combined. The densities for each selection are indicated with superscript symbols: $^*$ means a density of 0.001 $h^3$ cMpc $^{-3}$, $^{\dag}$ means 0.005 $h^3$ cMpc $^{-3}$. We note that for these parameters, rMSE errors for the LFI means around 1.0 indicate imprecise and inaccurate constraints, with error bars that span half the parameter space and predictions that are flat and around the mean of prior. Parameters with rMSE less than 0.8 tend to show a rough 1:1 relationship but with considerable 1$\sigma$ errors.
    We dive further into constraining the A$_{\text{AGN}}$ parameter in \textsection \ref{subsec:SCSAMfocus}.}
    \footnotesize
	\begin{tabular}{p{2.05cm}p{0.7cm}p{0.6cm}p{0.55cm}p{0.6cm}p{0.55cm}p{0.6cm}p{0.55cm}p{0.6cm}p{0.55cm}p{0.6cm}p{0.55cm}p{0.6cm}p{0.55cm}p{0.6cm}p{0.55cm}p{0.6cm}p{0.55cm}} 
	\hline \hline
	\multicolumn{2}{c}{Clustering Statistic:} & \multicolumn{4}{c}{2pt Correlation Function} & \multicolumn{4}{c}{Count-in-Cells} &  \multicolumn{4}{c}{Void Probability Function} & \multicolumn{4}{c}{All Statistics}\\
    \hline
	\multicolumn{2}{c}{Galaxy Selections} & \multicolumn{2}{c}{A$_{\text{SN1}}$} & \multicolumn{2}{c}{A$_{\text{SN2}}$} & \multicolumn{2}{c}{A$_{\text{SN1}}$} & \multicolumn{2}{c}{A$_{\text{SN2}}$} & \multicolumn{2}{c}{A$_{\text{SN1}}$} & \multicolumn{2}{c}{A$_{\text{SN2}}$} & \multicolumn{2}{c}{A$_{\text{SN1}}$} & \multicolumn{2}{c}{A$_{\text{SN2}}$} \\
    \hline
    Property & Value & rMSE & $\bar{\sigma}$ & rMSE & $\bar{\sigma}$ & rMSE & $\bar{\sigma}$ & rMSE & $\bar{\sigma}$ & rMSE & $\bar{\sigma}$ & rMSE & $\bar{\sigma}$ & rMSE & $\bar{\sigma}$ & rMSE & $\bar{\sigma}$ \\
    \hline
        M$_{\text{stellar}}$, $\log_{10}$M$_{\odot}$ & 9.0$^{\dag}$ & 0.94 & 0.853 & 0.815 & 0.725 & 0.948 & 0.866 & 0.873 & 0.768 & 0.906 & 0.491 & 0.749 & 0.476 & 0.908 & 0.777 & 0.812 & 0.522 \\
        ... & 10.0$^*$ & 0.923 & 0.878 & 1.029 & 1.079 & 0.924 & 0.893 & 1.052 & 1.083 & 0.791 & 0.683 & 0.988 & 1.034 & 0.793 & 0.717 & 0.965 & 0.969 \\
        ... & 10.3$^*$ & 0.916 & 0.919 & 0.916 & 0.919 & 1.09 & 1.02 & 0.932 & 0.924 & 1.104 & 1.022 & 1.02 & 0.931 & 0.753 & 0.621 & 1.013 & 0.881 \\
        SFR, M$_{\odot}$ yr$^{-1}$ & 1.25$^*$ & 0.981 & 0.928 & 0.991 & 0.973 & 0.885 & 0.775 & 1.082 & 1.024 & 0.795 & 0.777 & 0.995 & 1.009 & 0.821 & 0.721 & 1.008 & 0.932 \\
    \hline \hline	
    \label{table:ClusteringStats_AsnConstraints}
	\end{tabular}
	\end{center}
\end{table*}

% As representative examples, we plot the constraints from different clustering statistics under two mass selections:  the clustering of dark matter halos with halo mass greater than $2\times 10^{11}$ M$_{\odot}$, randomly sampled to a density of 0.005 $h^{3}$ cMpc$^{-3}$; and the clustering of SAM galaxies with stellar mass greater than $1\times 10^{9}$ M$_{\odot}$, randomly sampled to a density of 0.005 $h^{3}$ cMpc$^{-3}$. Figures \ref{fig:DMonlyBestOm_ClusteringStat}, \ref{fig:DMonlyBests8_ClusteringStat}, and \ref{fig:Mstar_ClusteringCompare} showing these constraints are shown in Appendix \ref{app:Zs_Clust_Focus}, as the exact constraints closely resemble Figures already shown. 

\begin{figure*}[b]
    \centering
     \begin{subfloat}[2-point correlation function.]{
        \includegraphics[width=0.23\textwidth]{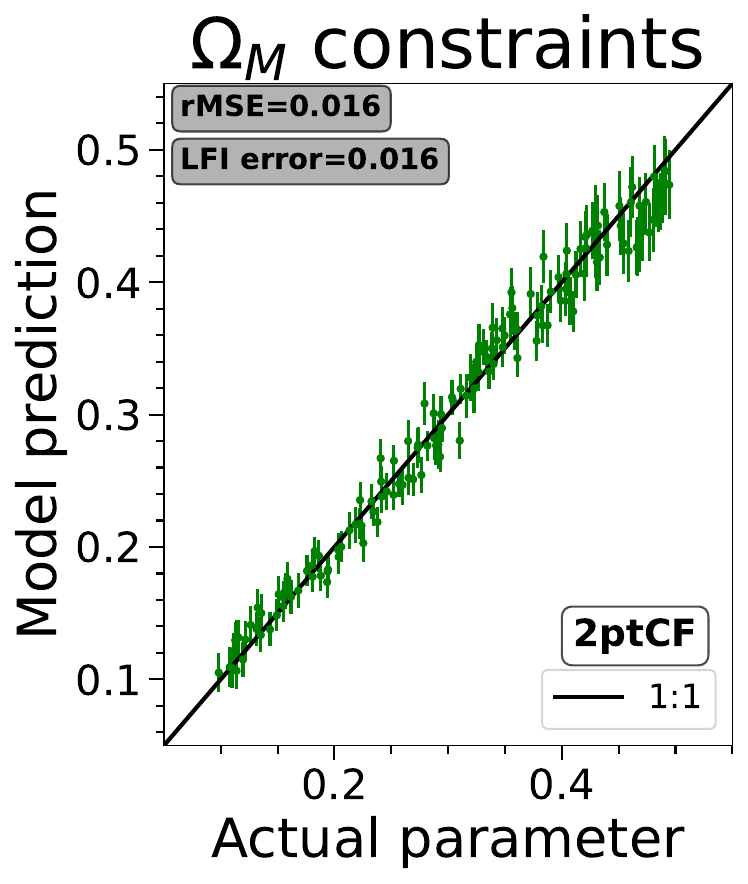}
         \label{fig:mhalog250N5k_2ptCFLFI_Om}
         }
     \end{subfloat}
     \begin{subfloat}[Count-in-Cells.]{
        \includegraphics[width=0.23\textwidth]{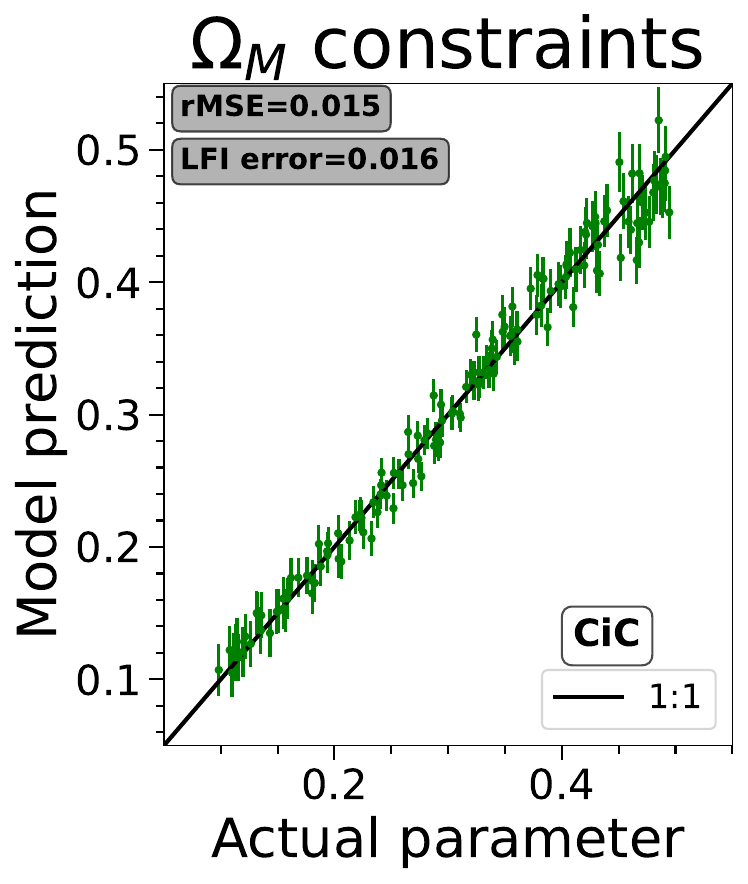}
         \label{fig:mhalogt250N5k_CiCLFI_Om}
         }
     \end{subfloat}
    \begin{subfloat}[Void Probability Function.]{
        \includegraphics[width=0.23\textwidth]{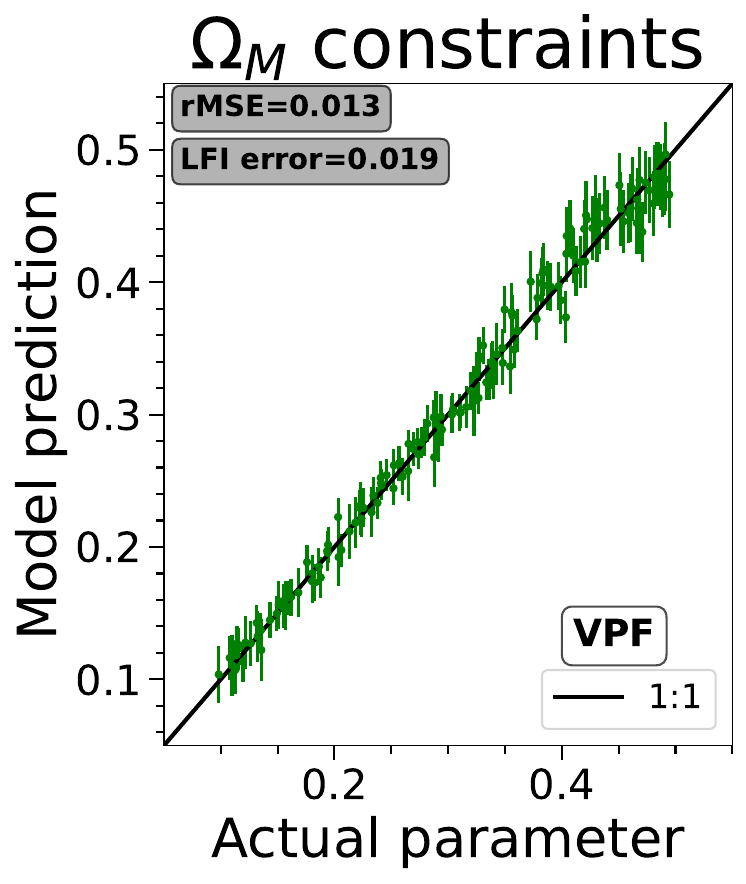}
         \label{fig:mhalogt250N5k_VPFLFI_Om}
         }
     \end{subfloat}
    \begin{subfloat}[`All' clustering statistics.]{
        \includegraphics[width=0.23\textwidth]{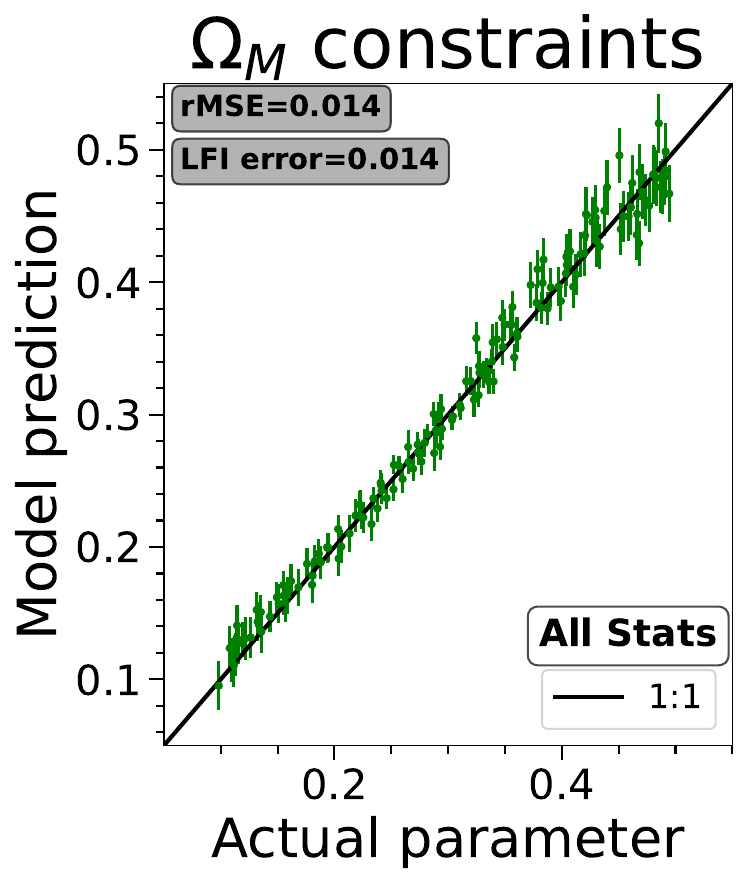}
         \label{fig:mhalogt250N5k_SampledLFI_Om}
         }
     \end{subfloat}
     \caption{Which clustering statistic is best at \textbf{constraining cosmology} through $\Omega_{\text{M}}$? We compare the constraints on $\Omega_{\text{M}}$ found by different clustering statistics based on the clustering of \textit{\textbf{dark matter halos}} with mass greater than $2\times 10^{11}$ M$_{\odot}$, randomly sampled to a density of 0.005 $h^{3}$ cMpc$^{-3}$. We combine the clustering at $z=\{0.0, 0.1, 0.5, 1.0\}$. The 2ptCF (a) is measured between  $1.1 < R < 36.1$ cMpc; CiC (b) is measured at at $R=16.0, 22.4, 28.8$ cMpc; and the VPF (c) is measured between $1.6 < R < 40$ cMpc. We combine `all' these statistics (d) for the best constraints.  Detailed quantitative comparisons can be found in Table \ref{table:ClusteringStats_CosmoConstraints}. }
    \label{fig:DMonlyBestOm_ClusteringStat}
\end{figure*}

\begin{figure*}[b]
    \centering
     \begin{subfloat}[2-point correlation function.]{
        \includegraphics[width=0.23\textwidth]{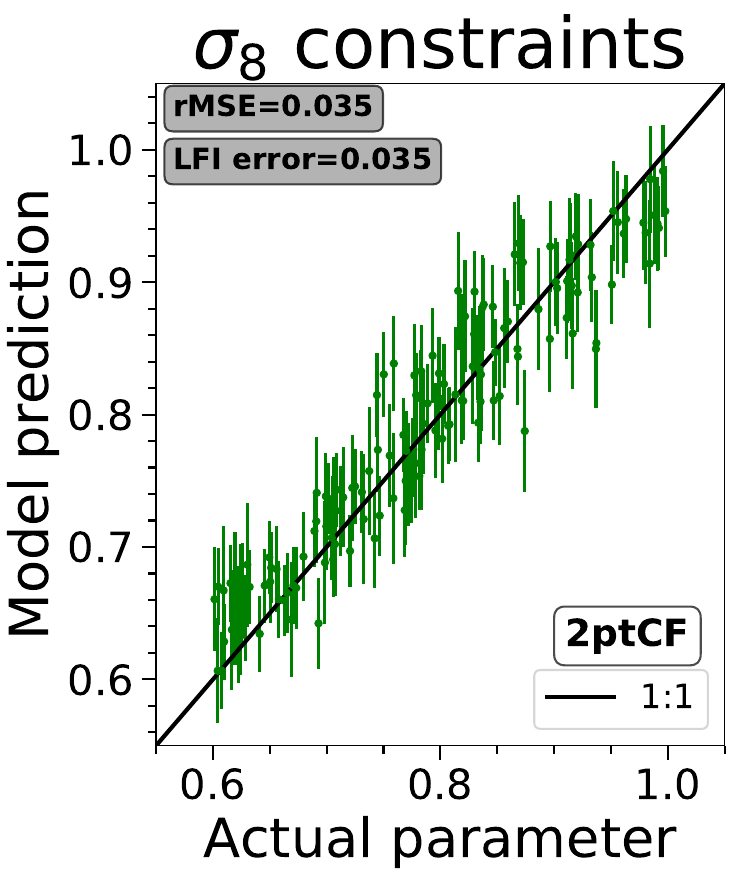}
         \label{fig:mhalog250N5k_2ptCFLFI_s8}
         }
     \end{subfloat}
     \begin{subfloat}[Count-in-Cells.]{
        \includegraphics[width=0.23\textwidth]{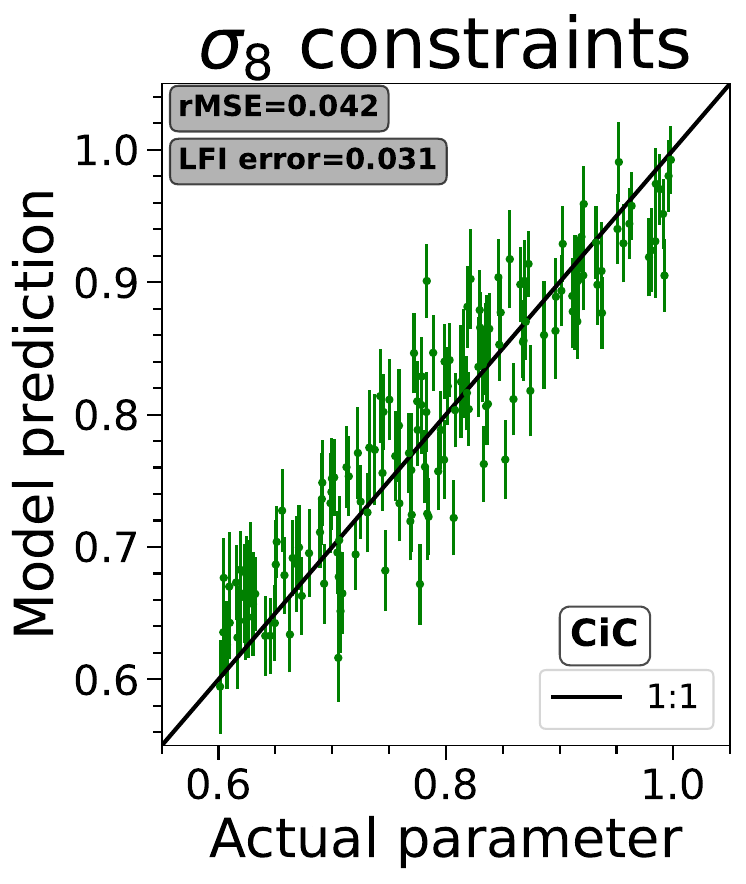}
         \label{fig:mhalogt250N5k_CiCLFI_s8}
         }
     \end{subfloat}
    \begin{subfloat}[Void Probability Function.]{
        \includegraphics[width=0.23\textwidth]{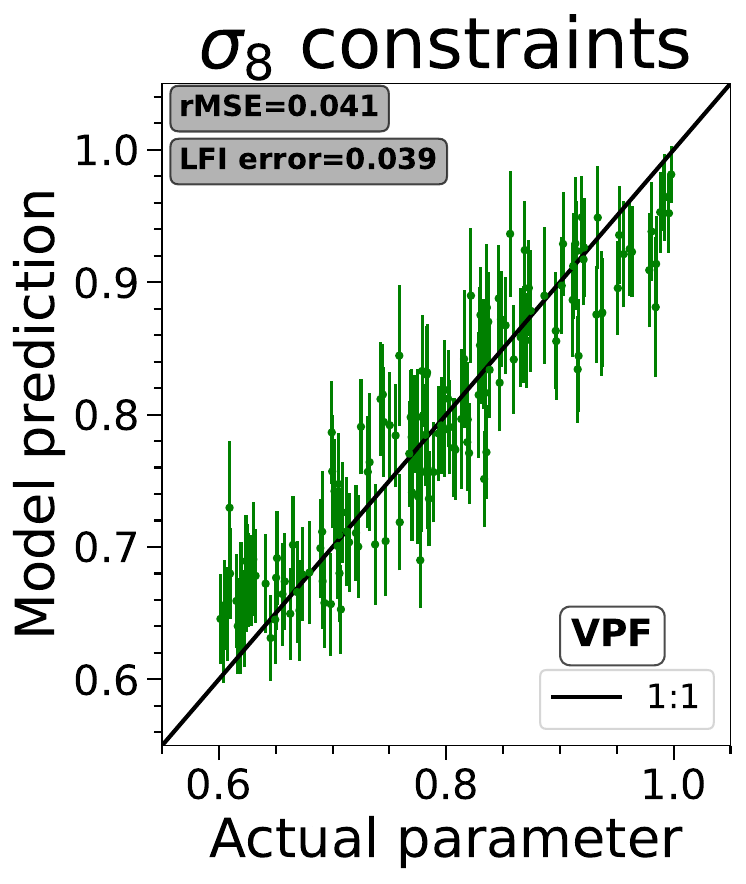}
         \label{fig:mhalogt250N5k_VPFLFI_s8}
         }
     \end{subfloat}
    \begin{subfloat}[`All' clustering statistics.]{
        \includegraphics[width=0.23\textwidth]{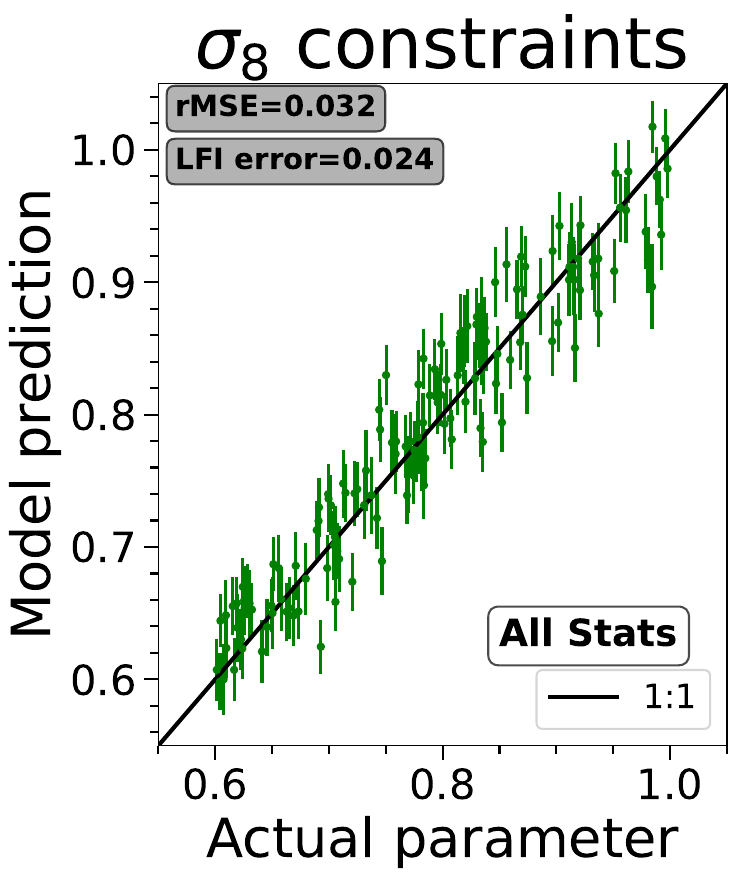}
         \label{fig:mhalogt250N5k_SampledLFI_s8}
         }
     \end{subfloat}
     \caption{Which clustering statistic is best at constraining $\sigma_{8}$? The same clustering and set-up as in Figure \ref{fig:DMonlyBestOm_ClusteringStat}, instead for $\sigma_{8}$.}
    \label{fig:DMonlyBests8_ClusteringStat}
\end{figure*}

\begin{figure*}
    \centering
    \begin{subfloat}[When training a neural network on the two-point correlation function between $1.1 < R < 36.1$ cMpc at $z=\{0.0, 0.1, 0.5, 1.0\}$.]{
        \includegraphics[width=0.97\textwidth]{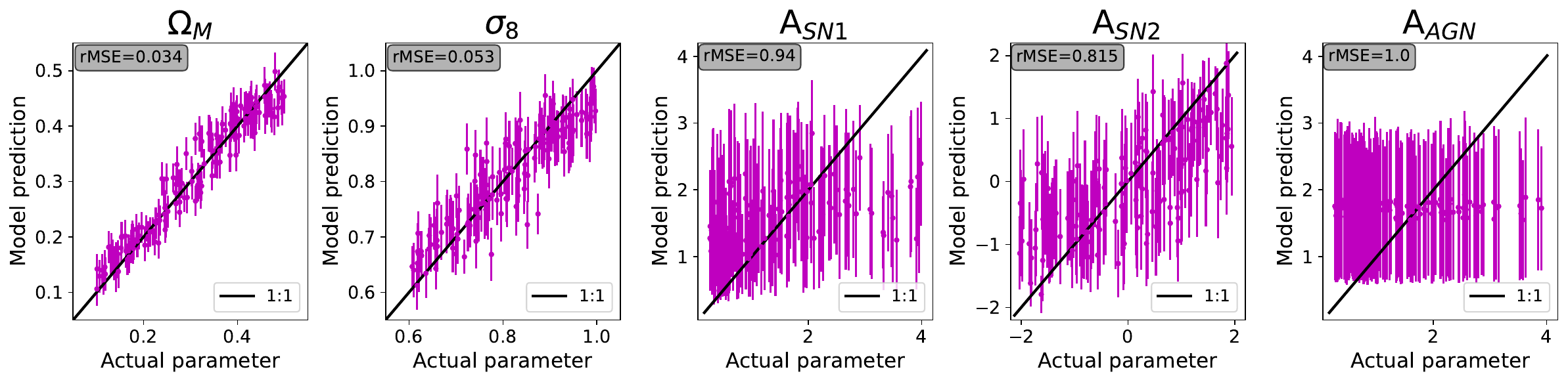}
         \label{fig:2ptCF_mstargt1N5k}
         }
     \end{subfloat}
     \begin{subfloat}[When training a neural network on the count-in-cells distribution at $R=16.0, 22.4, 28.8$ cMpc at $z=\{0.0, 0.1, 0.5, 1.0\}$.]{
        \includegraphics[width=0.97\textwidth]{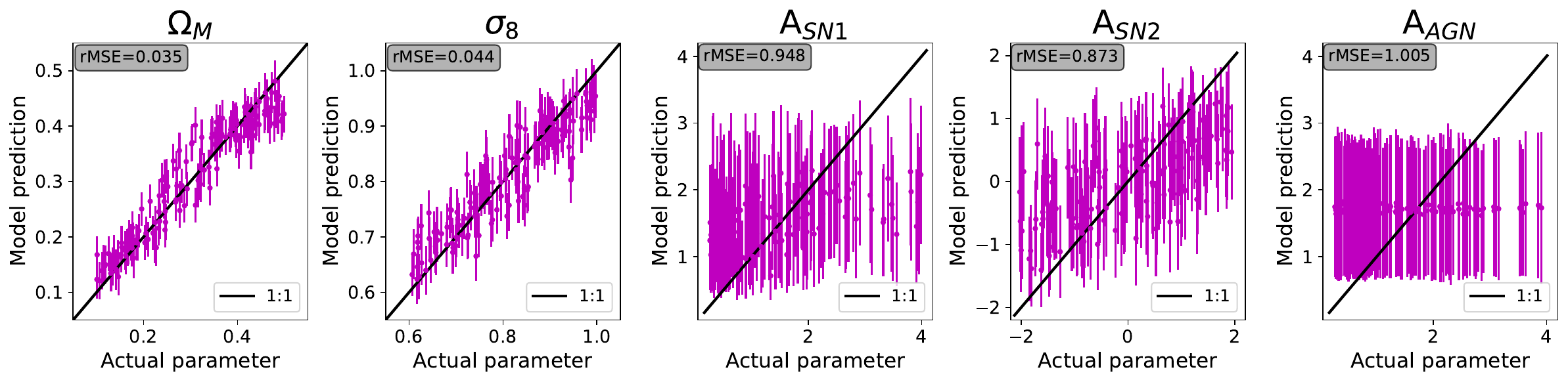}
         \label{fig:CiC_mstargt1N5k}
         }
     \end{subfloat}
     \begin{subfloat}[When training a neural network on the Void Probability Function between $1.6 < R < 40$ cMpc at $z=\{0.0, 0.1, 0.5, 1.0\}$.]{
        \includegraphics[width=0.97\textwidth]{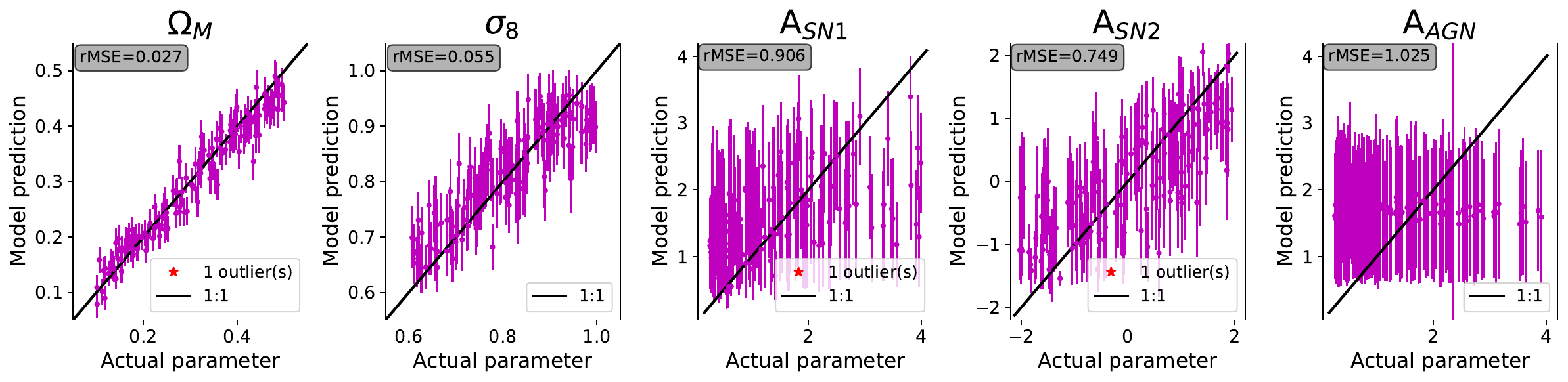}
         \label{fig:VPF_mstargt1N5k}
         }
     \end{subfloat}
    \begin{subfloat}[When training a neural network on `all' clustering statistics described above together.]{
        \includegraphics[width=0.97\textwidth]{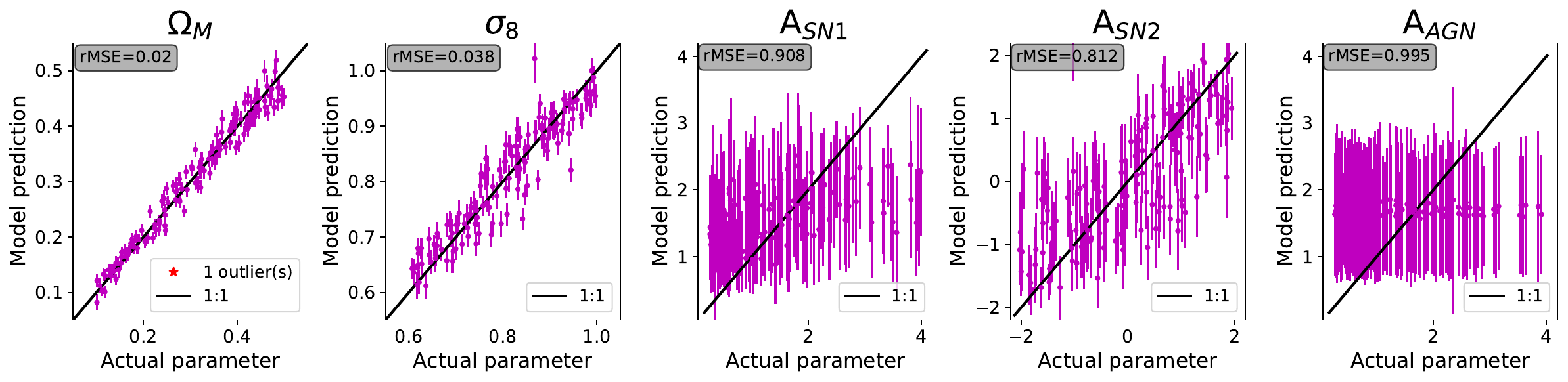}
         \label{fig:Sampled_mstargt1N5k}
         }
     \end{subfloat}
     \caption{ Which clustering statistic is best at \textbf{constraining cosmology \textit{and} astrophysical parameters?} We compare the constraints on $\Omega_{\text{M}}$, $\sigma_8$, A$_{\text{SN1}}$, A$_{\text{SN2}}$, and A$_{\text{AGN}}$ found by each of the clustering statistics that we use: (a) the 2ptCF , (b) CiC , (c) the VPF, and (d) all combined. We use the respective clustering of SAM galaxies with \textit{\textbf{stellar mass}} greater than $1\times 10^{9}$ M$_{\odot}$, down-sampled to a density of 0.005 $h^{3}$ cMpc$^{-3}$. Detailed quantitative comparisons can be found in Tables \ref{table:ClusteringStats_CosmoConstraints} and \ref{table:ClusteringStats_AsnConstraints}.}
    \label{fig:Mstar_ClusteringCompare}
\end{figure*}

\textbf{Comparing Clustering Statistics: }
In \textsection \ref{sec:CompareStats}, we compare the constraints that each independent clustering statistic is able to find for our five parameters. Though we tested several selections, we include two representative examples in this Appendix: cosmology-only constraints on $\Omega_{\text{M}}$ and $\sigma_8$ for a halo mass selection of $2\times 10^{11}$ M$_{\odot}$ down-sampled to a density of 0.005 $h^{3}$ cMpc$^{-3}$ (Figures \ref{fig:DMonlyBestOm_ClusteringStat} and \ref{fig:DMonlyBests8_ClusteringStat}), and constraints on all 5 parameters at once for a stellar mass selection of $1\times 10^{9}$ M$_{\odot}$ (Figure \ref{fig:Mstar_ClusteringCompare}). Outliers, marked in red stars in Figure \ref{fig:Mstar_ClusteringCompare}, are simulations in the test set whose `$Z$-value' (Eq. \ref{eq:Zvalue}) are greater than 6, and are excluded when calculating the rMSE (see \textsection \ref{subsubsec:Chi2} for more details).å

Finally, we remind readers that the true parameter distributions in the test set only \textit{appear} slightly skewed to lower values for the A$_{\text{SN1}}$ and A$_{\text{AGN}}$ parameters, due to their original generation in logarithmic space. This effect would disappear if plotted in log-scale, though we choose to keep all scales linear for consistency with plots in \citet{CAMELSannouncement}.

% \clearpage 

\textbf{Additional Information: Focused NNs.} As explored in \textsection \ref{subsec:SCSAMfocus}, we tested how our constraints fared if the neural network was told to focus on a single parameter at a time. When focusing on one parameter at a time, we find that the constraints either stay the same or slightly improve when comparing rMSE errors. We also find that the LFI errors from having all parameters trained at once occasionally match or even exceed those from focusing on one at a time. This indicates our LFI loss is performing as expected, and finding all possible available information for each parameter. 

\begin{figure*}
    \centering
    \begin{subfloat}[Neural network predicts \textit{only} A$_{\text{SN1}}$.]{
    \includegraphics[width=0.31\textwidth]{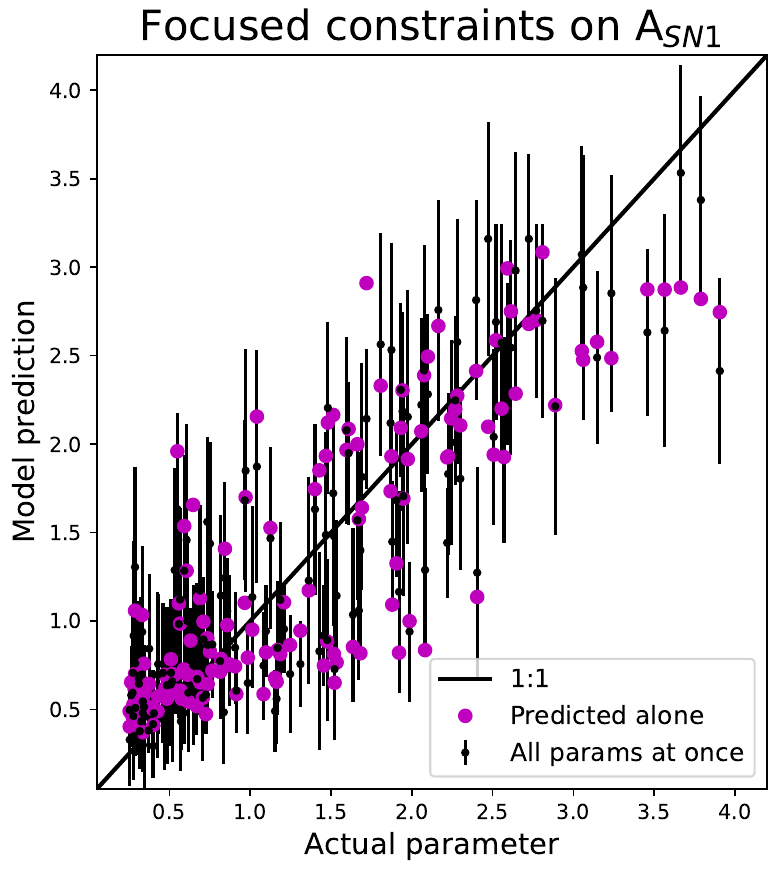}
    }
% 	\caption{}
    \end{subfloat}
    \hfill
    \begin{subfloat}[Neural network predicts \textit{only} A$_{\text{SN2}}$.]{
    \includegraphics[width=0.31\textwidth]{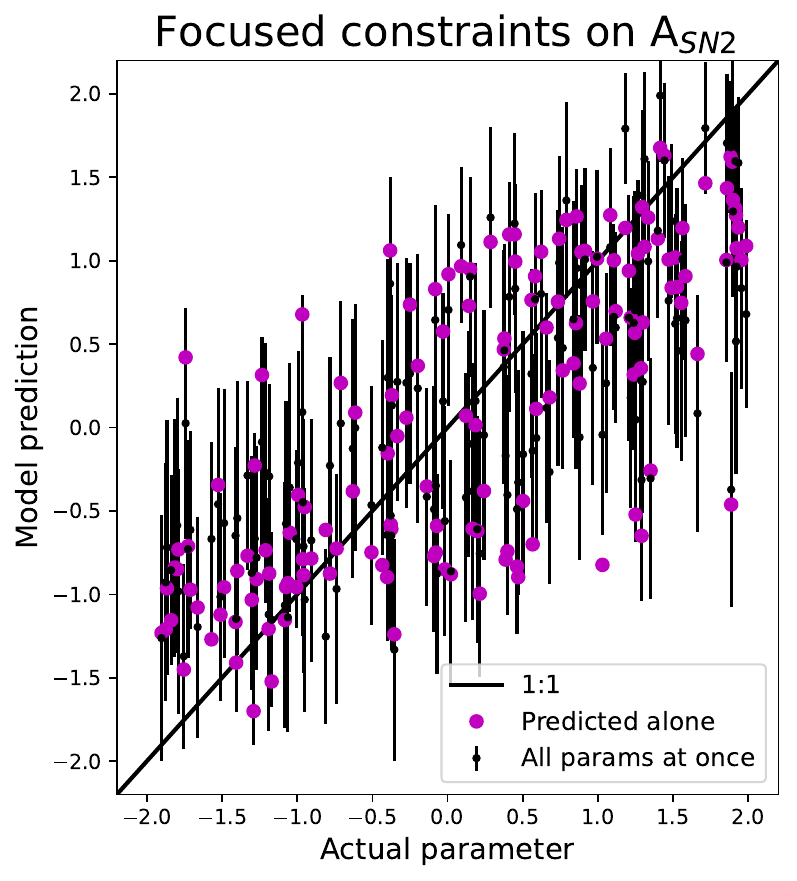}
    }
    % \caption{}
    \end{subfloat}
    \hfill
    \begin{subfloat}[Neural network predicts \textit{only} A$_{\text{AGN}}$.]{
    \includegraphics[width=0.31\textwidth]{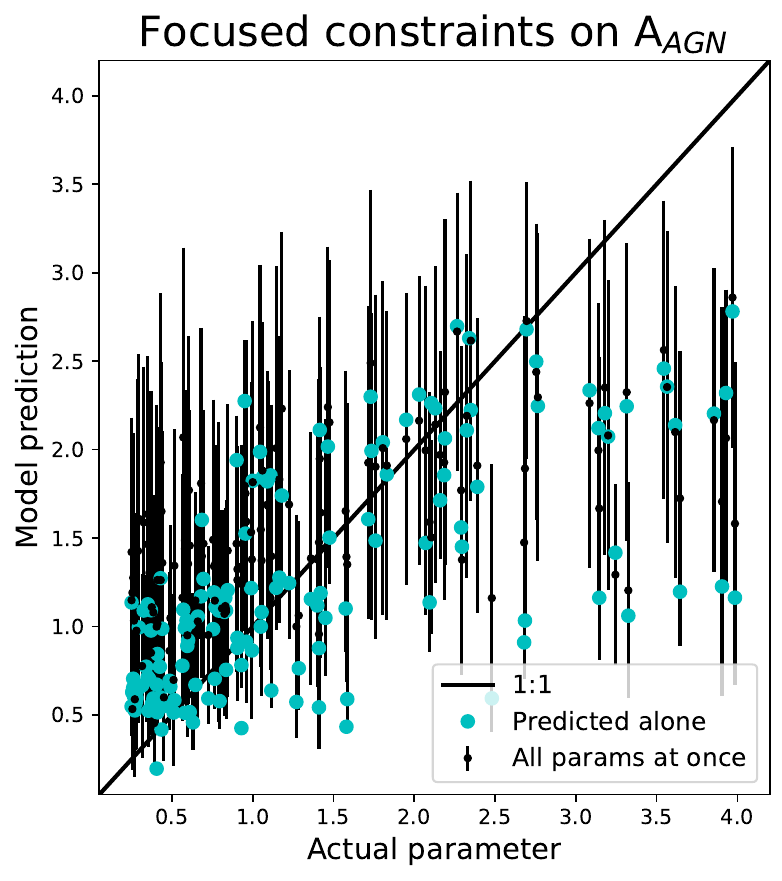}
    % \caption{}
    }
    \end{subfloat}
    \caption{Results from neural networks that attempt to predict one SAM parameter at a time. These networks were given `all' clustering statistics at $z=\{0.0, 0.1, 0.5, 1.0\}$ for all SAM galaxies with stellar mass greater than $2\times 10^{10}$ M$_{\odot}$ (a,b; magenta) or star formation rate greater than 1 M$_{\odot}$ yr$^{-1}$ (c; cyan), with no down-sampling to fixed number density. Predictions for each parameter when all 5 parameters are constrained at once are given in black with $\bar{\sigma}$ errors for comparison. The constraints do not demonstrably improve with a focused neural network---this improves our confidence that the neural networks fitting all 5 parameters at once are extracting close to the full available information. }
	\label{fig:SCSAMparamsalone}
\end{figure*}

For the cosmological parameters, the focused neural networks use `all' clustering of: SAM galaxies with halo mass greater than $2\times 10^{11}$ M$_{\odot}$ across 4 redshifts (down-sampled to $0.005\ h^{3}$ cMpc$^{-3}$); and SAM galaxies with stellar mass greater than $2\times 10^{10}$ M$_{\odot}$ across 4 redshifts (no down-sampling). For the A$_{\text{SN}}$ parameters, we select `all' clustering of SAM galaxies with stellar mass greater than $2\times 10^{10}$ M$_{\odot}$ across 4 redshifts (no down-sampling). For A$_{\text{AGN}}$, we select the `all' clustering of all SAM galaxies with star formation rate greater than 1 M$_{\odot}$ yr$^{-1}$ across 4 redshifts (also no down-sampling). Figure \ref{fig:SCSAMparamsalone} shows the focused NN results for the SC-SAM parameters.

\begin{table}[t]
	\begin{center}
    \caption{Constraints from learning the parameters \textbf{one at a time}. The neural networks are given `all' clustering across $0 < z < 1$ for the following selections on galaxy properties. We list the rMSE of the LFI posterior means when predicting the parameters alone of alongside the other 4.}
	\begin{tabular}{p{0.8cm}p{0.9cm}p{1.4cm}p{0.75cm}p{0.8cm}p{0.75cm}p{0.7cm}} 
	\hline \hline
	\multicolumn{4}{c}{Best-Performing Galaxy Selections} &  \multicolumn{1}{c}{Alone} 
	&  \multicolumn{2}{c}{All 5}\\
	\cline{1-4} \cline{6-7}
    % \hline
    Param. & Prop. & Value & $\mathcal{N}$ & rMSE & rMSE & $\bar{\sigma}$ \\
    \hline
		$\Omega_{\text{M}}$ & M$_{\text{halo}}$ & 2e11 M$_{\odot}$ & 0.005 & 0.014 & 0.014 & 0.014\\
		... & M$_{\text{star}}$ & 2e10 M$_{\odot}$ & N/A & 0.031 & 0.028 & 0.029 \\
		$\sigma_8$ & M$_{\text{halo}}$ & 2e11 M$_{\odot}$ & 0.005 & 0.031 & 0.032 & 0.024 \\
		... & M$_{\text{star}}$ & 2e10 M$_{\odot}$ & N/A & 0.039 & 0.036 & 0.043 \\
		A$_{\text{SN1}}$ & M$_{\text{star}}$ & 2e10 M$_{\odot}$ & N/A & 0.478 & 0.452 & 0.478 \\
        A$_{\text{SN2}}$ & M$_{\text{star}}$ & 2e10 M$_{\odot}$ & N/A & 0.745 & 0.757 & 0.717 \\
        A$_{\text{AGN}}$ & SFR & 1 M$_{\odot}$ yr$^{-1}$ & N/A & 0.766 & 0.889 & 0.758 \\
    \hline \hline	
    % \\
    \label{table:oneatatime}
	\end{tabular}
	\end{center}
\end{table}
 \clearpage

\section{How much information is lost to astrophysics?} \label{app:infolostAstro}

We have shown that our neural networks are able to constrain cosmology even when the SC-SAM is allowed to vary in several feedback parameters. However, just how much information is lost to these parameters? How well are the neural networks marginalizing over astrophysics? To answer this, we compare to results from neural networks trained and tested upon galaxy catalogs where the SC-SAM is held to the fiducial astrophysical parameters for all cosmologies. We focus on the case using the clustering of SAM galaxies with stellar mass greater than $1\times10^9$ M$_{\odot}$ down-sampled to 0.005 $h^{3}$ cMpc$^{-3}$.

We run the fiducial SC-SAM (e.g.\ best-fitting to many observations, as used in \citealt{Gabrielpillai2022}) over all 1000 LH N-body volumes. Therefore, we generate galaxy catalogs only varying in their cosmology and resulting halo merger history, with otherwise identical prescriptions for galaxy formation and evolution under the SC-SAM. We measure the clustering and train a neural network exactly as described in \textsection \ref{sec:Methods}.

\begin{figure*}[b]
     \centering
     \begin{subfloat}[With galaxy catalogs across the complete 5-parameter latin hypercube.]{
    \includegraphics[width=\textwidth]{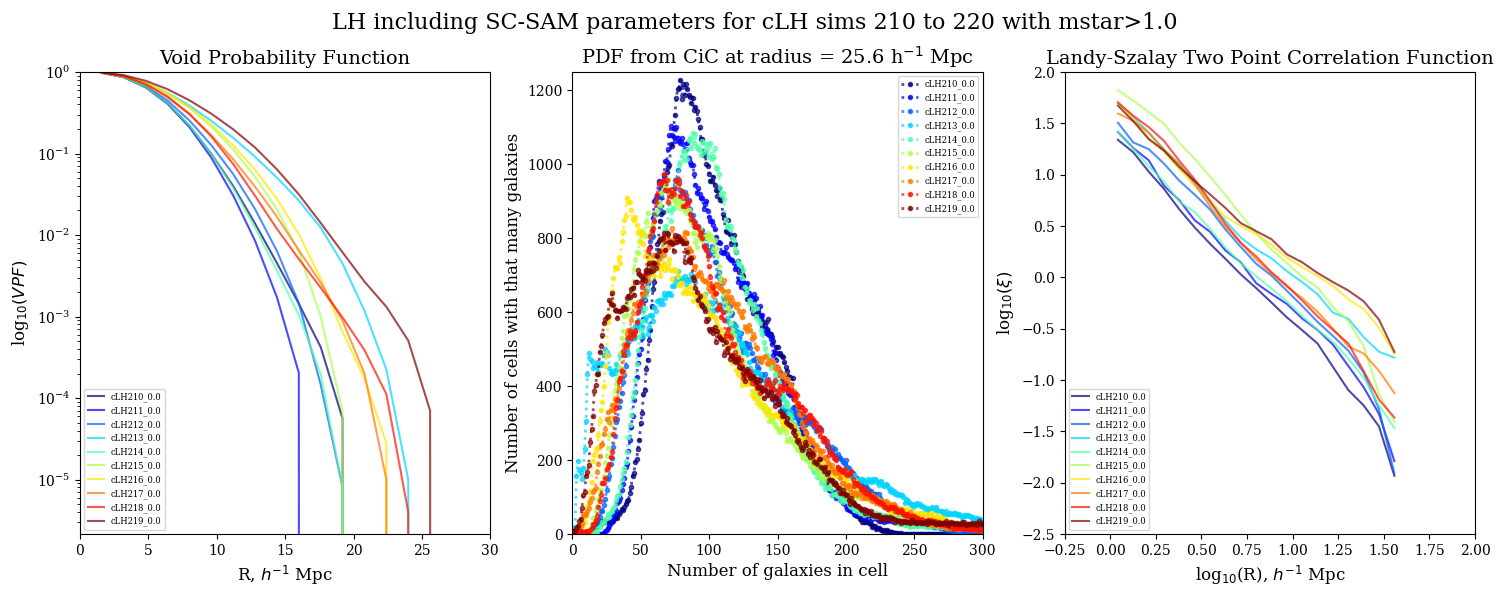}
     \label{fig:trueLHclustering}
     }
     \end{subfloat}
     \begin{subfloat}[With galaxy catalogs across only a cosmological latin hypercube, with the SC-SAM astrophysical parameters held at fiducial values.]{
    \includegraphics[width=\textwidth]{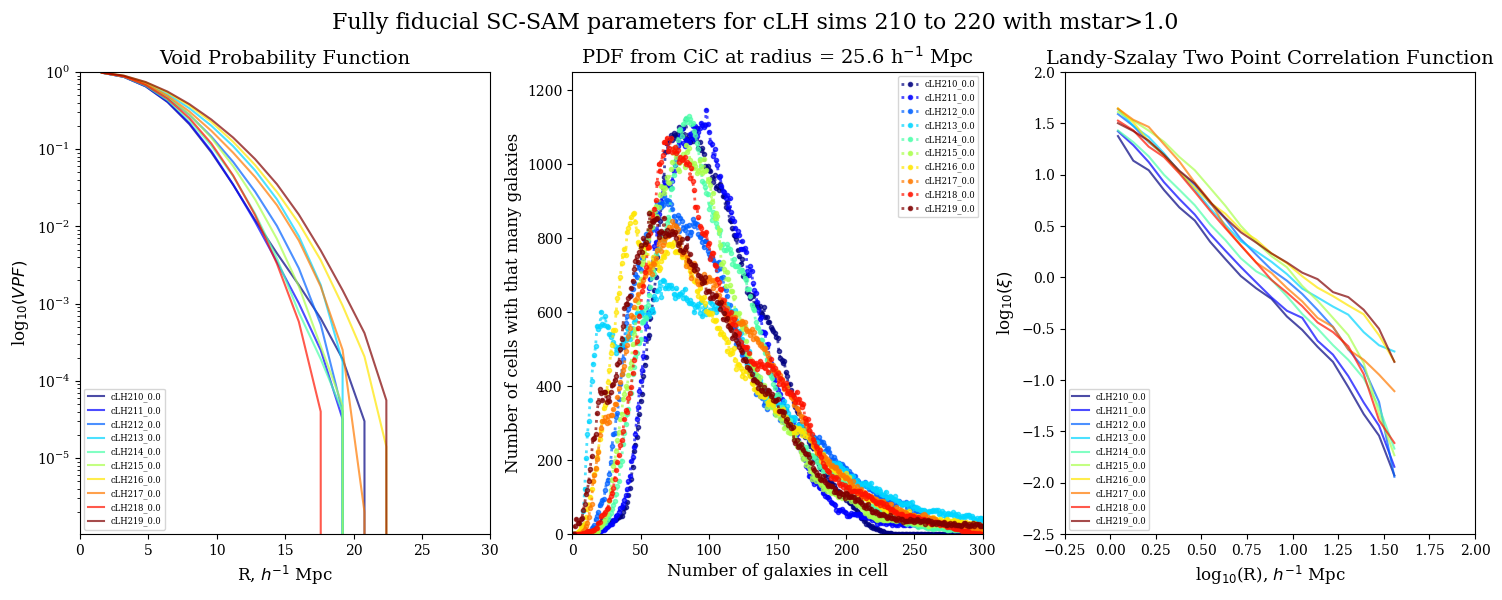}
     \label{fig:fidSAMclustering}
     }
    \end{subfloat}
    \caption{`All' clustering of $z=0$ SAM galaxies with stellar mass greater than $1\times10^9$ M$_{\odot}$ and randomly down-sampled to 0.005 $h^{3}$ cMpc$^{-3}$, with SAM parameters allowed to vary (top) or held at fiducial values (bottom). }
    \label{fig:example_LHvfidSAM}
\end{figure*}

Figure \ref{fig:example_LHvfidSAM} shows, for LH volumes 210 through 220, the clustering of $z=0$ SAM galaxies with stellar mass greater than $1\times10^9$ M$_{\odot}$ down-sampled to 0.005 $h^{3}$ cMpc$^{-3}$: with the SAM parameters allowed to vary within the latin hypercube (above), and when the SAM is held to fiducial values (below). We can visually see the effects that allowing the SC-SAM parameters to vary has on the clustering curves of these simulations, affirming the behavior the neural networks are learning.

In this experiment, the neural network is given the clustering of SC-SAM galaxies whose generation varied only in cosmology. The neural network trains while still expecting all 5 parameters to affect the clustering (including the astrophysical parameters that were not varied). However, the network immediately finds that there is no information to be learned about the astrophysical parameters, leading it to predict the mean of the prior for the SC-SAM parameters with an error encompassing each parameter's range. 

However, the cosmological predictions stay the same or improve for the fully fiducial SAM catalogs compared to the the latin hypercube SAM catalogs. Clustering using the wholly fiducial SAM catalogs yields constraints on $\Omega_{\text{M}}$ of 0.014/0.016 (LFI/rMSE losses); and constraints on $\sigma_8$ of 0.018/0.028 (LFI/rMSE). Compare these to the constraints when using the catalogs with all 5 parameters varied (in Figure \ref{fig:SAMmstarBest}): constraints of 0.014/0.02 (LFI/rMSE) for $\Omega_{\text{M}}$, and 0.021/0.038 for $\sigma_8$. Additionally, we remind readers of what our near-`best case' scenario constraints with dark matter only yielded in Figure \ref{fig:DMonlyBest}: constraints of 0.014/0.014 (LFI/rMSE) for $\Omega_{\text{M}}$, and 0.024/0.032 for $\sigma_8$. The possible improvement in $\sigma_8$ in the fully fiducial SAM catalogs compared to the halo catalogs may be attributable to the SC-SAM reinforcing or encoding cosmological information from the merger trees in its galaxy generation, but could be due to our selective testing of stellar and halo mass selections.s

The constraints with clustering of galaxy catalogs with no variation in the SC-SAM nearly matches the constraints from halo catalogs. This confirms a central tenet of our methodology: that neural networks are indeed capable of marginalizing over astrophysics to accurately and precisely measure cosmology. In this experiment, the neural network needed only to learn one mapping between the cosmological parameters and galaxy clustering in the presence of baryonic processes, and avoided possible scrambling in this mapping due to baryonic prescriptions (e.g.\ \citealt{Arico2021}).
% marginalize over one way of making galaxies given a cosmology and its resulting merger history. 
However, the consistently good results of the full latin hypercube including the SC-SAM parameters indicate that the neural networks are able to uncover much of the cosmological information still.
% \textit{so the variation of the SAM parameters introduces a very small amount of error in the rMSE, but none in the LFI for OmegaM; slightly more error in sigma8. With no SAM at all, essentially match the DMO results or do slightly better (which I suppose is to be expected, it's marginalizing over the single way of making galaxies given cosmo)}

% \textcolor{cyan}{do i actually need to talk about this experiment? or is it distracting? does remind us why want multiple glx formation models i guess}
As an additional experiment, we also compare constraints between clustering with and without varied astrophysics for SAM galaxies with stellar mass greater than $7\times 10^9$ M$_{\odot}$ with no density down-sampling (where the number density of the SAM galaxy catalogs enters as an additional and powerful tool for the neural network to leverage). The cosmological predictions from a neural network trained on the wholly fiducial SAM parameters yield constraints of 0.0029/0.008 (LFI/rMSE) for $\Omega_{\text{M}}$, and 0.0138/0.016 for $\sigma_8$. As Table \ref{table:GlxSelections_noDS} details, for the complete 5-parameter LH catalogs, the constraints we find are 0.021/0.025 (LFI/rMSE) for $\Omega_{\text{M}}$, and 0.03/0.036 for $\sigma_8$. The remarkable improvement in this scenario likely comes from the networks not having to work around the degenerate effects of the 5 parameters upon the number density of stellar-mass selected catalogs. However, this experiment should remind us of the importance of including multiple variations and prescriptions for astrophysics in cosmological parameter inference: the effects of astrophysics are many and poorly understood, and not accounting for them will lead to overly optimistic and possibly incorrect constraints. 

% \textit{what else to say? note that the LFI gets really good even with all 5 parameters? removing possible degeneracies and helping the network? or having some features of the cosmology that really affects the merger trees get amplified in SAM glx properties? not because jsut focusing on one param, i think the single param searches did just as well hmmmm}


\end{document}